\documentclass[a4paper,11pt,bold,enumerate,noupper,titlepage,nocenter,english,spanish]{book}
\usepackage{amssymb}

\usepackage{fancyheadings}
\usepackage[latin3]{inputenc}
\usepackage[english]{babel}
\usepackage{epsfig}
\usepackage{graphicx}
\usepackage{amsmath}

\def\zvs{
   \begin{picture}(1,1)(0,0)
   \put(0,0){\oval(0.5,0.5)[r]}
   \put(0,0.5){\oval(0.5,0.5)[l]}
   \end{picture}}
\newlength{\figwidth}
\begin{document}

\pagestyle{empty}

\centerline{\huge \bf  The electroweak matter sector from} \vspace{0.3cm}
\centerline{\huge \bf an effective theory
perspective}

\vspace{2cm} \centerline{\Large  Juli\'an \'Angel Manzano Flecha} \vspace{4cm%
} \centerline{\large \sf Barcelona, Juny 2002}

\vspace{1.5cm}
\begin{figure}[!h]
\centering \leavevmode \epsfysize=2cm \epsfbox{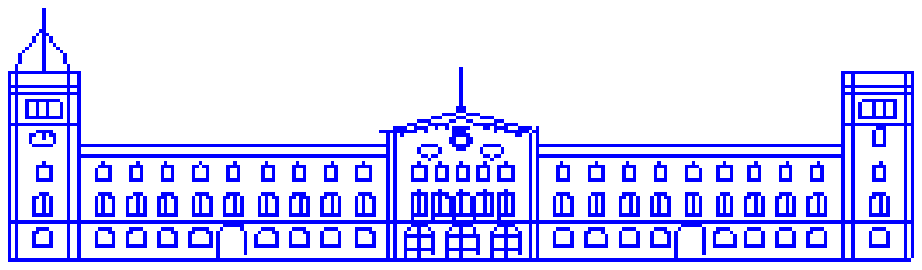}
\end{figure}

\vspace{0.5cm} \centerline{\large \bf Universitat de Barcelona} \vspace{0.5cm%
}
\centerline{\large Departament d'Estructura i
Constituents de la Mat\`eria}

\newpage \vspace*{2cm} \newpage \vspace*{1cm}

\centerline{\huge \bf  The electroweak matter sector from} \vspace{0.3cm}
\centerline{\huge \bf an effective theory
perspective}

\vspace{2cm} \centerline{\large Mem\`oria de la tesi presentada} %
\centerline{\large per Juli\'an \'Angel Manzano Flecha}
\centerline{\large per optar al grau de Doctor en Ci\`encies
F\'{\i}siques} \vspace{1cm}
\centerline{\large Director de tesi:
Dr. Dom\`enec Espriu} \vspace{2cm}
\centerline{\large Programa de
doctorat del Departament}
\centerline{\large d'Estructura i
Constituents de la Mat\`eria}
\centerline{\large
``Part\'{\i}cules, camps i fen\`omens qu\`antics col·lectius''} %
\centerline{\large Bienni 1997-99}
\centerline{\large \bf
Universitat de Barcelona} \vspace{4cm}
\centerline{Signat: Dr.
Dom\`enec Espriu} \newpage \vspace*{2cm} \newpage


\bigskip

\hspace{12cm}A Judith

\selectlanguage{english}

\newpage \vspace*{2cm} \newpage

\renewcommand{\thechapter}{\empty} \renewcommand{\chaptername}{\empty} %
\pagenumbering{roman} \pagestyle{plain}

\tableofcontents
\newpage

\renewcommand{\thesection}{\empty} \newpage \vspace*{2cm} \newpage %
\centerline{{\huge Preface}} \addcontentsline{toc}{section}{Preface}%
\vspace*{2cm}

\noindent This thesis deals with some theoretical and phenomenological
aspects of the electroweak matter sector with special emphasis on the
effective theory approach. This approach has been chosen for its versatility
when general conclusions are sought without entering in the details of the
currently available ``fundamental'' theories. Effective theories are present
in the description of almost all physical phenomena even though such
description is often not recognized as ``effective''. In particular,
effective theories in the context of quantum field theories are treated in
well known works in the literature and excellent introductions are
available. Because of that, I have chosen not to repeat what can be found
easily elsewhere but to indicate the reader the relevant references in the
introduction.

The thesis is structured in chapters that are almost in one to one
correspondence with my research articles. Namely Chapter \ref
{mattersectorchapter} is based on the article published in
Phys.Rev.D60: 114035, 1999 with some typos corrected and with some
notational modifications made in order to comply with the rest of
the thesis notation. Chapter \ref{cpviolationandmixing} is based
on the article published in Phys.Rev.D63: 073008, 2001 where again
some modifications have been made. In particular a whole section
was completely omitted in favor of the next chapter which is based
in a recent research article that extensively surpass the contents
of that section. This article, which is the is the groundwork of
Chapter \ref{LSZchapter}, has been accepted for publication in
Phys.Rev. and has E-Print archive number hep-ph/0204085 (in
http://xxx.lanl.gov/multi). Chapter \ref{LHCphenomenology} is
based on the publication Phys.Rev.D65: 073005, 2002 and finally
Chapter \ref {topdecaychapter} is based on a recent work not yet
published.

At the end of the thesis I have included a set of appendices that can be
useful for those interested in technical details of some sections. Even
though chapters are based on research articles some changes and new sections
have been inserted in order to make them more self-contained. Whenever
possible I have left some of the intermediate calculational steps to the
ease of those interested in reobtaining some results. In order to conform to
the University rules part of the thesis has been written in Spanish. In
particular the introduction and conclusion are presented in duplicate,
English and Spanish.

\newpage \selectlanguage{spanish}

\centerline{{\huge Prefacio}} \addcontentsline{toc}{section}{Prefacio}
\vspace*{2cm}

\noindent Esta tesis trata aspectos te\'{o}ricos y fenomenol\'{o}gicos del
sector de materia electrod\'{e}bil con especial \'{e}nfasis en el uso de
lagrangianos efectivos. Hemos utilizado la t\'{e}cnica de lagrangianos
efectivos debido a la versatilidad que nos brinda a la hora de obtener
resultados generales sin entrar en los detalles concretos de cada una de las
teor\'{i}as ``fundamentales'' actualmente utilizadas. Las teor\'{i}as
efectivas est\'{a}n presentes en la descripci\'{o}n de casi todos los
fen\'{o}menos f\'{i}sicos aun cuando muchas veces tal descripci\'{o}n no es
reconocida como tal. En particular el uso de teor\'{i}as efectivas en el
contexto de las teor\'{i}as cu\'{a}nticas de campos est\'{a} tratado en
reconocidos trabajos de la literatura cient\'{i}fica y se dispone de
excelentes introducciones. Es por ello que he preferido no repetir aqu\'{i}
los temas que f\'{a}cilmente pueden hallarse en dichos trabajos, optando por
dirigir al lector a las referencias adecuadas al comienzo de la
introducci\'{o}n.

Esta tesis se estructura en cap\'{i}tulos que han sido basados en
mis art\'{i}culos de investigaci\'{o}n. Concretamente, el
Cap\'{i}tulo \ref {mattersectorchapter} est\'{a} basado en el
art\'{i}culo publicado en la revista Phys.Rev.D60: 114035, 1999
con algunas correcciones tipogr\'{a}ficas y con algunas
modificaciones de notaci\'{o}n para adaptarlo al resto de la
tesis. El Cap\'{i}tulo \ref{cpviolationandmixing} est\'{a} basado
en el art\'{i}culo publicado en Phys.Rev.D63: 073008, 2001 donde
tambi\'{e}n se han efectuado algunas modificaciones. En particular
he omitido una secci\'{o}n completa ya que su antiguo contenido
est\'{a} ampliado y mejorado en el Cap\'{i}tulo \ref{LSZchapter}
basado en un art\'{i}culo reciente que trata el tema de manera
extensiva. Este art\'{i}culo ha sido aceptado para ser publicado
en la revista Phys.Rev. y tiene n\'{u}mero de archivo
electr\'{o}nico hep-ph/0204085 (en http://xxx.lanl.gov/multi). El
Cap\'{i}tulo \ref{LHCphenomenology} est\'{a} basado en la
publicaci\'{o}n en Phys.Rev.D65: 073005, 2002 y finalmente el
Cap\'{i}tulo \ref{topdecaychapter} est\'{a} basado en un trabajo
reciente que a\'{u}n no ha sido publicado.

Al final de la tesis he inclu\'{\i}do un conjunto de ap\'{e}ndices que
pueden ser \'{u}tiles para aquellos interesados en los detalles t\'{e}cnicos
de algunas secciones. A\'{u}n cuando los cap\'{\i}tulos est\'{a}n basados en
los art\'{\i}culos de investigaci\'{o}n he agregado algunas secciones para
hacerlos m\'{a}s independientes. Adem\'{a}s he intentado dejar pasos
intermedios en algunos c\'{a}lculos para facilitar la reproducci\'{o}n de
algunos de los resultados. Cumpliendo con las normas de la Universidad parte
de la tesis est\'{a} escrita en castellano. En particular la
introducci\'{o}n y las conclusiones se presentan por duplicado en ingl\'{e}s
y en castellano.

\newpage \centerline{{\huge Agradecimientos}} %
\addcontentsline{toc}{section}{Agradecimientos} \vspace*{2cm} \noindent Una
tesis es algo m\'{a}s que una colecci\'{o}n de trabajos, es por sobre todo
una colecci\'{o}n de experiencias. Es por ello que en esta secci\'{o}n
quisiera volcar al menos una \'{i}nfima parte del sentimiento de gratitud
que tengo hacia las personas con las que he tenido la suerte de relacionarme
en este proceso. En primer lugar quisiera comenzar con mi jefe, \textit{en}
Dom\`{e}nec. ``\textit{Yo hago fenomenolog\'{i}a}'' me dijo cuando
discut\'{i}amos las alternativas de tema de tesis. ``\textit{Uff,
fenomenolog\'{i}a..., no s\'{e}...}'' le contest\'{e} poco convencido. ``%
\textit{Mira, hay fenomenolog\'{i}a y FENOMENOLOGIA}'' me contest\'{o}. Fue
suficiente; era un alivio saber que no har\'{i}a \textit{fenomenolog\'{i}a}
sino \textit{FENOMENOLOGIA} y con esto comenc\'{e} a trabajar entusiasmado.
Por supuesto ahora que he acabado no se si he hecho \textit{fenomenolog\'{i}a%
} o \textit{FENOMENOLOGIA}, incluso la parte de \textit{TEORIA} quiz\'{a}s
sea s\'{o}lo \textit{teor\'{i}a}. Lo que si me queda claro es que en estos
a\~{n}os Dom\`{e}nec siempre me ha apoyado y me ha animado en mis frecuentes
desv\'{i}os del camino se\~{n}alado. Es este apoyo el que me ha hecho sentir
c\'{o}modo trabajando con \'{e}l y uno de los aspectos que m\'{a}s valoro de
su papel como director. \textit{Gracias Dom\`{e}nec, ha sido un placer
trabajar con vos}.

Y por supuesto no me puedo olvidar aqu\'{i} de mis compa\~{n}eros de
doctorado, los unos y los otros. Los unos, David, Guifr\'{e}, Joan, Ignasi,
Toni ahora ya doctores con los cuales comenc\'{e} a \textit{pelearme} con la
f\'{i}sica y con los cuales disfrut\'{e} de innumerables discusiones. Los
otros, Dani, Dolors, Enric, que empezaron mas tarde y que ya est\'{a}n
acabando! Quiero agradecerles aqu\'{i} los buenos momentos (\textexclamdown
malos no hubo!;) que pasamos entre estas venerables paredes. \textit{Enric i
Dolors, Gr\`{a}cies per l'ajuda en les meves cuites d'\'{u}ltima hora!!}
Tampoco me quiero olvidar de los \textit{benjamines} (\textexclamdown que
nadie se ofenda!), Alex, Aleix, Llu\'{i}s, Toni, Luca, a todos les deseo que
disfruten de la tesis al m\'{a}ximo, que todo se acaba aunque no lo parezca!
\textit{Luca, Gracias por tus ideas revolucionarias, en f\'{i}sica,
pol\'{i}tica y otros asuntos que no nombrar\'{e} aqu\'{i}, he disfrutado
mucho de tu compa\~{n}\'{i}a}.

Algo que no quiero olvidarme de agradecer es el buen ambiente de trabajo del
departamento, Joan, Jos\'{e} Ignacio, Quim, Pere, Rolf (al menos antes de su
paso a las altas esferas), Josep,... les quiero agradecer especialmente
haber estado all\'{i} siempre dispuestos a aguantar y a pasarlo bien con
nuestras dudas y certezas.

\textit{Pere, gr\`{a}cies per l'entusiasme contagi\'{o}s!}

Finalmente quiero acordarme de mi familia, M\'{a}, P\'{a}, Pablo,
\textexclamdown qu\'{e} les voy a agradecer! \textexclamdown \textexclamdown
Qu\'{e} los quiero!! Y a vos Ju, que sos mi joya en esta vida, \textit{%
aquesta tesi \'{e}s per a tu}.

\newpage \vspace*{2cm} \newpage




\renewcommand{\chaptername}{\empty} \renewcommand{\thesection}{%
\arabic{section}} 
\setcounter{chapter}{0} \renewcommand{\thepage}{\arabic{page}} %
\renewcommand{\thechapter}{\empty}

\pagestyle{myheadings} \pagestyle{fancyplain}

\renewcommand{\chaptermark}[1]           {\markboth{#1}{}} %
\renewcommand{\sectionmark}[1]           {\markright{\thesection\ #1}}
\lhead[\fancyplain{}{\bfseries \thepage}]%
           {\fancyplain{}{\bfseries\rightmark}}
\rhead[\fancyplain{}{\bfseries \leftmark}]%
           {\fancyplain{}{\bfseries\thepage}}

\chapter{Resumen de la Tesis}

\setcounter{page}{1} 

\section{Introducci\'{o}n}

\bigskip \renewcommand{\thetable}{\arabic{section}.\arabic{table}} %
\setcounter{table}{0}

\renewcommand{\thefigure}{\arabic{section}.\arabic{figure}} %
\setcounter{figure}{0}



Las Teor\'{i}as Cu\'{a}nticas de Campos (QFT) se definen utilizando el grupo
de renormalizaci\'{o}n. La idea b\'{a}sica tiene sus or\'{i}genes en el
mundo de la materia condensada \cite{Kadanoff:1966wm} y b\'{a}sicamente se
puede expresar diciendo que en el l\'{i}mite termodin\'{a}mico (un
n\'{u}mero infinito de grados de libertad) la integraci\'{o}n de los grados
de libertad de alta frecuencia es equivalente a una redefinici\'{o}n de los
operadores que aparecen en la teor\'{i}a. Cuando el n\'{u}mero de dichos
operadores es finito decimos que la teor\'{i}a es `renormalizable' y cuando
no lo es decimos que es `no renormalizable' o efectiva \cite
{efftheointro,Donoghue}. Las teor\'{i}as renormalizables pueden ser
consideradas como Teor\'{i}as Cu\'{a}nticas de Campos (QFT) `fundamentales'
ya que el l\'{i}mite al continuo es posible.

En cualquier caso, los operadores renormalizados poseen una dependencia en
el \textit{cut-off} que regulariza la teor\'{i}a. Esta dependencia est\'{a}
dictada principalmente por la dimensi\'{o}n \textit{naive} del operador.
Cuanto mayor es dicha dimensi\'{o}n, mayor es la supresi\'{o}n dictada por
el \textit{cut-off}. Por ello, las teor\'{i}as no renormalizables pueden ser
analizadas en la pr\'{a}ctica truncando el n\'{u}mero de operadores que se
ordenan por dimensi\'{o}n creciente. Los operadores de dimensi\'{o}n menor
dan las contribuciones m\'{a}s importantes a los observables de baja
energ\'{i}a, lo cual hace que estas teor\'{i}as tengan poder de
predicci\'{o}n si nos restringimos a dicho r\'{e}gimen energ\'{e}tico. A
medida que incrementamos la energ\'{i}a o el orden en teor\'{i}a de
perturbaciones (relacionado con el orden en energ\'{i}a por el teorema de
Weinberg \cite{Weinberg}), se necesitan m\'{a}s y m\'{a}s operadores en los
c\'{a}lculos, y por lo tanto el poder de predicci\'{o}n se reduce y
eventualmente la teor\'{i}a se vuelve ineficaz. Esta caracter\'{i}stica (o
inconveniente) de las teor\'{i}as efectivas est\'{a} compensada por sus
ventajas en t\'{e}rminos de generalidad. Como diferentes teor\'{i}as de
altas energ\'{i}as pertenecen a la misma clase de universalidad (la misma
fenomenolog\'{i}a a bajas energ\'{i}as) las teor\'{i}as efectivas se pueden
considerar como una forma compacta de probar diversas teor\'{i}as sin entrar
en sus peculiaridades irrelevantes de altas energ\'{i}as. Podemos resumir
estas consideraciones en la Tabla (\ref{tablaefectiva})

Aparte de consideraciones dimensionales, las simetr\'{\i}as son el otro
ingrediente b\'{a}sico que clasifica operadores y restringe la mezcla de los
mismos generada por el grupo de renormalizaci\'{o}n.

\newpage

\begin{table}[tbp]
\centering
\begin{tabular}{|c|c|}
\hline
QFT renormalizables & QFT efectivas \\ \hline
n\'{u}mero finito de operadores &
\begin{tabular}{c}
n\'{u}mero infinito de operadores \\
\multicolumn{1}{l}{{(truncaci\'{o}n controlada por la dimensi\'{o}n)}}
\end{tabular}
\\ \hline
poder de predicci\'{o}n a energ\'{i}as arbitrarias & poder de predicci\'{o}n
a bajas energ\'{i}as \\ \hline
proliferaci\'{o}n de modelos & generalidad \\ \hline
\end{tabular}
\caption{QFT renormalizables vs. QFT efectivas}
\label{tablaefectiva}
\end{table}

El objetivo de esta tesis es el estudio de algunos problemas abiertos en el
sector de materia electrod\'{e}bil. Los temas estudiados incluyen:

\begin{itemize}
\item  Aspectos generales de modelos de ruptura din\'{a}mica de
simetr\'{\i}a donde estudiamos posibles trazas que estos mecanismos pueden
dejar a bajas energ\'{\i}as.

\item  Un tratamiento general de la violaci\'{o}n de la simetr\'{\i}a $CP$ y
la mezcla de familias en el \'{a}mbito de una teor\'{\i}a efectiva y la
determinaci\'{o}n de algunos de los coeficientes efectivos involucrados.

\item  Aspectos te\'{o}ricos conectando el grupo de renormalizaci\'{o}n, la
invariancia \textit{gauge}, $CP$, $CPT$, y los observables f\'{\i}sicos.

\item  La posibilidad de acotar experimentalmente algunos de los acoplos
efectivos involucrados en el futuro acelerador de protones LHC.
\end{itemize}

En lo que sigue presentaremos un resumen detallado de los temas tratados en
esta tesis.\bigskip

A pesar de que la estructura b\'{a}sica del Modelo Est\'{a}ndar (SM) de las
interacciones electrod\'{e}biles ya ha sido bien verificada gracias a un
gran n\'{u}mero de experimentos, su sector de ruptura de simetr\'{i}a no ha
sido firmemente establecido a\'{u}n, tanto desde el punto de vista
te\'{o}rico como experimental.

En la versi\'{o}n m\'{\i}nima del SM de interacciones electrod\'{e}biles, el
mismo mecanismo (un \'{u}nico doblete escalar complejo) da masa
simult\'{a}neamente a los bosones de \textit{gauge} $W$ y $Z$ y a los campos
de materia fermi\'{o}nicos (con la posible excepci\'{o}n del neutrino). Este
mecanismo est\'{a}, sin embargo, basado en una aproximaci\'{o}n
perturbativa. Desde el punto de vista no perturbativo el sector escalar del
SM m\'{\i}nimo se supone trivial, que a su vez es equivalente a considerar a
dicho modelo como una truncaci\'{o}n de una teor\'{\i}a efectiva. Esto
implica que a una escala $\sim 1$ TeV nuevas interacciones deber\'{\i}an
aparecer si el Higgs no se encuentra a m\'{a}s bajas energ\'{\i}as \cite
{triviality}. El \textit{cut-off} de 1 TeV est\'{a} determinado por estudios
no perturbativos y sugerido por la falta de validez del esquema perturbativo
a esa escala. Por otro lado, en el SM m\'{\i}nimo es completamente
antinatural tener un Higgs ligero ya que su masa no est\'{a} protegida por
ninguna simetr\'{\i}a (el as\'{\i} denominado problema de jerarqu\'{\i}as).

Esta contradicci\'{o}n se resuelve utilizando extensiones
supersim\'{e}tricas del SM, donde esencialmente tenemos el mismo mecanismo,
aunque el sector escalar es mucho m\'{a}s rico en este caso con preferencia
de escalares relativamente ligeros. En realidad, si la supersimetr\'{i}a
resulta ser una idea \'{u}til en fenomenolog\'{i}a, es crucial que el Higgs
se encuentre con una masa $M_{H}\leq 125$ GeV, ya que si esto no ocurre los
problemas te\'{o}ricos que motivaron la introducci\'{o}n de la
supersimetr\'{i}a reaparecer\'{i}an \cite{hierarchy}. C\'{a}lculos a dos
\textit{loops} \cite{twoloop} elevan este l\'{i}mite a alrededor de los 130
GeV.

Una tercera posibilidad es la dada por modelos de ruptura din\'{a}mica de la
simetr\'{i}a (tales como la teor\'{i}as de \textit{technicolor} (TC) \cite
{TC}). En este caso existen interacciones que se vuelven fuertes,
t\'{i}picamente a la escala $\Lambda _{\chi }\simeq 4\pi v$ ($v=250$ GeV),
rompiendo la simetr\'{i}a global $SU(2)_{L}\times SU(2)_{R}$ a su subgrupo
diagonal $SU(2)_{V}$ y produciendo bosones de Goldstone que eventualmente
pasan a ser los grados de libertad longitudinales de $W^{\pm }$ y $Z$. Para
transmitir esta ruptura de simetr\'{i}a a los campos ordinarios de materia
se requiere de interacciones adicionales, usualmente denominadas \textit{%
technicolor} extendido (ETC) y caracterizado por una escala diferente $M$.
Generalmente, se asume que $M\gg 4\pi v$ para mantener bajo control a
posibles corrientes neutras de cambio de sabor (FCNC) \cite{FCNC}. As\'{i},
una caracter\'{i}stica distintiva de estos modelos es que el mecanismo
responsable de dar masas a los bosones $W^{\pm }$ y $Z$ y a los campos de
materia es diferente.

\textquestiondown D\'{o}nde estamos actualmente? Algunos irian tan lejos
como para decir que un Higgs elemental (supersim\'{e}trico o de otro tipo)
ha sido `visto' a trav\'{e}s de correcciones radiativas y que su masa es
menor que 200 GeV, o incluso que ha sido descubierto en los \'{u}ltimos
d\'{i}as del LEP con una masa $\simeq $115 GeV \cite{LEPC}. Otros descreen
de estas afirmaciones (ver por ejemplo \cite{chanowitz} para un estudio
cr\'{i}tico sobre las actuales afirmaciones acerca de un Higgs ligero).

El enfoque basado en los Lagrangianos efectivos ha sido notablemente
\'{u}til a la hora de fijar restricciones al tipo de nueva f\'{i}sica
detr\'{a}s del mecanismo de ruptura de simetr\'{i}a del SM tomando como
datos b\'{a}sicamente los resultados experimentales de LEP \cite{aleph} (y
SLC \cite{SLD}). Hasta ahora ha sido aplicado principalmente al sector \emph{%
bos\'{o}nico}, las as\'{i} denominadas correcciones 'oblicuas'. La idea es
considerar el Lagrangiano m\'{a}s general que describe las interacciones
entre el sector de \textit{gauge} y los bosones de Goldstone que aparecen
luego de que la ruptura $SU(2)_{L}\times SU(2)_{R}\rightarrow SU(2)_{V}$
tiene lugar. Ya que no se asume ning\'{u}n mecanismo especial para esta
ruptura, el procedimiento es completamente general asumiendo, por supuesto,
que las part\'{i}culas no expl\'{i}citamente incluidas en el Lagrangiano
efectivo son mucho m\'{a}s pesadas que las que s\'{i} lo est\'{a}n. La
dependencia en el modelo espec\'{i}fico tiene que estar contenida en los
coeficientes de los operadores de dimensi\'{o}n m\'{a}s alta.

Con la idea de extender este enfoque que ha sido tan eficaz, en el
Cap\'{i}tulo \ref{mattersectorchapter} parametrizamos, independientemente
del modelo, posibles desviaciones de las predicciones del Modelo
Est\'{a}ndar m\'{i}nimo en el sector de \emph{materia}. Como ya hemos dicho,
esto se realiza asumiendo s\'{o}lo el esquema de ruptura de simetr\'{i}a del
Modelo Est\'{a}ndar y que las part\'{i}culas a\'{u}n no observadas son
suficientemente pesadas, de manera que la simetr\'{i}a est\'{a} realizada de
manera no lineal. Tambi\'{e}n reexaminamos, dentro del lenguaje de las
teor\'{i}as efectivas, hasta que punto los modelos m\'{a}s simples de
ruptura din\'{a}mica est\'{a}n realmente acotados y las hip\'{o}tesis
utilizadas en la comparaci\'{o}n con el experimento. Ya que los modelos de
ruptura din\'{a}mica de simetr\'{i}a pueden ser aproximados a energ\'{i}as
intermedias $\Lambda _{\chi }<E<M$ por operadores de cuatro fermiones,
presentamos una clasificaci\'{o}n completa de los mismos cuando las nuevas
part\'{i}culas aparecen en la representaci\'{o}n usual del grupo $%
SU(2)_{L}\times SU(3)_{c}$ y tambi\'{e}n una clasificaci\'{o}n parcial en el
caso general. Luego discutimos la precisi\'{o}n de la descripci\'{o}n basada
en operadores de cuatro fermiones efectuando el \textit{matching} con una
teor\'{i}a `fundamental' en un ejemplo simple. Los coeficientes del
Lagrangiano efectivo en el sector de materia para los modelos de ruptura
din\'{a}mica de simetr\'{i}a (expresados en t\'{e}rminos de los coeficientes
de los operadores de cuatro fermiones) son luego comparados con aquellos
provenientes de modelos con escalares elementales (como el Modelo
Est\'{a}ndar m\'{i}nimo). Contrariamente a lo cre\'{i}do com\'{u}nmente,
observamos que el signo de las correcciones de v\'{e}rtice no est\'{a}n
fijadas en los modelos de ruptura din\'{a}mica de simetr\'{i}a. Resumiendo,
sin analizar los temas de violaci\'{o}n de $CP$ o fenomenolog\'{i}a de
mezcla de familias, el trabajo de este cap\'{i}tulo proporciona las
herramientas te\'{o}ricas requeridas para analizar en t\'{e}rminos generales
restricciones en el sector de materia del Modelo Est\'{a}ndar.

Hasta aqu\'{i} nada definitivo se ha dicho acerca de la violaci\'{o}n de $CP$
o la mezcla de familias. Sin embargo, tal como sucede en el SM, estos
fen\'{o}menos est\'{a}n probablemente relacionados con el sector de ruptura
de simetr\'{i}a.\bigskip

La violaci\'{o}n de $CP$ y la mezcla de familias se encuentran entre los
enigmas m\'{a}s intrigantes del SM. La comprensi\'{o}n del origen de la
violaci\'{o}n de $CP$ es en realidad uno de los objetivos m\'{a}s
importantes de los experimentos actuales y futuros. Esto est\'{a}
completamente justificado ya que dicha comprensi\'{o}n puede no s\'{o}lo
revelar caracter\'{i}sticas inesperadas de sectores de nueva f\'{i}sica,
sino tambi\'{e}n dar pistas en el entendimiento de aspectos
fenomenol\'{o}gicos complejos como la bariog\'{e}nesis en cosmolog\'{i}a.

En el Modelo Est\'{a}ndar m\'{i}nimo la informaci\'{o}n sobre las cantidades
que describen esta fenomenolog\'{i}a est\'{a} codificada en la matriz de
mezcla de Cabibbo-Kobayashi-Maskawa (CKM) (aqu\'{i} denotada $K$). En este
contexto, aunque la matriz de masas m\'{a}s general posee, en principio, un
gran n\'{u}mero de fases, s\'{o}lo las matrices de diagonalizaci\'{o}n de
fermiones de quiralidad \textit{left} sobreviven combinadas en una \'{u}nica
matriz CKM. Esta matriz contiene s\'{o}lo una fase compleja observable. Si
esta \'{u}nica fuente de violaci\'{o}n de $CP$ es suficiente o no para
explicar nuestro mundo es, actualmente, una inc\'{o}gnita.

Como es bien sabido, algunas de las entradas de esta matriz est\'{a}n muy
bien medidas, mientras que otras (tales como $K_{tb}$, $K_{ts}$ y $K_{td}$)
son poco conocidas y la \'{u}nica restricci\'{o}n experimental real viene
dada por los requerimientos de unitariedad. En este problema en particular
se ha invertido un gran esfuerzo en la \'{u}ltima d\'{e}cada y esta
dedicaci\'{o}n continuar\'{a} en el futuro inmediato destinada a lograr en
el sector cargado una precisi\'{o}n comparable con la lograda en el sector
neutro. Como gu\'{i}a, mencionamos que la precisi\'{o}n en $\sin 2\beta $ se
espera que sea superior al 1\% en el futuro LHCb, y una precisi\'{o}n
semejante se espera para ese momento en los experimentos actualmente en
curso (BaBar, Belle) \cite{Amato:1998xt}.

Unos de los prop\'{o}sitos de los experimentos de nueva generaci\'{o}n es
testear la `unitariedad de la matriz CKM'. Puesto de esta forma, dicho
prop\'{o}sito no parece tener mucho sentido. Por supuesto si s\'{o}lo
mantenemos las tres generaciones conocidas, la mezcla ocurre a trav\'{e}s de
una matriz de $3\times 3$ que es, por construcci\'{o}n, necesariamente
unitaria. Lo que realmente se quiere decir con la afirmaci\'{o}n anterior es
que se quiere verificar si los elementos de matriz $S$ observables, que a
nivel \'{a}rbol son proporcionales a elementos de CKM, cuando son medidos en
decaimientos d\'{e}biles est\'{a}n o no de acuerdo con las relaciones de
unitariedad a nivel \'{a}rbol predichas por el Modelo Est\'{a}ndar. Si
escribimos por ejemplo
\begin{equation}
\left\langle q_{j}\left| W_{\mu }^{+}\right| q_{i}\right\rangle
=U_{ij}V_{\mu },  \label{eqa1introsp}
\end{equation}
a nivel \'{a}rbol, est\'{a} claro que la unitariedad de la matriz CKM
implica
\begin{equation}
\sum_{k}U_{ik}U_{jk}^{\ast }=\delta _{ij},  \label{eqa2introsp}
\end{equation}
Sin embargo, incluso si no existe nueva f\'{i}sica m\'{a}s all\'{a} del
Modelo Est\'{a}ndar las correcciones radiativas contribuyen a los elementos
de matriz relevantes en los decaimientos d\'{e}biles y arruinan la
unitariedad de la `matriz CKM' $U$, en el sentido de que los
correspondientes elementos de matriz $S$ no estar\'{a}n restringidos a
obedecer las relaciones de unitariedad indicadas arriba. Obviamente, las
desviaciones de unitariedad debidas a las correcciones radiativas
electrod\'{e}biles ser\'{a}n necesariamente peque\~{n}as. Despu\'{e}s
veremos a que nivel debemos esperar violaciones de unitariedad debidas a
correcciones radiativas.

Pero por supuesto, las violaciones de unitariedad que realmente son
interesantes son las causadas por nueva f\'{\i}sica. La f\'{\i}sica m\'{a}s
all\'{a} del Modelo Est\'{a}ndar se puede manifestar de diferentes maneras y
a diferentes escalas. Otra vez, tal como hemos hecho con el caso sin mezcla
ni violaci\'{o}n de $CP$ asumiremos que la nueva f\'{\i}sica puede aparecer
a una escala $\Lambda $ que es relativamente grande comparada con $M_{Z}$.
Esta observaci\'{o}n incluye al sector escalar tambi\'{e}n; es decir,
asumimos que el Higgs ---si es que existe--- es suficientemente pesado. Con
estas hip\'{o}tesis trataremos de extraer algunas conclusiones acerca de la
mezcla de familias y la violaci\'{o}n de $CP$ utilizando t\'{e}cnicas de
Lagrangianos efectivos.

Ilustremos esta idea con un ejemplo simple: Supongamos el caso en el que hay
una nueva generaci\'{o}n pesada. En ese caso podemos proceder de dos
maneras. Una posibilidad consiste en tratar a todos los fermiones, ligeros o
pesados, al mismo nivel. Terminar\'{\i}amos entonces con una matriz de
mezcla de $4\times 4$ unitaria, cuya submatriz de $3\times 3$,
correspondiente a los fermiones ligeros, no necesitar\'{\i}a ser ---y en
realidad no ser\'{\i}a--- unitaria. Puesto de esta manera, las desviaciones
de unitariedad (\textexclamdown incluso a nivel \'{a}rbol!) podr\'{\i}an ser
considerables. La manera alternativa de proceder consistir\'{\i}a, de
acuerdo a la filosof\'{\i}a de los Lagrangianos efectivos, en integrar
completamente a la generaci\'{o}n pesada. Nos quedar\'{\i}amos entonces, al
nivel m\'{a}s bajo en la expansi\'{o}n en la inversa de la masa pesada, con
los t\'{e}rminos cin\'{e}ticos y de masa ordinarios para los fermiones
ligeros y una matriz de mezcla ordinaria de $3\times 3$ que ser\'{\i}a
obviamente unitaria. Naturalmente no existe contradicci\'{o}n l\'{o}gica
entre ambos procedimientos ya que lo que realmente importa es el elemento
matriz $S$ y este adquiere, si seguimos el segundo procedimiento
(integraci\'{o}n de campos pesados), dos clases de contribuciones: una de
los operadores de dimensi\'{o}n m\'{a}s baja, que contienen s\'{o}lo
fermiones ligeros, y otra de los de dimensi\'{o}n m\'{a}s alta obtenidos
despu\'{e}s de integrar los campos pesados. El resultado para el elemento de
matriz $S$ observable debe ser el mismo sea cual sea el procedimiento
aplicado, pero del segundo m\'{e}todo aprendemos que las violaciones de
unitariedad en el tri\'{a}ngulo de tres generaciones est\'{a}n suprimidas
por una masa pesada. Este simple ejercicio ilustra las ventajas del enfoque
basado en los Lagrangianos efectivos.

En el Cap\'{i}tulo \ref{cpviolationandmixing} extendemos el Lagrangiano
efectivo presentado en el Cap\'{i}tulo \ref{mattersectorchapter} para
considerar mezcla de familias y violaci\'{o}n de $CP$. Este Lagrangiano
contiene los operadores efectivos que dan la contribuci\'{o}n dominante en
teor\'{i}as donde la f\'{i}sica m\'{a}s all\'{a} del Modelo Est\'{a}ndar
aparece a la escala $\Lambda >>M_{W}$. Como en el Cap\'{i}tulo \ref
{mattersectorchapter} aqu\'{i} mantenemos s\'{o}lo los operadores efectivos
no universales dominantes, o sea los de dimensi\'{o}n cuatro. Como no
hacemos otras suposiciones aparte de las de simetr\'{i}a, consideramos
t\'{e}rminos cin\'{e}tico y de masa no diagonales y efectuamos con toda
generalidad la diagonalizaci\'{o}n y el paso a la base f\'{i}sica. Esta
diagonalizaci\'{o}n no deja trazas en el SM aparte de la matriz CKM. Sin
embargo, veremos aqu\'{i} que mucha m\'{a}s informaci\'{o}n de la base
d\'{e}bil queda en los operadores efectivos escritos en la base diagonal.
Luego determinaremos la contribuci\'{o}n en diferentes observables y
discutiremos las posibles nuevas fuentes de violaci\'{o}n de $CP$, la idea
es extraer conclusiones sobre nueva f\'{i}sica m\'{a}s all\'{a} del Modelo
Est\'{a}ndar de consideraciones generales, sin tener que calcular en cada
modelo. En el mismo cap\'{i}tulo presentamos los valores de los coeficientes
del Lagrangiano efectivo calculados en algunas teor\'{i}as, incluido el
Modelo Est\'{a}ndar con un Higgs pesado, y tratamos de obtener conclusiones
generales sobre el esquema general exhibido por la f\'{i}sica m\'{a}s
all\'{a} del Modelo Est\'{a}ndard en lo que concierne a la violaci\'{o}n de $%
CP$.

En el proceso tenemos que tratar un problema te\'{o}rico que es interesante
por s\'{i} mismo: la renormalizaci\'{o}n de la matriz CKM y de la
funci\'{o}n de onda (wfr.) en el esquema \textit{on-shell} en presencia de
mezcla de familias. Pero, \textquestiondown por qu\'{e} tenemos que
preocuparnos de la wfr. o de los contra-t\'{e}rminos de CKM si aqu\'{i}
trabajamos a nivel \'{a}rbol? La respuesta es bastante simple: incluso a
nivel \'{a}rbol uno de los operadores efectivos contribuye a las
autoenerg\'{i}as fermi\'{o}nicas y por lo tanto a las wfr. Esto implica que
esta contribuci\'{o}n ``indirecta'' tiene que ser tenida en cuenta ya que
para calcular observables f\'{i}sicos las wfr. est\'{a}n dictadas por los
requerimientos de LSZ que a su vez son equivalentes a los requerimientos del
esquema \textit{on-shell}. Adem\'{a}s, se puede ver que los
contra-t\'{e}rminos de CKM est\'{a}n tambi\'{e}n relacionados con las wfr.
(aunque no con las f\'{i}sicas o ``externas'') y por lo tanto otra
contribuci\'{o}n potencial puede aparecer a trav\'{e}s de este
contra-t\'{e}rmino.\bigskip

En este punto descubrimos que algunas preguntas acerca de la correcta
implementaci\'{o}n del esquema \textit{on-shell} en presencia de mezcla de
familias quedaban por contestar. Algunas de estas preguntas fueron hechas
por primera vez en \cite{Grassi} donde se presentaron supuestas
inconsistencias entre el esquema \textit{on-shell} y la invariancia \textit{%
gauge}. Motivados por estos resultados decidimos investigar el tema del
esquema \textit{on-shell} en presencia de mezcla de familias y su
relaci\'{o}n con la invariancia \textit{gauge}. Nuestro trabajo en
relaci\'{o}n con este tema est\'{a} presentado en el Cap\'{i}tulo \ref
{LSZchapter} y los resultados de este cap\'{i}tulo se utilizan en el caso
mucho m\'{a}s simple de la contribuci\'{o}n de teor\'{i}a efectiva a primer
orden. Aqu\'{i} vale la pena remarcar que los resultados obtenidos en el
Cap\'{i}tulo \ref{LSZchapter} van mucho m\'{a}s all\'{a} que su
aplicaci\'{o}n en el Cap\'{i}tulo \ref{cpviolationandmixing} y son
relevantes en los c\'{a}lculos de violaci\'{o}n de $CP$ en futuros
experimentos de alta precisi\'{o}n.

Hagamos aqu\'{i} una breve introducci\'{o}n al problema: Cuando calculamos
una amplitud f\'{i}sica de v\'{e}rtice a nivel \textit{1-loop} tenemos que
considerar las contribuciones de nivel \'{a}rbol m\'{a}s correcciones de
varios tipos. O sea, necesitamos contra-t\'{e}rminos para la carga
el\'{e}ctrica, \'{a}ngulo de Weinberg y renormalizaci\'{o}n de la
funci\'{o}n de onda del bos\'{o}n de \textit{gauge} $W$. Tambi\'{e}n
necesitamos la wfr. de los fermiones externos y los contra-t\'{e}rminos de
CKM. Estas \'{u}ltimas renormalizaciones est\'{a}n relacionadas en una forma
que veremos en el Cap\'{i}tulo \ref{LSZchapter} \cite{Balzereit:1999id}.
Finalmente necesitamos calcular los diagramas 1PI correspondientes al
v\'{e}rtice en cuesti\'{o}n.

Hasta aqu\'{i} todo lo dicho es.est\'{a}ndar. Sin embargo, una controversia
relativamente antigua existe en la literatura con respecto a cu\'{a}l es la
manera adecuada de definir las wfr. externas y los contra-t\'{e}rminos de
CKM. La cuesti\'{o}n es bastante compleja ya que estamos tratando con
part\'{i}culas que son inestables (y por lo tanto las autoenerg\'{i}as,
relacionadas con las wfr., desarrollan cortes en el plano complejo que en
general dependen de la fijaci\'{o}n de \textit{gauge}) y con la cuesti\'{o}n
de mezcla de familias.

Varias propuestas han aparecido en la literatura tratando de definir los
contra-t\'{e}rminos adecuados tanto para las patas externas (wfr.) como para
los elementos de matriz de CKM. Las condiciones \textit{on-shell} que
diagonalizan el propagador fermi\'{o}nico \textit{on-shell} fueron
introducidas originalmente en \cite{Aoki}. En \cite{DennerSack} las wfr. que
``satisfac\'{i}an'' las condiciones de \cite{Aoki} fueron derivadas. Sin
embargo en \cite{DennerSack} no se ten\'{i}a en cuenta la presencia de
cortes en las autoenerg\'{i}as, un hecho que entra en conflicto con las
condiciones en \cite{Aoki}. M\'{a}s tarde esto fue reconocido en \cite
{Denner}. El problema se puede resumir diciendo que las condiciones \textit{%
on-shell} definidas en \cite{Aoki} son en realidad imposibles de satisfacer
por un conjunto m\'{i}nimo de constantes de renormalizaci\'{o}n\footnote{%
Por un conjunto m\'{i}nimo queremos decir un conjunto de wfr. de $\bar{\Psi}%
_{0}=\bar{\Psi}\bar{Z}^{\frac{1}{2}}$ y $\Psi _{0}=Z^{\frac{1}{2}}\Psi $
relacionadas por $\bar{Z}^{\frac{1}{2}}=\gamma ^{0}Z^{\frac{1}{2}\dagger
}\gamma ^{0}.$} debido a la presencia de partes absortivas en las
autoenerg\'{i}as. El autor de \cite{Denner} evita este problema
introduciendo una prescripci\'{o}n que elimina \textit{de facto} estas
partes absortivas, pero pagando el precio de no diagonalizar el propagador
fermi\'{o}nico en sus \'{i}ndices de familia.

Las identidades de Ward basadas en la simetr\'{i}a de \textit{gauge} SU(2)$%
_{L}$ relacionan las wfr. y los contra-t\'{e}rminos de CKM \cite
{Balzereit:1999id}. En \cite{Grassi} se muestra que si la prescripci\'{o}n
de \cite{DennerSack} se utiliza en los contra-t\'{e}rminos de CKM, el
resultado del c\'{a}lculo de un observable f\'{i}sico resulta dependiente
del par\'{a}metro de \textit{gauge}. Como ya hemos mencionado, los
resultados en \cite{DennerSack} no tratan adecuadamente las partes
absortivas presentes en las autoenerg\'{i}as; que a su vez resultan ser
dependientes del par\'{a}metro de \textit{gauge}. En el Cap\'{i}tulo \ref
{LSZchapter} veremos que a pesar de los problemas existentes en la
prescripci\'{o}n dada en \cite{DennerSack}, las conclusiones dadas en \cite
{Grassi} son correctas: una condici\'{o}n necesaria para la invariancia
\textit{gauge} de las amplitudes f\'{i}sicas es que el contra-t\'{e}rmino de
CKM sea independiente del par\'{a}metro de \textit{gauge}. Tanto el
contra-t\'{e}rmino de CKM propuesto \cite{Grassi} como los propuestos en
\cite{Balzereit:1999id}, \cite{Diener:2001qt} satisfacen dicha condici\'{o}n.

Existen en la literatura otras propuestas para definir la
renormalizaci\'{o}n de CKM, \cite{Diener:2001qt}, \cite{Barroso} y \cite
{Yamada}. En todos estos trabajos, o se utilizan las wfr. propuestas
originalmente en \cite{DennerSack} o las dadas en \cite{Denner}, o la
cuesti\'{o}n de la correcta definici\'{o}n de la wfr. externas se evita
completamente. En cualquier caso las partes absortivas de las
autoenerg\'{i}as no son tenidas en cuenta (incluso las partes absortivas de
los diagramas 1PI son evitadas en \cite{Barroso}). Como veremos, hacer esto
conduce a amplitudes f\'{i}sicas ---elementos de matriz $S$ --- que son
dependientes del par\'{a}metro de \textit{gauge}, independientemente del
m\'{e}todo utilizado para renormalizar $K_{ij}$ siempre que la
redefinici\'{o}n de $K_{ij}$ sea independiente del \textit{gauge} y preserve
unitariedad.

Debido a la estructura de los cortes absortivos resulta que, sin embargo, la
dependencia en el par\'{a}metro de \textit{gauge} en la amplitud ---elemento
de matriz $S$--- , usando la prescripci\'{o}n de \cite{Denner}, cancela en
el modulo cuadrado de la misma en el SM. Esta cancelaci\'{o}n ha sido
verificada num\'{e}ricamente por los autores de \cite{Kniehl}. En el
Cap\'{\i}tulo \ref{LSZchapter} presentaremos los resultados anal\'{\i}ticos
que muestran que esta cancelaci\'{o}n es exacta. Sin embargo la dependencia
en el par\'{a}metro de \textit{gauge} permanece en la amplitud.

\textquestiondown Es esto aceptable? Creemos que no. Los diagramas que
contribuyen al mismo proceso f\'{\i}sico fuera del sector electrod\'{e}bil
del SM pueden interferir con la amplitud del SM y revelar la inaceptable
dependencia \textit{gauge}. M\'{a}s a\'{u}n, las partes absortivas
independientes del \textit{gauge} est\'{a}n tambi\'{e}n eliminadas en la
prescripci\'{o}n en \cite{Denner}. Sin embargo, estas partes, a diferencia
de las dependientes del \textit{gauge}, no desaparecen de la amplitud al
cuadrado tal como veremos. Adem\'{a}s, no debemos olvidar que el esquema en
\cite{Denner} no diagonaliza correctamente los propagadores en sus
\'{\i}ndices de familia. El Cap\'{\i}tulo \ref{LSZchapter} est\'{a} dedicado
a respaldar las afirmaciones anteriores.

En resumen, en el Cap\'{i}tulo \ref{LSZchapter}, con la ayuda de un uso
extensivo de las identidades de Nielsen \cite{Nielsen,Piguet,Sibold}
complementadas con c\'{a}lculos expl\'{i}citos, corroboramos que el
contra-t\'{e}rmino de CKM tiene que ser independiente del par\'{a}metro de
\textit{gauge} y demostramos que la prescripci\'{o}n com\'{u}nmente
utilizada para la renormalizaci\'{o}n de la funci\'{o}n de onda conduce a
amplitudes f\'{i}sicas dependientes del par\'{a}metro de \textit{gauge},
incluso si el contra-t\'{e}rmino de CKM no depende del par\'{a}metro de
\textit{gauge} tal como se requiere. Para aquellos lectores no
familiarizados con las identidades de Nielsen presentamos un resumen
pedag\'{o}gico de las mismas indicando las referencias relevantes. Usando
esta tecnolog\'{i}a mostramos que una prescripci\'{o}n que cumple los
requerimientos de LSZ conduce a amplitudes independientes del par\'{a}metro
de \textit{gauge}. Las renormalizaciones de funci\'{o}n de onda resultantes
necesariamente poseen partes absortivas. Por ello verificamos
expl\'{i}citamente que dicha presencia no altera los requerimientos
esperados en cuanto a $CP$ y $CPT$. Los resultados obtenidos utilizando esta
prescripci\'{o}n son diferentes (incluso a nivel del m\'{o}dulo cuadrado de
la amplitud) de los que se obtienen despreciando las partes absortivas en el
caso del decaimiento del quark top. Mostramos asimismo que esta diferencia
es num\'{e}ricamente relevante.\bigskip

Una vez que estos aspectos te\'{o}ricos est\'{a}n aclarados pasamos al
estudio de la fenomenolog\'{i}a capaz de probar la f\'{i}sica del sector de
corrientes cargadas que es el sector sensible a la violaci\'{o}n de $CP$ en
el Modelo Est\'{a}ndar. Cuando nos centramos en interacciones que involucran
a los bosones $W,Z$, los operadores presentes en el Lagrangiano efectivo
electrod\'{e}bil inducen v\'{e}rtices efectivos que acoplan los bosones de
\textit{gauge} con los campos de materia \cite{burgess}
\begin{equation}
-\frac{e}{4c_{W}s_{W}}\bar{f}\gamma ^{\mu }\left( \kappa _{L}^{NC}L+\kappa
_{R}^{NC}R\right) Z_{\mu }f-\frac{e}{s_{W}}\bar{f}\gamma ^{\mu }\left(
\kappa _{L}^{CC}L+\kappa _{R}^{CC}R\right) \frac{\tau ^{-}}{2}W_{\mu
}^{+}f+h.c.
\end{equation}
Otros posibles efectos no son f\'{i}sicamente observables, tal como veremos
en el Cap\'{i}tulo \ref{LHCphenomenology}. En t\'{e}rminos pr\'{a}cticos,
LHC establecer\'{a} restricciones en los acoplos efectivos del v\'{e}rtice
del $W$, y por lo tanto en la nueva f\'{i}sica que contribuye a los mismos.
Nuestros resultados son tambi\'{e}n relevantes en un contexto
fenomenol\'{o}gico m\'{a}s amplio como una manera de restringir $\kappa _{L}$
y $\kappa _{R}$ (incluyendo nueva f\'{i}sica y correcciones radiativas), sin
necesidad de apelar a un Lagrangiano efectivo subyacente que describa un
modelo espec\'{i}fico de ruptura de simetr\'{i}a. Por supuesto en ese caso
se pierde el poder de un Lagrangiano efectivo, es decir, se pierde el
conjunto bien definido de reglas de contaje y la capacidad de relacionar
diferentes procesos.

Como ya hemos destacado, incluso en el Modelo Est\'{a}ndar m\'{\i}nimo, las
correcciones radiativas inducen modificaciones en los v\'{e}rtices.
Asumiendo una dependencia suave en los momentos externos estos factores de
forma pueden ser expandidos en potencias de momentos. Al orden m\'{a}s bajo
en la expansi\'{o}n en derivadas, el efecto de las correcciones radiativas
puede ser codificado en los v\'{e}rtices efectivos $\kappa _{L}$ y $\kappa
_{R}$. As\'{\i}, estos v\'{e}rtices efectivos toman valores bien definidos,
valores calculables en el Modelo Est\'{a}ndar m\'{\i}nimo, y cualquier
desviaci\'{o}n de los mismos (que, incidentalmente, no han sido determinados
completamente en el Modelo Est\'{a}ndar a\'{u}n) indicar\'{\i}a la presencia
de nueva f\'{\i}sica en el sector de materia. La capacidad que LHC tiene
para fijar restricciones directas en los v\'{e}rtices efectivos, en
particular en aquellos que involucran a la tercera generaci\'{o}n, es de
vital importancia para acotar los posibles modelos de f\'{\i}sica m\'{a}s
all\'{a} del Modelo Est\'{a}ndard. El trabajo del Cap\'{\i}tulo \ref
{LHCphenomenology} est\'{a} dedicado a este an\'{a}lisis en procesos
cargados involucrando al quark top en el LHC.

A la energ\'{i}a de LHC (14 Tev) el mecanismo dominante en la producci\'{o}n
de tops, con una secci\'{o}n eficaz de 800 pb \cite{catani}, es el mecanismo
de fusi\'{o}n gluon-gluon. Este mecanismo no tiene nada que ver con el
sector electrod\'{e}bil y por lo tanto no el m\'{a}s adecuado para nuestros
prop\'{o}sitos. Aunque es el mecanismo que m\'{a}s tops produce y por lo
tanto es importante considerarlo a la hora de estudiar los acoplos del top a
trav\'{e}s de su decaimiento, que ser\'{a} nuestro principal inter\'{e}s en
el Cap\'{i}tulo \ref{topdecaychapter}, y tambi\'{e}n como \textit{background}
al proceso que nos ocupar\'{a} en este cap\'{i}tulo.
\begin{figure}[!hbp]
\begin{center}
\includegraphics[width=8cm]{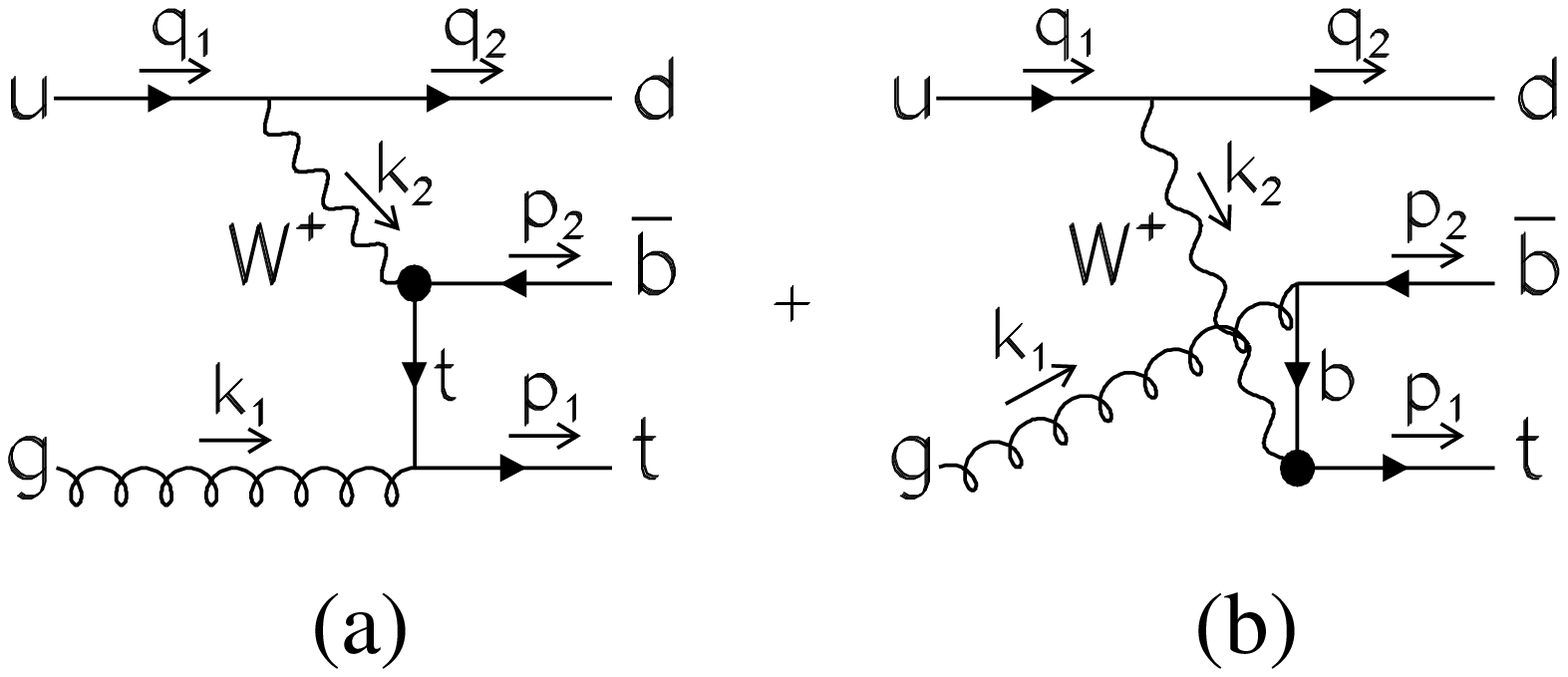}
\end{center}
\caption{Diagramas de Feynman que contribuyen al subproceso de
producci\'{o}n de un \textit{single-top}. En este caso tenemos un quark $d$
como quark espectador}
\label{u+gt+b-d+totintrosp}
\end{figure}
\begin{figure}[!hbp]
\begin{center}
\includegraphics[width=8cm]{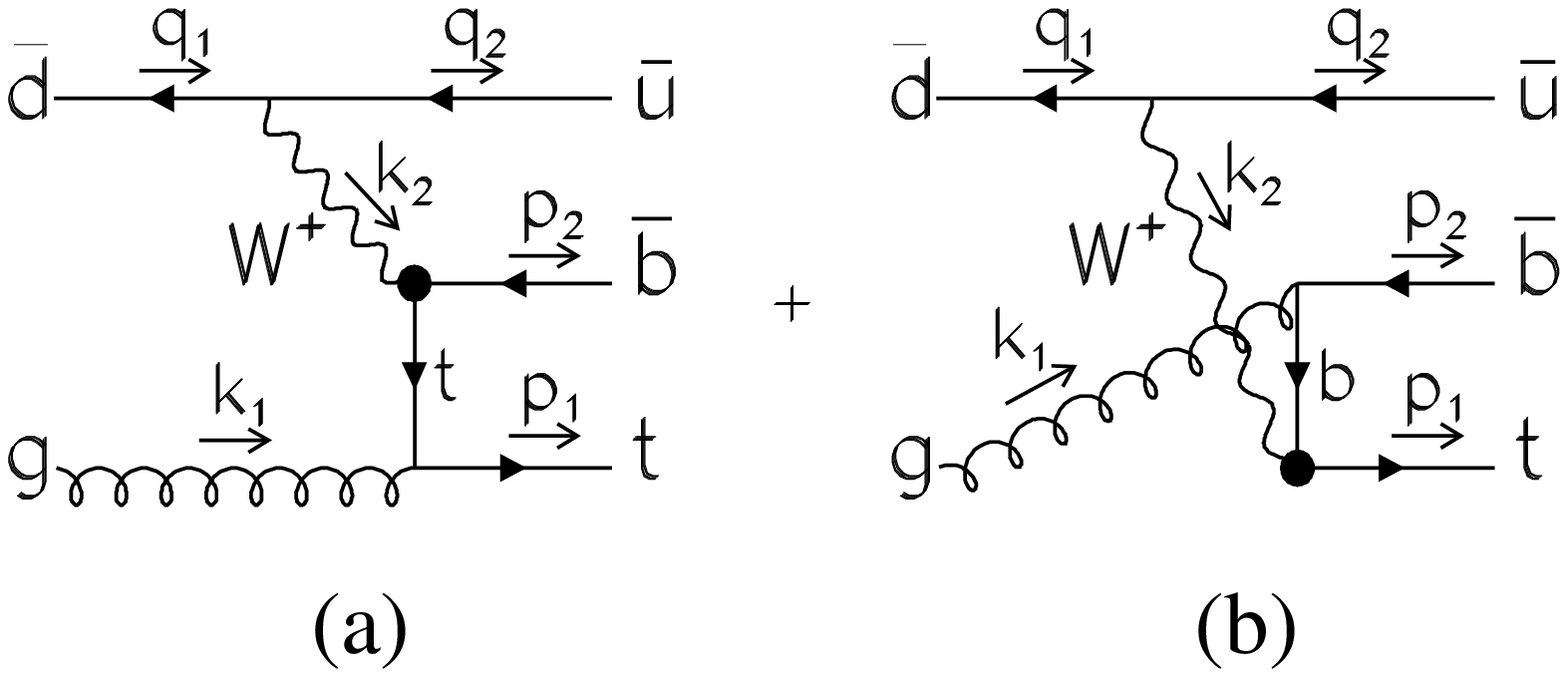}
\end{center}
\caption{Diagramas de Feynman que contribuyen al subproceso de
producci\'{o}n de un \textit{single-top}. En este caso tenemos un quark $%
\bar{u}$ como quark espectador}
\label{d-gt+b-u-totintrosp}
\end{figure}

La f\'{i}sica electrod\'{e}bil entra en juego en la producci\'{o}n de
\textit{single-top} (un \'{u}nico top). (para una revisi\'{o}n reciente ver
e.g. \cite{Tait}.) A las energ\'{i}as de LHC el subproceso electrod\'{e}bil
dominante (de lejos) que contribuye a la producci\'{o}n de \textit{single-top%
} est\'{a} dado por un gluon ($g$) viniendo de un prot\'{o}n y un quark o
anti-quark ligero viniendo del otro (este proceso tambi\'{e}n se denomina de
producci\'{o}n en canal t \cite{tchannel,SSW}). Este proceso est\'{a}
graficado en las Figs. \ref{u+gt+b-d+totintrosp} y \ref{d-gt+b-u-totintrosp}%
, donde quarks ligeros de tipo $u$ o antiquarks ligeros de tipo $\bar{d}$
son extra\'{i}dos del prot\'{o}n, respectivamente. Estos quarks luego radian
un $W$ cuyo acoplo efectivo es el objeto de nuestro inter\'{e}s. La
secci\'{o}n eficaz total para este proceso en el LHC ha sido calculada en
250 pb \cite{SSW}, a ser comparada con los 50 pb para la asociada a la
producci\'{o}n con un bos\'{o}n $W^{+}$ y un quark $b$ extra\'{i}do del mar
de prot\'{o}n, y 10 pb que corresponden a la fusi\'{o}n quark-quark
(producci\'{o}n en canal s que ser\'{a} analizada en el Cap\'{i}tulo \ref
{topdecaychapter}). En el Tevatron (2 GeV) la secci\'{o}n eficaz de
producci\'{o}n para fusi\'{o}n $W$-gluon es de 2.5 pb, y por lo tanto, en
comparaci\'{o}n, la producci\'{o}n de tops en este subproceso en particular
es realmente copiosa en LHC. La simulaciones de Monte Carlo incluyendo el
an\'{a}lisis de los productos de decaimiento del top indican que este
proceso puede ser analizado en detalle en LHC y tradicionalmente ha sido
considerado como el m\'{a}s importante para nuestros prop\'{o}sitos.

En una colisi\'{o}n prot\'{o}n prot\'{o}n tambi\'{e}n se produce un par
bottom anti-top a trav\'{e}s de un subproceso an\'{a}logo. En cualquier caso
los resultados cualitativos son muy similares a aquellos correspondientes a
la producci\'{o}n de tops, de donde las secciones eficaces pueden ser
f\'{a}cilmente derivadas haciendo los cambios adecuados.

En el contexto de teor\'{i}as efectivas, la contribuci\'{o}n de operadores
de dimension cinco a la producci\'{o}n de tops a trav\'{e}s de fusi\'{o}n de
bosones vectoriales longitudinales fue estimada hace alg\'{u}n tiempo en
\cite{LY}, aunque el estudio no fue de ning\'{u}n modo completo. Debe ser
mencionado que la producci\'{o}n de un par $t,\bar{t}$ a trav\'{e}s de este
mecanismo est\'{a} muy enmascarada por el mecanismo dominante que es la
fusi\'{o}n gluon-gluon, mientras que la producci\'{o}n de \textit{single-top}%
, a trav\'{e}s de fusi\'{o}n $WZ$, se supone mucho m\'{a}s suprimida
comparada con el mecanismo presentado en este trabajo. Esto se debe a que
los dos v\'{e}rtices son electrod\'{e}biles en el proceso discutido en \cite
{LY}, y a que los operadores de dimensi\'{o}n cinco se suponen suprimidos
por una escala elevada. La contribuci\'{o}n de operadores de dimensi\'{o}n
cuatro no ha sido, por lo que sabemos, considerada anteriormente, aunque la
capacidad de la producci\'{o}n de \textit{single-top} para medir el elemento
de matriz de CKM $K_{tb}$, ha sido hasta cierto punto analizado en el pasado
(ver por ejemplo \cite{SSW,mandp}).

Para resumir, en el Cap\'{i}tulo \ref{LHCphenomenology} analizamos la
sensibilidad de diferentes observables a la magnitud de los coeficientes
efectivos que parametrizan la nueva f\'{i}sica m\'{a}s all\'{a} del Modelo
Est\'{a}ndar. Tambi\'{e}n mostramos que los observables relevantes para la
distinci\'{o}n de los acoplos quirales \textit{left} y \textit{right}
involucra, en la pr\'{a}ctica, la medici\'{o}n del esp\'{i}n del top que
s\'{o}lo puede ser realizada de forma indirecta midiendo la distribuci\'{o}n
angular de sus productos de decaimiento. Mostramos que la presencia de
acoplos efectivos de quiralidad \textit{right} implican que el top no se
encuentra en un estado puro y que existe una \'{u}nica base de esp\'{i}n
\'{u}til para conectar la distribuci\'{o}n de los productos de decaimiento
del top con la secci\'{o}n eficaz diferencial de producci\'{o}n de tops
polarizados. Presentamos adem\'{a}s las expresiones anal\'{i}ticas
completas, incluyendo acoplos efectivos generales, de las secciones eficaces
diferenciales correspondientes a los subprocesos de producci\'{o}n de
\textit{single-top} polarizado en canal t. La masa del quark bottom, que
resulta ser m\'{a}s relevante de lo que se puede esperar, se mantiene en
todo el c\'{a}lculo. Finalmente analizamos diferentes aspectos de la
secci\'{o}n eficaz total relevantes para la detecci\'{o}n de nueva
f\'{i}sica a trav\'{e}s de los acoplos efectivos. Tambi\'{e}n hemos
desarrollado la aproximaci\'{o}n llamada de \textit{W} efectivo para este
proceso pero los resultados no se presentan en esta tesis \cite{effW}%
.\bigskip

\begin{figure}[!htp]
\begin{center}
\includegraphics[width=6cm]{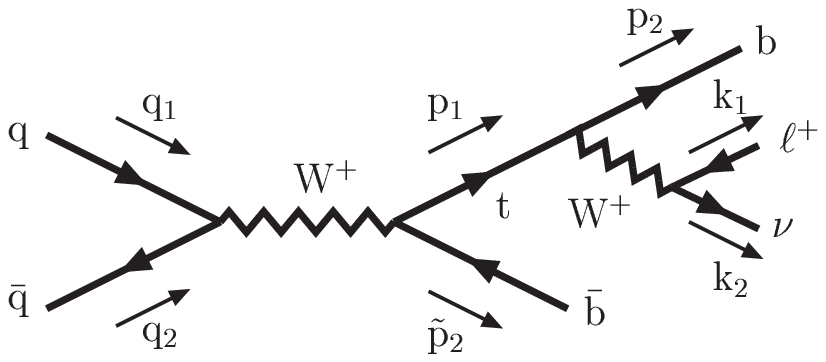}
\end{center}
\par
\caption{Diagrama de Feynman correspondiente a la producci\'{o}n y
decaimiento de \textit{single-top} en el canal s.}
\label{singletopschannelanddecayintrosp}
\end{figure}

Finalmente en el Cap\'{i}tulo \ref{topdecaychapter} estudiamos un aspecto de
la producci\'{o}n de tops que no fue finalizado en el cap\'{i}tulo anterior;
la ``medici\'{o}n'' del esp\'{i}n del top a trav\'{e}s de sus productos de
decaimiento. El an\'{a}lisis num\'{e}rico de la sensibilidad de los
diferentes observables al acoplo \textit{right} $g_{R}$ se realiza aqu\'{i}
incluyendo los productos de decaimiento del top. Ya que el principal
objetivo de este cap\'{i}tulo es aclarar el rol del esp\'{i}n del top cuando
el decaimiento del top tambi\'{e}n se considera, estudiamos la
producci\'{o}n de \textit{single-top} a trav\'{e}s del canal s, m\'{a}s
simple de analizar desde el punto de vista te\'{o}rico. La producci\'{o}n y
decaimiento del top en este canal se grafica en la Fig. (\ref
{singletopschannelanddecayintrosp})

En el Cap\'{i}tulo \ref{topdecaychapter} mostramos como la secci\'{o}n
eficaz diferencial correspondiente al proceso de la Fig. (\ref
{singletopschannelanddecayintrosp}) se calcula en dos pasos usando la
aproximaci\'{o}n resonancia estrecha teniendo en cuenta el esp\'{i}n del
top. O sea, en primer lugar calculamos la probabilidad de producir tops con
una dada polarizaci\'{o}n y luego convolucionamos dicha probabilidad con la
probabilidad de decaimiento, sumando sobre las dos polarizaciones del top.
Exponemos los argumentos que permiten demostrar que los efectos de
interferencia cu\'{a}nticos pueden ser minimizados con una elecci\'{o}n
adecuada de la base de esp\'{i}n. Presentamos expresiones expl\'{i}citas
tanto para el canal s como para el canal t de la base de esp\'{i}n que
diagonaliza la matriz densidad del top. En el caso del canal s utilizamos
esta base en nuestro programa de integraci\'{o}n de Monte Carlo analizando
num\'{e}ricamente la sensibilidad de nuestros resultados ante cambios de la
base de esp\'{i}n o incluso ante la posibilidad de prescindir del esp\'{i}n
completamente. Estos estudios num\'{e}ricos muestran que la
implementaci\'{o}n de la base correcta de esp\'{i}n es importante a nivel
del 4\%. Adem\'{a}s de la cuesti\'{o}n del esp\'{i}n del top, nuestros
resultados num\'{e}ricos muestran claramente el papel crucial de elegir
configuraciones cinem\'{a}ticas concretas para los productos de decaimiento
del top que maximicen la sensibilidad al acoplo $g_{R}$ tanto en magnitud
como en fase.\bigskip

En los ap\'{e}ndices de esta tesis hemos inclu\'{i}do material t\'{e}cnico
que complementa los contenidos de los cap\'{i}tulos y algunos c\'{a}lculos
que pueden servir al lector interesado en reproducir los resultados. En
particular hemos inclu\'{i}do el c\'{a}lculo completo de todas las
autoenerg\'{i}as fermi\'{o}nicas en un \textit{gauge} arbitrario $R_{\xi }$ .

\newpage


\section{Resultados y Conclusiones}

\bigskip

En lo que sigue presentamos un sumario de los principales resultados y
conclusiones de esta tesis.

\begin{itemize}
\item  {En el Cap\'{\i}tulo \ref{mattersectorchapter}: }

\begin{itemize}
\item  Ofrecemos una clasificaci\'{o}n completa de los operadores de cuatro
campos fermi\'{o}nicos responsables de dar masa a fermiones f\'{i}sicos y a
bosones \textit{gauge} vectoriales en modelos con rotura din\'{a}mica de
simetr\'{i}a. Dicha clasificaci\'{o}n se realiza cuando las nuevas
part\'{i}culas aparecen en las representaciones usuales del grupo $%
SU(2)_{L}\times SU(3)_{c}$. En el caso general discutimos, adem\'{a}s, una
clasificaci\'{o}n parcial. Debido a que se ha tomado \'{u}nicamente el caso
de una sola familia, el problema de mezcla no ha sido aqu\'{i} considerado.

\item  Investigamos las consecuencias fenomenol\'{o}gicas para el sector
electrod\'{e}bil neutro en dicha clase de modelos. Para ello realizamos el
\textit{matching} entre la descripci\'{o}n mediante t\'{e}rminos de cuatro
fermiones y una teor\'{i}a a m\'{a}s bajas energ\'{i}as que contiene
s\'{o}lamente los grados de libertad del SM (a excepci\'{o}n del Higgs). Los
coeficientes de este Lagrangiano efectivo de bajas energ\'{i}as para modelos
con rotura din\'{a}mica de simetr\'{i}a son, a continuaci\'{o}n, comparados
con los de modelos con escalares elementales (como por ejemplo, en el Modelo
Est\'{a}ndar m\'{i}nimo).

\item  Determinamos el valor del acoplamiento efectivo de $Zb\bar{b}$ en
modelos con rotura din\'{a}mica de simetr\'{i}a verificando que su
contribuci\'{o}n es importante, pero su signo no est\'{a} determinado
contrariamente a afirmaciones anteriores. El valor experimental actual se
desv\'{i}a del predicho por el SM en casi 3 $\sigma $. Estimamos tambi\'{e}n
los efectos en los fermiones ligeros, a pesar de que no son observables
actualmente. Algunas consideraciones generales concernientes al mecanismo de
rotura din\'{a}mica de simetr\'{i}a son presentadas.
\end{itemize}

\item  En el Cap\'{\i}tulo \ref{cpviolationandmixing}:

\begin{itemize}
\item  Analizamos la estructura de los operadores efectivos de cuatro
dimensiones para el sector de materia de la teor\'{\i}a electrod\'ebil
cuando se permiten violaciones $CP$ y mezcla de familias.

\item  Realizamos la diagonalizaci\'on de los t\'erminos de masa y
cin\'eticos demostrando que, adem\'as de la presencia de la matriz CKM en el
v\'ertice cargado del SM, aparecen nuevas estructuras en los operadores
efectivos constru\'{\i}dos con fermiones de quiralidad \textit{left}. En
particular la matriz CKM se encuentra tambi\'en presente en el sector neutro.

\item  Calculamos tambi\'en la contribuci\'on de los operadores efectivos en
el SM m\'{\i}nimo con un Higgs pesado y en el SM con un doblete de fermiones
pesados adicional.

\item  En general, incluso si la f\'{\i}sica responsable de la generaci\'on
de los operadores efectivos adicionales conserva $CP$, las fases presentes
en los acoplamientos Yukawa y cin\'eticos se hacen observables en los
operadores efectivos tras su diagonalizaci\'on.
\end{itemize}

\item  En el Cap\'{\i}tulo \ref{LSZchapter}:

\begin{itemize}
\item  Presentamos y resolvemos la cuesti\'{o}n sobre la definici\'{o}n de
un conjunto de constantes wfr. a \textit{1 loop} consistentes con los
requerimientos de \textit{on-shell} y la invariancia gauge de las amplitudes
f\'{i}sicas electrod\'{e}biles. Demostramos, utilizando las identidades de
Nielsen, que con nuestro conjunto de constantes wfr. y una
renormalizaci\'{o}n del CKM independiente del gauge, se obtienen unas
amplitudes f\'{i}sicas para el decaimiento del top y del W independientes
del gauge.

\item  Mostramos que la prescripci\'{o}n \textit{on-shell} dada en \cite
{Denner} no diagonaliza el propagador en los \'{i}ndices de familia y que
dicha prescripci\'{o}n origina amplitudes que dependen del \textit{gauge},
aunque dicha dependencia desaparece en m\'{o}dulo de la amplitud
correspondiente al v\'{e}rtice cargado electrod\'{e}bil. El hecho de que
s\'{o}lo el m\'{o}dulo de las amplitudes electrod\'{e}biles no dependa del
\textit{gauge} no es satisfactorio, ya que la interferencia con fases
fuertes puede, por ejemplo, originar una dependencia \textit{gauge}
inaceptable. En el caso del decaimiento del top encontramos que la
diferencia num\'{e}rica entre nuestro resultado para el m\'{o}dulo al
cuadrado de la amplitud y el mismo obtenido con la prescripci\'{o}n dada en
\cite{Denner} llega al $0.5\%$. Esta diferencia ser\'{a} relevante en los
futuros experimentos de precisi\'{o}n dise\~{n}ados para determinar el
v\'{e}rtice $tb$.

\item  Comprobamos la consistencia de nuestro esquema con el teorema $CPT$.
Dicha comprobaci\'{o}n se hace mostrando que, aunque nuestras constantes
wfr. no verifican la condici\'{o}n de pseudo-hermiticidad ($\bar{Z}\neq
\gamma ^{0}Z^{\dagger }\gamma ^{0}$), la anchura total de part\'{\i}culas y
anti-part\'{\i}culas coincide.
\end{itemize}

\item  En el Cap\'{\i}tulo \ref{LHCphenomenology}:

\begin{itemize}
\item  Presentamos un c\'{a}lculo completo de las secciones eficaces en el
canal t para tops o anti-tops polarizados incluyendo acoplamientos efectivos
\textit{right} y contribuciones a la masa del quark bottom.

\item  Realizamos una simulaci\'{o}n Monte Carlo de la producci\'{o}n de {%
\textit{single-top}} polarizado en el LHC para una colecci\'{o}n de
distribuciones en $p_{T}$ y distribuciones angulares para los quarks $t$ y $%
\bar{b}$. Mostramos, sin tener en cuenta \textit{backgrounds} o el efecto
del decaimiento del top, que podemos esperar una sensibilidad de 2
desviaciones est\'{a}ndar para variaciones de $g_{R}$ del orden de $5\times
10^{-2}$.

\item  Mostramos, bas\'{a}ndonos en consideraciones te\'{o}ricas, que el top
no puede producirse en un estado de esp\'{i}n puro si $g_{R}\neq 0$. M\'{a}s
a\'{u}n, indicamos cual es la base de esp\'{i}n adecuada para convolucionar
la secci\'{o}n eficaz de producci\'{o}n del top con la secci\'{o}n eficaz de
decaimiento del mismo. Dicha convoluci\'{o}n se efect\'{u}a para poder
calcular el proceso completo en el marco de la aproximaci\'{o}n de
resonancia estrecha.
\end{itemize}

\item  En el Cap\'{\i}tulo \ref{topdecaychapter}:

\begin{itemize}
\item  Presentamos un c\'{a}lculo completo de la secci\'{o}n eficaz en el
canal s de producci\'{o}n de\textit{\ single-top} incluyendo su decaimiento.
Los c\'{a}lculos incluyen acoplamientos efectivos \textit{right} y
contribuciones de la masa del quark bottom.

\item  Efectuamos una simulaci\'{o}n Monte Carlo de la producci\'{o}n y
decaimiento de tops polarizados en el LHC en el canal s. Representamos
gr\'{a}ficamente diferentes distribuciones de $p_{T}$, masa invariante y
distribuciones angulares constru\'{i}das con los momentos del
anti-lept\'{o}n y el momento de los jets del bottom y del anti-bottom.
Encontramos que las variaciones de $g_{R}$ del orden $5\times 10^{-2}$ son
visibles con se\~{n}ales comprendidas entre 3 y 1 desviaciones est\'{a}ndar
dependiendo de la fase de $g_{R}$ y de los observables elegidos.

\item  Presentamos expresiones expl\'{i}citas para los canales t y s de la
base de esp\'{i}n del top que diagonaliza su matriz densidad. Comprobamos
num\'{e}ricamente que para el canal s dicha base minimiza los t\'{e}rminos
de interferencia ignorados en la aproximaci\'{o}n de resonancia estrecha.
\end{itemize}
\end{itemize}

\newpage

\selectlanguage{english}

\pagestyle{myheadings} \pagestyle{fancyplain}

\renewcommand{\chaptermark}[1]           {\markboth{#1}{}} %
\renewcommand{\sectionmark}[1]           {\markright{\thesection\ #1}}
\lhead[\fancyplain{}{\bfseries \thepage}]%
           {\fancyplain{}{\bfseries\rightmark}}
\rhead[\fancyplain{}{\bfseries \leftmark}]%
           {\fancyplain{}{\bfseries\thepage}}

\renewcommand{\thesection}{\arabic{section}} 
\setcounter{chapter}{0} \renewcommand{\thepage}{\arabic{page}} %
\setcounter{page}{1} \renewcommand{\thechapter}{\arabic{chapter}}

\renewcommand{\chaptername}{Chapter}

\renewcommand{\thetable}{\arabic{chapter}.\arabic{table}} %
\setcounter{table}{0}

\renewcommand{\thefigure}{\arabic{chapter}.\arabic{figure}} %
\setcounter{figure}{0}

\chapter{Introduction}

Quantum field theories (QFT) are defined through the renormalization group.
The basic idea has its origins in the condensed matter world \cite
{Kadanoff:1966wm} and briefly can be stated by saying that in the
thermodinamic limit (an infinite number of degrees of freedom) the
integration of high frequency degrees of freedom can be seen as a
redefinition of the operators appearing in the theory. When the number of
such operators is finite we call this theory `renormalizable' and when it is
not we call it non-renormalizable or effective theory \cite
{efftheointro,Donoghue}. Renormalizable theories are in principle capable of
being considered as `fundamental' QFT since the continuum limit is feasible.

In any case renormalized operators bear dependence on the cut-off that
regularizes the theory. Such dependence is dictated mainly by the naive
dimension of the operator. The bigger the dimension the bigger the cut-off
suppression. Because of that, non renormalizable theories can be analyzed in
practice truncating the number of operators which are ordered by increasing
dimensionality. Lower dimensional operators provide the leading contribution
to observables at low energies and because of that these theories still have
predictive power if we restrict ourselves to such regime. As we increase the
energy or the order in perturbation theory (related to the energy counting
by the Weinberg theorem \cite{Weinberg}) more and more operators are needed
in calculations and therefore the predictive power reduces and eventually
the theory becomes worthless. This inconvenient feature of effective
theories is compensated by their advantage in terms of generality. Since
different high energy models belong to the same universality class (the same
phenomenology at low energies) effective field theories provide a way to
probe theories in a compact way without entering in irrelevant high energy
features. In total we can summarize these considerations in Table (\ref
{effectivetable})
\begin{table}[tbp]
\centering
\begin{tabular}{|c|c|}
\hline
renormalizable QFT & effective QFT \\ \hline
finite number of operators &
\begin{tabular}{c}
infinite number of operators \\
\multicolumn{1}{l}{(truncation controlled by dimensionality)}
\end{tabular}
\\ \hline
predictability power at all energies & predictability power at low enegies
\\ \hline
model proliferation & generality \\ \hline
\end{tabular}
\caption{renormalizable vs. effective QFT's}
\label{effectivetable}
\end{table}

Besides dimensionality considerations, symmetry is the other basic
ingredient that classifies operators and restricts the renormalization group
mixing between operators.\bigskip

The object of this thesis is the study of some open problems in the
electroweak matter sector from an effective theory perspective. The topics
studied include:

\begin{itemize}
\item  General aspects of dynamical symmetry breaking models, studying what
traces these mechanisms may leave at low energies.

\item  A treatment of $CP$ violation and family mixing in the framework of
an effective theory and the determination of some of the effective couplings
involved.

\item  Theoretical issues connecting the renormalization group, gauge
invariance, $CP$, $CPT$ and physical observables.

\item  The possibility of experimentally constraining some of the effective
couplings involved at the LHC.
\end{itemize}

In what follows we will provide a more detailed picture of the scope of this
thesis.\bigskip

Even though the basic structure of Standard Model (SM) of electroweak
interactions has already been well tested thanks to a number of accurate
experiments, its symmetry breaking sector is not firmly established yet,
both from the theoretical and the experimental point of view.

In the minimal version of the SM of electroweak interactions the same
mechanism (a one-doublet complex scalar field) gives masses simultaneously
to the $W$ and $Z$ gauge bosons and to the fermionic matter fields (with the
possible exception of the neutrino). This mechanism is, however, based in a
perturbative approximation. From the non-perturbative point of view the
minimal SM scalar sector is believed to be trivial, which in turn is
equivalent to considering such model as a truncation of an effective theory.
This implies that at a scale $\sim 1$ TeV new interactions should appear if
the Higgs particle is not found by then \cite{triviality}. The 1 TeV cut-off
is determined from non-perturbative studies and hinted by the breakdown of
perturbative unitarity. On the other hand, in the minimal SM it is
completely unnatural to have a light Higgs particle since its mass is not
protected by any symmetry (the so-called hierarchy problem).

This contradiction is solved by supersymmetric extensions of the SM, where
essentially the same symmetry breaking mechanism is at work, although the
scalar sector becomes much richer in this case with relatively light scalars
preferred. In fact, if supersymmetry is to remain a useful idea in
phenomenology, it is crucial that the Higgs particle is found with a mass $%
M_{H}\leq 125$ GeV, or else the theoretical problems, for which
supersymmetry was invoked in the first place, will reappear \cite{hierarchy}%
. Two-loop calculations \cite{twoloop} raise this limit somewhat to 130 GeV
or thereabouts.

A third possibility is the one provided by models of dynamical symmetry
breaking (such as technicolor (TC) theories \cite{TC}). Here there are
interactions that become strong, typically at the scale $\Lambda _{\chi
}\simeq 4\pi v$ ($v=250$ GeV), breaking the global $SU(2)_{L}\times
SU(2)_{R} $ symmetry to its diagonal subgroup $SU(2)_{V}$ and producing
Goldstone bosons which eventually become the longitudinal degrees of freedom
of the $W^{\pm }$ and $Z$. In order to transmit this symmetry breaking to
the ordinary matter fields one requires additional interactions, usually
called extended technicolor (ETC) and characterized by a different scale $M$%
. Generally, it is assumed that $M\gg 4\pi v$ to keep possible
flavor-changing neutral currents (FCNC) under control \cite{FCNC}. Thus a
distinctive characteristic of these models is that the mechanism giving
masses to the $W^{\pm }$ and $Z$ bosons and to the matter fields is
different.

Where do we stand at present? Some will go as far as saying that an
elementary Higgs (supersymmetric or otherwise) has been `seen' through
radiative corrections and that its mass is below 200 GeV, or even discovered
in the last days of LEP with a mass $\simeq $115 GeV \cite{LEPC}. Others
dispute this fact (see for instance \cite{chanowitz} for a critical review
of current claims of a light Higgs).

The effective Lagrangian approach has proven to be remarkably useful in
setting very stringent bounds on the type of new physics behind the symmetry
breaking mechanism of the SM taking as input basically the LEP \cite{aleph}
(and SLC \cite{SLD}) experimental results. So far it has been applied mostly
to the \emph{bosonic} sector, the so-called `oblique' corrections. The idea
is to consider the most general Lagrangian which describes the interactions
between the gauge sector and the Goldstone bosons appearing after the $%
SU(2)_{L}\times SU(2)_{R}\rightarrow SU(2)_{V}$ breaking takes place. Since
no special mechanism is assumed for this breaking the procedure is
completely general, assuming of course that particles not explicitly
included in the effective Lagrangian are much heavier than those appearing
in it. The dependence on the specific model must be contained in the
coefficients of higher dimensional operators.

With the idea of extending this successful approach, in Chapter \ref
{mattersectorchapter} we parametrize in a model-independent way possible
departures from the minimal Standard Model predictions in the \emph{matter}
sector. As we have said that is done assuming only the symmetry breaking
pattern of the Standard Model and that new particles are sufficiently heavy
so that the symmetry is non-linearly realized. We also review in the
effective theory language to what extent the simplest models of dynamical
breaking are actually constrained and the assumptions going into the
comparison with experiment. Since dynamical symmetry breaking models can be
approximated at intermediate energies $\Lambda _{\chi }<E<M$ by four-fermion
operators we present a complete classification of the latter when new
particles appear in the usual representations of the $SU(2)_{L}\times
SU(3)_{c}$ group as well as a partial classification in the general case.
Then we discuss the accuracy of the four-fermion description by matching to
a simple `fundamental' theory. The coefficients of the effective Lagrangian
in the matter sector for dynamical symmetry breaking models (expressed in
terms of the coefficients of the four-quark operators) are then compared to
those of models with elementary scalars (such as the minimal Standard
Model). Contrary to a somewhat widespread belief, we see that the sign of
the vertex corrections is not fixed in dynamical symmetry breaking models.
Summing up, without dealing with $CP$ violating or mixing phenomenology, the
work of this chapter provides the theoretical tools required to analyze in a
rather general setting constraints on the matter sector of the Standard
Model.

Up to this point nothing definite has been said about $CP$ violation or
mixing. However as is the case in the SM these phenomena are probably
related to the symmetry breaking sector.\bigskip

$CP$ violation and family mixing are among the most intriguing puzzles of
the SM. Understanding the origin of $CP$ violation is in fact one of the
important objectives of ongoing and future experiments. This is fairly
justified since such understanding may not only reveal unexpected features
of physics beyond the SM but also add clues to the comprehension of more
complex phenomena such as baryogenesis in cosmology.

In the minimal Standard Model the information about quantities describing
these phenomena is encoded in the Cabibbo-Kobayashi-Maskawa (CKM) mixing
matrix (here denoted $K$). In this context, although the most general mass
matrix does, in principle, contain a large number of phases, only the left
handed diagonalization matrices survive combined in a single CKM mixing
matrix. This matrix contains only one observable complex phase. Whether this
source of $CP$ violation is enough to explain our world is, at present, an
open question.

As it is well known, some of the entries of this matrix are remarkably well
measured, while others (such as the $K_{tb}$, $K_{ts}$ and $K_{td}$
elements) are poorly known and the only real experimental constraint come
from the unitarity requirements. A lot of effort in the last decade has been
invested in this particular problem and this dedication will continue in the
foreseeable future aiming to a precision in the charged current sector
comparable to the one already reached in the neutral sector. As a guidance,
let us mention that the accuracy in $\sin 2\beta $ after LHCb is expected to
be just beyond the 1\% level, and a comparable accuracy might be expected by
that time from the ongoing generation of experiments (BaBar, Belle) \cite
{Amato:1998xt}.

One of the commonly stated purposes of the new generation of experiments is
to check the `unitarity of the CKM matrix'. Stated this way, the purpose
sounds rather meaningless. Of course, if one only retains the three known
generations, mixing occurs through a $3\times 3$ matrix that is, by
construction, necessarily unitary. What is really meant by the above
statement is whether the observable $S$-matrix elements, which at tree level
are proportional to a CKM matrix element, when measured in charged weak
decays, turn out to be in good agreement with the tree-level unitarity
relations predicted by the Standard Model. If we write, for instance,
\begin{equation}
\left\langle q_{j}\left| W_{\mu }^{+}\right| q_{i}\right\rangle
=U_{ij}V_{\mu },  \label{eqa1intro}
\end{equation}
at tree level, it is clear that unitarity of the CKM matrix implies
\begin{equation}
\sum_{k}U_{ik}U_{jk}^{\ast }=\delta _{ij},  \label{eqa2intro}
\end{equation}
However, even if there is no new physics at all beyond the Standard Model
radiative corrections contribute to the matrix elements relevant for weak
decays and spoil the unitarity of the `CKM matrix' $U$, in the sense that
the corresponding $S$-matrix elements are no longer constrained to verify
the above relation. Obviously, departures from unitarity due to the
electroweak radiative corrections are bound to be small. Later we shall see
at what level are violations of unitarity due to radiative corrections to be
expected.

But of course, the violations of unitarity that are really interesting are
those caused by new physics. Physics beyond the Standard Model can manifest
itself in several ways and at several scales. Again as we have done with the
case without mixing or $CP$ violation we shall assume that new physics may
appear at a scale $\Lambda $ which is relatively large compared to the $%
M_{Z} $ scale. This remark again includes the scalar sector too; i.e. we
assume that the Higgs particle ---if it exists at all--- it is sufficiently
heavy. With these assumptions we will try to derive some conclusions about
mixing and $CP$ violation using effective Lagrangian techniques.

Let us illustrate the idea with a simple example: Suppose we consider the
case of a new heavy generation. In that case we can proceed in two ways. One
possibility is to treat all fermions, light or heavy, on the same footing.
We would then end up with a $4\times 4$ unitary mixing matrix, the one
corresponding to the light fermions being a $3\times 3$ submatrix which, of
course need not be ---and in fact, will not be--- unitary. Stated this way
the departures from unitarity (already at tree level!) could conceivably be
sizeable. The alternative way to proceed would be, in the philosophy of
effective Lagrangians, to integrate out completely the heavy generation. One
is then left, at lowest order in the inverse mass expansion, with just the
ordinary kinetic and mass terms for light fermions, leading ---obviously---
to an ordinary $3\times 3$ mixing matrix, which is of course unitary.
Naturally, there is no logical contradiction between the two procedures
because what really matters is the physical $S$-matrix element and this
gets, if we follow the second procedure (integrating out the heavy fields),
two type of contributions: from the lowest dimensional operators involving
only light fields and from the higher dimensional operators obtained after
integrating out the heavy fields. The result for the observable $S$-matrix
element should obviously be the same whatever procedure we follow, but in
using the second method we learn that the violations of unitarity in the
(three generation) unitarity triangle are suppressed by some heavy mass.
This simple consideration illustrates the virtues of the effective
Lagrangian approach.

In Chapter \ref{cpviolationandmixing} we extend the effective Lagrangian
presented in Chapter \ref{mattersectorchapter} in order to consider mixing
and $CP$ violating terms. Such Lagrangian contains the effective operators
that give the leading contribution in theories where the physics beyond the
Standard Model shows at a scale $\Lambda >>M_{W}$. Like in Chapter \ref
{mattersectorchapter} we keep here only the leading non-universal effective
operators, that is dimension four ones. Since we make no assumptions besides
symmetries, we take non-diagonal kinetic and mass terms and we perform the
diagonalization and passage to the physical basis in full generality. Such
diagonalization leaves no traces in the SM besides the CKM matrix, however
we shall see here that a lot more information of the weak basis remains in
the effective operators written in the diagonal basis. Then we shall
determine the contribution to different observables and discuss the possible
new sources of $CP$ violation, the idea being to be able to gain some
knowledge about new physics beyond the Standard Model from general
considerations, without having to compute model by model. In this same
chapter the values of the coefficients of the effective Lagrangian in some
theories, including the Standard Model with a heavy Higgs, are presented and
we try to draw some general conclusions about the general pattern exhibited
by physics beyond the Standard Model in what concerns $CP$ violation.

In the process we have to deal with two theoretical problems which are very
interesting in their own: the renormalization of the CKM matrix elements and
the wave function renormalization (wfr.) in the on-shell scheme when mixing
is present. But why should we care about wfr. or CKM counterterms if here we
work at tree level? The answer is quite simple: even at tree level one of
the effective operators contribute to the fermionic self-energies and
therefore to the wfr. constants. This implies that this ``indirect''
contribution must also be taken into account since in order to calculate
physical observables the wfr. constants are constrained by the LSZ
requirements which in turn are equivalent to the requirements of the
on-shell scheme. Moreover, it can be shown that CKM counterterm is also
related to the wfr. constants (although not to the physical or ``external''
ones) so another potential contribution may arise through this
counterterm.\bigskip

At this point one discovers that some questions remained to be answered
regarding the correct implementation of the on-shell scheme in the presence
of mixing. Some of these questions were raised in \cite{Grassi} where
supposed inconsistencies between the on-shell scheme and gauge invariance
were put forward. Spurred by these results we decided to investigate the
issue of the on-shell scheme in the presence of mixing and its relation to
gauge invariance. Our work with respect to this issue is condensed in
Chapter \ref{LSZchapter} and the results of this chapter are applied in the
much more simple case of the effective theory contribution at leading order.
Here, it is worthwhile to point out that the results obtained in Chapter \ref
{LSZchapter} go far beyond their application in Chapter \ref
{cpviolationandmixing} and are sure to be relevant in forthcoming high
precision experiments to compare with theoretical expectations.

Let us here make a brief introduction to the problem: When calculating a
vertex physical amplitude at 1-loop level we have to consider tree level
contributions plus corrections of several types. That is, we need counter
terms for the electric charge, Weinberg angle and wave-function
renormalization for the $W$ gauge boson. We also require wfr. for the
external fermions and counter terms for the entries of the CKM matrix. The
latter are in fact related in a way that will be described in Chapter \ref
{LSZchapter} \cite{Balzereit:1999id}. Finally one needs to compute the 1PI
diagrams correspoding to the given vertex.

So far everything is clear. However, a long standing controversy exists in
the literature concerning what is the appropriate way to define both an
external wfr. and CKM counter terms. The issue becomes involved because we
are dealing with particles which are unstable (and therefore the
self-energies, that are related to the wfr. constants, develop branch cuts;
even gauge dependent ones) and because of mixing.

Several proposals have been put forward in the literature to define
appropriate counter terms both for the external legs and for the CKM matrix
elements in the on-shell scheme. The original conditions diagonalizing the
fermionic on-shell propagator were introduced in \cite{Aoki}. In \cite
{DennerSack} the wfr. ``satisfying'' the conditions of \cite{Aoki} were
derived. However in \cite{DennerSack} no care was taken about the presence
of branch cuts in the self-energies, a fact that enters into conflict with
conditions in \cite{Aoki}. That was later realized in \cite{Denner}. The
problem can be stated saying that the on-shell conditions defined in \cite
{Aoki} are in fact impossible to satisfy for a minimal set of
renormalization constants\footnote{%
By minimal set we mean a set where the wfr. of $\bar{\Psi}_{0}=\bar{\Psi}%
\bar{Z}^{\frac{1}{2}}$ and $\Psi _{0}=Z^{\frac{1}{2}}\Psi $ are related by $%
\bar{Z}^{\frac{1}{2}}=\gamma ^{0}Z^{\frac{1}{2}\dagger }\gamma ^{0}.$} due
to the absorptive parts present in the self-energies. The author of \cite
{Denner} circumvented this problem by introducing a prescription that
\textit{de facto} eliminates such absorptive parts, but at the price of not
diagonalizing the fermionic propagators in family space.

Ward identities based on the SU(2)$_{L}$ gauge symmetry relate wfr. and
counter terms for the CKM matrix elements \cite{Balzereit:1999id}. In \cite
{Grassi} it was seen that if the prescription of \cite{DennerSack} was used
in the counter terms for the CKM matrix elements, the result of a
calculation of a given vertex observable is gauge dependent. As we have just
mentioned, the results in \cite{DennerSack} do not deal properly with the
absorptive terms appearing in the self-energies; which in addition happen to
be gauge dependent. In Chapter \ref{LSZchapter} we will see that in spite of
the problems with the prescription for the wfr. given in \cite{DennerSack},
the conclusions reached in \cite{Grassi} are correct: a necessary condition
for gauge invariance of the physical amplitudes is that counter terms for
the CKM matrix elements $K_{ij}$ are by themselves gauge independent. This
condition is fulfilled by the CKM counter term proposed in \cite{Grassi} as
it is in minimal subtraction \cite{Balzereit:1999id}, \cite{Diener:2001qt}.

Other proposals to handle CKM renormalization exist in the literature \cite
{Diener:2001qt}, \cite{Barroso} and \cite{Yamada}. In all these works either
the external wfr. proposed originally in \cite{DennerSack} or \cite{Denner}
are used, or the issue of the correct definition of the external wfr. is
sidestepped altogether. In any case the absorptive part of the self-energies
(and even the absorptive part of the 1PI vertex part in one particular
instance \cite{Barroso}) are not taken into account. As we shall see doing
so leads to physical amplitudes --- $S$-matrix elements--- which are gauge
dependent, and this irrespective of the method one uses to renormalize $%
K_{ij}$ provided the redefinition of $K_{ij}$ is gauge independent and
preserves unitarity.

Due to the structure of the imaginary branch cuts it turns out however, that
the gauge dependence present in the amplitude using the prescription of \cite
{Denner} cancels in the modulus squared of the physical $S$-matrix element
in the SM. This cancellation has been checked numerically by the authors in
\cite{Kniehl}. In Chapter \ref{LSZchapter} we shall provide analytical
results showing that this cancellation is exact. However the gauge
dependence remains at the level of the amplitude.

Is this acceptable? We do not think so. Diagrams contributing to the same
physical process outside the SM electroweak sector may interfere with the SM
amplitude and reveal the unwanted gauge dependence. Furthermore, gauge
independent absorptive parts are also discarded by the prescription in \cite
{Denner}. These parts, contrary to the gauge dependent ones, do not drop in
the squared amplitude as we shall show. In addition, one should not forget
that the scheme in \cite{Denner} does not deliver on-shell renormalized
propagators that are diagonal in family space. Chapter \ref{LSZchapter} is
dedicated to substantiate the above claims.

Briefly, in Chapter \ref{LSZchapter} with the aid of an extensive use of the
Nielsen identities \cite{Nielsen,Piguet,Sibold} complemented by explicit
calculations we corroborate that the counter term for the CKM mixing matrix
must be explicitly gauge independent and demonstrate that the commonly used
prescription for the wave function renormalization constants leads to gauge
parameter dependent amplitudes, even if the CKM counter term is gauge
invariant as required. For those not familiar with Nielsen identities we
provide a brief, and hopefully pedagogical, introduction and indicate the
relevant references. Using that technology we show that a proper
LSZ-compliant prescription leads to gauge independent amplitudes. The
resulting wave function renormalization constants necessarily possess
absorptive parts, but we verify that they comply with the expected
requirements concerning $CP$ and $CPT$. The results obtained using this
prescription are different (even at the level of the modulus squared of the
amplitude) from the ones neglecting the absorptive parts in the case of top
decay. We show that the difference is numerically relevant.\bigskip

Once those theoretical aspects are settled we move onto the study of the
phenomenology capable of probing the physics of the charged current sector
which is the one sensible to the electroweak $CP$ violation in the SM. When
particularizing to interactions involving the $W,Z$ bosons, the operators
present in the effective electroweak Lagrangian induce effective vertices
coupling the gauge bosons to the matter fields \cite{burgess}
\begin{equation}
-\frac{e}{4c_{W}s_{W}}\bar{f}\gamma ^{\mu }\left( \kappa _{L}^{NC}L+\kappa
_{R}^{NC}R\right) Z_{\mu }f-\frac{e}{s_{W}}\bar{f}\gamma ^{\mu }\left(
\kappa _{L}^{CC}L+\kappa _{R}^{CC}R\right) \frac{\tau ^{-}}{2}W_{\mu
}^{+}f+h.c.
\end{equation}
Other possible effects are not physically observable, as we shall see in
Chapter \ref{LHCphenomenology}. In practical terms, LHC will set bounds on
these effective $W$ vertices, and therefore on the new physics contributing
to them. Our results are also relevant in a broader phenomenological context
as a way to bound $\kappa _{L}$ and $\kappa _{R}$ (including both new
physics and universal radiative corrections), without any need to appeal to
an underlying effective Lagrangian describing a specific model of symmetry
breaking. Of course one then looses the power of an effective Lagrangian,
namely a well defined set of counting rules and the ability to relate
different processes.

As already remarked, even in the minimal Standard Model, radiative
corrections induce modifications in the vertices. Assuming a smooth
dependence in the external momenta these form factors can be expanded in
powers of momenta. At the lowest order in the derivative expansion, the
effect of radiative corrections can be encoded in the effective vertices $%
\kappa _{L}$ and $\kappa _{R}$. Thus these effective vertices take well
defined, calculable values in the minimal Standard Model, and any deviation
from these values (which, incidentally, have not been fully determined in
the Standard Model yet) would indicate the presence of new physics in the
matter sector. The extent to what LHC can set direct bounds on the effective
vertices, in particular on those involving the third generation, is highly
relevant to constraint physics beyond the Standard Model in a direct way.
The work in Chapter \ref{LHCphenomenology} is devoted to such an analysis in
charged processes involving a top quark at the LHC.

At the LHC energy (14 TeV) the dominant mechanism of top production, with a
cross section of 800 pb \cite{catani}, is gluon-gluon fusion. This mechanism
has nothing to do with the electroweak sector and thus is not the most
adequate for our purposes. Although it is the one producing most of the tops
and thus its consideration becomes necessary in order to study the top
couplings through their decay, which will our main interest in Chapter \ref
{topdecaychapter}, and also as a background to the process we shall be
interested in.
\begin{figure}[!hbp]
\begin{center}
\includegraphics[width=7cm]{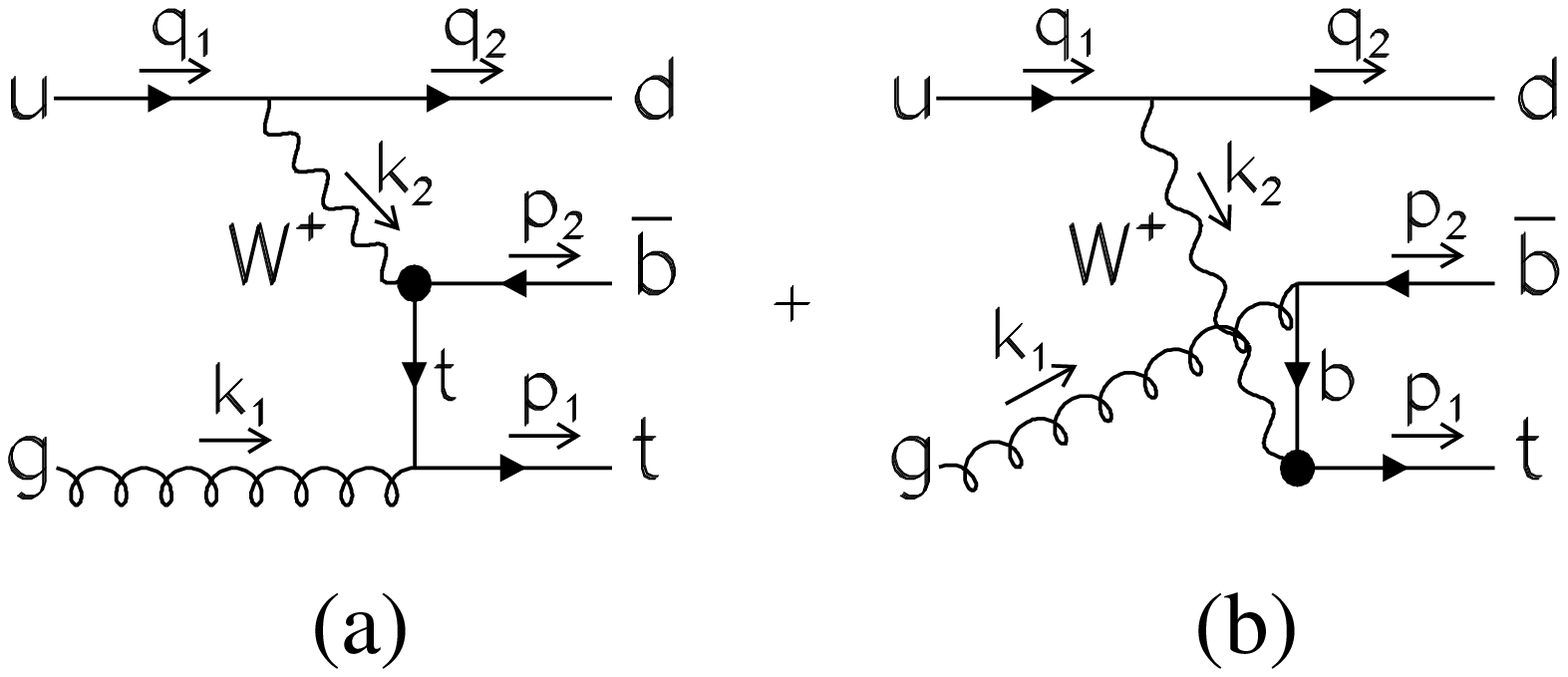}
\end{center}
\caption{Feynman diagrams contributing single top production subprocess. In
this case we have a $d$ as spectator quark}
\label{u+gt+b-d+totintro}
\end{figure}
\begin{figure}[!hbp]
\begin{center}
\includegraphics[width=8cm]{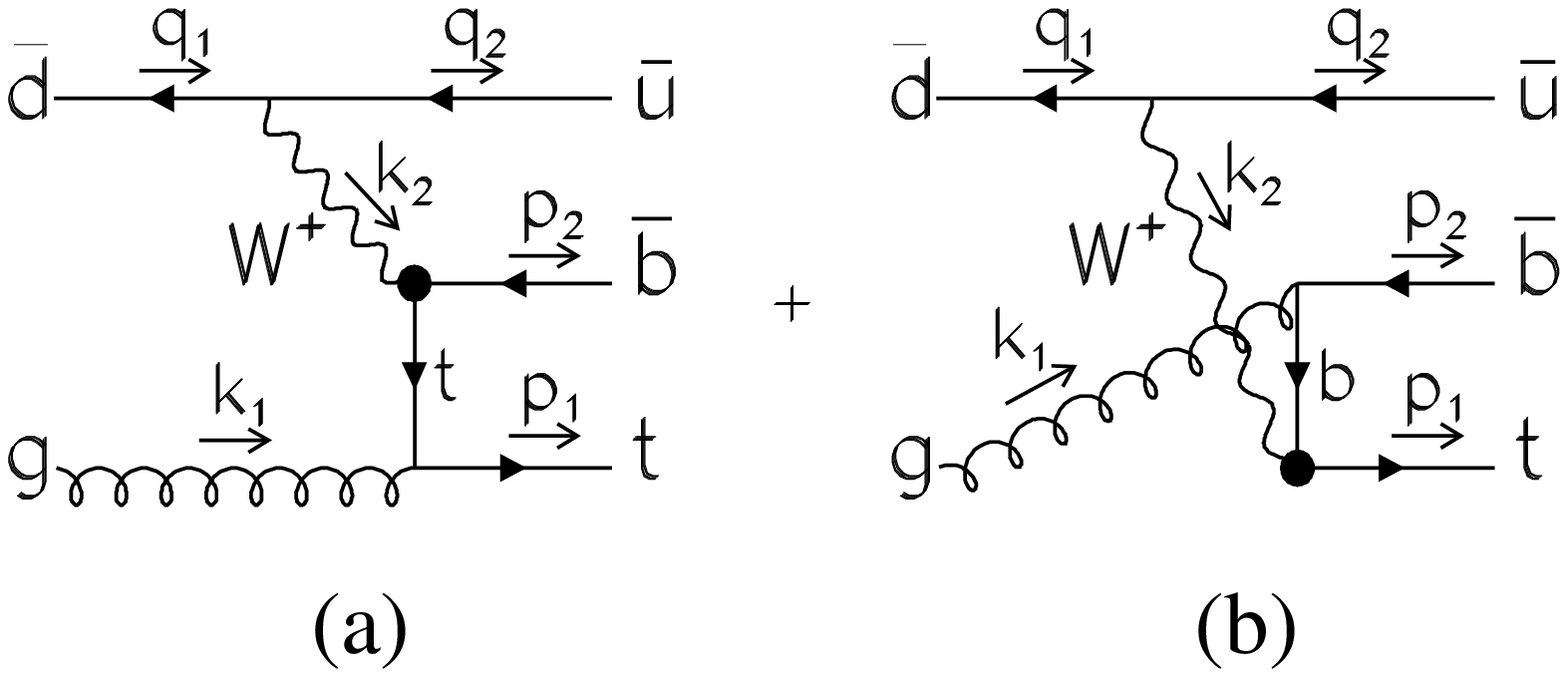}
\end{center}
\caption{Feynman diagrams contributing single top production subprocess. In
this case we have a $\bar{u}$ as spectator quark}
\label{d-gt+b-u-totintro}
\end{figure}
Electroweak physics enters the game in single top production. (for a recent
review see e.g. \cite{Tait}.) At LHC energies the (by far) dominant
electroweak subprocess contributing to single top production is given by a
gluon ($g$) coming from one proton and a light quark or anti-quark coming
from the other (this process is also called \textit{t}-channel production
\cite{tchannel,SSW}). This process is depicted in Figs. \ref
{u+gt+b-d+totintro} and \ref{d-gt+b-u-totintro}, where light $u$-type quarks
or $\bar{d}$-type antiquarks are extracted from the proton, respectively.
These quarks then radiate a $W$ whose effective couplings are the object of
our interest. The cross total section for this process at the LHC is
estimated to be 250 pb \cite{SSW}, to be compared to 50 pb for the
associated production with a $W^{+}$ boson and a $b$-quark extracted from
the sea of the proton, and 10 pb corresponding to quark-quark fusion (%
\textit{s}-channel production to be analyzed in chapter \ref{topdecaychapter}%
). For comparison, at the Tevatron (2 GeV) the cross section for $W$-gluon
fusion is 2.5 pb, so the production of tops through this particular
subprocess is copious at the LHC. Monte Carlo simulations including the
analysis of the top decay products indicate that this process can be
analyzed in detail at the LHC and traditionally has been regarded as the
most important one for our purposes.

In a proton-proton collision a bottom-anti-top pair is also produced,
through analogous subprocesses. At any rate qualitative results are very
similar to those corresponding to top production, from where the cross
sections can be easily derived doing the appropriate changes.

In the context of effective theories, the contribution from operators of
dimension five to top production via longitudinal vector boson fusion was
estimated some time ago in \cite{LY}, although the study was by no means
complete. It should be mentioned that $t,\bar{t}$ pair production through
this mechanism is very much masked by the dominant mechanism of gluon-gluon
fusion, while single top production, through $WZ$ fusion, is expected to be
much suppressed compared to the mechanism presented in this work, the reason
being that both vertices are electroweak in the process discussed in \cite
{LY}, and that operators of dimension five are expected to be suppressed, at
least at moderate energies, by some large mass scale. The contribution from
dimension four operators as such has not, to our knowledge, been considered
before, although the potential for single top production for measuring the
CKM matrix element $K_{tb}$, has to some extent been analyzed in the past
(see e.g. \cite{SSW,mandp}).

To summarize, in Chapter \ref{LHCphenomenology} we analyze the sensitivity
of different observables to the magnitude of the effective couplings that
parametrize new physics beyond the Standard Model. We also show that the
observables relevant to the distinction between left and right effective
couplings involve in practice the measurement of the spin of the top that
only can be achieved indirectly by measuring the angular distribution of its
decay products. We show that the presence of effective right-handed
couplings implies that the top is not in a pure spin state and that a unique
spin basis is singled out which allows one to connect top decay products
angular distribution with the polarized top differential cross section. We
present a complete analytical expression of the differential polarized cross
section of the relevant perturbative subprocess including general effective
couplings. The mass of the bottom quark, which actually turns out to be more
relevant than naively expected, is retained. Finally we analyze different
aspects the total cross section relevant to the measurement of new physics
through the effective couplings. We have also worked out the effective-W
approximation for this process but results are not presented here \cite{effW}%
.\bigskip

\begin{figure}[!hbp]
\begin{center}
\includegraphics[width=6cm]{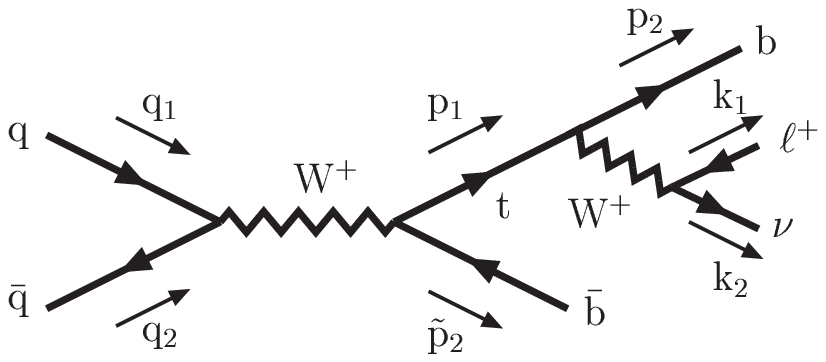}
\end{center}
\par
\caption{Feynman diagram contributing to single top production and decay
process in the s-channel.}
\label{singletopschannelanddecayintro}
\end{figure}

Finally in Chapter \ref{topdecaychapter} we address an aspect of single top
production that was not finished in the previous chapter, namely the
``measurement'' of the top spin via its decay products. Here, the numerical
analysis of the sensitivity of different observables to the right coupling $%
g_{R}$ is performed including the top decay products. Since the main
objective of this chapter is to clarify the role of the top spin when the
top decay is also considered we study single top production through the
theoretically simpler \textit{s}-channel. Single top production and decay in
this channel is depicted in Fig. (\ref{singletopschannelanddecayintro})

In Chapter \ref{topdecaychapter} we show how the differential cross section
corresponding to the process of Fig. (\ref{singletopschannelanddecayintro})
is calculated in a two step process using the narrow-width approximation
with the top spin taken into account. That is, we first calculate the
probability of producing tops with a given polarization and then we
convolute it with the probability of decay summing over both polarizations.
We argue how quantum interference effects can be minimized by the
appropriate choice of spin basis. We present explicit expressions for the
top spin basis that diagonalizes top density matrix both for the \textit{t}-
and \textit{s}- channels. In the case of the \textit{s}-channel we use this
basis in our Monte Carlo integration and we check numerically how sensitive
our results are to a change of spin-basis or even to disregarding top spin
altogether This numerical study shows that the implementation of the correct
spin basis is numerically important at the 4\% level. Besides the top-spin
issue, our simulations clearly show the crucial role of selecting specific
kinematical configurations for the top decay products in order to achieve
maximal sensitivity to $g_{R}$ both in magnitude and phase.\bigskip

In the appendices of this thesis we have included technical material that
complement the contents of the chapters and some calculations that can be a
useful reference for those interested in some technical results. In
particular we have included the complete calculation of all fermionic
self-energies in an arbitrary $R_{\xi }$ gauges.

\chapter{The effective Lagrangian approach in the matter sector}

\label{mattersectorchapter}The Standard Model of electroweak interactions
has by now been impressively tested up to one part in a thousand level
thanks to the formidable experimental work of LEP, SLC and other experiments
in recent years. However, when it comes to the symmetry breaking mechanism
clouds remain in this otherwise bright horizon and the mechanism giving
masses to $W^{\pm }$, $Z,$ and fermions remains largely veiled.

The effective Lagrangian approach has already proven remarkably useful in
setting very stringent bounds on some types of new physics taking as input
basically the LEP \cite{aleph} (and SLC \cite{SLD}) experimental results.
One writes the most general Lagrangian describing the interactions between
the gauge sector and the Goldstone bosons appearing after the $%
SU(2)_{L}\times SU(2)_{R}\rightarrow SU(2)_{V}$ breaking . Since nothing is
assumed for this breaking, the procedure is completely universal. The
dependence on the specific model underlying the symmetry breaking is
contained in the coefficients of higher dimensional operators. These kind of
techniques ---inherited from pion physics--- have been already used to
analyze contributions to the $S$, $T$ and $U$ parameters \cite{PT} and
extract useful constraints on the models of symmetry breaking from them.

Our purpose in this chapter is to extend these techniques to the matter
sector of the Standard Model. We shall write the leading non-universal
operators, determine how their coefficients affect different physical
observables and then determine their value in two very general families of
models: those containing elementary scalars and those with dynamical
symmetry breaking. Since the latter become non-perturbative at the $M_{Z}$
scale, effective Lagrangian techniques are called for anyway. In short, we
would like to provide the theoretical tools required to test ---at least in
principle--- whether the mechanism giving masses to quarks and fermions is
the same as that which makes the intermediate vector bosons massive or not
without having to get involved in the nitty-gritty details of particular
models.

\section{The effective Lagrangian approach}

\label{theefflangappr}Let us start by briefly recalling the salient features
of the effective Lagrangian analysis of the oblique corrections.

Including only those operators which are relevant for oblique corrections,
the effective Lagrangian reads (see e.g. \cite{DEH,eff-lag} for the complete
Lagrangian)
\begin{equation}
\mathcal{L}_{\mathrm{eff}}=\frac{v^{2}}{4}\mathrm{tr}D_{\mu }UD^{\mu
}U^{\dagger }+a_{0}{g^{\prime }}^{2}\frac{v^{2}}{4}(\mathrm{tr}TD_{\mu
}UU^{\dagger })^{2}+a_{1}gg^{\prime }\mathrm{tr}UB_{\mu \nu }U^{\dagger
}W^{\mu \nu }-a_{8}\frac{g^{2}}{4}(\mathrm{tr}TW^{\mu \nu })^{2},
\label{effobl}
\end{equation}
where $U=\exp (i\tau \cdot \chi /v)$ contains the 3 Goldstone bosons
generated after the breaking of the global symmetry $SU(2)_{L}\times
SU(2)_{R}\rightarrow SU(2)_{V}$. The covariant derivative is defined by
\begin{equation}
D_{\mu }U=\partial _{\mu }U+ig\frac{\tau }{2}\cdot W_{\mu }U-ig^{\prime }U%
\frac{\tau ^{3}}{2}B_{\mu },.  \label{30.6a}
\end{equation}
$B_{\mu \nu }$ and $W^{\mu \nu }$ are the field-strength tensors
corresponding to the right and left gauge groups, respectively
\begin{equation}
W_{\mu \nu }=\frac{\vec{\tau}}{2}\cdot \vec{W}_{\mu \nu },\qquad B_{\mu \nu
}=\frac{\tau ^{3}}{2}(\partial _{\mu }B_{\nu }-\partial _{\nu }B_{\mu }),
\end{equation}
and $T=U\tau ^{3}U^{\dagger }$. Only terms up to order $\mathcal{O}(p^{4})$
have been included. The reason is that dimensional counting arguments
suppress, at presently accessible energies, higher dimensional terms, in the
hypothesis that all undetected particles are much heavier than those
included in the effective Lagrangian. While the first term on the r.h.s. of (%
\ref{effobl}) is universal (in the unitary gauge it is just the mass term
for the $W^{\pm }$ and $Z$ bosons), the coefficients $a_{0}$, $a_{1}$ and $%
a_{8}$ are non-universal. In other words, they depend on the specific
mechanism responsible for the symmetry breaking. (Throughout this chapter
the term `universal' means `independent of the specific mechanism triggering
$SU(2)_{L}\times SU(2)_{R}\rightarrow SU(2)_{V}$ breaking'.)

Most $Z$-physics observables relevant for electroweak physics can be
parametrized in terms of vector and axial couplings $g_{V}$ and $g_{A}$ (see
section \ref{Zdecayobs}). These are, in practice, flavor-dependent since
they include vertex corrections which depend on the specific final state.
Oblique corrections are however the same for all final states. The
non-universal (but generation-independent) contributions to $g_{V}$ and $%
g_{A}$ coming from the effective Lagrangian (\ref{effobl}) are
\begin{eqnarray}
\bar{g}_{V} &=&a_{0}g^{\prime \,2}\left[ \frac{\tau ^{3}}{2}+2Q_{f}\left(
2c_{W}^{2}-s_{W}^{2}\right) \right]
+2a_{1}Q_{f}g^{2}s_{W}^{2}+2a_{8}Q_{f}g^{2}c_{W}^{2},  \label{gagvobl} \\
\bar{g}_{A} &=&a_{0}\frac{\tau ^{3}}{2}g^{\prime \,2}.  \label{gagvvv}
\end{eqnarray}
They do depend on the specific underlying breaking mechanism through the
values of the $a_{i}$. It should be noted that these coefficients depend
logarithmically on some unknown scale. In the minimal Standard Model the
characteristic scale is the Higgs boson mass, $M_{H}$. In other theories the
scale $M_{H}$ will be replaced by some other scale $\Lambda $. A crucial
prediction of chiral perturbation theory is that the dependence on these
different scales is logarithmic and actually the same. It is thus possible
to eliminate this dependence by building suitable combinations of $g_{V}$
and $g_{A}$ \cite{EH,EM} determined by the condition of absence of logs.
Whether this line intersects or not the experimentally allowed region is a
direct test of the nature of the symmetry breaking sector, independently of
the precise value of Higgs mass (in the minimal Standard Model) or of the
scale of new interactions (in other scenarios)\footnote{%
Notice that, contrary to a somewhat widespread belief, the limit $%
M_{H}\rightarrow \infty $ does not correspond a Standard Model `without the
Higgs'. There are some non-trivial non-decoupling effects}.

One could also try to extract information about the individual coefficients $%
a_{0}$, $a_{1}$ and $a_{8}$ themselves, and not only on the combinations
cancelling the dependence on the unknown scale. This necessarily implies
assuming a specific value for the scale $\Lambda $ and one should be aware
that when considering these scale dependent quantities there are finite
uncertainties of order $1/16\pi ^{2}$ associated to the subtraction
procedure ---an unavoidable consequence of using an effective theory, that
is often overlooked. (And recall that using an effective theory is almost
mandatory in dynamical symmetry breaking models.) Only finite combinations
of coefficients have a universal meaning. The subtraction scale uncertainty
persists when trying to find estimates of the above coefficients via
dispersion relations and the like \cite{PT}.

In the previous analysis it is assumed that the hypothetical new physics
contributions from vertex corrections are completely negligible. But is it
so? The way to analyze such vertex corrections in a model-independent way is
quite similar to the one outlined for the oblique corrections. We shall
introduce in the next section the most general effective Lagrangian
describing the matter sector. In this sector there is one universal operator
(playing a role analogous to that of the first operator on the r.h.s. of (%
\ref{effobl}) in the purely bosonic sector)
\begin{equation}
\mathcal{L}_{\mathrm{eff}}=-v\mathrm{\bar{f}}Uy_{f}R\mathrm{f}+h.c.,\qquad
y_{f}=y\mathbf{1}+y_{3}\tau _{3}.  \label{mass}
\end{equation}
It is an operator of dimension 3. In the unitary gauge $U=1$, it is just the
mass term for the matter fields. For instance if $\bar{q}_{L}$ is the
doublet $(\bar{t},\bar{b})$
\begin{equation}
m_{t}=v(y+y_{3})=vy_{t},\qquad m_{b}=v(y-y_{3})=vy_{b}.
\end{equation}
Non-universal operators carrying in their coefficients the information on
the mechanism giving masses to leptons and quarks will be of dimension 4 and
higher.

We shall later derive the values of the coefficients corresponding to
operators in the effective Lagrangian of dimension 4 within the minimal
Standard Model in the large $M_{H}$ limit and see how the effective
Lagrangian provides a convenient way of tracing the Higgs mass dependence in
physical observables. We shall later argue that non-decoupling effects
should be the same in other theories involving elementary scalars, such as
e.g. the two-Higgs doublet model, replacing $M_{H}$ by the appropriate mass.

Large non-decoupling effects appear in theories of dynamical symmetry
breaking and thus they are likely to produce large contributions to the
dimension 4 coefficients. If the scale characteristic of the extended
interactions (i.e. those responsible of the fermion mass generation) is much
larger than the scale characteristic of the electroweak breaking, it makes
sense to parametrize the former, at least at low energies, via effective
four-fermion operators\footnote{%
While using an effective theory description based on four-fermion operators
alone frees us from having to appeal to any particular model it is obvious
that some information is lost. This issue turns out to be a rather subtle
one and shall be discussed and quantified in turn.}. We shall assume here
that this clear separation of scales does take place and only in this case
are the present techniques really accurate. The appearance of pseudo
Goldstone bosons (abundant in models of dynamical breaking) may thus
jeopardize our conclusions, as they bring a relatively light scale into the
game (typically even lighter than the Fermi scale). In fact, for the
observables we consider their contribution is not too important, unless they
are extremely light. For instance a pseudo-Goldstone boson of 100 GeV can be
accommodated without much trouble, as we shall later see.

The four-fermion operators we have just alluded to can involve either four
ordinary quarks or leptons (but we will see that dimensional counting
suggests that their contribution will be irrelevant at present energies with
the exception of those containing the top quark), or two new (heavy)
fermions and two ordinary ones. This scenario is quite natural in several
extended technicolor (ETC) or top condensate (TopC) models \cite{ETC,TopC},
in which the underlying dynamics is characterized by a scale $M$. At scales $%
\mu <M$ the dynamics can be modelled by four-fermion operators (of either
technifermions in ETC models, or ordinary fermions of the third family in
TopC models). We perform a classification\footnote{%
In the case of ordinary fermions and leptons, four-fermion operators have
been studied in \cite{4FC}. To our knowledge a complete analysis when
additional fields beyond those present in the Standard Model are present has
not been presented in the literature before.} of these operators. We shall
concentrate in the case where technifermions appear in ordinary
representations of $SU(2)_{L}\times SU(3)_{c}$ (hypercharge can be
arbitrary). The classification will then be exhaustive. We shall discuss
other representations as well, although we shall consider custodially
preserving operators only, and only those operators which are relevant for
our purposes.

As a matter of principle we have tried not to make any assumptions regarding
the actual way different generations are embedded in the extended
interactions. In practice, when presenting our numerical plots and figures,
we are assuming that the appropriate group-theoretical factors are similar
for all three generations of physical fermions.

It has been our purpose in this chapter to be as general as possible, not
advocating or trying to put forward any particular theory. Thus, the
analysis may, hopefully, remain useful beyond the models we have just used
to motivate the problem. We hope to convey to the reader our belief that a
systematic approach based on four-fermion operators and the effective
Lagrangian treatment can be very useful.

\section{The matter sector}

\label{S-17.6a}Appelquist, Bowick, Cohler and Hauser established some time
ago a list of $d=4$ operators \cite{ABCH}. These are the operators of lowest
dimensionality which are non-universal. In other words, their coefficients
will contain information on whatever mechanism Nature has chosen to make
quarks and leptons massive. Of course operators of dimensionality 5, 6 and
so on will be generated at the same time. We shall turn to these later. We
have reanalyzed all possible independent operators of $d=4$ (see the
discussion in appendix \ref{Msectorapp}.\ref{MsectorappA}) and we find the
following ones
\begin{eqnarray}
\mathcal{L}_{L}^{1} &=&i\mathrm{\bar{f}}M_{L}^{1}\gamma ^{\mu }U\left(
D_{\mu }U\right) ^{\dagger }L\mathrm{f}+h.c.,  \label{2.6b} \\
\mathcal{L}_{L}^{2} &=&i\mathrm{\bar{f}}M_{L}^{2}\gamma ^{\mu }\left( D_{\mu
}U\right) \tau ^{3}U^{\dagger }L\mathrm{f}+h.c., \\
\mathcal{L}_{L}^{3} &=&i\mathrm{\bar{f}}M_{L}^{3}\gamma ^{\mu }U\tau
^{3}U^{\dagger }\left( D_{\mu }U\right) \tau ^{3}U^{\dagger }L\mathrm{f}%
+h.c., \\
\mathcal{L}_{L}^{4} &=&i\mathrm{\bar{f}}M_{L}^{4}\gamma ^{\mu }U\tau
^{3}U^{\dagger }D_{\mu }^{L}L\mathrm{f}+h.c.,  \label{2.6c} \\
\mathcal{L}_{R}^{1} &=&i\mathrm{\bar{f}}M_{R}^{1}\gamma ^{\mu }U^{\dagger
}\left( D_{\mu }U\right) R\mathrm{f}+h.c., \\
\mathcal{L}_{R}^{2} &=&i\mathrm{\bar{f}}M_{R}^{2}\gamma ^{\mu }\tau
^{3}U^{\dagger }\left( D_{\mu }U\right) R\mathrm{f}+h.c., \\
\mathcal{L}_{R}^{3} &=&i\mathrm{\bar{f}}M_{R}^{3}\gamma ^{\mu }\tau
^{3}U^{\dagger }\left( D_{\mu }U\right) \tau ^{3}R\mathrm{f}+h.c., \\
\mathcal{L}_{R}^{\prime } &=&i\mathrm{\bar{f}}M_{R}^{\prime }\tau ^{3}\gamma
^{\mu }D_{\mu }^{L}R\mathrm{f}+h.c..  \label{2.6d}
\end{eqnarray}
Each operator is accompanied by a coefficient $M_{L,R}^{i}$. In this chapter
we will not consider mixing and therefore these coefficient are pure
numbers. In Chapter \ref{cpviolationandmixing} mixing is considered and
therefore we will allow the $M_{L,R}^{i}$ to have family indices. Thus, up
to $\mathcal{O}(p^{4})$, our effective Lagrangian is\footnote{%
Although there is only one derivative in (\ref{30.6b}) and thus this is a
misname, we stick to the same notation here as in the purely bosonic
effective lagrangian}

\begin{equation}
\mathcal{L}_{\mathrm{eff}}=\mathcal{L}_{R}^{\prime }+\sum_{i=1}^{4}\mathcal{L%
}_{L}^{i}+\sum_{i=1}^{3}\mathcal{L}_{R}^{i}.  \label{30.6b}
\end{equation}
In the above, $D_{\mu }U$ is defined in~(\ref{30.6a}) whereas
\begin{eqnarray*}
D_{\mu }^{L}f_{L} &=&\left[ \partial _{\mu }+ig\frac{\tau }{2}\cdot W_{\mu
}+ig^{\prime }\left( Q-\frac{\tau ^{3}}{2}\right) B_{\mu }+ig_{s}\frac{%
\lambda }{2}\cdot G_{\mu }\right] f_{L}, \\
D_{\mu }^{R}f_{R} &=&\left[ \partial _{\mu }+ig^{\prime }QB_{\mu }+ig_{s}%
\frac{\lambda }{2}\cdot G_{\mu }\right] f_{R},
\end{eqnarray*}
where $Q$ is the electric charge given by
\begin{equation*}
Q=\frac{\tau ^{3}}{2}+z,
\end{equation*}
with $z=1/6$ for quarks and $z=-1/2$ for leptons and therefore with the
hypercharge given by
\begin{equation*}
Y=\left\{
\begin{tabular}{ll}
$z$ & $\mathrm{for\ lefts.}$ \\
$\frac{\tau ^{3}}{2}+z$ & $\mathrm{for\ rights.}$%
\end{tabular}
\right.
\end{equation*}
This list differs from the one in \cite{ABCH} by the presence of the last
operator (\ref{2.6d}). It will turn out, however, that $M_{R}^{\prime }$
does not contribute to any observable. All these operators are invariant
under local $SU(2)_{L}\times U(1)_{Y}$ transformations. 

This list includes both the custodially preserving operators $\mathcal{L}%
_{L}^{1}$ and $\mathcal{L}_{R}^{1}$ and the rest of operators that are
custodially breaking ones. In the purely bosonic part of the effective
Lagrangian (\ref{effobl}), the first (universal) operator and the one
accompanying $a_{1}$ are custodially preserving, while those going with $%
a_{0}$ and $a_{8}$ are custodially breaking. E.g., $a_{0}$ parametrizes the
contribution of the new physics to the $\Delta \rho $ parameter. If the
underlying physics is custodially preserving only $M_{L,R}^{1}$ will get
non-vanishing contributions\footnote{%
Of course hypercharge $Y$ breaks custodial symmetry, since only a subgroup
of $SU(2)_{R}$ is gauged. Therefore, \textit{all} operators involving
right-handed fields break custodial symmetry. However, there is still a
distinction between those operators whose structure is formally custodially
invariant (and custodial symmetry is broken only through the coupling to the
external gauge field) and those which would not be custodially preserving
even if the full $SU(2)_{R}$ were gauged.}.

The operator $\mathcal{L}_{L}^{4}$ deserves some comments. By using the
equations of motion it can be reduced to the mass term (\ref{mass})
\begin{equation*}
vM_{L}^{4}\mathrm{\bar{f}}U\tau ^{3}y_{f}R\mathrm{f}+h.c.,
\end{equation*}
However this procedure is, generally speaking, only justified if the matter
fields appear only as external legs. For the time being we shall keep $%
\mathcal{L}_{L}^{4}$ as an independent operator and in the next section we
shall determine its value in the minimal Standard Model after integrating
out a heavy Higgs. We shall see that, after imposing that physical on-shell
fields have unit residue, $M_{L}^{4}$ does drop from all physical
predictions.

What is the expected size of the $M_{L,R}^{i}$ coefficients in the minimal
Standard Model? This question is easily answered if we take a look at the
diagrams that have to be computed to integrate out the Higgs field (Fig. (%
\ref{Figmsector-3})). Notice that the calculation is carried out in the
non-linear variables $U$, hence the appearance of the unfamiliar diagram e).
Diagram d) is actually of order $1/M_{H}^{2}$, which guarantees the gauge
independence of the effective Lagrangian coefficients. The diagrams are
obviously proportional to $y^{2}$, $y$ being a Yukawa coupling, and also to $%
1/16\pi ^{2}$, since they originate from a one-loop calculation. Finally,
the screening theorem shows that they may depend on the Higgs mass only
logarithmically, therefore
\begin{equation}
M_{L,R}^{i\left( \mathrm{SM}\right) }\sim {\frac{y^{2}}{{16\pi ^{2}}}}\log {%
\frac{M_{H}^{2}}{M_{Z}^{2}}}.
\end{equation}
These dimensional considerations show that the vertex corrections are only
sizeable for third generation quarks.

\begin{figure}[!hbp]
\begin{center}
\includegraphics[width= 6cm]{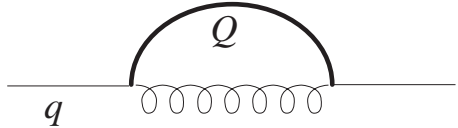}
\end{center}
\caption{Mechanism generating quark masses through the exchange of a ETC
particle.}
\label{Figmsector-4}
\end{figure}

In models of dynamical symmetry breaking, such as TC or ETC, we shall have
new contributions to the $M_{L,R}^{i}$ from the new physics (which we shall
later parametrize with four-fermion operators). We have several new scales
at our disposal. One is $M$, the mass normalizing dimension six four-fermion
operators. The other can be either $m_{b}$ (negligible, since $M$ is large),
$m_{t}$, or the dynamically generated mass of the techniquarks $m_{Q}$
(typically of order $\Lambda _{TC}$, the scale associated to the
interactions triggering the breaking of the electroweak group). Thus we can
get a contribution of order
\begin{equation}
M_{L,R}^{i\left( \mathrm{Q}\right) }\sim \frac{1}{16\pi ^{2}}\frac{m_{Q}^{2}%
}{M^{2}}\log \frac{m_{Q}^{2}}{M^{2}}.
\end{equation}
While $m_{Q}$ is, at least naively, expected to be $\simeq \Lambda _{TC}$
and therefore similar for all flavors, there should be a hierarchy for $M$.
As will be discussed in the following sections, the scale $M$ which is
relevant for the mass generation (encoded in the only dimension 3 operator
in the effective Lagrangian), via techniquark condensation and ETC
interaction exchange (Fig. (\ref{Figmsector-4})), is the one normalizing
chirality flipping operators. On the contrary, the scale normalizing
dimension 4 operators in the effective theory is the one that normalizes
chirality preserving operators. Both scales need not be exactly the same,
and one may envisage a situation with relatively light scalars present where
the former can be much lower. However, it is natural to expect that $M$
should at any rate be smallest for the third generation. Consequently the
contribution to the $M_{L,R}^{i}$'s from the third generation should be
largest.

We should also discuss dimension 5, 6, etc. operators and why we need not
include them in our analysis. Let us write some operators of dimension 5:
\begin{eqnarray*}
&&\mathrm{\bar{f}}\hat{W}UR\mathrm{f}+h.c., \\
&&\mathrm{\bar{f}}U\hat{B}R\mathrm{f}+h.c., \\
&&\mathrm{\bar{f}}\sigma ^{\mu \nu }\left( D_{\mu }\left( D_{\nu }U\right)
\right) ^{\dagger }R\mathrm{f}+h.c., \\
&&\mathrm{\bar{f}}\sigma ^{\mu \nu }\left( D_{\mu }U\right) ^{\dagger
}D_{\nu }R\mathrm{f}+h.c., \\
&&\mathrm{\bar{f}}UD^{2}R\mathrm{f}+h.c.,
\end{eqnarray*}
where we use the notation $\hat{W}\equiv ig\sigma ^{\mu \nu }W_{\mu \nu }$, $%
\hat{B}\equiv ig^{\prime }\sigma ^{\mu \nu }B_{\mu \nu }$. These are a few
of a long list of about 25 operators, and this including only the ones
contributing to the $ffZ$ vertex. All these operators are however chirality
flipping and thus their contribution to the amplitude must be suppressed by
one additional power of the fermion masses. This makes their study
unnecessary at the present level of precision. Similar considerations apply
to operators of dimensionality 6 or higher.

\section{The effective theory of the Standard Model}

\label{31.7-1}In this section we shall obtain the values of the coefficients
$M_{L,R}^{i}$ in the minimal Standard Model. The appropriate effective
coefficients for the oblique corrections $a_{i}$ have been obtained
previously by several authors \cite{EH,EM,all}. Their values are
\begin{eqnarray}
a_{0} &=&\frac{1}{16\pi ^{2}}\frac{3}{8}\left( \frac{1}{\hat{\epsilon}}-\log
\frac{M_{H}^{2}}{\mu ^{2}}+\frac{5}{6}\right) , \\
a_{1} &=&\frac{1}{16\pi ^{2}}\frac{1}{12}\left( \frac{1}{\hat{\epsilon}}%
-\log \frac{M_{H}^{2}}{\mu ^{2}}+\frac{5}{6}\right) , \\
a_{8} &=&0.
\end{eqnarray}
where $1/\hat{\epsilon}\equiv 1/\epsilon -\gamma _{E}+\log 4\pi $. We use
dimensional regularization with a space-time dimension $4-2\epsilon $.

We begin by writing the Standard Model in terms of the non-linear variables $%
U$. The matrix
\begin{equation}
\mathcal{M}=\sqrt{2}(\tilde{\Phi},\Phi ),
\end{equation}
constructed with the Higgs doublet, $\Phi $ and its conjugate, $\tilde{\Phi}%
\equiv i\tau ^{2}\Phi ^{\ast }$, is rewritten in the form
\begin{equation}
\mathcal{M}=(v+\rho )U,\qquad U^{-1}=U^{\dagger },
\end{equation}
where $\rho $ describe the `radial' excitations around the v.e.v. $v$.
Integrating out the field $\rho $ produces an effective Lagrangian of the
form (\ref{effobl}) with the values of the $a_{i}$ given above (as well as
some other pieces not shown there). This functional integration also
generates the vertex corrections (\ref{30.6b}).

We shall determine the $M_{L,R}^{i}$ by demanding that the renormalized
one-particle irreducible Green functions (1PI), $\hat{\Gamma}$, are the same
(up to some power in the external momenta and mass expansion) in both, the
minimal Standard Model and the effective Lagrangian. In other words, we
require that
\begin{equation}
\Delta \hat{\Gamma}=0,  \label{match-2}
\end{equation}
where throughout this section
\begin{equation}
\Delta \Gamma \equiv \Gamma _{\mathrm{SM}}-\Gamma _{\mathrm{eff}},
\end{equation}
and the hat denotes renormalized quantities. This procedure is known as
matching. It goes without saying that in doing so the same renormalization
scheme must be used. The on-shell scheme is particularly well suited to
perform the matching and will be used throughout this work.

One only needs to worry about SM diagrams that are not present in the
effective theory; namely, those containing the Higgs. The rest of the
diagrams give exactly the same result, thus dropping from the matching. In
contrast, the diagrams containing a Higgs propagator are described by local
terms (such as $\mathcal{L}_{L}^{1}$ through $\mathcal{L}_{L}^{4}$) in the
effective theory, they involve the coefficients $M_{L,R}^{i}$, and give rise
to the Feynman rules collected in appendix \ref{Msectorapp}.\ref{MsectorappB}%
.

\begin{figure}[!hbp]
\begin{center}
\includegraphics[width=\figwidth]{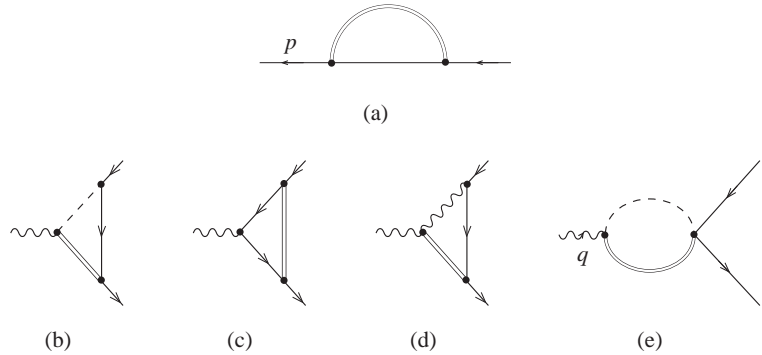}
\end{center}
\caption{The diagrams relevant for the matching of the fermion self-energies
and vertices (counterterm diagrams are not included). Double lines represent
the Higgs, dashed lines the Goldstone bosons, and wiggly lines the gauge
bosons.}
\label{Figmsector-3}
\end{figure}

Let us first consider the fermion self-energies. There is only one 1PI
diagram with a Higgs propagator (see Fig. (\ref{Figmsector-3})).

A straightforward calculation gives
\begin{equation}
\Sigma _{\mathrm{SM}}^{f}=-{\frac{y_{f}^{2}}{16\pi ^{2}}}\left\{ \not{p}%
\left[ {\frac{1}{2}}\frac{1}{\hat{\epsilon}}-{\frac{1}{2}}\log {\frac{%
M_{H}^{2}}{\mu ^{2}}}+{\frac{1}{4}}\right] +m_{f}\left[ \frac{1}{\hat{%
\epsilon}}-\log {\frac{M_{H}^{2}}{\mu ^{2}}}+1\right] \right\} .  \label{SM5}
\end{equation}
$\Delta \Sigma ^{f}$ can be computed by subtracting Eqs. (\ref{new-1}) and (%
\ref{new-2}) from Eq. (\ref{SM5}).

Next, we have to renormalize the fermion self-energies. We introduce the
following notation
\begin{equation}
\Delta Z\equiv Z_{\mathrm{SM}}-Z_{\mathrm{eff}}=\delta Z_{\mathrm{SM}%
}-\delta Z_{\mathrm{eff}},
\end{equation}
where $Z_{\mathrm{SM}}$ ($Z_{\mathrm{eff}}$) stands for any renormalization
constant of the SM (effective theory). To compute $\Delta \hat{\Sigma}^{f}$,
we simply add to $\Delta \Sigma ^{f}$ the counterterm diagram (\ref{17.6a})
with the replacements $\delta Z_{V,A}^{f}\rightarrow \Delta Z_{V,A}^{f}$ and
$\delta m_{f}\rightarrow \Delta m_{f}$. This, of course, amounts to Eqs. (%
\ref{17.6d}), (\ref{17.6e}) and (\ref{17.6f}) with the same replacements.
From Eqs. (\ref{9.6a}), (\ref{17.6b}) and (\ref{17.6c}) (which also hold for
$\Delta Z$, $\Delta m$ and $\Delta \Sigma $) one can express $\Delta
Z_{V,A}^{f}$ and $\Delta m_{f}/m_{f}$ in terms of the bare fermion
self-energies and finally obtain $\Delta \hat{\Sigma}^{f}$. The result is
\begin{eqnarray}
\Delta \hat{\Sigma}_{A,V,S}^{d} &=&0, \\
\Delta \hat{\Sigma}_{A}^{u} &=&0, \\
\Delta \hat{\Sigma}_{V,S}^{u} &=&4M_{L}^{4}-{\frac{1}{16\pi ^{2}}}{\frac{%
y_{u}^{2}-y_{d}^{2}}{2}}\left[ \frac{1}{\hat{\epsilon}}-\log {\frac{M_{H}^{2}%
}{\mu ^{2}}}+{\frac{1}{2}}\right] .  \label{17.6g}
\end{eqnarray}
We see from Eq. (\ref{17.6g}) that the matching conditions, $\Delta \hat{%
\Sigma}_{V,S}^{u}=0$, imply
\begin{equation}
M_{L}^{4}={\frac{1}{16\pi ^{2}}}{\frac{y_{u}^{2}-y_{d}^{2}}{8}}\left[ \frac{1%
}{\hat{\epsilon}}-\log {\frac{M_{H}^{2}}{\mu ^{2}}}+{\frac{1}{2}}\right] .
\label{SM-de7}
\end{equation}
The other matchings are satisfied automatically and do not give any
information.

Let us consider the vertex $ffZ$. The relevant diagrams are shown in Fig. (%
\ref{Figmsector-3}) (diagrams b--e). We shall only collect the contributions
proportional to $\gamma _{\mu }$ and $\gamma _{\mu }\gamma _{5}$.
Remembering
\begin{eqnarray}
s_{W} &\equiv &\sin \theta _{W}\equiv \frac{g^{\prime }}{\sqrt{%
g^{2}+g^{\prime \,2}}},\qquad c_{W}\equiv \cos \theta _{W}\equiv \frac{g}{%
\sqrt{g^{2}+g^{\prime \,2}}},  \notag \\
e &\equiv &gs_{W}=g^{\prime }c_{W},\qquad W_{\mu }^{3}=s_{W}A_{\mu
}+c_{W}Z_{\mu },\qquad B_{\mu }=c_{W}A_{\mu }-s_{W}Z_{\mu },  \label{phybase}
\end{eqnarray}
and Eq. (\ref{vfaf}) the result is
\begin{equation}
\Gamma _{\mu }^{ffZ}={\frac{-1}{16\pi ^{2}}}{\frac{y_{f}^{2}}{2}}\gamma
_{\mu }\left\{ v_{f}\left( \frac{1}{\hat{\epsilon}}-\log {\frac{M_{H}^{2}}{%
\mu ^{2}}}+{\frac{1}{2}}\right) -3a_{f}\,\gamma _{5}\left( \frac{1}{\hat{%
\epsilon}}-\log {\frac{M_{H}^{2}}{\mu ^{2}}}+{\frac{11}{6}}\right) \right\} .
\end{equation}
By subtracting the diagrams (\ref{newN1d1}) and (\ref{newN1d2}) from $\Gamma
_{\mu }^{ffZ}$ one gets $\Delta \Gamma _{\mu }^{ffZ}$. Renormalization
requires that we add the counterterm diagram (\ref{17.6h}) where, again, $%
\delta Z\rightarrow \Delta Z$. One can check that both $\Delta
Z_{1}^{Z}-\Delta Z_{2}^{Z}$ and $\Delta Z_{1}^{Z\gamma }-\Delta
Z_{2}^{Z\gamma }$ are proportional to $\Delta \Sigma _{Z\gamma }(0)$, which
turns out to be zero. Hence the only relevant renormalization constants are $%
\Delta Z_{V}^{f}$ and $\Delta Z_{A}^{f}$. These renormalization constant
have already been determined. One obtains for $\Delta \hat{\Gamma}_{\mu
}^{ffZ}$ the result
\begin{eqnarray*}
\Delta \hat{\Gamma}_{\mu }^{ddZ} &=&{\frac{-e}{2s_{W}c_{W}}}\gamma _{\mu
}\left\{ \left[ M_{L}^{1}-M_{L}^{3}-M_{R}^{1}-M_{R}^{3}+M_{L}^{2}+M_{R}^{2}%
\right] \right. \\
&-&\left. \gamma _{5}\left[ {\frac{1}{16\pi ^{2}}}{\frac{y_{d}^{2}}{2}}%
\left( \frac{1}{\hat{\epsilon}}-\log {\frac{M_{H}^{2}}{\mu ^{2}}}+{\frac{5}{2%
}}\right) +M_{L}^{1}-M_{L}^{3}+M_{R}^{1}+M_{R}^{3}+M_{L}^{2}-M_{R}^{2}\right]
\right\} \\
\Delta \hat{\Gamma}_{\mu }^{uuZ} &=&{\frac{e}{2s_{W}c_{W}}}\gamma _{\mu
}\left\{ \left[ M_{L}^{1}-M_{L}^{3}-M_{R}^{1}-M_{R}^{3}-M_{L}^{2}-M_{R}^{2}%
\right] \right. \\
&-&\left. \gamma _{5}\left[ {\frac{1}{16\pi ^{2}}}{\frac{y_{u}^{2}}{2}}%
\left( \frac{1}{\hat{\epsilon}}-\log {\frac{M_{H}^{2}}{\mu ^{2}}}+{\frac{5}{2%
}}\right) +M_{L}^{1}-M_{L}^{3}+M_{R}^{1}+M_{R}^{3}-M_{L}^{2}+M_{R}^{2}\right]
\right\} ,
\end{eqnarray*}
where use has been made of Eq. (\ref{SM-de7}). The matching condition, $%
\Delta \hat{\Gamma}_{\mu }^{ffZ}=0$ implies
\begin{eqnarray}
M_{L}^{1}-M_{L}^{3} &=&-{\frac{1}{16\pi ^{2}}}{\frac{y_{u}^{2}+y_{d}^{2}}{8}}%
\left( \frac{1}{\hat{\epsilon}}-\log {\frac{M_{H}^{2}}{\mu ^{2}}}+{\frac{5}{2%
}}\right) ,  \label{SMmt1} \\
M_{R}^{1}+M_{R}^{3} &=&-{\frac{1}{16\pi ^{2}}}{\frac{y_{u}^{2}+y_{d}^{2}}{8}}%
\left( \frac{1}{\hat{\epsilon}}-\log {\frac{M_{H}^{2}}{\mu ^{2}}}+{\frac{5}{2%
}}\right) ,  \label{SMmt2} \\
M_{L}^{2} &=&{\frac{1}{16\pi ^{2}}}{\frac{y_{u}^{2}-y_{d}^{2}}{8}}\left(
\frac{1}{\hat{\epsilon}}-\log {\frac{M_{H}^{2}}{\mu ^{2}}}+{\frac{5}{2}}%
\right) ,  \label{SMmt3} \\
M_{R}^{2} &=&-{\frac{1}{16\pi ^{2}}}{\frac{y_{u}^{2}-y_{d}^{2}}{8}}\left(
\frac{1}{\hat{\epsilon}}-\log {\frac{M_{H}^{2}}{\mu ^{2}}}+{\frac{5}{2}}%
\right) .  \label{SMmt4}
\end{eqnarray}

To determine completely the $M_{L,R}^{i}$ coefficients we need to consider
the vertex $udW$. The relevant diagrams are analogous to those of Fig. \ref
{Figmsector-3}. A straightforward calculation gives
\begin{eqnarray*}
\Delta \hat{\Gamma}_{\mu }^{udW} &=&{\frac{e}{4\sqrt{2}s_{W}}}\;\gamma _{\mu
}\left\{ \left[ {\frac{y_{u}y_{d}}{16\pi ^{2}}}\left( \frac{1}{\hat{\epsilon}%
}-\log {\frac{M_{H}^{2}}{\mu ^{2}}}+{\frac{5}{2}}\right) +4\left(
M_{R}^{1}-M_{R}^{3}\right) \right] (1+\gamma _{5})\right. \\
&&\left. -\left[ {\frac{y_{u}^{2}+y_{d}^{2}}{16\pi ^{2}}}\frac{1}{2}\left(
\frac{1}{\hat{\epsilon}}-\log {\frac{M_{H}^{2}}{\mu ^{2}}}+{\frac{5}{2}}%
\right) +4\left( M_{L}^{1}+M_{L}^{3}\right) \right] (1-\gamma _{5})\right\} .
\end{eqnarray*}
The matching condition $\Delta \hat{\Gamma}_{\mu }^{udW}=0$ amounts to the
following set of equations
\begin{eqnarray*}
M_{R}^{1}-M_{R}^{3} &=&-{\frac{1}{16\pi ^{2}}}{\frac{y_{u}y_{d}}{4}}\left(
\frac{1}{\hat{\epsilon}}-\log {\frac{M_{H}^{2}}{\mu ^{2}}}+{\frac{5}{2}}%
\right) , \\
M_{L}^{1}+M_{L}^{3} &=&-{\frac{1}{16\pi ^{2}}}{\frac{y_{u}^{2}+y_{d}^{2}}{8}}%
\left( \frac{1}{\hat{\epsilon}}-\log {\frac{M_{H}^{2}}{\mu ^{2}}}+{\frac{5}{2%
}}\right) ,
\end{eqnarray*}
Combining these equations with Eqs. (\ref{SMmt1}, \ref{SMmt2}) we finally
get
\begin{eqnarray}
M_{L}^{1} &=&-{\frac{1}{16\pi ^{2}}}{\frac{y_{u}^{2}+y_{d}^{2}}{8}}\left(
\frac{1}{\hat{\epsilon}}-\log {\frac{M_{H}^{2}}{\mu ^{2}}}+{\frac{5}{2}}%
\right) , \\
M_{R}^{1} &=&-{\frac{(y_{u}+y_{d})^{2}}{\left( 16\pi \right) ^{2}}}\left(
\frac{1}{\hat{\epsilon}}-\log {\frac{M_{H}^{2}}{\mu ^{2}}}+{\frac{5}{2}}%
\right) , \\
M_{L}^{3} &=&0, \\
M_{R}^{3} &=&-{\frac{(y_{u}-y_{d})^{2}}{\left( 16\pi \right) ^{2}}}\left(
\frac{1}{\hat{\epsilon}}-\log {\frac{M_{H}^{2}}{\mu ^{2}}}+{\frac{5}{2}}%
\right) ,
\end{eqnarray}
This, along with Eqs. (\ref{SMmt3}, \ref{SMmt4}) and Eq. (\ref{SM-de7}), is
our final answer. These results coincide, where the comparison is possible,
with those obtained in \cite{DG} by functional methods. It is interesting to
note that it has not been necessary to consider the matching of the vertex $%
ff\gamma $.

We shall show explicitly that $M_{L}^{4}$ drops from the $S$ matrix element
corresponding to $Z\rightarrow f\bar{f}$. It is well known that the
renormalized $u$-fermion propagator has residue $1+\delta _{res}$, where $%
\delta _{res}$ is given in Eq. (\ref{17.6j}) of appendix \ref{Msectorapp}.%
\ref{MsectorappD}. Therefore, in order to evaluate $S$-matrix elements
involving external $u$ lines at one-loop, one has to multiply the
corresponding amputated Green functions by a factor $1+n\,\delta _{res}/2$,
where $n$ is the number on external $u$-lines (in the case under
consideration $n=2$). One can check that when this factor is taken into
account, the $M_{L}^{4}$ appearing in the renormalized S-matrix vertex are
cancelled.

We notice that $M_{L}^{1}$ and $M_{R}^{1}$ indeed correspond to custodially
preserving operators, while $M_{L,R}^{2}$ and $M_{L,R}^{3}$ do not. All
these coefficients (just as $a_{0}$, $a_{1}$ and $a_{8}$) are ultraviolet
divergent (with the exception of $M_{L}^{3}$). This is so because the Higgs
particle is an essential ingredient to guarantee the renormalizability of
the Standard Model. Once this is removed, the usual renormalization process
(e.g. the on-shell scheme) is not enough to render all ``renormalized''
Green functions finite. This is why the bare coefficients of the effective
Lagrangian (which contribute to the renormalized Green functions either
directly or via counterterms) have to be proportional to $1/\epsilon $ to
cancel the new divergences. The coefficients of the effective Lagrangian are
manifestly gauge invariant.

What is the value of these coefficients in other theories with elementary
scalars and Higgs-like mechanism? This issue has been discussed in some
detail in \cite{ciafaloni} in the context of the two-Higgs doublet model,
but it can actually be extended to supersymmetric theories (provided of
course scalars other than the $CP$-even Higgs can be made heavy enough, see
e.g. \cite{mjh}). It was argued there that non-decoupling effects are
exactly the same as in the minimal Standard Model, including the constant
non-logarithmic piece. Since the $M_{L,R}^{i}$ coefficients contain all the
non-decoupling effects associated to the Higgs particle at the first
non-trivial order in the momentum or mass expansion, the low energy
effective theory will be exactly the same.

\section{Z decay observables}

\label{Zdecayobs}The decay width of $Z\rightarrow f\bar{f}$ is described by
\begin{equation}
\Gamma _{f}\equiv \Gamma \left( Z\rightarrow f\bar{f}\right) =4n_{c}\Gamma
_{0}\left[ \left( g_{V}^{f}\right) ^{2}R_{V}^{f}+\left( g_{A}^{f}\right)
^{2}R_{A}^{f}\right] ,
\end{equation}
where $g_{V}^{f}$ and $g_{A}^{f}$ are the effective electroweak couplings as
defined in \cite{yellow} and $n_{c}$ is the number of colors of fermion $f$.
The radiation factors $R_{V}^{f}$ and $R_{A}^{f}$ describe the final state
QED and QCD interactions \cite{yellow2}. For a charged lepton we have
\begin{eqnarray*}
R_{V}^{l} &=&1+\frac{3\bar{\alpha}}{4\pi }+\mathcal{O}\left( \bar{\alpha}%
^{2},\left( \frac{m_{l}}{M_{Z}}\right) ^{4}\right) , \\
R_{A}^{l} &=&1+\frac{3\bar{\alpha}}{4\pi }-6\left( \frac{m_{l}}{M_{Z}}%
\right) ^{2}+\mathcal{O}\left( \bar{\alpha}^{2},\left( \frac{m_{l}}{M_{Z}}%
\right) ^{4}\right) ,
\end{eqnarray*}
where $\bar{\alpha}$ is the electromagnetic coupling constant at the scale $%
M_{Z}$ and $m_{l}$ is the final state lepton mass

The tree-level width $\Gamma _{0}$ is given by
\begin{equation}
\Gamma _{0}=\frac{G_{\mu }M_{Z}^{3}}{24\sqrt{2}\pi }.
\end{equation}
If we define
\begin{eqnarray}
\rho _{f} &\equiv &4\left( g_{A}^{f}\right) ^{2}, \\
\bar{s}_{W}^{2} &\equiv &\frac{\tau ^{3}}{4Q_{f}}\left( 1-\frac{g_{V}^{f}}{%
g_{A}^{f}}\right) ,
\end{eqnarray}
we can write
\begin{equation}
\Gamma _{f}=n_{c}\Gamma _{0}\rho _{f}\left[ 4\left( \frac{\tau ^{3}}{2}%
-2Q_{f}\bar{s}_{W}^{2}\right) ^{2}R_{V}^{f}+R_{A}^{f}\right] .
\end{equation}
Other quantities which are often used are $\Delta \rho _{f}$, defined
through
\begin{equation}
\rho _{f}\equiv \frac{1}{1-\Delta \rho _{f}},
\end{equation}
the forward-backward asymmetry $A_{FB}^{f}$
\begin{equation}
A_{FB}^{f}=\frac{3}{4}A^{e}A^{f},
\end{equation}
and $R_{b}$
\begin{equation}
R_{b}=\frac{\Gamma _{b}}{\Gamma _{h}},
\end{equation}
where
\begin{equation*}
A^{f}\equiv \frac{2g_{V}^{f}g_{A}^{f}}{\left( g_{A}^{f}\right) ^{2}+\left(
g_{V}^{f}\right) ^{2}},
\end{equation*}
and $\Gamma _{b}$, $\Gamma _{h}$ are the b-partial width and total hadronic
width, respectively (each of them, in turn, can be expressed in terms of the
appropriate effective couplings). As we see, nearly all of $Z$ physics can
be described in terms of $g_{A}^{f}$ and $g_{V}^{f}$. The box contributions
to the process $e^{+}e^{-}\rightarrow f\bar{f}$ are not included in the
analysis because they are negligible and they cannot be incorporated as
contributions to effective electroweak neutral current couplings anyway.

We shall generically denote these effective couplings by $g^{f}$. If we
express the value they take in the Standard Model by $g^{f\left( \mathrm{SM}%
\right) }$, we can write a perturbative expansion for them in the following
way
\begin{equation}
g^{f\left( \mathrm{SM}\right) }=g^{f\left( 0\right) }+g^{f\left( 2\right) }+%
\bar{g}^{f}(a_{i}^{\mathrm{(SM)}})+\hat{g}^{f}(M_{L,R}^{i\mathrm{(SM)}}),
\label{gsm}
\end{equation}
where $g^{f\left( 0\right) }$ are the tree-level expressions for these form
factors, $g^{f\left( 2\right) }$ are the one-loop contributions which do not
contain any Higgs particle as internal line in the Feynman graphs. In the
effective Lagrangian language they are generated by the quantum corrections
computed by operators such as (\ref{mass}) or the first operator on the
r.h.s. of (\ref{effobl}). On the other hand, the Feynman diagrams containing
the Higgs particle contribute to $g^{f\left( \mathrm{SM}\right) }$ in a
twofold way. One is via the $\mathcal{O}(p^{2})$ and $\mathcal{O}(p^{4})$
Longhitano effective operators (\ref{effobl}) which depend on the $a_{i}$
coefficients, which are Higgs-mass dependent, and thus give a
Higgs-dependent oblique correction to $g^{f\left( \mathrm{SM}\right) }$,
which is denoted by $\bar{g}^{f}$. The other one is via genuine vertex
corrections which depend on the $M_{L,R}^{i}$. This contribution is denoted
by $\hat{g}^{f}$.

The tree-level value for the form factors are
\begin{equation}
g_{V}^{f\left( 0\right) }=\frac{\tau ^{3}}{2}-2s_{W}^{2}Q_{f},\qquad
g_{A}^{f\left( 0\right) }=\frac{\tau ^{3}}{2}.
\end{equation}
In a theory X, different from the minimal Standard Model, the effective form
factors will take values $g^{f\left( \mathrm{X}\right) }$, where
\begin{equation}
g^{f\left( \mathrm{X}\right) }=g^{f\left( 0\right) }+g^{f\left( 2\right) }+%
\bar{g}^{f}(a_{i}^{\mathrm{(X)}})+\hat{g}^{f}(M_{L,R}^{i\mathrm{(X)}}),
\label{g4q}
\end{equation}
and the $a_{i}^{\mathrm{(X)}}$ and $M_{L,R}^{i\mathrm{(X)}}$ are effective
coefficients corresponding to theory X.

Within one-loop accuracy in the symmetry breaking sector (but with arbitrary
precision elsewhere), $\bar{g}^{f}$ and $\hat{g}^{f}$ are linear functions
of their arguments and thus we have
\begin{equation}
g^{f\left( \mathrm{X}\right) }=g^{f\left( \mathrm{SM}\right) }+\bar{g}%
^{f}(a_{i}^{\mathrm{(X)}}-a_{i}^{\mathrm{(SM)}})+\hat{g}^{f}(M_{L,R}^{i%
\mathrm{(X)}}-M_{L,R}^{i\mathrm{(SM)}}).
\end{equation}

The expression for $\bar{g}^{f}$ in terms of $a_{i}$ was already given in
Eqs. (\ref{gagvobl}) and (\ref{gagvvv}). On the other hand from appendix \ref
{Msectorapp}.\ref{MsectorappB} we learn that
\begin{eqnarray*}
\hat{g}_{V}^{f}\left( M_{L,R}^{i}\right) &=&M_{L}^{2}+M_{R}^{2}-\tau
^{3}\left( M_{L}^{1}-M_{L}^{3}-M_{R}^{1}-M_{R}^{3}\right) , \\
\hat{g}_{A}^{f}\left( M_{L,R}^{i}\right) &=&M_{L}^{2}-M_{R}^{2}-\tau
^{3}\left( M_{L}^{1}-M_{L}^{3}+M_{R}^{1}+M_{R}^{3}\right) ,
\end{eqnarray*}

In the minimal Standard Model all the Higgs dependence at the one loop level
(which is the level of accuracy assumed here) is logarithmic and is
contained in the $a_{i}$ and $M_{L,R}^{i}$ coefficients. Therefore one can
easily construct linear combinations of observables where the leading Higgs
dependence cancels. These combinations allow for a test of the minimal
Standard Model independent of the actual value of the Higgs mass.

Let us now review the comparison with current electroweak data for theories
with dynamical symmetry breaking. Some confusion seem to exist on this point
so let us try to analyze this issue critically.

A first difficulty arises from the fact that at the $M_Z$ scale perturbation
theory is not valid in theories with dynamical breaking and the contribution
from the symmetry breaking sector must be estimated in the framework of the
effective theory, which is non-linear and non-renormalizable. Observables
will depend on some subtraction scale. (Estimates based on dispersion
relations and resonance saturation amount, in practice, to the same,
provided that due attention is paid to the scale dependence introduced by
the subtraction in the dispersion relation.)

A somewhat related problem is that, when making use of the variables $S,T$
and $U$ \cite{PT}, or $\epsilon _{1},\epsilon _{2}$ and $\epsilon _{3}$ \cite
{epsilon}, one often sees in the literature bounds on possible ``new
physics'' in the symmetry breaking sector without actually removing the
contribution from the Standard Model Higgs that the ``new physics'' is
supposed to replace (this is not the case e.g. in \cite{PT} where this issue
is discussed with some care). Unless the contribution from the ``new
physics'' is enormous, this is a flagrant case of double counting, but it is
easy to understand why this mistake is made: removing the Higgs makes the
Standard Model non-renormalizable and the observables of the Standard Model
without the Higgs depend on some arbitrary subtraction scale.

In fact the two sources of arbitrary subtraction scales (the one originating
from the removal of the Higgs and the one from the effective action
treatment) are one an the same and the problem can be dealt with the help of
the coefficients of higher dimensional operators in the effective theory
(i.e. the $a_{i}$ and $M_{L,R}^{i}$). The dependence on the unknown
subtraction scale is absorbed in the coefficients of higher dimensional
operators and traded by the scale of the ``new physics''. Combinations of
observables can be built where this scale (and the associated
renormalization ambiguities) drops. These combinations allow for a test of
the ``new physics'' independently of the actual value of its characteristic
scale. In fact they are the same combinations of observables where the Higgs
dependence drops in the minimal Standard Model.

A third difficulty in making a fair comparison of models of dynamical
symmetry breaking with experiment lies in the vertex corrections. If we
analyze the lepton effective couplings $g_A^l$ and $g_V^l$, the minimal
Standard Model predicts very small vertex corrections arising from the
symmetry breaking sector anyway and it is consistent to ignore them and
concentrate in the oblique corrections. However, this is not the situation
in dynamical symmetry breaking models. We will see in the next sections that
for the second and third generation vertex corrections can be sizeable. Thus
if we want to compare experiment to oblique corrections in models of
dynamical breaking we have to concentrate on electron couplings only.

\begin{figure}[!hbp]
\hspace{2.5cm} \includegraphics[width=\figwidth]{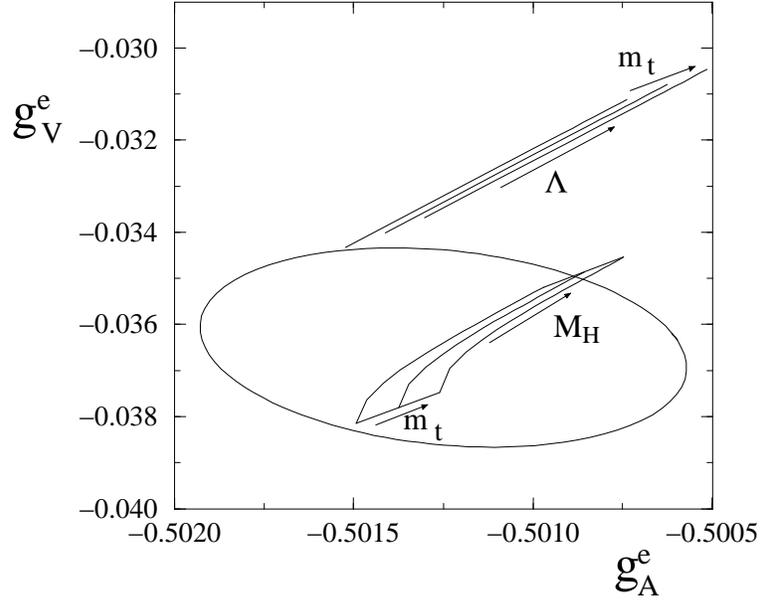}
\caption{The $1-\protect\sigma $ experimental region in the $%
g_{A}^{e}-g_{V}^{e}$ plane. The Standard Model predictions as a function of $%
m_{t}$ ($170.6\leq m_{t}\leq 180.6$ GeV) and $M_{H}$ ($70\leq M_{H}\leq 1000$
GeV) are shown (the middle line corresponds to the central value $%
m_{t}=175.6 $ GeV). The predictions of a QCD-lke technicolor theory with $%
n_{TC}n_{D}=8$ and degenerate technifermion masses are shown as straight
lines (only oblique corrections are included). One moves along the straight
lines by changing the scale $\Lambda $. The three lines correspond to the
extreme and central values for $m_{t}$. Recall that the precise location
anywhere on the straight lines (which definitely do intersect the $1-\protect%
\sigma $ region) depends on the renormalization procedure and thus is not
predictable within the non-renormalizable effective theory. In addition the
technicolor prediction should be considered accurate only at the 15\% level
due to the theoretical uncertainties discussed in the text (this error is at
any rate smaller than the one associated to the uncertainty in $\Lambda $).
Notice that the oblique corrections, in the case of degenerate masses, are
independent of the value of the technifermion mass. Assuming universality of
the vertex corrections reduces the error bars by about a factor one-half and
leaves technicolor predictions outside the $1-\protect\sigma $ region.}
\label{Figmsector-1}
\end{figure}

In Fig. \ref{Figmsector-1} we see the prediction of the minimal Standard
Model for $170.6<m_{t}<180.6$ GeV and $70<M_{H}<1000$ GeV including the
leading two-loop corrections \cite{yellow2}, falling nicely within the
experimental $1-\sigma $ region for the electron effective couplings. In
this and in subsequent plots we present the data from the combined four LEP
experiments only. What is the actual prediction for a theory with dynamical
symmetry breaking? The straight solid lines correspond to the prediction of
a QCD-like technicolor model with $n_{TC}=2$ and $n_{D}=4$ (a one-generation
model) in the case where all technifermion masses are assumed to be equal
(we follow \cite{DEH}, see \cite{other} for related work) allowing the same
variation for the top mass as in the Standard Model. We do not take into
account here the contribution of potentially present pseudo Goldstone
bosons, assuming that they can be made heavy enough. The corresponding
values for the $a_{i}$ coefficients in such a model are given in appendix
\ref{Msectorapp}.\ref{MsectorappE} and are derived using chiral quark model
techniques and chiral perturbation theory. They are scale dependent in such
a way as to make observables finite and unambiguous, but of course
observables depend in general on the scale of ``new physics'' $\Lambda $.

We move along the straight lines by changing the scale $\Lambda $. It would
appear at first sight that one needs to go to unacceptably low values of the
new scale to actually penetrate the $1-\sigma $ region, something which
looks unpleasant at first sight (we have plotted the part of the line for $%
100\leq \Lambda \leq 1500$ GeV), as one expects $\Lambda \sim \Lambda _{\chi
}$. In fact this is not necessarily so. There is no real prediction of the
effective theory \emph{along} the straight lines, because only combinations
which are $\Lambda $-independent are predictable. As for the location not
\emph{along} the line, but \emph{of} the line itself it is in principle
calculable in the effective theory, but of course subject to the
uncertainties of the model one relies upon, since we are dealing with a
strongly coupled theory. (We shall use chiral quark model estimates in this
work as we believe that they are quite reliable for QCD-like theories, see
the discussion below.)

\begin{figure}[!hbp]
\begin{center}
\includegraphics[width=\figwidth]{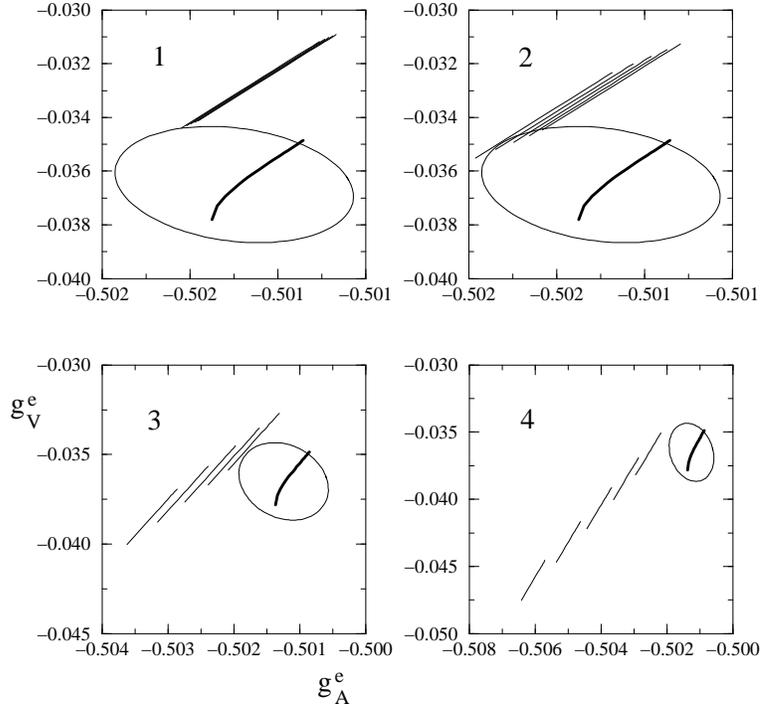}
\end{center}
\caption{The effect of isospin breaking in the oblique corrections in
QCD-like technicolor theories. The $1-\protect\sigma$ region for the $%
g_A^e-g_V^e$ couplings and the SM prediction (for $m_t=175.6$ GeV, and $%
70\le M_H\le 1000$ GeV) are shown. The different straight lines correspond
to setting the technifermion masses in each doublet ($m_1$, $m_2$) to the
value $m_2=$ 250, 300, 350, 400 and 450 GeV (larger masses are the ones
deviating more from the SM predictions), and $m_1=1.05 m_2$ (plot 1), $%
m_1=1.1 m_2$ (plot 2), $m_1=1.2 m_2$ (plot 3), and $m_1=1.3 m_2$ (plot 4).
The results are invariant under the exchange of $m_1$ and $m_2$. As in
figure \ref{Figmsector-1} the prediction of the effective theory is the
whole straight line and not any particular point on it, as we move along the
line by varying the unknown scale $\Lambda$. Clearly isospin breakings
larger than 20 \% give very poor agreement with the data, even for low
values of the dynamically generated mass.}
\label{Figmsector-2}
\end{figure}

If we allow for a splitting in the technifermion masses the comparison with
experiment improves very slightly. The values of the effective Lagrangian
coefficients relevant for the oblique corrections in the case of unequal
masses are also given in appendix \ref{Msectorapp}.\ref{MsectorappE}. Since $%
a_{1}$ is independent of the technifermion dynamically generated masses
anyway, the dependence is fully contained in $a_{0}$ (the parameter $T$ of
Peskin and Takeuchi \cite{PT}) and $a_{8}$ (the parameter $U$). This is
shown in Fig. \ref{Figmsector-2}. We assume that the splitting is the same
for all doublets, which is not necessarily true\footnote{%
In fact it can be argued that QCD corrections may, in some cases \cite
{holdom}, enhance techniquark masses.}.

If other representations of the $SU(2)_{L}\times SU(3)_{c}$ gauge group are
used, the oblique corrections have to be modified in the form prescribed in
section \ref{heavyfermions}. Larger group theoretical factors lead to larger
oblique corrections and, from this point of view, the restriction to weak
doublets and color singlets or triplets is natural.

Let us close this section by justifying the use of chiral quark model
techniques, trying to assess the errors involved, and at the same time
emphasizing the importance of having the scale dependence under control. A
parameter like $a_{1}$ (or $S$ in the notation of Peskin and Takeuchi \cite
{PT}) contains information about the long-distance properties of a strongly
coupled theory. In fact, $a_{1}$ is nothing but the familiar $L_{10}$
parameter of the strong chiral Lagrangian of Gasser and Leutwyler \cite{GL}
translated to the electroweak sector. This strong interaction parameter can
be measured and it is found to be $L_{10}=(-5.6\pm 0.3)\times 10^{-3}$ (at
the $\mu =M_{\eta }$ scale, which is just the conventional reference value
and plays no specific role in the Standard Model.) This is almost twice the
value predicted by the chiral quark model \cite{AA,ERT} ($L_{10}=-1/32\pi
^{2}$), which is the estimate plotted in Fig. \ref{Figmsector-1}. Does this
mean that the chiral quark model grossly underestimates this observable? Not
at all. Chiral perturbation theory predicts the running of $L_{10}$. It is
given by
\begin{equation}
L_{10}(\mu )=L_{10}(M_{\eta })+\frac{1}{128\pi ^{2}}\log \frac{\mu ^{2}}{%
M_{\eta }^{2}}.
\end{equation}
According to our current understanding (see e.g. \cite{AET}), the chiral
quark model gives the value of the chiral coefficients at the chiral
symmetry breaking scale ($4\pi f_{\pi }$ in QCD, $\Lambda _{\chi }$ in the
electroweak theory). Then the coefficient $L_{10}$ (or $a_{1}$ for that
matter) predicted within the chiral quark model agrees with QCD at the 10\%
level.

Let us now turn to the issue of vertex corrections in theories with
dynamical symmetry breaking and the determination of the coefficients $%
M_{L,R}^{i}$ which are, after all, the focal point of this chapter.

\section{New physics and four-fermion operators}

\label{new-phys}In order to have a picture in our mind, let us assume that
at sufficiently high energies the symmetry breaking sector can be described
by some renormalizable theory, perhaps a non-abelian gauge theory. By some
unspecified mechanism some of the carriers of the new interaction acquire a
mass. Let us generically denote this mass by $M$. One type of models that
comes immediately to mind is the extended technicolor scenario. $M$ would
then be the mass of the ETC bosons. Let us try, however, not to adhere to
any specific mechanism or model.

Below the scale $M$ we shall describe our underlying theory by four-fermion
operators. This is a convenient way of parametrizing the new physics below $%
M $ without needing to commit oneself to a particular model. Of course the
number of all possible four-fermion operators is enormous and one may think
that any predictive power is lost. This is not so because of two reasons: a)
The size of the coefficients of the four fermion operators is not arbitrary.
They are constrained by the fact that at scale $M$ they are given by
\begin{equation}
-\xi_{\mathrm{CG}} \frac{G^2}{M^2}  \label{4qcoef}
\end{equation}
where $\xi_{\mathrm{CG}}$ is built out of Clebsch-Gordan factors and $G$ a
gauge coupling constant, assumed perturbative of $\mathcal{O}(1)$ at the
scale $M$. The $\xi_{\mathrm{CG}}$ being essentially group-theoretical
factors are probably of similar size for all three generations, although not
necessarily identical as this would assume a particular style of embedding
the different generations into the large ETC (for instance) group. Notice
that for four-fermion operators of the form $\mathbf{J}\cdot \mathbf{J}%
^\dagger$, where $\mathbf{J}$ is some fermion bilinear, $\xi_{\mathrm{CG}}$
has a well defined sign, but this is not so for other operators. b) It turns
out that only a relatively small number of combinations of these
coefficients do actually appear in physical observables at low energies.

Matching to the fundamental physical theory at $\mu =M$ fixes the value of
the coupling constants accompanying the four-fermion operators to the value (%
\ref{4qcoef}). In addition contact terms, i.e. non-zero values for the
effective coupling constants $M_{L,R}^{i}$, are generally speaking required
in order for the fundamental and four-fermion theories to match. These will
later evolve under the renormalization group due to the presence of the
four-fermion interactions. Because we expect that $M\gg \Lambda _{\chi }$,
the $M_{L,R}^{i}$ will be typically logarithmically enhanced. Notice that
there is no guarantee that this is the case for the third generation, as we
will later discuss. In this case the TC and ETC dynamics would be tangled up
(which for most models is strongly disfavored by the constraints on oblique
corrections). For the first and second generation, however, the logarithmic
enhancement of the $M_{L,R}^{i}$ is a potentially large correction and it
actually makes the treatment of a fundamental theory via four-fermion
operators largely independent of the particular details of specific models,
as we will see.

Let us now get back to four-fermion operators and proceed to a general
classification. A first observation is that, while in the bosonic sector
custodial symmetry is just broken by the small $U(1)_{Y}$ gauge
interactions, which is relatively small, in the matter sector the breaking
is not that small. We thus have to assume that whatever underlying new
physics is present at scale $M$ it gives rise both to custodially preserving
and custodially non-preserving four-fermion operators with coefficients of
similar strength. Obvious requirements are hermiticity, Lorentz invariance
and $SU(3)_{c}\times SU(2)_{L}\times U(1)_{Y}$ symmetry. Neither $C$ nor $P$
invariance are imposed, but invariance under $CP$ is assumed.

We are interested in $d=6$ four-fermion operators constructed with two
ordinary fermions (either leptons or quarks), denoted by $q_L$, $q_R$, and
two fermions $Q^{A}_L$, $Q^{A}_R$. Typically $A$ will be the technicolor
index and the $Q_L$, $Q_R$ will therefore be techniquarks and technileptons,
but we may be as well interested in the case where the $Q$ may be ordinary
fermions. In this case the index $A$ drops (in our subsequent formulae this
will correspond to taking $n_{TC}=1$). We shall not write the index $A$
hereafter for simplicity, but this degree of freedom is explicitly taken
into account in our results.

As we have already mentioned we shall discuss in detail the case where the
additional fermions fall into ordinary representations of $SU(2)_{L}\times
SU(3)_{c}$ and will discuss other representations later. The fields $Q_{L}$
will therefore transform as $SU(2)_{L}$ doublets and we shall group the
right-handed fields $Q_{R}$ into doublets as well, but then include suitable
insertions of $\tau ^{3}$ to consider custodially breaking operators. In
order to determine the low energy remnants of all these four-fermion
operators (i.e. the coefficients $M_{L,R}^{i}$) it is enough to know their
couplings to $SU(2)_{L}$ and no further assumptions about their electric
charges (or hypercharges) are needed. Of course, since the $Q_{L}$, $Q_{R}$
couple to the electroweak gauge bosons they must not lead to new anomalies.
The simplest possibility is to assume they reproduce the quantum numbers of
one family of quarks and leptons (that is, a total of four doublets $n_{D}=4$%
), but other possibilities exist (for instance $n_{D}=1$ is also possible
\cite{1doub}, although this model presents a global $SU(2)_{L}$ anomaly).

We shall first be concerned with the $Q_{L}$, $Q_{R}$ fields belonging to
the representation $\mathbf{3}$ of $SU(3)_{c}$ and afterwards, focus in the
simpler case where the $Q_{L}$, $Q_{R}$ are color singlet (technileptons).
Colored $Q_{L}$, $Q_{R}$ fermions can couple to ordinary quarks and leptons
either via the exchange of a color singlet or of a color octet. In addition
the exchanged particle can be either an $SU(2)_{L}$ triplet or a singlet,
thus leading to a large number of possible four-fermion operators. More
important for our purposes will be whether they flip or not the chirality.
We use Fierz rearrangements in order to write the four-fermion operators as
product of either two color singlet or two color octet currents. A complete
list is presented in table \ref{table-2} and table \ref{table-3} for the
chirality preserving and chirality flipping operators, respectively.
\begin{table}[tbp]
\centering
\begin{tabular}{|c|c|}
\hline
$L^2=(\bar Q_{L} \gamma_{\mu} Q_{L})(\bar q_{L}\gamma^\mu q_{L})$ &  \\
\hline
$R^2=(\bar Q_{R}\gamma_{\mu} Q_{R})(\bar q_{R}\gamma^\mu q_{R})$ & $%
R_{3}R=(\bar Q_{R}\gamma_{\mu}\tau^3 Q_{R})(\bar q_{R}\gamma^\mu q_{R})$ \\
\cline{2-2}
& $RR_{3}=(\bar Q_{R}\gamma_{\mu} Q_{R})(\bar q_{R}\gamma^\mu\tau^3 q_{R})$
\\ \cline{2-2}
& $R_{3}^2=(\bar Q_{R}\gamma_{\mu}\tau^3 Q_{R})(\bar q_{R}\gamma^\mu\tau^3
q_{R})$ \\ \hline\hline
$RL=(\bar Q_{R} \gamma_{\mu} Q_{R})(\bar q_{L}\gamma^\mu q_{L})$ & $%
R_{3}L=(\bar Q_{R}\gamma_{\mu} \tau^3 Q_{R})(\bar q_{L}\gamma^\mu q_{L})$ \\
\hline
$LR=(\bar Q_{L}\gamma_{\mu} Q_{L})(\bar q_{R}\gamma^\mu q_{R})$ & $%
LR_{3}=(\bar Q_{L}\gamma_{\mu} Q_{L})(\bar q_{R}\gamma^\mu \tau^3 q_{R})$ \\
\hline
$rl=(\bar Q_{R} \gamma_{\mu}\vec\lambda Q_{R})\cdot (\bar
q_{L}\gamma^\mu\vec\lambda q_{L}) $ & $r_{3}l= (\bar
Q_{R}\gamma_{\mu}\vec\lambda \tau^3 Q_{R})\cdot (\bar
q_{L}\gamma^\mu\vec\lambda q_{L})$ \\ \hline
$lr=(\bar Q_{L} \gamma_{\mu}\vec\lambda Q_{L})\cdot (\bar
q_{R}\gamma^\mu\vec\lambda q_{R}) $ & $lr_{3}= (\bar
Q_{L}\gamma_{\mu}\vec\lambda Q_{L})\cdot (\bar q_{R}\gamma^\mu\vec\lambda
\tau^3 q_{R})$ \\ \hline\hline
$(\bar Q_{L}\gamma_{\mu} q_{L})(\bar q_{L}\gamma^\mu Q_{L})$ &  \\ \hline
$(\bar Q_{R}\gamma_{\mu} q_{R})(\bar q_{R}\gamma^\mu Q_{R}) $ & $(\bar
Q_{R}\gamma_{\mu}\tau^3 q_{R})(\bar q_{R}\gamma^\mu Q_{R})+ (\bar
Q_{R}\gamma_{\mu} q_{R})(\bar q_{R}\gamma^\mu \tau^3 Q_{R})$ \\ \cline{2-2}
& $(\bar Q_{R}\gamma_{\mu}\tau^3 q_{R})(\bar q_{R}\gamma^\mu\tau^3 Q_{R})$
\\ \hline\hline
$(\bar Q_{L}^i\gamma_{\mu} Q^j_{L} )(\bar q_{L}^j\gamma^\mu q_{L}^i)$ &  \\
\hline
$(\bar Q_{R}^i\gamma_{\mu} Q^j_{R} )(\bar q_{R}^j\gamma^\mu q_{R}^i)$ &  \\
\hline
$(\bar Q_{L}^i\gamma_{\mu} q_{L}^j)(\bar q_{L}^j\gamma^\mu Q_{L}^i)$ &  \\
\hline
$(\bar Q_{R}^i\gamma_{\mu}q_{R}^j) (\bar q_{R}^j\gamma^\mu Q_{R}^i)$ & $%
(\bar Q_{R}^i\gamma_{\mu} q_{R}^j)(\bar q_{R}^j\gamma^\mu [\tau^3 Q_{R}]^i)$
\\ \hline
\end{tabular}
\caption{Four-fermion operators which do not change the fermion chirality.
The first (second) column contains the custodially preserving (breaking)
operators.}
\label{table-2}
\end{table}
\begin{table}[tbp]
\centering
\begin{tabular}{|c|c|}
\hline
\mbox{  $(\bar Q_{L}\gamma^\mu q_{L})(\bar
q_{R}\gamma_{\mu} Q_{R})$} &
\mbox{  $(\bar Q_{L}\gamma^\mu
q_{L})(\bar q_{R}\gamma_{\mu}\tau^3 Q_{R})$} \\ \hline
\mbox{
$(\bar
          q_{L}^i q_{R}^j)(\bar Q_{L}^k Q_{R}^l)\epsilon_{ik}\epsilon_{jl}$}
&
\mbox{ $(\bar
          q_{L}^i  [\tau^3q_{R}]^j)(\bar Q_{L}^k
Q_{R}^l)\epsilon_{ik}\epsilon_{jl}$} \\ \hline
\mbox{ $(\bar
          q_{L}^i Q_{R}^j) (\bar Q_{L}^k q_{R}^l)\epsilon_{ik}\epsilon_{jl}$}
&
\mbox{ $(\bar
          q_{L}^i Q_{R}^j) (\bar Q_{L}^k [\tau^3
q_{R}]^l)\epsilon_{ik}\epsilon_{jl}$} \\ \hline
\mbox{  $(\bar
Q_{L}\gamma^\mu\vec\lambda q_{L})\cdot(\bar
q_{R}\gamma_{\mu}\vec\lambda Q_{R})$} &
\mbox{  $(\bar
Q_{L}\gamma^\mu\vec\lambda q_{L})\cdot(\bar
q_{R}\gamma_{\mu}\vec\lambda\tau^3 Q_{R})$} \\ \hline
\mbox{ $(\bar
          q_{L}^i\vec\lambda q_{R}^j)\cdot( \bar Q_{L}^k\vec\lambda
Q_{R}^l)\epsilon_{ik}\epsilon_{jl}$} &
\mbox{ $(\bar
          q_{L}^i\vec\lambda [\tau^3q_{R}]^j) \cdot(\bar Q_{L}^k\vec\lambda
Q_{R}^l)\epsilon_{ik}\epsilon_{jl}$} \\ \hline
\mbox{ $(\bar
          q_{L}^i\vec\lambda Q_{R}^j)\cdot( \bar Q_{L}^k\vec\lambda
q_{R}^l)\epsilon_{ik}\epsilon_{jl}$} &
\mbox{ $(\bar
          q_{L}^i\vec\lambda Q_{R}^j)\cdot( \bar Q_{L}^k\vec\lambda
[\tau^3q_{R}]^l)\epsilon_{ik}\epsilon_{jl}$} \\ \hline
\end{tabular}
\caption{Chirality-changing four-fermion operators. To each entry, the
corresponding hermitian conjugate operator should be added. The left (right)
column contains custodially preserving (breaking) operators.}
\label{table-3}
\end{table}
\begin{table}[tbp]
\centering
\begin{tabular}{|c|c|}
\hline
$l^2=(\bar Q_{L}\gamma_{\mu}\vec\lambda Q_{L})\cdot(\bar q_{L}\gamma^\mu\vec
\lambda q_{L}) $ &  \\ \hline
$r^2=(\bar Q_{R}\gamma_{\mu} \vec\lambda Q_{R})\cdot(\bar
q_{R}\gamma^\mu\vec\lambda q_{R}) $ & $r_{3}r=(\bar
Q_{R}\gamma_{\mu}\vec\lambda\tau^3 Q_{R})\cdot(\bar
q_{R}\gamma^\mu\vec\lambda q_{R})$ \\ \cline{2-2}
& $rr_{3}=(\bar Q_{R}\gamma_{\mu}\vec\lambda Q_{R})\cdot(\bar
q_{R}\gamma^\mu\vec\lambda \tau^3 q_{R})$ \\ \cline{2-2}
& $r_{3}^2=(\bar Q_{R}\gamma_{\mu}\vec\lambda\tau^3 Q_{R}) \cdot(\bar
q_{R}\gamma^\mu\vec\lambda\tau^3 q_{R})$ \\ \hline\hline
$\vec L^2=(\bar Q_{L}\gamma_{\mu}\vec\tau Q_{L})\cdot (\bar
q_{L}\gamma^\mu\vec\tau q_{L})$ &  \\ \hline
$\vec R^2=(\bar Q_{R}\gamma_{\mu}\vec\tau Q_{R})\cdot (\bar
q_{R}\gamma^\mu\vec\tau q_{R})$ &  \\ \hline
$\vec l^2=(\bar Q_{L}\gamma_{\mu}\vec\lambda \vec\tau Q_{L})\cdot (\bar
q_{L}\gamma^\mu\vec\lambda \vec\tau q_{L})$ &  \\ \hline
$\vec r^2=(\bar Q_{R}\gamma_{\mu}\vec\lambda \vec\tau Q_{R})\cdot (\bar
q_{R}\gamma^\mu\vec\lambda \vec\tau q_{R})$ &  \\ \hline
\end{tabular}
\caption{New four-fermion operators of the form $\mathbf{J}\cdot \mathbf{j}$
obtained after fierzing. The left (right) column contains custodially
preserving (breaking) operators. In addition those written in the two upper
blocks of table \ref{table-2} should also be considered. Together with the
above they form a complete set of chirality preserving operators.}
\label{table-4}
\end{table}

Note that the two upper blocks of table \ref{table-2} contain operators of
the form $\mathbf{J}\cdot \mathbf{j}$, where ($\mathbf{J}$) $\mathbf{j}$
stands for a (heavy) fermion current with well defined color and flavor
numbers; namely, belonging to an irreducible representation of $SU(3)_{c}$
and $SU(2)_{L}$. In contrast, those in the two lower blocks are not of this
form. In order to make their physical content more transparent, we can
perform a Fierz transformation and replace the last nine operators (two
lower blocks) in table \ref{table-2} by those in table \ref{table-4}. These
two basis are related by
\begin{eqnarray}
(\bar{Q}_{L}\gamma _{\mu }q_{L})(\bar{q}_{L}\gamma ^{\mu }Q_{L}) &=&\frac{1}{%
4}l^{2}+\frac{1}{6}L^{2}+\frac{1}{4}\vec{l}~{}^{2}+\frac{1}{6}\vec{L}^{2} \\
(\bar{Q}_{L}^{j}\gamma _{\mu }Q_{L}^{i})(\bar{q}_{L}^{i}\gamma ^{\mu
}q_{L}^{j}) &=&\frac{1}{2}L^{2}+\frac{1}{2}\vec{L}^{2} \\
(\bar{Q}_{L}^{j}\gamma _{\mu }q_{L}^{i})(\bar{q}_{L}^{i}\gamma ^{\mu
}Q_{L}^{j}) &=&\frac{1}{2}l^{2}+\frac{1}{3}L^{2} \\
(\bar{Q}_{R}\gamma _{\mu }q_{R})(\bar{q}_{R}\gamma ^{\mu }Q_{R}) &=&\frac{1}{%
4}r^{2}+\frac{1}{6}R^{2}+\frac{1}{4}\vec{r}^{2}+\frac{1}{6}\vec{R}^{2} \\
(\bar{Q}_{R}\gamma _{\mu }q_{R})(\bar{q}_{R}\gamma ^{\mu }\tau ^{3}Q_{R})%
\kern1.5cm && \\
+(\bar{Q}_{R}\gamma _{\mu }\tau ^{3}q_{R})(\bar{q}_{R}\gamma ^{\mu }Q_{R})
&=&\frac{1}{2}rr_{3}+\frac{1}{3}RR_{3}+\frac{1}{2}r_{3}r+\frac{1}{3}R_{3}R \\
(\bar{Q}_{R}\gamma _{\mu }\tau ^{3}q_{R})(\bar{q}_{R}\gamma ^{\mu }\tau
^{3}Q_{R}) &=&\frac{1}{4}r^{2}+\frac{1}{6}R^{2}-\frac{1}{4}\vec{r}^{2}-\frac{%
1}{6}\vec{R}^{2}+\frac{1}{2}r_{3}^{2}+\frac{1}{3}R_{3}^{2} \\
(\bar{Q}_{R}^{j}\gamma _{\mu }Q_{R}^{i})(\bar{q}_{R}^{i}\gamma ^{\mu
}q_{R}^{j}) &=&\frac{1}{2}R^{2}+\frac{1}{2}\vec{R}^{2} \\
(\bar{Q}_{R}^{j}\gamma _{\mu }q_{R}^{i})(\bar{q}_{R}^{i}\gamma ^{\mu
}Q_{R}^{j}) &=&\frac{1}{2}r^{2}+\frac{1}{3}R^{2} \\
(\bar{Q}_{R}^{j}\gamma _{\mu }q_{R}^{i})(\bar{q}_{R}^{i}\gamma ^{\mu }[\tau
^{3}Q_{R}]^{j}) &=&\frac{1}{2}r_{3}r+\frac{1}{3}R_{3}R
\end{eqnarray}
for colored techniquarks. Notice the appearance of some minus signs due to
the fierzing and that operators such as $L^{2}$ (for instance) get
contributions from four fermions operators which do have a well defined sign
as well as from others which do not.

The use of this basis simplifies the calculations considerably as the Dirac
structure is simpler. Another obvious advantage of this basis, which will
become apparent only later, is that it will make easier to consider the long
distance contributions to the $M_{L,R}^{i}$, from the region of momenta $\mu
<\Lambda _{\chi }$.

The classification of the chirality preserving operator involving
technileptons is of course simpler. Again we use Fierz rearrangements to
write the operators as $\mathbf{J}\cdot \mathbf{j}$. However, in this case
only a color singlet $\mathbf{J}$ (and, thus, also a color singlet $\mathbf{j%
}$) can occur. Hence, the complete list can be obtained by crossing out from
table \ref{table-4} and from the first eight rows of table \ref{table-2} the
operators involving $\vec{\lambda}$. Namely, those designated by lower-case
letters. We are then left with the two operators $\vec{L}^{2}$, $\vec{R}^{2}$
from table \ref{table-4} and with the first six rows of table \ref{table-2}:
$L^{2}$, $R^{2}$, $R_{3}R$, $RR_{3}$, $R_{3}^{2}$, $RL$, $R_{3}L$, $LR$ and $%
LR_{3}$. If we choose to work instead with the original basis of chirality
preserving operators in table \ref{table-2}, we have to supplement these
nine operators in the first six rows of the table with $(\bar{Q}_{L}\gamma
_{\mu }q_{L})(\bar{q}_{L}\gamma ^{\mu }Q_{L})$ and $(\bar{Q}_{R}\gamma _{\mu
}q_{R})(\bar{q}_{R}\gamma ^{\mu }Q_{R})$, which are the only independent
ones from the last seven rows. These two basis are related by
\begin{eqnarray}
(\bar{Q}_{L}\gamma _{\mu }q_{L})(\bar{q}_{L}\gamma ^{\mu }Q_{L}) &=&\frac{1}{%
2}L^{2}+\frac{1}{2}\vec{L}^{2} \\
(\bar{Q}_{R}\gamma _{\mu }q_{R})(\bar{q}_{R}\gamma ^{\mu }Q_{R}) &=&\frac{1}{%
2}R^{2}+\frac{1}{2}\vec{R}^{2}
\end{eqnarray}
for technileptons.

It should be borne in mind that Fierz transformations, as presented in the
above discussion, are strictly valid only in four dimensions. In $%
4-2\epsilon $ dimensions for the identities to hold we need `evanescent'
operators \cite{evan}, which vanish in 4 dimensions. However the replacement
of some four-fermion operators in terms of others via the Fierz identities
is actually made inside a loop of technifermions and therefore a finite
contribution is generated. Thus the two basis will eventually be equivalent
up to terms of order
\begin{equation}
\frac{1}{16\pi ^{2}}\frac{G^{2}}{M^{2}}m_{Q}^{2}  \label{fier}
\end{equation}
where $m_{Q}$ is the mass of the technifermion (this estimate will be
obvious only after the discussion in the next sections). In particular no
logarithms can appear in (\ref{fier}).

Let us now discuss how the appearance of other representations might enlarge
the above classification. We shall not be completely general here, but
consider only those operators that may actually contribute to the
observables we have been discussing (such as $g_{V}$ and $g_{A}$).
Furthermore, for reasons that shall be obvious in a moment, we shall
restrict ourselves to operators which are $SU(2)_{L}\times SU(2)_{R}$
invariant.

The construction of the chirality conserving operators for fermions in
higher dimensional representations of $SU(2)$ follows essentially the same
pattern presented in appendix \ref{Msectorapp}.\ref{MsectorappC} for doublet
fields, except for the fact that operators such as
\begin{equation}
(\bar{Q}_{L}\gamma _{\mu }q_{L})(\bar{q}_{L}\gamma ^{\mu }Q_{L}),\qquad (%
\bar{Q}_{L}^{i}\gamma _{\mu }Q_{L}^{j})(\bar{q}_{L}^{j}\gamma ^{\mu
}q_{L}^{i}),  \label{impossible}
\end{equation}
and their right-handed versions, which appear on the right hand side of
table \ref{table-2}, are now obviously not acceptable since $Q_{L}$ and $%
q_{L}$ are in different representations. Those operators, restricting
ourselves to color singlet bilinears (the only ones giving a non-zero
contribution to our observables) can be replaced in the fundamental
representation by
\begin{equation}
(\bar{Q}_{L}\gamma _{\mu }Q_{L})(\bar{q}_{L}\gamma ^{\mu }q_{L}),\qquad (%
\bar{Q}_{L}\gamma _{\mu }\vec{\tau}Q_{L})(\bar{q}_{L}\gamma ^{\mu }\vec{\tau}%
q_{L}),  \label{possible}
\end{equation}
when we move to the $\mathbf{J}\cdot \mathbf{j}$ basis. Now it is clear how
to modify the above when using higher representations for the $Q$ fields.
The first one is already included in our set of custodially preserving
operators, while the second one has to be modified to
\begin{equation}
{\vec{L}}^{2}\ \equiv \ (\bar{Q}_{L}\gamma _{\mu }\vec{T}Q_{L})(\bar{q}%
_{L}\gamma ^{\mu }\vec{\tau}q_{L}),  \label{higherrep}
\end{equation}
where $\vec{T}$ are the $SU(2)$ generators in the relevant representation.
In addition we have the right-handed counterpart, of course. We could in
principle now proceed to construct custodially violating operators by
introducing suitable $T^{3}$ and $\tau ^{3}$ matrices. Unfortunately, it is
not possible to present a closed set of operators of this type, as the
number of independent operators does obviously depend on the dimensionality
of the representation. For this reason we shall only consider custodially
preserving operators when moving to higher representations, namely $L^{2}$, $%
R^{2}$, $RL$, $LR$, ${\vec{L}}^{2}$ and ${\vec{R}}^{2}$.

If we examine tables 1, 2 and 3 we will notice that both chirality violating
and chirality preserving operators appear. It is clear that at the leading
order in an expansion in external fermion masses only the chirality
preserving operators (tables \ref{table-2} and \ref{table-4}) are important,
those operators containing both a $q_{L}$ and a $q_{R}$ field will be
further suppressed by additional powers of the masses of the fermions and
thus subleading. Furthermore, if we limit our analysis to the study of the
effective $W^{\pm }$ and $Z$ couplings, such as $g_{V}$ and $g_{A}$, as we
do here, chirality-flipping operators can contribute only through a two-loop
effect. Thus the contribution from the chirality flipping operators
contained in table \ref{table-3} is suppressed both by an additional $%
1/16\pi ^{2}$ loop factor and by a $m_{Q}^{2}/M^{2}$ chirality factor. If
for the sake of the argument we take $m_{Q}$ to be 400 GeV, the correction
will be below or at the 10\% level for values of $M$ as low as 100 GeV. This
automatically eliminates from the game operators generated through the
exchange of a heavy scalar particle, but of course the presence of light
scalars, below the mentioned limit, renders their neglection unjustified. It
is not clear where simple ETC models violate this limit (see e.g. \cite{mao}%
). We just assume that all scalar particles can be made heavy enough.

Additional light scalars may also appear as pseudo Goldstone bosons at the
moment the electroweak symmetry breaking occurs due to $\bar{Q}Q$
condensation. We had to assume somehow that their contribution to the
oblique correction was small (e.g. by avoiding their proliferation and
making them sufficiently heavy). They also contribute to vertex corrections
(and thus to the $M_{L,R}^{i}$), but here their contribution is naturally
suppressed. The coupling of a pseudo Goldstone boson $\omega $ to ordinary
fermions is of the form
\begin{equation}
\frac{1}{4\pi }\frac{m_{Q}^{2}}{M^{2}}\omega \bar{q}_{L}q_{R},  \label{psse}
\end{equation}
thus their contribution to the $M_{L,R}^{i}$ will be or order
\begin{equation}
g\frac{G^{4}}{(16\pi ^{2})^{2}}(\frac{m_{Q}^{2}}{M^{2}})^{2}\log \frac{%
\Lambda _{\chi }^{2}}{m_{\omega }^{2}}.  \label{contrismall}
\end{equation}
Using the same reference values as above a pseudo Goldstone boson of 100 GeV
can be neglected.

If the operators contained in table \ref{table-3} are not relevant for the $%
W^{\pm }$ and $Z$ couplings, what are they important for? After electroweak
breaking (due to the strong technicolor forces or any other mechanism) a
condensate $\langle \bar{Q}Q\rangle $ emerges. The chirality flipping
operators are then responsible for generating a mass term for ordinary
quarks and leptons. Their low energy effects are contained in the only $d=3$
operator appearing in the matter sector, discussed in section \ref
{theefflangappr}. We thus see that the four fermion approach allows for a
nice separation between the operators responsible for mass generation and
those that may eventually lead to observable consequences in the $W^{\pm }$
and $Z$ couplings. One may even entertain the possibility that the relevant
scale is, for some reason, different for both sets of operators (or, at
least, for some of them). It could, at least in principle, be the case that
scalar exchange enhances the effect of chirality flipping operators,
allowing for large masses for the third generation, without giving
unacceptably large contributions to the $Z$ effective coupling. Whether one
is able to find a satisfactory fundamental theory where this is the case is
another matter, but the four-fermion approach allows, at least, to pose the
problem.

We shall now proceed to determine the constants $M_{L,R}^{i}$ appearing in
the effective Lagrangian after integration of the heavy degrees of freedom.
For the sake of the discussion we shall assume hereafter that technifermions
are degenerate in mass and set their masses equal to $m_{Q}$. The general
case is discussed in appendix \ref{Msectorapp}.\ref{MsectorappE}.

\section{Matching to a fundamental theory (ETC)}

\begin{figure}[!hbp]
\begin{center}
\includegraphics[width=\figwidth]{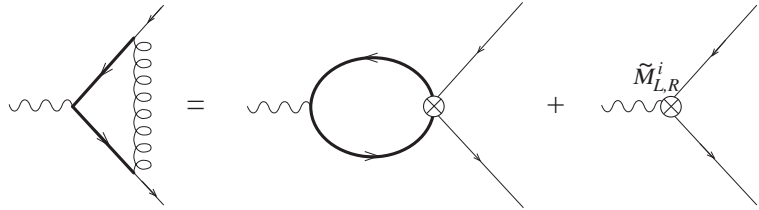}
\end{center}
\caption{The matching at the scale $\protect\mu=M$.}
\label{Figmsector-5}
\end{figure}

At the scale $\mu =M$ we integrate out the heavier degrees of freedom by
matching the renormalized Green functions computed in the underlying
fundamental theory to a four-fermion interaction. This matching leads to the
values (\ref{4qcoef}) for the coefficients of the four-fermion operators as
well as to a purely short distance contribution for the $M_{L,R}^{i}$, which
shall be denoted by $\tilde{M}_{L,R}^{i}$. The matching procedure is
indicated in Fig. \ref{Figmsector-5}.

It is perhaps useful to think of the $\tilde{M}_{L,R}^{i}$ as the value that
the coefficients of the effective Lagrangian take at the matching scale, as
they contain the information on modes of frequencies $\mu >M$. The $\tilde{M}%
_{L,R}^{i}$ will be, in general, divergent, i.e. they will have a pole in $%
1/\epsilon $. Let us see how to obtain these coefficients $\tilde{M}%
_{L,R}^{i}$ in a particular case.

As discussed in the previous section we understand that at very high
energies our theory is described by a gauge theory. Therefore we have to add
to the Standard Model Lagrangian (already extended with technifermions) the
following pieces
\begin{equation}
-\frac{1}{4}E_{\mu \nu }E^{\mu \nu }-\frac{1}{2}M^{2}E_{\mu }E^{\mu }+G\bar{Q%
}\gamma ^{\mu }E_{\mu }q+\mathrm{h.c.}.  \label{Eboson}
\end{equation}
The $E_{\mu }$ vector boson (of mass $M$) acts in a large flavor group space
which mixes ordinary fermions with heavy ones. (The notation in (\ref{Eboson}%
) is somewhat symbolic as we are not implying that the theory is
vector-like, in fact we do not assume anything at all about it.)

At energies $\mu <M$ we can describe the contribution from this sector to
the effective Lagrangian coefficients either using the degrees of freedom
present in (\ref{Eboson}) or via the corresponding four quark operator and a
non-zero value for the $\tilde{M}_{L,R}^{i}$ coefficients. Demanding that
both descriptions reproduce the same renormalized $ffW$ vertex fixes the
value of the $\tilde{M}_{L,R}^{i}$.

Let us see this explicitly in the case where the intermediate vector boson $%
E_{\mu }$ is a $SU(3)_{c}\times SU(2)_{L}$ singlet. For the sake of
simplicity, we take the third term in (\ref{Eboson}) to be
\begin{equation}
G\bar{Q}_{L}\gamma ^{\mu }E_{\mu }q_{L}.  \label{Eboson'}
\end{equation}
At energies below $M$, the relevant four quark operator is then
\begin{equation}
-{\frac{G^{2}}{M^{2}}}(\bar{Q}_{L}\gamma ^{\mu }q_{L})(\bar{q}_{L}\gamma
_{\mu }Q_{L}).  \label{4q-1}
\end{equation}
In the limit of degenerate techniquark masses, it is quite clear that only $%
\tilde{M}_{L}^{1}$ can be different from zero. Thus, one does not need to
worry about matching quark self-energies. Concerning the vertex (Fig. \ref
{Figmsector-5}), we have to impose Eq. (\ref{match-2}), where now
\begin{equation}
\Delta \Gamma \equiv \Gamma _{E}-\Gamma _{4Q}.  \label{match-1}
\end{equation}
Namely, $\Delta \Gamma $ is the difference between the vertex computed using
Eq. (\ref{Eboson}) and the same quantity computed using the four quark
operators as well as non zero $\tilde{M}_{L,R}^{i}$ coefficients (recall
that the hat in Eq. (\ref{match-2}) denotes renormalized quantities). A
calculation analogous to that of section \ref{31.7-1} (now the leading terms
in $1/M^{2}$ are retained) leads to
\begin{equation}
\tilde{M}_{L}^{1}=-{\frac{G^{2}}{16\pi ^{2}}}{\frac{m_{Q}^{2}}{M^{2}}}{\frac{%
1}{\hat{\epsilon}}}.  \label{de-1}
\end{equation}

\section{Integrating out heavy fermions}

\label{heavyfermions}As we move down in energies we can integrate lower and
lower frequencies with the help of the four-fermion operators (which do
accurately describe physics below $M$). This modifies the value of the $%
M_{L,R}^{i}$
\begin{equation}
M_{L,R}^{i}(\mu )=\tilde{M}_{L,R}^{i}+\Delta M_{L,R}^{i}(\mu /M),\qquad \mu
<M.  \label{short}
\end{equation}
The quantity $\Delta M_{L,R}^{i}(\mu /M)$ can be computed in perturbation
theory down to the scale $\Lambda _{\chi }$ where the residual interactions
labelled by the index $A$ becomes strong and confine the technifermions. The
leading contribution is given by a loop of technifermions.

\begin{figure}[!hbp]
\begin{center}
\includegraphics[width=\figwidth]{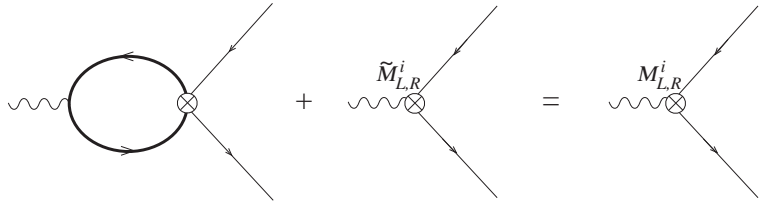}
\end{center}
\caption{Matching at the scale $\protect\mu=\Lambda_\protect\protect\chi$.}
\label{Figmsector-6}
\end{figure}

To determine such contribution it is necessary to demand that the
renormalized Green functions match when computed using explicitly the
degrees of freedom $Q_{L}$, $Q_{R}$ and when their effect is described via
the effective Lagrangian coefficients $M_{L,R}^{i}$. The matching procedure
is illustrated in Fig. \ref{Figmsector-6}.

The scale $\mu $ of the matching must be such that $\mu <M$, but such that $%
\mu >\Lambda _{\chi }$, where perturbation theory in the technicolor
coupling constant starts being questionable.

The result of the calculation in the case of degenerate masses is
\begin{equation}
\Delta M_{L,R}^{i}(\mu /M)=-\bar{M}_{L,R}^{i}\left( 1-\hat{\epsilon}\log {%
\frac{\mu ^{2}}{M^{2}}}\right) ,  \label{logg}
\end{equation}
where we have kept the logarithmically enhanced contribution only and have
neglected any other possible constant pieces. $\bar{M}_{L,R}^{i}$ is the
singular part of $\tilde{M}_{L,R}^{i}$. The finite parts of $\tilde{M}%
_{L,R}^{i}$ are clearly very model dependent (cf for instance the previous
discussion on evanescent operators) and we cannot possibly take them into
account in a general analysis. Accordingly, we ignore all other terms in (%
\ref{logg}) as well as those finite pieces generated through the fierzing
procedure (see discussion in previous section). Keeping the logarithmically
enhanced terms therefore sets the level of accuracy of our calculation. We
will call (\ref{short}) the short-distance contribution to the coefficient $%
M_{L,R}^{i}$. General formulae for the case where the two technifermions are
not degenerate in masses can be found in appendix \ref{Msectorapp}.\ref
{MsectorappE}.

Notice that the final short distance contribution to the $M_{L,R}^{i}$ is
ultraviolet finite, as it should. The divergences in $\tilde{M}_{L,R}^{i}$
are exactly matched by those in $\Delta M_{L,R}^{i}$. The pole in $\tilde{M}%
_{L,R}^{i}$ combined with singularity in $\Delta M_{L,R}^{i}$ provides a
finite contribution.

There is another potential source of corrections to the $M_{L,R}^{i}$
stemming from the renormalization of the four fermion coupling constant $%
G^{2}/M^{2}$ (similar to the renormalization of the Fermi constant in the
electroweak theory due to gluon exchange). This effect is however subleading
here. The reason is that we are considering technigluon exchange only for
four-fermion operators of the form $\mathbf{J}\cdot \mathbf{j}$, where,
again, $\mathbf{j}$ ($\mathbf{J}$) stands for a (heavy) fermion current
(which give the leading contribution, as discussed). The fields carrying
technicolor have the same handedness and thus there is no multiplicative
renormalization and the effect is absent.

Of course in addition to the short distance contribution there is a
long-distance contribution from the region of integration of momenta $\mu
<\Lambda _{\chi }$. Perturbation theory in the technicolor coupling constant
is questionable and we have to resort to other methods to determine the
value of the $M_{L,R}^{i}$ at the $Z$ mass.

There are two possible ways of doing so. One is simply to mimic the
constituent chiral quark model of QCD. There one loop of chiral quarks with
momentum running between the scale of chiral symmetry breaking and the scale
of the constituent mass of the quark, which acts as infrared cut-off,
provide the bulk of the contribution \cite{ERT,AET} to $f_{\pi }$, which is
the equivalent of $v$. Making the necessary translations we can write for
QCD-like theories
\begin{equation}
v^{2}\simeq n_{TC}n_{D}\frac{m_{Q}^{2}}{4\pi ^{2}}\log \frac{\Lambda _{\chi
}^{2}}{m_{Q}^{2}}.  \label{v2}
\end{equation}

Alternatively, we can use chiral Lagrangian techniques \cite{HG} to write a
low-energy bosonized version of the technifermion bilinears $\bar{Q}%
_{L}\Gamma Q_{L}$ and $\bar{Q}_{R}\Gamma Q_{R}$ using the chiral currents $%
\mathbf{J}_{L}$ and $\mathbf{J}_{R}$. The translation is
\begin{eqnarray}
\bar{Q}_{L}\gamma ^{\mu }Q_{L} &\rightarrow &\frac{v^{2}}{2}\mathrm{tr}%
U^{\dagger }iD_{\mu }U \\
\bar{Q}_{L}\gamma ^{\mu }\tau ^{i}Q_{L} &\rightarrow &\frac{v^{2}}{2}\mathrm{%
tr}U^{\dagger }\tau ^{i}iD_{\mu }U \\
\bar{Q}_{R}\gamma ^{\mu }Q_{R} &\rightarrow &\frac{v^{2}}{2}\mathrm{tr}%
UiD_{\mu }U^{\dagger } \\
\bar{Q}_{R}\gamma ^{\mu }\tau ^{i}Q_{R} &\rightarrow &\frac{v^{2}}{2}\mathrm{%
tr}U\tau ^{i}iD_{\mu }U^{\dagger }
\end{eqnarray}
Other currents do not contribute to the effective coefficients. Both methods
agree.

Finally, we collect all contributions to the coefficients $M_{L,R}^{i}$ of
the effective Lagrangian. For fields in the usual representations of the
gauge group
\begin{eqnarray}
2M_{L}^{1} &=&a_{\vec{L}^{2}}\frac{G^{2}}{M^{2}}(v^{2}+n_{TC}n_{D}\frac{%
m_{Q}^{2}}{4\pi ^{2}}\log \frac{M^{2}}{\Lambda _{\chi }^{2}})-\frac{1}{16\pi
^{2}}\frac{y_{u}^{2}+y_{d}^{2}}{4}(\frac{1}{\hat{\epsilon}}-\log \frac{%
\Lambda ^{2}}{\mu ^{2}}), \\
2M_{R}^{1} &=&(a_{\vec{R}^{2}}+\frac{1}{2}a_{R_{3}^{2}})\frac{G^{2}}{M^{2}}%
(v^{2}+n_{TC}n_{D}\frac{m_{Q}^{2}}{4\pi ^{2}}\log \frac{M^{2}}{\Lambda
_{\chi }^{2}})-\frac{1}{16\pi ^{2}}\frac{(y_{u}+y_{d})^{2}}{8}(\frac{1}{\hat{%
\epsilon}}-\log \frac{\Lambda ^{2}}{\mu ^{2}}), \\
M_{L}^{2} &=&\frac{1}{2}a_{R_{3}L}\frac{G^{2}}{M^{2}}(v^{2}+n_{TC}n_{D}\frac{%
m_{Q}^{2}}{4\pi ^{2}}\log \frac{M^{2}}{\Lambda _{\chi }^{2}})+\frac{1}{16\pi
^{2}}\frac{y_{u}^{2}-y_{d}^{2}}{8}(\frac{1}{\hat{\epsilon}}-\log \frac{%
\Lambda ^{2}}{\mu ^{2}}), \\
2M_{L}^{3} &=&0, \\
M_{R}^{2} &=&\frac{1}{2}a_{R_{3}R}\frac{G^{2}}{M^{2}}(v^{2}+n_{TC}n_{D}\frac{%
m_{Q}^{2}}{4\pi ^{2}}\log \frac{M^{2}}{\Lambda _{\chi }^{2}})-\frac{1}{16\pi
^{2}}\frac{y_{u}^{2}-y_{d}^{2}}{8}(\frac{1}{\hat{\epsilon}}-\log \frac{%
\Lambda ^{2}}{\mu ^{2}}), \\
2M_{R}^{3} &=&\frac{1}{2}a_{R_{3}^{2}}\frac{G^{2}}{M^{2}}(v^{2}+n_{TC}n_{D}%
\frac{m_{Q}^{2}}{4\pi ^{2}}\log \frac{M^{2}}{\Lambda _{\chi }^{2}})-\frac{1}{%
16\pi ^{2}}\frac{(y_{u}-y_{d})^{2}}{8}(\frac{1}{\hat{\epsilon}}-\log \frac{%
\Lambda ^{2}}{\mu ^{2}}),
\end{eqnarray}
while in the case of higher representations, where only custodially
preserving operators have been considered, only $M_{L}^{1}$ and $M_{R}^{1}$
get non-zero values (through $a_{{\vec{L}}^{2}}$ and $a_{{\vec{R}}^{2}}$).
The long distance contribution is, obviously, universal (see section \ref
{theefflangappr}), while we have to modify the short distance contribution
by replacing the Casimir of the fundamental representation of $SU(2)$ for
the appropriate one ($1/2\rightarrow c(R)$), the number of doublets by the
multiplicity of the given representation, and $n_{c}$ by the appropriate
dimensionality of the $SU(3)_{c}$ representation to which the $Q$ fields
belong.

These expressions require several comments. First of all, they contain the
same (universal) divergences as their counterparts in the minimal Standard
Model. The scale $\Lambda $ should, in principle, correspond to the matching
scale $\Lambda _{\chi }$, where the low-energy non-linear effective theory
takes over. However, we write an arbitrary scale just to remind us that the
finite part accompanying the log is regulator dependent and cannot be
determined within the effective theory. Recall that the leading $\mathcal{O}%
(n_{TC}n_{D})$ term is finite and unambiguous, and that the ambiguity lies
in the formally subleading term (which, however, due to the log is
numerically quite important). Furthermore only logarithmically enhanced
terms are included in the above expressions. Finally one should bear in mind
that the chiral quark model techniques that we have used are accurate only
in the large $n_{TC}$ expansion (actually $n_{TC}n_{D}$ here). The same
comments apply of course to the oblique coefficients $a_{i}$ presented in
appendix \ref{Msectorapp}.\ref{MsectorappE}.

The quantities $a_{\vec{L}^{2}}$, $a_{\vec{R}^{2}}$, $a_{R_{3}^{2}}$, $%
a_{R_{3}L}$ and $a_{R_{3}R}$ are the coefficients of the four-fermion
operators indicated by the sub-index (a combination of Clebsch-Gordan and
fierzing factors). They depend on the specific model. As discussed in
previous sections these coefficients can be of either sign. This observation
is important because it shows that the contribution to the effective
coefficients has no definite sign \cite{noncomm} indeed. It is nice that
there is almost a one-to-one correspondence between the effective Lagrangian
coefficients (all of them measurable, at least in principle) and
four-fermion coefficients.

Apart from these four-fermion coefficients, the $M_{L,R}^{i}$ depend on a
number of quantities ($v$, $m_{Q}$, $\Lambda _{\chi }$, $G$ and $M$). Let us
first discuss those related to the electroweak symmetry breaking, ($m_{Q}$
and $\Lambda _{\chi }$) and postpone the considerations on $M$ to the next
section ($G$ will be assumed to be of $\mathcal{O}(1)$). $v$ is of course
the Fermi scale and hence not an unknown at all ($v\simeq 250$ GeV). The
value of $m_{Q}$ can be estimated from (\ref{v2}) since $v^{2}$ is known and
$\Lambda _{\chi }$, for QCD-like technicolor theories is $\sim 4\pi v$.
Solving for $m_{Q}$ one finds that if $n_{D}=4$, $m_{Q}\simeq v$, while if $%
n_{D}=1$, $m_{Q}\simeq 2.5v$. Notice that $m_{Q}$ and $v$ depend differently
on $n_{TC}$ so it is not correct to simply assume $m_{Q}\simeq v$. In
theories where the technicolor $\beta $ function is small (and it is pretty
small if $n_{D}=4$ and $n_{TC}=2$) the characteristic scale of the breaking
is pushed upwards, so we expect $\Lambda _{\chi }\gg 4\pi v$. This brings $%
m_{Q}$ somewhat downwards, but the decrease is only logarithmic. We shall
therefore take $m_{Q}$ to be in the range 250 to 450 GeV. We shall allow for
a mass splitting within the doublets too. The splitting within each doublet
cannot be too large, as Fig. \ref{Figmsector-2} shows. For simplicity we
shall assume an equal splitting of masses for all doublets.

\section{Results and discussion}

Let us first summarize our results so far. The values of the effective
Lagrangian coefficients encode the information about the symmetry breaking
sector that is (and will be in the near future) experimentally accessible.
The $M_{L,R}^{i}$ are therefore the counterpart of the oblique corrections
coefficients $a_{i}$ and they have to be taken together in precision
analysis of the Standard Model, even if they are numerically less
significant.

These effective coefficients apply to $Z$-physics at LEP, top production at
the Next Linear Collider, measurements of the top decay at CDF, or indeed
any other process involving the third generation (where their effect is
largest), provided the energy involved is below $4\pi v$, the limit of
applicability of chiral techniques. (Of course chiral effective Lagrangian
techniques fails well below $4\pi v$ if a resonance is present in a given
channel, see also \cite{gd}.)

In the Standard model the $M_{L,R}^{i}$ are useful to keep track of the $%
\log M_{H}$ dependence in all processes involving either neutral or charged
currents. They also provide an economical description of the symmetry
breaking sector, in the sense that they contain the relevant information in
the low-energy regime, the only one testable at present. Beyond the Standard
model the new physics contribution is parametrized by four-fermion
operators. By choosing the number of doublets, $m_{Q}$, $M$, and $\Lambda
_{\chi }$ suitably, we are in fact describing in a single shot a variety of
theories: extended technicolor (commuting and non-commuting), walking
technicolor \cite{walking} or top-assisted technicolor, provided that all
remaining scalars and pseudo-Goldstone bosons are sufficiently heavy.

The accuracy of the calculation is limited by a number of approximations we
have been forced to make and which have been discussed at length in previous
sections. In practice we retain only terms which are logarithmically
enhanced when running from $M$ to $m_{Q}$, including the long distance part,
below $\Lambda _{\chi }$. The effective Lagrangian coefficients $M_{L,R}^{i}$
are all finite at the scale $\Lambda _{\chi }$, the lower limit of
applicability of perturbation theory. Below that scale they run following
the renormalization group equations of the non-linear theory and new
divergences have to be subtracted\footnote{%
The divergent contribution coming from the Standard Model $M_{L,R}^{i}$'s
has to be removed, though, as discussed in section \ref{Zdecayobs}, so the
difference is finite and would be fully predictable, had we good theoretical
control on the subleading corrections. At present only the $\mathcal{O}%
(n_{TC}n_{D})$ contribution is under reasonable control.}. These
coefficients contain finally the contribution from scales $M>\mu >m_{Q}$,
the dynamically generated mass of the technifermion (expected to be of $%
\mathcal{O}(\Lambda _{TC})$). In view of the theoretical uncertainties, to
restrict oneself to logarithmically enhanced terms is a very reasonable
approximation which should capture the bulk of the contribution.

Let us now proceed to a more detailed discussion of the implications of our
analysis. Let us begin by discussing the value that we should take for $M$,
the mass scale normalizing four-fermion operators. Fermion condensation
gives a mass to ordinary fermions via chirality-flipping operators of order
\begin{equation}
m_{f}\simeq \frac{G^{2}}{M^{2}}\langle \bar{Q}Q\rangle ,  \label{mfer}
\end{equation}
through the operators listed in table \ref{table-3}. A chiral quark model
calculation shows that
\begin{equation}
\langle \bar{Q}Q\rangle \simeq v^{2}m_{Q}.
\end{equation}
Thus, while $\langle \bar{Q}Q\rangle $ is universal, there is an inverse
relation between $M^{2}$ and $m_{f}$. In QCD-like theories this leads to the
following rough estimates for the mass $M$ (the subindex refers to the
fermion which has been used in the l.h.s. of (\ref{mfer}))
\begin{equation}
M_{e}\sim 150\mathrm{TeV},\qquad M_{\mu }\sim 10\mathrm{TeV},\qquad
M_{b}\sim 3\mathrm{TeV}.
\end{equation}
If taken at face value, the scale for $M_{b}$ is too low, even the one for $%
M_{\mu }$ may already conflict with current bounds on FCNC, unless they are
suppressed by some other mechanism in a natural way. Worse, the top mass
cannot be reasonably reproduced by this mechanism. This well-known problem
can be partly alleviated in theories where technicolor walks or invoking
top-color or a similar mechanism \cite{topcolor}). Then $M$ can be made
larger and $m_{Q}$, as discussed, somewhat smaller. For theories which are
not vector-like the above estimates become a lot less reliable.

However one should not forget that none of the four-fermion operators
playing a role in the vertex effective couplings participates at all in the
fermion mass determination. In principle we can then entertain the
possibility that the relevant mass scale for the latter should be lower
(perhaps because they get a contribution through scalar exchange, as some of
them can be generated this way). Even in this case it seems just natural
that $M_{b}$ (the scale normalizing chirality preserving operators for the
third generation, that is) is low and not too different from $\Lambda _{\chi
}$. Thus the logarithmic enhancement is pretty much absent in this case and
some of the approximations made become quite questionable in this case.
(Although even for the $b$ couplings there is still a relatively large
contribution to the $M_{L,R}^{i}$'s coming from long distance
contributions.) Put in another words, unless an additional mechanism is
invoked, it is not really possible to make definite estimates for the $b$%
-effective couplings without getting into the details of the underlying
theory. The flavor dynamics and electroweak breaking are completely
entangled in this case. If one only retains the long distance part (which is
what we have done in practice) we can, at best, make order-of-magnitude
estimates. However, what is remarkable in a way is that this does not happen
for the first and second generation vertex corrections. The effect of flavor
dynamics can then be encoded in a small number of coefficients.

We shall now discuss in some detail the numerical consequences of our
assumptions. We shall assume the above values for the mass scale $M$; in
other words, we shall place ourselves in the most disfavored situation. We
shall only present results for QCD-like theories and $n_{D}=4$ exclusively.
For other theories the appropriate results can be very easily obtained from
our formulae. For the coefficients $a_{\vec{L}^{2}}$, $a_{R_{3}R}$, $%
a_{R_{3}L}$, etc. we shall use the range of variation [-2, 2] (since they
are expected to be of $\mathcal{O}(1)$). Of course larger values of the
scale, $M$, would simply translate into smaller values for those
coefficients, so the results can be easily scaled down.

\begin{figure}[!hbp]
\begin{center}
\includegraphics[width=12cm]{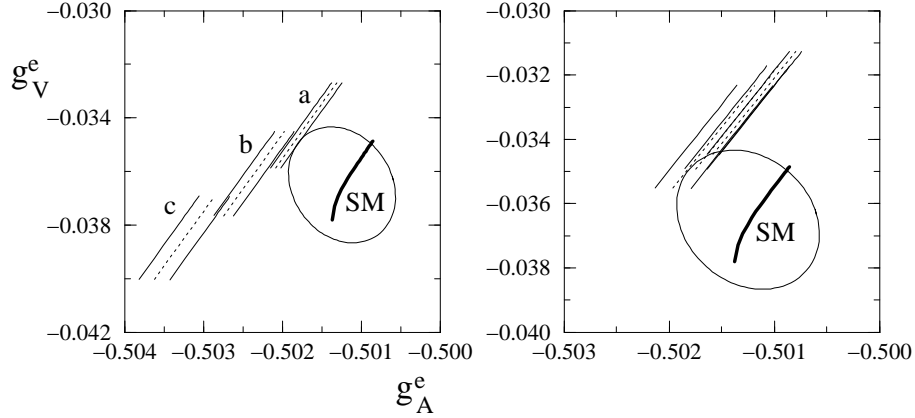}
\end{center}
\caption{Oblique and vertex corrections for the electron effective
couplings. The elipse indicate the 1-$\protect\sigma$ experimental region.
Three values of the effective mass $m_2$ are considered: 250 (a), 350 (b)
and 450 GeV (c), and two splittings: 10\% (right) and 20\% (left). The
dotted lines correspond to including the oblique corrections only. The
coefficients of the four-fermion operators vary in the range [-2,2] and this
spans the region between the two solid lines. The Standard Model prediction
(thick solid line) is shown for $m_t=175.6$ GeV and $70\le M_H\le 1500$ GeV.}
\label{Figmsector-7}
\end{figure}

Fig. \ref{Figmsector-7} shows the $g_{A}^{e},g_{V}^{e}$ electron effective
couplings when vertex corrections are included and allowed to vary within
the stated limits. To avoid clutter, the top mass is taken to the central
value 175.6 GeV. The Standard Model prediction is shown as a function of the
Higgs mass. The dotted lines in Fig. \ref{Figmsector-7} correspond to
considering the oblique corrections only. Vertex corrections change these
results and, depending on the values of the four-fermion operator
coefficients, the prediction can take any value in the strip limited by the
two solid lines (as usual we have no specific prediction in the direction
along the strip due to the dependence on $\Lambda $, inherited from the
non-renormalizable character of the effective theory). A generic
modification of the electron couplings is of $\mathcal{O}(10^{-5})$, small
but much larger than in the Standard Model and, depending on its sign, may
help to bring a better agreement with the central value.

The modifications are more dramatic in the case of the second generation,
for the muon, for instance. Now, we expect changes in the $M_{L,R}^{i}$'s
and, eventually, in the effective couplings of $\mathcal{O}(10^{-3})$ These
modifications are just at the limit of being observable. They could even
modify the relation between $M_{W}$ and $G_{\mu }$ (i.e. $\Delta r$).

\begin{figure}[!hbp]
\begin{center}
\includegraphics[width=8cm]{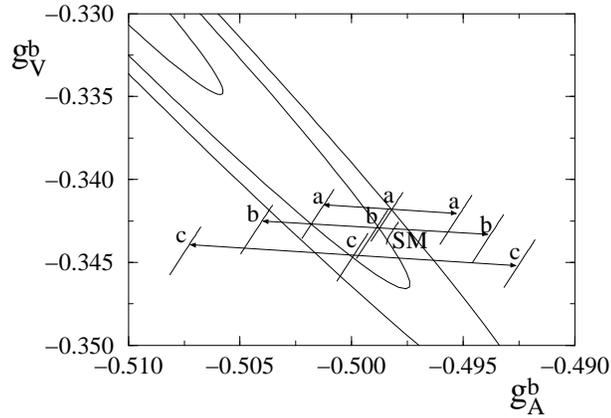}
\end{center}
\caption{Bottom effective couplings compared to the SM prediction for $%
m_t=175.6$ as a function of the Higgs mass (in the range [70,1500] GeV). The
elipses indicate 1, 2, and 3-$\protect\sigma$ experimental regions. The
dynamically generated masses are 250 (a), 350 (b) and 400 GeV (c) and we
show a 20\% splitting between the masses in the heavy doublet. The
degenerate case does not present quantitative differencies if we consider
the experimental errors. The central lines correspond to including only the
oblique corrections. When we include the vertex corrections (depending on
the size of the four-fermion coefficients) we predict the regions between
lines indicated by the arrows. The four-fermion coefficients in this case
take values in the range [-0.1,0.1].}
\label{Figmsector-8}
\end{figure}

Fig. \ref{Figmsector-8} shows a similar plot for the bottom effective
couplings $g_{A}^{b},g_{V}^{b}$. It is obvious that taking generic values
for the four-fermion operators (of $\mathcal{O}(1)$) leads to enormous
modifications in the effective couplings, unacceptably large in fact. The
corrections become more manageable if we allow for a smaller variation of
the four-fermion operator coefficients (in the range [-0.1,0.1]). This
suggests that the natural order of magnitude for the mass $M_{b}$ is $\sim
10 $ TeV, at least for chirality preserving operators. As we have discussed
the corrections can be of either sign.

One could, at least in the case of degenerate masses, translate the
experimental constraints on the $M_{L,R}^{i}$ (recall that their
experimental determination requires a combination of charged and neutral
processes, since there are six of them) to the coefficients of the
four-fermion operators. Doing so would provide us with a four-fermion
effective theory that would exactly reproduce all the available data. It is
obvious however that the result would not be very satisfactory. While the
outcome would, most likely, be coefficients of $\mathcal{O}(1)$ for the
electron couplings, they would have to be of $\mathcal{O}(10^{-1})$, perhaps
smaller for the bottom. Worse, the same masses we have used lead to
unacceptably low values for the top mass (\ref{mfer}). Allowing for a
different scale in the chirality flipping operators would permit a large top
mass without affecting the effective couplings. Taking this as a tentative
possibility we can pose the following problem: measure the effective
couplings $M_{L,R}^{i}$ for all three generations and determine the values
of the four-fermion operator coefficients and the characteristic mass scale
that fits the data best. In the degenerate mass limit we have a total of 8
unknowns (5 of them coefficients, expected to be of $\mathcal{O}(1)$) and 18
experimental values (three sets of the $M_{L,R}^{i}$). A similar exercise
could be attempted in the chirality flipping sector. If the solution to this
exercise turned out to be mathematically consistent (within the experimental
errors) it would be extremely interesting. A negative result would certainly
rule out this approach. Notice that dynamical symmetry breaking predicts the
pattern $M_{L,R}^{i}\sim m_{f}$, while in the Standard Model $%
M_{L,R}^{i}\sim m_{f}^{2}$.

We should end with some words of self-criticism. It may seem that the
previous discussion is not too conclusive and that we have managed only to
rephrase some of the long-standing problems in the symmetry breaking sector.
However, the \textit{raison d'\^{e}tre} of the present work is not really to
propose a solution to these problems, but rather to establish a theoretical
framework to treat them systematically. Experience from the past shows that
often the effects of new physics are magnified and thus models are ruled out
on this basis, only to find out that a careful and rigorous analysis leaves
some room for them. We believe that this may be the case in dynamical
symmetry breaking models and we believe too that only through a detailed and
careful comparison with the experimental data will progress take place.

The effective Lagrangian provides the tools to look for an `existence proof'
(or otherwise) of a phenomenologically viable, mathematically consistent
dynamical symmetry breaking model. We hope that there is any time soon
sufficient experimental data to attempt to determine the four-fermion
coefficients, at lest approximately.

\chapter{$CP$ violation and mixing}

\label{cpviolationandmixing}One of the pressing open problems in particle
physics is to understand the origin of $CP$ violation and family mixing. In
the minimal Standard Model we have only two possible sources of $CP$
violation, one is the strong $CP$ phase controlling the gluonic topological
term and the other is the single phase present in the so-called
Cabibbo-Kobayashi-Maskawa (CKM) mixing matrix (here denoted $K$) in the
electroweak sector$.$ In this chapter we will deal only with the electroweak
sector even though both sectors are related \cite{Donoghue}.

Our main purpose here is to parametrize all possible sources of $CP$
violation and family mixing that may arise in the electroweak sector when
considering new physics beyond the SM using an effective Lagrangian
approach. Like in the previous chapter we consider only leading
four-dimensional operators keeping all fields of the SM, except the not yet
observed Higgs field. We start with a general classification of
four-dimensional operators respecting the $SU(3)_{c}\times SU(2)_{L}\times
U(1)_{Y}$ gauge symmetry with matter and gauge fields in the standard
representations. This classification includes non-diagonal kinetic and mass
terms along with Appelquist et al. effective operators. We perform a
diagonalization in section \ref{physical} showing that, besides the presence
of the CKM matrix in the SM charged vertex, new structures show up in
effective operators constructed with left handed fermions. The rest of the
chapter is dedicated to the study of the contribution of the effective
operators to the physical observables in the neutral and charged vertices.
Care is taken to ensure that all contributions to the observable quantities
are taken into account, including wave function and CKM renormalization that
are present even at tree level. The final section is devoted to the analysis
of the SM supplemented with an additional heavy fermion doublet and the case
of the SM with a heavy Higgs.

\section{Effective Lagrangian and $CP$ violation}

\label{efflag}Let us first state the assumptions behind the present
framework. We shall assume that the scale of any new physics beyond the
standard model is sufficiently high so that an inverse mass expansion is
granted, and we shall organize the effective Lagrangian accordingly. We
shall also assume that the Higgs field either does not exist or is massive
enough to permit an effective Lagrangian treatment by expanding in inverse
powers of its mass, $M_{H}$. In short, we assume that all as yet undetected
new particles are heavy, with a mass much larger than the energy scale at
which the effective Lagrangian is to be used. Thus it is natural to use a
non-linear realization of the $SU(2)_{L}\times U(1)$ symmetry where the
unphysical scalar fields are collected in a unitary $2\times 2$ matrix $U(x)$
(see e.g. \cite{eff-lag}).

An additional assumption that we shall later make is that whatever is the
source of $CP$-violation beyond the Standard Model, when compared to the $CP$
conserving part, is `small'. This statement does need qualification. What we
actually mean is that \emph{observable} $CP$ violating deviations must me
small. This does not mean that $CP$-violating operators are always
suppressed. We can have the situation where lots of $CP$ violating phases
appear or disappear when we pass to the physical basis. Since \emph{this
basis} is the most directly related one to observable quantities it is the
chosen one to make a qualitative and quantitative analysis of $CP$ violating
effects. However, we always have to remember that observable quantities
include in their calculation renormalization contributions along with finite
renormalizations of the involved external fields which generically alter the
weight of the different $CP$ violating operators as we shall see.

Let us commence our classification of the operators present in the matter
sector of the effective electroweak Lagrangian. We shall use the following
projectors
\begin{equation}
R=\frac{1+\gamma ^{5}}{2},\qquad L=\frac{1+\gamma ^{5}}{2}\qquad \tau ^{u}=%
\frac{1+\tau ^{3}}{2},\qquad \tau ^{d}=\frac{1-\tau ^{3}}{2},  \label{eqa3}
\end{equation}
where $R$ is the right projector and $L$ the left projector in chirality
space, and $\tau ^{u}$ is the up projector and $\tau ^{d}$ the down
projector in $SU(2)$ space. The different gauge groups act on the scalar, $%
U(x)$, and fermionic, $f_{L}(x),f_{R}(x)$, fields in the following way
\begin{eqnarray}
D_{\mu }U &=&\partial _{\mu }U+ig\frac{\tau }{2}\cdot W_{\mu }U-ig^{\prime }U%
\frac{\tau ^{3}}{2}B_{\mu },  \notag \\
D_{\mu }^{L}f_{L} &=&\left[ \partial _{\mu }+ig\frac{\tau }{2}\cdot W_{\mu
}+ig^{\prime }\left( Q-\frac{\tau ^{3}}{2}\right) B_{\mu }+ig_{s}\frac{%
\lambda }{2}\cdot G_{\mu }\right] f_{L},  \notag \\
D_{\mu }^{R}f_{R} &=&\left[ \partial _{\mu }+ig^{\prime }QB_{\mu }+ig_{s}%
\frac{\lambda }{2}\cdot G_{\mu }\right] f_{R},  \label{deriv1}
\end{eqnarray}
with
\begin{equation*}
Q=\left\{
\begin{array}{ccc}
\frac{\tau ^{3}}{2}+\frac{1}{6} &  & \mathrm{quarks} \\
\frac{\tau ^{3}}{2}-\frac{1}{2} &  & \mathrm{leptons}
\end{array}
\right. .
\end{equation*}

The following terms are universal. They must be present in any effective
theory whose long-distance properties are those of the Standard Model. They
correspond to the Standard Model kinetic and mass terms (we use the notation
$\mathrm{f}$ to describe both left and right degrees of freedom
simultaneously)
\begin{eqnarray}
\mathcal{L}_{kin}^{L} &=&i\mathrm{\bar{f}}X_{L}\gamma ^{\mu }D_{\mu }^{L}L%
\mathrm{f},  \notag \\
\mathcal{L}_{kin}^{R} &=&i\mathrm{\bar{f}}\left( \tau ^{u}X_{R}^{u}+\tau
^{d}X_{R}^{d}\right) \gamma ^{\mu }D_{\mu }^{R}R\mathrm{f},  \notag \\
\mathcal{L}_{m} &=&-\mathrm{\bar{f}}\left( U\left( \tau ^{u}\tilde{y}%
_{u}^{f}+\tau ^{d}\tilde{y}_{d}^{f}\right) R+\left( \tau ^{u}\tilde{y}%
_{u}^{f\dagger }+\tau ^{d}\tilde{y}_{d}^{f\dagger }\right) U^{\dagger
}L\right) \mathrm{f},  \label{lm}
\end{eqnarray}
where $X_{L}$, $X_{R}^{u}$ and $X_{R}^{d}$ are non-singular Hermitian
matrices having only family indices, and $\tilde{y}_{u}^{f}$ and $\tilde{y}%
_{d}^{f}$ are arbitrary matrices and have only family indices too. Note that
in general $X_{R}^{d}\neq X_{R}^{u}$ and therefore the operator $\mathcal{L}%
_{R}^{\prime }$ presented in the previous chapter is automatically
incorporated in these terms . In the Standard Model, these matrices can
always be reabsorbed by an appropriate redefinition of the fields (we shall
see this explicitly later), so one does not even contemplate the possibility
that left and right `kinetic' terms are differently normalized, but this is
perfectly possible in an effective theory, and the transformations required
to bring these kinetic terms to the standard form do leave some fingerprints.

In order to write the above terms in the familiar form in the Standard Model
we shall perform a series of chiral changes of variables. In general, due to
the axial anomaly, these changes will modify the topological $CP$ violating $%
\theta $-terms
\begin{equation}
\mathcal{L}_{\theta }=\epsilon ^{\alpha \beta \mu \nu }\left( \theta
_{1}B_{\alpha \beta }B_{\mu \nu }+\theta _{2}W_{\alpha \beta }^{a}W_{\mu \nu
}^{a}+\theta _{3}G_{\alpha \beta }^{a}G_{\mu \nu }^{a}\right) .  \label{eqa4}
\end{equation}
From those terms only the gluonic $\theta $-term is observable \cite
{Anselm:1994uj} but we will not deal here with this issue.

Notice the appearance of the unitary matrix $U$ collecting the (unphysical)
Goldstone bosons. The Higgs field ---as emphasized above--- should it exist,
has been integrated out. Since the global symmetries are non-linearly
realized the above Lagrangian is non-renormalizable and additional operators
are required to absorb the additional divergences which are generated due to
the non-linear nature of the theory.

In addition to (\ref{lm}) a number of operators of dimension four should be
included in the matter sector of the effective electroweak Lagrangian. They
are, to begin with, necessary as counterterms to remove some ultraviolet
divergences that appear at the quantum level due to the non-linear nature of
(\ref{lm}). Moreover, physics beyond the Standard Model does in general
contribute to the coefficients of those operators, as it may do to $X_{L}$, $%
X_{Ru}$ $X_{Rd}$, $\tilde{y}_{u}$ and $\tilde{y}_{d}$. The dimension 4
operators can be written generically as
\begin{eqnarray}
\mathcal{L}_{L} &=&\mathrm{\bar{f}}\gamma _{\mu }M_{L}O_{L}^{\mu }L\mathrm{f}%
+h.c.,  \notag \\
\mathcal{L}_{R} &=&\mathrm{\bar{f}}\gamma _{\mu }M_{R}O_{R}^{\mu }R\mathrm{f}%
+h.c.,  \label{eqa5}
\end{eqnarray}
where $M_{L}$ and $M_{R}$ are matrices having family indices only and $%
O_{L}^{\mu }$ and $O_{R}^{\mu }$ are operators of dimension one having weak
indices (u,d) only. These operators were first written by \cite{eff-lag} in
the case where mixing between families is absent and they have been recently
considered in \cite{nomix}. The extension to the three-generation case is
new and presented here.

The complete list of the dimension four operators is
\begin{eqnarray}
\mathcal{L}_{L}^{1} &=&i\mathrm{\bar{f}}M_{L}^{1}\gamma ^{\mu }U\left(
D_{\mu }U\right) ^{\dagger }L\mathrm{f}+h.c.,  \notag \\
\mathcal{L}_{L}^{2} &=&i\mathrm{\bar{f}}M_{L}^{2}\gamma ^{\mu }\left( D_{\mu
}U\right) \tau ^{3}U^{\dagger }L\mathrm{f}+h.c.  \notag \\
\mathcal{L}_{L}^{3} &=&i\mathrm{\bar{f}}M_{L}^{3}\gamma ^{\mu }U\tau
^{3}U^{\dagger }\left( D_{\mu }U\right) \tau ^{3}U^{\dagger }L\mathrm{f}%
+h.c.,  \notag \\
\mathcal{L}_{L}^{4} &=&i\mathrm{\bar{f}}M_{L}^{4}\gamma ^{\mu }U\tau
^{3}U^{\dagger }D_{\mu }^{L}L\mathrm{f}+h.c..  \notag \\
\mathcal{L}_{R}^{1} &=&i\mathrm{\bar{f}}M_{R}^{1}\gamma ^{\mu }U^{\dagger
}\left( D_{\mu }U\right) R\mathrm{f}+h.c.,  \notag \\
\mathcal{L}_{R}^{2} &=&i\mathrm{\bar{f}}M_{R}^{2}\gamma ^{\mu }\tau
^{3}U^{\dagger }\left( D_{\mu }U\right) R\mathrm{f}+h.c.,  \notag \\
\mathcal{L}_{R}^{3} &=&i\mathrm{\bar{f}}M_{R}^{3}\gamma ^{\mu }\tau
^{3}U^{\dagger }\left( D_{\mu }U\right) \tau ^{3}R\mathrm{f}+h.c.,
\label{effec}
\end{eqnarray}
Without any loss of generality we take the matrices in family space $%
M_{L}^{1}$, $M_{R}^{1}$, $M_{L}^{3}$ and $M_{R}^{3}$ Hermitian, while $%
M_{L}^{2}$, $M_{R}^{2}$ and $M_{L}^{4}$ are completely general. If we
require the above operators to be $CP$ conserving, the matrices $M_{LR}^{i}$
must be real (see section \ref{cptranssection}).

In addition to the above ones, physics beyond the Standard Model generates,
in general, an infinite tower of higher-dimensional operators with $d\geq 5$
(these operators are eventually required as counterterms too due to the
non-linear nature of the Lagrangian (\ref{lm}) ). On dimensional grounds
these operators shall be suppressed by powers of the scale $\Lambda $
characterizing new physics or by powers of $4\pi v$ ($v$ being the scale of
the breaking ---250 GeV). Therefore, if the scale of new physics is
sufficiently high the contribution of higher dimensional operators can be
neglected as compared to those of $d=4$. Of course for this to be true the
later must be non-vanishing and sizeable. Thanks to the violation of the
Appelquist-Carazzone decoupling theorem\cite{AC} in spontaneously broken
theories, this is generically the case, unless the new physics is tuned so
as to be decoupling as is the case in the minimal supersymmetric Standard
Model. Our results do not apply in this case (see e.g. \cite{DHP} for a
recent discussion on this matter).

\section{Passage to the physical basis}

\label{physical}Let us first consider the operators which are already
present in the Standard Model, Eq. (\ref{lm}). The diagonalization and
passage to the physical basis are of course well known, but some
modifications are required when one considers the general case in (\ref{lm})
so it is worth going through the discussion with some detail.

We perform first the unitary change of variables
\begin{equation}
\mathrm{f}=\left[ \tilde{V}_{L}L+\left( \tilde{V}_{Ru}\tau ^{u}+\tilde{V}%
_{Rd}\tau ^{d}\right) R\right] \mathrm{f},  \label{1}
\end{equation}
with the help of the unitary matrices $\tilde{V}_{L}$ ,$\tilde{V}_{Ru}$ and $%
\tilde{V}_{Rd}$. Hence
\begin{equation}
\left( \tilde{y}_{u}^{f}\tau ^{u}+\tilde{y}_{d}^{f}\tau ^{d}\right)
\rightarrow \left( \tilde{V}_{L}^{\dagger }\tilde{y}_{u}^{f}\tilde{V}%
_{Ru}\tau ^{u}+\tilde{V}_{L}^{\dagger }\tilde{y}_{d}^{f}\tilde{V}_{Rd}\tau
^{d}\right) ,  \label{eqa6}
\end{equation}
and
\begin{eqnarray}
X_{L} &\rightarrow &\tilde{V}_{L}^{\dagger }X_{L}\tilde{V}_{L}=D^{L},  \notag
\\
X_{R}^{u} &\rightarrow &\tilde{V}_{Ru}^{\dagger }X_{R}^{u}\tilde{V}%
_{Ru}=D_{u}^{R},  \notag \\
X_{R}^{d} &\rightarrow &\tilde{V}_{Rd}^{\dagger }X_{R}^{d}\tilde{V}%
_{Rd}=D_{d}^{R},  \label{eqa8}
\end{eqnarray}
where $D^{L}$ $D_{u}^{R}$ and $D_{d}^{R}$ are diagonal matrices with
eigenvalues different from zero. Then, with the help of a non-unitary
transformation
\begin{equation}
\mathrm{f}\rightarrow \left[ \left( D^{L}\right) ^{-\frac{1}{2}}L+\left(
\left( D_{u}^{R}\right) ^{-\frac{1}{2}}\tau ^{u}+\left( D_{d}^{R}\right) ^{-%
\frac{1}{2}}\tau ^{d}\right) R\right] \mathrm{f},  \label{2}
\end{equation}
we obtain
\begin{eqnarray}
D^{L} &\rightarrow &\left( \left( D^{L}\right) ^{-\frac{1}{2}}\right) ^{\ast
}D^{L}\left( D^{L}\right) ^{-\frac{1}{2}}=I,  \notag \\
D_{u}^{R} &\rightarrow &\left( \left( D_{u}^{R}\right) ^{-\frac{1}{2}%
}\right) ^{\ast }D_{u}^{R}\left( D_{u}^{R}\right) ^{-\frac{1}{2}}=I,  \notag
\\
D_{d}^{R} &\rightarrow &\left( \left( D_{d}^{R}\right) ^{-\frac{1}{2}%
}\right) ^{\ast }D_{d}^{R}\left( D_{d}^{R}\right) ^{-\frac{1}{2}}=I,
\label{eqa9}
\end{eqnarray}
and the matrix $\tilde{y}_{u}^{f}\tau ^{u}+\tilde{y}_{d}^{f}\tau ^{d}$
transforms into
\begin{equation}
\left( \left( D^{L}\right) ^{-\frac{1}{2}}\right) ^{\ast }\tilde{V}%
_{L}^{\dagger }\tilde{y}_{u}^{f}\tilde{V}_{Ru}\left( D_{u}^{R}\right) ^{-%
\frac{1}{2}}\tau ^{u}+\left( \left( D^{L}\right) ^{-\frac{1}{2}}\right)
^{\ast }\tilde{V}_{L}^{\dagger }\tilde{y}_{d}^{f}\tilde{V}_{Rd}\left(
D_{d}^{R}\right) ^{-\frac{1}{2}}\tau ^{d}\equiv y_{u}^{f}\tau
^{u}+y_{d}^{f}\tau ^{d},  \label{eqa10}
\end{equation}
where $y_{u}^{f}$ and $y_{d}^{f}$ are the Yukawa couplings. Thus, the left
and right kinetic terms can be brought to the canonical form at the sole
expense of redefining the Yukawa couplings. Since this is all there is in
the Standard Model, we see that the effect of considering more general
coefficients for the right-handed kinetic terms is irrelevant. This will not
be the case when additional operators are considered. Fermions transform, up
to this point, in irreducible representations of the gauge group.

We now perform the unitary change of variables
\begin{equation}
\mathrm{f}\rightarrow \left[ \left( V_{Lu}\tau ^{u}+V_{Ld}\tau ^{d}\right)
L+\left( V_{Ru}\tau ^{u}+V_{Rd}\tau ^{d}\right) R\right] \mathrm{f},
\label{3}
\end{equation}
with unitary matrices $V_{Lu}$, $V_{Ru}$, $V_{Ld}$ and $V_{Rd}$ and having
family indices only. They are chosen so that the Yukawa terms become
diagonal and definite positive (see \cite{peskin})
\begin{equation}
\left( V_{Lu}^{\dagger }\tau ^{u}+V_{Ld}^{\dagger }\tau ^{d}\right) \left(
y_{u}^{f}\tau ^{u}+y_{d}^{f}\tau ^{d}\right) \left( V_{Ru}\tau
^{u}+V_{Rd}\tau ^{d}\right) =d_{u}^{f}\tau ^{u}+d_{d}^{f}\tau ^{d}.
\label{eqa11}
\end{equation}
After all these transformations $\mathcal{L}_{m}$ transforms into
\begin{equation}
\mathcal{L}_{m}=-\mathrm{\bar{f}}\left\{ \left( \tau ^{u}U+K^{\dagger }\tau
^{d}U\right) \tau ^{u}d_{u}^{f}+\left( \tau ^{d}U+K\tau ^{u}U\right) \tau
^{d}d_{d}^{f}\right\} R\mathrm{f}+h.c.,  \label{6}
\end{equation}
where $K\equiv V_{Lu}^{\dagger }V_{Ld}$ is well known
Cabibbo-Kobayashi-Maskawa matrix. Note in Eq. (\ref{6}) that when we set $%
U=I $ we obtain
\begin{equation*}
\mathcal{L}_{m}\rightarrow -\mathrm{\bar{f}}\left( \tau ^{u}d_{u}^{f}+\tau
^{d}d_{d}^{f}\right) R\mathrm{f}+h.c.,
\end{equation*}
which is a diagonal mass term. Fermions now transform in reducible
representations of the gauge group.

The left and right kinetic terms now read
\begin{equation}
\mathcal{L}_{kin}^{R}=i\mathrm{\bar{f}}\gamma ^{\mu }D_{\mu }^{R}R\mathrm{f},
\label{eqa12}
\end{equation}
and
\begin{eqnarray}
\mathcal{L}_{kin}^{L} &=&i\mathrm{\bar{f}}\gamma ^{\mu }L\left\{ \partial
_{\mu }+ig^{\prime }\left( Q-\frac{\tau ^{3}}{2}\right) B_{\mu }+ig\frac{%
\tau ^{3}}{2}W_{\mu }^{3}\right.  \notag \\
&&\left. +ig\left( K\frac{\tau ^{-}}{2}W_{\mu }^{+}+K^{\dagger }\frac{\tau
^{+}}{2}W_{\mu }^{-}\right) +ig_{s}\frac{\lambda }{2}\cdot G_{\mu }\right\}
\mathrm{f}.  \label{eqa13}
\end{eqnarray}
$CP$ violation is obtained if and only if $K\neq K^{\ast }$. In total the SM
kinetic term is written using physical gauge bosons is given by
\begin{eqnarray}
\mathcal{L}_{kin} &=&\mathcal{L}_{kin}^{L}+\mathcal{L}_{kin}^{R}=i\bar{f}%
\gamma ^{\mu }\left\{ \partial _{\mu }+ig_{s}\frac{\mathbf{\lambda }}{2}%
\mathbf{\cdot G}_{\mu }+ieQA_{\mu }\right.  \notag \\
&&+\left. \frac{ie}{s_{W}c_{W}}\left[ \left( \frac{\tau ^{3}}{2}%
-Qs_{W}^{2}\right) L-Qs_{W}^{2}R\right] Z_{\mu }\right\} f  \notag \\
&&-\frac{e}{\sqrt{2}s_{W}}\left[ \bar{u}\gamma ^{\mu }KW_{\mu }^{+}Ld+\bar{d}%
\gamma ^{\mu }K^{\dagger }W_{\mu }^{-}Lu\right] .  \label{SMkin}
\end{eqnarray}
As is well known, some freedom for additional phase redefinitions is left.
If we make the replacement
\begin{equation}
\mathrm{f}\rightarrow \left[ \left( W_{Lu}\tau ^{u}+W_{Ld}\tau ^{d}\right)
L+\left( W_{Ru}\tau ^{u}+W_{Rd}\tau ^{d}\right) R\right] \mathrm{f},
\label{eqa14}
\end{equation}
we have to change
\begin{equation}
K=V_{Lu}^{\dagger }V_{Ld}\rightarrow W_{Lu}^{\dagger }V_{Lu}^{\dagger
}V_{Ld}W_{Ld}=W_{Lu}^{\dagger }KW_{Ld},  \label{eqa15}
\end{equation}
and
\begin{eqnarray}
d_{u} &=&V_{Lu}^{\dagger }y_{u}^{f}V_{Ru}\rightarrow W_{Lu}^{\dagger
}V_{Lu}^{\dagger }y_{u}^{f}V_{Ru}W_{Ru}=W_{Lu}^{\dagger }d_{u}^{f}W_{Ru},
\notag \\
d_{d} &=&V_{Ld}^{\dagger }y_{d}^{f}V_{Rd}\rightarrow W_{Ld}^{\dagger
}V_{Ld}^{\dagger }y_{d}^{f}V_{Rd}W_{Rd}=W_{Ld}^{\dagger }d_{d}^{f}W_{Rd},
\label{eqa16}
\end{eqnarray}
but if we want to keep $d_{u}^{f}$ and $d_{d}^{f}$ diagonal real and
definite positive, and if we suppose that they do not have degenerate
eigenstates the only possibility for the unitary matrices $W$ is to be
diagonal. This freedom can be used, for example, to extract five phases from
$K$. After this no further redefinitions are possible neither in the left
nor in the right handed sector. Henceforth, without any loss of generality,
we will absorb matrices $W$ in the definition of matrices $V$.

So much for the Standard Model. Let us now move to the more general case
represented at low energies by the $d=4$ operators listed in the previous
section. We have to analyze the effect of the transformations given by Eqs. (%
\ref{1}) (\ref{2}) and (\ref{3}) on the operators (\ref{effec}). The
composition of those transformations is given by
\begin{eqnarray}
\mathrm{f} &\rightarrow &\tilde{V}_{L}\left( D^{L}\right) ^{\frac{-1}{2}%
}\left( V_{Lu}\tau ^{u}+V_{Ld}\tau ^{d}\right) L\mathrm{f}  \notag \\
&&+\left( \tilde{V}_{Ru}\left( D_{u}^{R}\right) ^{\frac{-1}{2}}V_{Ru}\tau
^{u}+\tilde{V}_{Rd}\left( D_{d}^{R}\right) ^{\frac{-1}{2}}V_{Rd}\tau
^{d}\right) R\mathrm{f}  \notag \\
&\equiv &\left( C_{L}^{u}\tau ^{u}+C_{L}^{d}\tau ^{d}\right) L\mathrm{f}%
+\left( C_{R}^{u}\tau ^{u}+C_{R}^{d}\tau ^{d}\right) R\mathrm{f},
\label{totalt}
\end{eqnarray}
Note that because of the presence of matrices $D$, matrices $C$ are in
general non-unitary. We begin with the effective operators involving left
handed fields. In this case when we perform transformation (\ref{totalt}) we
obtain
\begin{equation}
\mathcal{L}_{L}\rightarrow \mathrm{\bar{f}}\gamma _{\mu }Q_{L}^{\mu }L%
\mathrm{f}+h.c.,  \label{eqa 17}
\end{equation}
with the operator $Q_{L}^{\mu }$ containing family and weak indices given by
\begin{eqnarray}
Q_{L}^{\mu } &=&\hat{M}_{L}\tau ^{u}O_{L}^{\mu }\tau ^{u}+\hat{M}_{L}K\tau
^{u}O_{L}^{\mu }\tau ^{d}  \notag \\
&&+K^{\dagger }\hat{M}_{L}K\tau ^{d}O_{L}^{\mu }\tau ^{d}+K^{\dagger }\hat{M}%
_{L}\tau ^{d}O_{L}^{\mu }\tau ^{u},  \label{7}
\end{eqnarray}
where we have defined
\begin{equation}
\hat{M}_{L}\equiv C_{L}^{u\dagger }M_{L}C_{L}^{u}.  \label{eqa18}
\end{equation}
Thus new structures do appear involving the CKM matrix $K$ and left-handed
fields. The former cannot be reduced to our starting set of operators by a
simple redefinition of the original couplings $M_{L}$.

The case of the effective operators involving right handed fields ($\mathcal{%
L}_{R}$) is, in this sense, simpler because transformation (\ref{totalt})
only redefine the matrices $M_{R}$. The operators involving right-handed
fields are $\mathcal{L}_{R}^{1}$, $\mathcal{L}_{R}^{2}$, and $\mathcal{L}%
_{R}^{3}$ and can be written generically as (see next section)
\begin{equation}
\mathcal{L}_{R}^{p}=i\mathrm{\bar{f}}\gamma _{\mu }M_{R}^{p}O_{p}^{\mu }R%
\mathrm{f}+h.c.,  \label{eqa19}
\end{equation}
with
\begin{equation}
O_{1}^{\mu }=U^{\dagger }\left( D_{\mu }U\right) ,\qquad O_{2}^{\mu }=\tau
^{3}U^{\dagger }\left( D_{\mu }U\right) ,\qquad O_{3}^{\mu }=\tau
^{3}U^{\dagger }\left( D_{\mu }U\right) \tau ^{3}.  \label{eqa20}
\end{equation}
Note that because of the $h.c.$ in $\mathcal{L}_{R}^{p}$ we can change $%
O_{2}^{\mu }$ by $U^{\dagger }\left( D_{\mu }U\right) \tau ^{3}$ along with $%
M_{R}^{2}$ by $M_{R}^{2\dagger }$. So under the transformation (\ref{totalt}%
) we obtain
\begin{equation*}
\mathcal{L}_{R}^{p}\rightarrow i\mathrm{\bar{f}}\gamma _{\mu }Q_{pR}^{\mu }R%
\mathrm{f}+h.c.,
\end{equation*}
with the operators $Q_{pR}^{\mu }$ containing family and weak indices given
by
\begin{eqnarray}
Q_{pR}^{\mu } &=&C_{R}^{u\dagger }M_{R}^{p}C_{R}^{u}\tau ^{u}O_{p}^{\mu
}\tau ^{u}+C_{R}^{u\dagger }M_{R}^{p}C_{R}^{d}\tau ^{u}O_{p}^{\mu }\tau ^{d}
\notag \\
&&+C_{R}^{u\dagger }M_{R}^{p}C_{R}^{d}\tau ^{d}O_{p}^{\mu }\tau
^{d}+C_{R}^{d\dagger }M_{R}^{p}C_{R}^{u}\tau ^{d}O_{p}^{\mu }\tau ^{u},
\label{9}
\end{eqnarray}
hence
\begin{eqnarray}
\sum_{p=1}^{3}\mathcal{L}_{R}^{p} &\rightarrow &\sum_{p=1}^{3}\left( i%
\mathrm{\bar{f}}\gamma _{\mu }Q_{pR}^{\mu }R\mathrm{f}+h.c.\right)  \notag \\
&=&\sum_{p=1}^{3}\left( i\mathrm{\bar{f}}\gamma _{\mu }\tilde{M}%
_{R}^{p}O_{p}^{\mu }R\mathrm{f}+h.c.\right) ,  \label{eqa21}
\end{eqnarray}
with
\begin{eqnarray}
\hat{M}_{R}^{1} &=&C_{+}^{\dagger }M_{R}^{1}C_{+}+C_{-}^{\dagger
}M_{R}^{2}C_{+}+C_{-}^{\dagger }M_{R}^{3}C_{-},  \notag \\
\hat{M}_{R}^{2} &=&C_{-}^{\dagger }M_{R}^{1}C_{+}+C_{+}^{\dagger
}M_{R}^{2}C_{+}+C_{+}^{\dagger }M_{R}^{3}C_{-},  \notag \\
\hat{M}_{R}^{3} &=&C_{-}^{\dagger }M_{R}^{1}C_{-}+C_{+}^{\dagger
}M_{R}^{2}C_{-}+C_{+}^{\dagger }M_{R}^{3}C_{+},  \label{eqa22}
\end{eqnarray}
where $C_{\pm }=(C_{R}^{u}\pm C_{R}^{d})/2$. Hence, transformations (\ref
{totalt}) can be absorbed by a mere redefinition of the matrices$\
M_{R}^{1}, $ $M_{R}^{2}$ and $M_{R}^{3}$.

\section{Effective couplings and $CP$ violation}

\label{effective}After the transformations discussed in the previous section
we are now in the physical basis and in a position to discuss the physical
relevance of the couplings in the effective Lagrangian. On dimensional
grounds the contribution of all possible dimension four operators to the
vertices can be parametrized in terms of effective couplings
\begin{eqnarray}
\mathcal{L}_{eff} &=&-g_{s}\mathrm{\bar{f}}\gamma ^{\mu }\left(
a_{L}L+a_{R}R\right) \lambda \cdot G_{\mu }\mathrm{f}  \notag \\
&&-e\mathrm{\bar{f}}\gamma ^{\mu }\left( b_{L}L+b_{R}R\right) A_{\mu }%
\mathrm{f}  \notag \\
&&-\frac{e}{2c_{W}s_{W}}\mathrm{\bar{f}}\gamma ^{\mu }\left(
g_{L}L+g_{R}R\right) Z_{\mu }\mathrm{f}  \notag \\
&&-\frac{e}{s_{W}}\mathrm{\bar{f}}\gamma ^{\mu }\left( h_{L}L+h_{R}R\right)
\frac{\tau ^{-}}{2}W_{\mu }^{+}\mathrm{f}  \notag \\
&&-\frac{e}{s_{W}}\mathrm{\bar{f}}\gamma ^{\mu }\left( h_{L}^{\dagger
}L+h_{R}^{\dagger }R\right) \frac{\tau ^{+}}{2}W_{\mu }^{-}\mathrm{f},
\label{vert1}
\end{eqnarray}
where we define
\begin{equation}
a_{LR}=a_{LR}^{u}\tau ^{u}+a_{LR}^{d}\tau ^{d},\qquad b_{LR}=b_{LR}^{u}\tau
^{u}+b_{LR}^{d}\tau ^{d},\qquad g_{LR}=g_{LR}^{u}\tau ^{u}+g_{LR}^{d}\tau
^{d}.  \label{eqa23}
\end{equation}
After rewriting the effective operators (\ref{effec}) in the physical basis,
their contribution to the couplings $a_{R},a_{L},b_{R},\ldots $ can be found
out by setting $U=I$.

The operators involving right-handed fields give rise to ($c_{W}$ and $s_{W}$
are the cosine and sine of the Weinberg angle, respectively)
\begin{eqnarray}
\sum_{p=1}^{3}\mathcal{L}_{R}^{p} &=&-\bar{f}\gamma ^{\mu }\left( \hat{M}%
_{R}^{1}+\hat{M}_{R}^{2}\tau ^{3}\right) \left[ \frac{e}{s_{W}}\left( \frac{%
\tau ^{-}}{2}W_{\mu }^{+}+\frac{\tau ^{+}}{2}W_{\mu }^{-}\right) +\frac{e}{%
c_{W}s_{W}}\frac{\tau ^{3}}{2}Z_{\mu }\right] Rf+h.c.  \notag \\
&&-\bar{f}\gamma ^{\mu }\hat{M}_{R}^{3}\tau ^{3}\left[ \frac{e}{s_{W}}\left(
\frac{\tau ^{-}}{2}W_{\mu }^{+}+\frac{\tau ^{+}}{2}W_{\mu }^{-}\right) +%
\frac{e}{c_{W}s_{W}}\frac{\tau ^{3}}{2}Z_{\mu }\right] \tau ^{3}Rf+h.c
\label{l0fv}
\end{eqnarray}

For the operators involving left-handed fields we have instead
\begin{eqnarray}
\mathcal{L}_{L}^{1} &=&\bar{\mathrm{\ f}}\gamma ^{\mu }\left\{ \frac{e}{%
c_{W}s_{W}}\left( \hat{M}_{L}^{1}\frac{\tau ^{u}}{2}-K^{\dagger }\hat{M}%
_{L}^{1}K\frac{\tau ^{d}}{2}\right) Z_{\mu }\right.  \notag \\
&&+\left. \frac{e}{s_{W}}\left( \hat{M}_{L}^{1}K\frac{\tau ^{-}}{2}W_{\mu
}^{+}+K^{\dagger }\hat{M}_{L}^{1}\frac{\tau ^{+}}{2}W_{\mu }^{-}\right)
\right\} L\mathrm{\ f}+h.c.,  \label{l1fv} \\
\mathcal{L}_{L}^{2} &=&-\bar{\mathrm{\ f}}\gamma ^{\mu }\left\{ \frac{e}{%
c_{W}s_{W}}\left( \hat{M}_{L}^{2}\frac{\tau ^{u}}{2}+K^{\dagger }\hat{M}%
_{L}^{2}K\frac{\tau ^{d}}{2}\right) Z_{\mu }\right.  \notag \\
&&+\left. \frac{e}{s_{W}}\left( -\hat{M}_{L}^{2}K\frac{\tau ^{-}}{2}W_{\mu
}^{+}+K^{\dagger }\hat{M}_{L}^{2}\frac{\tau ^{+}}{2}W_{\mu }^{-}\right)
\right\} L\mathrm{f}+h.c.,  \label{13fv} \\
\mathcal{L}_{L}^{3} &=&-\bar{\mathrm{f}}\gamma ^{\mu }\left\{ \frac{e}{%
c_{W}s_{W}}\left( \hat{M}_{L}^{3}\frac{\tau ^{u}}{2}-K^{\dagger }\hat{M}%
_{L}^{3}K\frac{\tau ^{d}}{2}\right) Z_{\mu }\right.  \notag \\
&&+\left. \frac{e}{s_{W}}\left( -\hat{M}_{L}^{3}K\frac{\tau ^{-}}{2}W_{\mu
}^{+}-K^{\dagger }\hat{M}_{L}^{3}\frac{\tau ^{+}}{2}W_{\mu }^{-}\right)
\right\} L\mathrm{f}+h.c..  \label{14fv}
\end{eqnarray}

The contribution from $\mathcal{L}_{L}^{4}$ is a little bit different and
deserves some additional comments. Let us first see how this effective
operator looks in the physical basis and after setting $U=I$
\begin{eqnarray}
\mathcal{L}_{L}^{4} &=&-\bar{\mathrm{f}}\gamma ^{\mu }\left\{ \left( \hat{M}%
_{L}^{4}\tau ^{u}-K^{\dagger }\hat{M}_{L}^{4}K\tau ^{d}\right) \left[
-i\partial _{\mu }+eQA_{\mu }\right. \right.  \notag \\
&&+\left. \frac{e}{c_{W}s_{W}}\left( \frac{\tau ^{3}}{2}-Qs_{W}^{2}\right)
Z_{\mu }+g_{s}\frac{\mathbf{\lambda }}{2}\mathbf{\cdot G}_{\mu }\right]
\notag \\
&&+\left. \frac{e}{s_{W}}\left( \hat{M}_{L}^{4}K\frac{\tau ^{-}}{2}W_{\mu
}^{+}-K^{\dagger }\hat{M}_{L}^{4}\frac{\tau ^{+}}{2}W_{\mu }^{-}\right)
\right\} L\mathrm{f}+h.c..  \label{l7fv}
\end{eqnarray}
One sees that $\mathcal{L}_{L}^{4}$ is the only operator potentially
contributing to the gluon and photon effective couplings. This is of course
surprising since both the photon and the gluon are associated to currents
which are exactly conserved and radiative corrections (including those from
new physics) are prohibited at zero momentum transfer. However one should
note that the effective couplings listed in (\ref{vert1}) are not directly
observable yet because one must take into account the renormalization of the
external legs. In fact $\mathcal{L}_{L}^{4}$ is the only operator that can
possibly contribute to such renormalization at the order we are working.
This issue will be discussed in detail in the next section. When the
contribution from the external legs is taken into account one observes that $%
\mathcal{L}_{L}^{4}$ can be eliminated altogether from the neutral gauge
bosons couplings (and this includes the $Z$ couplings).

Another way of seeing this (as pointed out in \cite{nomix}) is by realizing
that after use of the equations of motion $\mathcal{L}_{L}^{4}$ transforms
into a mass term, so the effect of $\mathcal{L}_{L}^{4}$ can be absorbed by
a redefinition of the fermion masses, if the fermions are on-shell, as it
will be the case in the present discussion. Then it is clear that $\mathcal{L%
}_{L}^{4}$ may possibly contribute to the renormalization of the CKM matrix
elements only (i.e. to the charged current sector).

All this considered, from Eqs. (\ref{vert1}) and (\ref{l0fv}-\ref{l7fv}),
and from the results presented in section \ref{dk} concerning wave function
renormalization, we obtain
\begin{equation}
a_{L}=a_{R}=b_{L}=b_{R}=0,  \label{eqa24}
\end{equation}
both for the up and down components. For the $Z$ couplings we get
\begin{eqnarray}
g_{L}^{u} &=&-\hat{M}_{L}^{1}-\hat{M}_{L}^{1\dagger }+\hat{M}_{L}^{2\dagger
}+\hat{M}_{L}^{2}+\hat{M}_{L}^{3}+\hat{M}_{L}^{3\dagger }  \notag \\
g_{L}^{d} &=&K^{\dagger }\left( \hat{M}_{L}^{1}+\hat{M}_{L}^{1\dagger }+\hat{%
M}_{L}^{2\dagger }+\hat{M}_{L}^{2}-\hat{M}_{L}^{3}-\hat{M}_{L}^{3\dagger
}\right) K,  \notag \\
g_{R}^{u} &=&\hat{M}_{R}^{1}+\hat{M}_{R}^{1\dagger }+\hat{M}_{R}^{2}+\hat{M}%
_{R}^{2\dagger }+\hat{M}_{R}^{3}+\hat{M}_{R}^{3\dagger },  \notag \\
g_{R}^{d} &=&\hat{M}_{R}^{2}+\hat{M}_{R}^{2\dagger }-\hat{M}_{R}^{1}-\hat{M}%
_{R}^{1\dagger }-\hat{M}_{R}^{3}-\hat{M}_{R}^{3\dagger }.  \label{Zcoupl1}
\end{eqnarray}
To compare this results with the ones presented in section \ref{Zdecayobs}
regarding $Z$ decay we have to use
\begin{eqnarray*}
g_{V}^{f} &=&\frac{g_{R}^{f}+g_{L}^{f}}{2}, \\
g_{A}^{f} &=&\frac{g_{L}^{f}-g_{R}^{f}}{2}.
\end{eqnarray*}
The contribution from wave-function renormalization cancels the dependence
from the vertices on the Hermitian combination $\hat{M}_{L}^{4}+\hat{M}%
_{L}^{4\dagger }$, which is the only one that appears from the vertices
themselves.

As for the effective charged couplings we give here the contribution coming
from the vertices only. So in order to get the full effective couplings one
must still add the contribution from wave-function renormalization and from
the renormalization of the CKM matrix elements. Actually we will see in
section \ref{dk} that these contributions cancel out at tree level so in
fact the following results include the full dependence on $\hat{M}_{L}^{4}$
\begin{eqnarray}
h_{L} &=&\left( -\hat{M}_{L}^{1}-\hat{M}_{L}^{1\dagger }+\hat{M}_{L}^{2}-%
\hat{M}_{L}^{2\dagger }-\hat{M}_{L}^{3}-\hat{M}_{L}^{3\dagger }+\hat{M}%
_{L}^{4}-\hat{M}_{L}^{4\dagger }\right) K,  \notag \\
h_{R} &=&\hat{M}_{R}^{1}+\hat{M}_{R}^{1\dagger }+\hat{M}_{R}^{2}-\hat{M}%
_{R}^{2\dagger }-\hat{M}_{R}^{3}-\hat{M}_{R}^{3\dagger },  \label{vertices1}
\end{eqnarray}

The above effective couplings thus summarize all effects due to the mixing
of families in the low energy theory caused by the presence of new physics
at some large scale $\Lambda $. Our aim now is to investigate the possible
new sources of $CP$ violation in the above effective couplings. Let us first
give a brief account of $C$ and $P$ transformations.

\section{$CP$ transformations}

Under $P$ we have
\begin{eqnarray*}
f\left( x\right) &\rightarrow &\gamma ^{0}f\left( \tilde{x}\right) , \\
\bar{f}\left( x\right) &\rightarrow &\bar{f}\left( \tilde{x}\right) \gamma
^{0}, \\
U\left( x\right) &\rightarrow &U^{\dagger }\left( \tilde{x}\right) , \\
B_{\mu }\left( x\right) &\rightarrow &B^{\mu }\left( \tilde{x}\right) , \\
\mathbf{W}_{\mu }\left( x\right) &\rightarrow &\mathbf{W}^{\mu }\left(
\tilde{x}\right) , \\
\partial _{\mu } &\rightarrow &\partial ^{\mu },
\end{eqnarray*}
where
\begin{equation*}
\tilde{x}^{\mu }=x_{\mu }=\left( x^{0},-x^{i}\right) .
\end{equation*}
Under $C$ we have
\begin{eqnarray*}
f\left( x\right) &\rightarrow &i\gamma ^{2}\gamma ^{0}\bar{f}^{T}, \\
\bar{f}\left( x\right) &\rightarrow &f^{T}i\gamma ^{2}\gamma ^{0}, \\
U\left( x\right) &\rightarrow &U^{\intercal }\left( x\right) , \\
B_{\mu }\left( x\right) &\rightarrow &-B_{\mu }\left( x\right) , \\
\left( W_{\mu }^{1},W_{\mu }^{2},W_{\mu }^{3}\right) &\rightarrow &\left(
-W_{\mu }^{1},W_{\mu }^{2},-W_{\mu }^{3}\right) ,
\end{eqnarray*}
The transformation for $\mathbf{W}_{\mu }$ can be written as
\begin{equation*}
\mathbf{\tau \cdot W}_{\mu }\rightarrow -\mathbf{\tau }^{\intercal }\mathbf{%
\cdot W}_{\mu },
\end{equation*}
accordingly the $SU\left( 3\right) _{c}$ gauge bosons transforms so as to
satisfy
\begin{equation*}
\mathbf{\lambda \cdot G}_{\mu }\rightarrow -\mathbf{\lambda }^{\intercal }%
\mathbf{\cdot G}_{\mu },
\end{equation*}
Let us investigate the effects of these transformations in some kinetic
terms. The kinetic term of the Goldstone bosons is given by
\begin{equation*}
\mathcal{L}_{0}=Tr\left( D_{\mu }UD^{\mu }U^{\dagger }\right)
\end{equation*}
transforming under $P$ as
\begin{eqnarray*}
\mathcal{L}_{0} &\rightarrow &Tr\left( D^{\mu }U^{\dagger }D_{\mu }U\right)
\\
&=&Tr\left( D_{\mu }UD^{\mu }U^{\dagger }\right) =\mathcal{L}_{0},
\end{eqnarray*}
where the change $x\rightarrow \tilde{x}$ has no effect since the Lagrangian
is integrated over space-time. Under $C$ we have
\begin{equation*}
D_{\mu }U\rightarrow \partial _{\mu }U^{\intercal }-ig\frac{\mathbf{\tau }%
^{\intercal }}{2}\mathbf{\cdot W}_{\mu }U^{\intercal }+ig^{\prime
}U^{\intercal }\frac{\tau ^{3}}{2}B_{\mu },
\end{equation*}
and under $CP$ we have
\begin{equation*}
D_{\mu }U\rightarrow \partial ^{\mu }U^{\ast }-ig\frac{\mathbf{\tau }%
^{\intercal }}{2}\mathbf{\cdot W}^{\mu }U^{\ast }+ig^{\prime }U^{\ast }\frac{%
\tau ^{3}}{2}B^{\mu }=\left( D^{\mu }U\right) ^{\ast },
\end{equation*}
so
\begin{eqnarray*}
\left( D_{\mu }U\right) ^{\intercal } &\rightarrow &\left( \partial _{\mu
}U^{\intercal }-ig\frac{\mathbf{\tau }^{\intercal }}{2}\mathbf{\cdot W}_{\mu
}U^{\intercal }+ig^{\prime }U^{\intercal }\frac{\tau ^{3}}{2}B_{\mu }\right)
^{\intercal } \\
&=&\left( \partial _{\mu }U-igU\frac{\mathbf{\tau }}{2}\mathbf{\cdot W}_{\mu
}+ig^{\prime }\frac{\tau ^{3}}{2}B_{\mu }U\right) \\
&=&-U\left( \partial _{\mu }U^{\dagger }+ig\frac{\mathbf{\tau }}{2}\mathbf{%
\cdot W}_{\mu }U^{\dagger }-ig^{\prime }U^{\dagger }\frac{\tau ^{3}}{2}%
B_{\mu }\right) U \\
&=&-U\left( D_{\mu }U^{\dagger }\right) U,
\end{eqnarray*}
and
\begin{equation*}
\left( D^{\mu }U^{\dagger }\right) ^{\intercal }\rightarrow -U^{\dagger
}\left( D^{\mu }U\right) U^{\dagger },
\end{equation*}
from the above we obtain
\begin{equation*}
\mathcal{L}_{0}=Tr\left( \left( D^{\mu }U^{\dagger }\right) ^{\intercal
}\left( D_{\mu }U\right) ^{\intercal }\right) \rightarrow Tr\left( D^{\mu
}UD_{\mu }U^{\dagger }\right) =\mathcal{L}_{0},
\end{equation*}
so $\mathcal{L}_{0}$ is invariant under $C$ and $P$ separately. Finally it
will be useful to keep for future use the following $CP$ transformations
\begin{eqnarray}
f &\rightarrow &-i\gamma ^{2}\bar{f}^{T},  \notag \\
\bar{f} &\rightarrow &if^{T}\gamma ^{2},  \notag \\
U &\rightarrow &U^{\ast }  \notag \\
D_{\mu }U &\rightarrow &\left( D^{\mu }U\right) ^{\ast },  \notag \\
B_{\mu } &\rightarrow &-B^{\mu },  \notag \\
\mathbf{\tau \cdot W}_{\mu } &\rightarrow &-\mathbf{\tau }^{\intercal }%
\mathbf{\cdot W}_{\mu },  \notag \\
\mathbf{\lambda \cdot G}_{\mu } &\rightarrow &-\mathbf{\lambda }^{\intercal }%
\mathbf{\cdot G}_{\mu },  \notag \\
\partial _{\mu } &\rightarrow &\partial ^{\mu },  \label{cptrans}
\end{eqnarray}
where the change $x\rightarrow \tilde{x}$ is understood

\section{Dimension 4 operators under $CP$ transformations}

\label{cptranssection} In this section we will test necessary and sufficient
conditions to have $CP$ invariant operators under transformations (\ref
{cptrans}). Let us start with the kinetic term defining
\begin{eqnarray*}
\mathcal{L}_{L} &=&\mathcal{O}_{L}+\mathcal{O}_{L}^{\dagger } \\
\mathcal{O}_{L} &=&i\bar{f}M\gamma ^{\mu }D_{\mu }^{L}\frac{1-\gamma ^{5}}{2}%
f \\
&=&i\bar{f}M\gamma ^{\mu }\frac{1-\gamma ^{5}}{2}\left( \partial _{\mu }+ig%
\frac{\mathbf{\tau }}{2}\mathbf{\cdot W}_{\mu }+ig^{\prime }zB_{\mu }+ig_{s}%
\frac{\mathbf{\lambda }}{2}\mathbf{\cdot G}_{\mu }\right) f,
\end{eqnarray*}
with the matrix $M$ having only mixing family indices. Then we have
\begin{eqnarray*}
\mathcal{O}_{L}^{\dagger } &=&if^{\dagger }\left( \partial _{\mu }+ig\frac{%
\mathbf{\tau }}{2}\mathbf{\cdot W}_{\mu }+ig^{\prime }zB_{\mu }+ig_{s}\frac{%
\mathbf{\lambda }}{2}\mathbf{\cdot G}_{\mu }\right) \frac{1-\gamma ^{5}}{2}%
\gamma ^{\mu \dagger }\gamma ^{0}f \\
&=&i\bar{f}\gamma ^{0}\gamma ^{\mu \dagger }\gamma ^{0}\left( \partial _{\mu
}+ig\frac{\mathbf{\tau }}{2}\mathbf{\cdot W}_{\mu }+ig^{\prime }zB_{\mu
}+ig_{s}\frac{\mathbf{\lambda }}{2}\mathbf{\cdot G}_{\mu }\right) \frac{%
1-\gamma ^{5}}{2}M^{\dagger }f \\
&=&i\bar{f}M^{\dagger }\gamma ^{\mu }D_{\mu }^{L}\frac{1-\gamma ^{5}}{2}f,
\end{eqnarray*}
so the complete term is
\begin{eqnarray*}
\mathcal{L}_{L} &=&i\bar{f}\left( M+M^{\dagger }\right) \gamma ^{\mu }D_{\mu
}^{L}\frac{1-\gamma ^{5}}{2}f \\
&=&i\bar{f}A\gamma ^{\mu }D_{\mu }^{L}\frac{1-\gamma ^{5}}{2}f,
\end{eqnarray*}
with $A$ an arbitrary $3\times 3$ Hermitian matrix. Under $CP$ we have
\begin{eqnarray*}
\mathcal{L}_{L} &\rightarrow &if^{\intercal }\gamma ^{2}\gamma ^{0}\gamma
^{0}A\gamma ^{\mu }\left( \partial ^{\mu }-ig\frac{\mathbf{\tau }^{\intercal
}}{2}\mathbf{\cdot W}^{\mu }+ig^{\prime }zB^{\mu }-ig_{s}\frac{\mathbf{%
\lambda }^{\intercal }}{2}\mathbf{\cdot G}^{\mu }\right) \frac{1-\gamma ^{5}%
}{2}\gamma ^{2}\gamma ^{0}f^{\ast } \\
&=&if^{\intercal }A\frac{1-\gamma ^{5}}{2}\gamma ^{\mu \ast }\left( \partial
^{\mu }-ig\frac{\mathbf{\tau }^{\intercal }}{2}\mathbf{\cdot W}^{\mu
}+ig^{\prime }zB^{\mu }-ig_{s}\frac{\mathbf{\lambda }^{\intercal }}{2}%
\mathbf{\cdot G}^{\mu }\right) \gamma ^{0}f^{\ast } \\
&=&-i\bar{f}A^{\intercal }\gamma ^{\mu \dagger }\left( \overset{\leftarrow }{%
\partial ^{\mu }}-ig\frac{\mathbf{\tau }}{2}\mathbf{\cdot W}^{\mu
}+ig^{\prime }zB^{\mu }-ig_{s}\frac{\mathbf{\lambda }}{2}\mathbf{\cdot G}%
^{\mu }\right) \frac{1-\gamma ^{5}}{2}f \\
&=&i\bar{f}A^{\intercal }\gamma _{\mu }\left( \partial ^{\mu }+ig\frac{%
\mathbf{\tau }}{2}\mathbf{\cdot W}^{\mu }+ig^{\prime }zB^{\mu }+ig_{s}\frac{%
\mathbf{\lambda }}{2}\mathbf{\cdot G}^{\mu }\right) \frac{1-\gamma ^{5}}{2}f.
\end{eqnarray*}
where a minus sign is present due to the commutation of grassman variables
and where a 'by parts' integration was performed. Hence $\int d^{4}x\mathcal{%
L}_{L}$ is invariant under $CP$ if and only if
\begin{equation*}
A=A^{\intercal },
\end{equation*}
Due to the fact that $A$ is Hermitian, this is the same as asking
\begin{equation*}
A=A^{\ast },
\end{equation*}
in other words $A$ must be \emph{real symmetric}.

Now we take
\begin{equation*}
\mathcal{L}_{L}^{1}=\mathcal{O}_{1}+\mathcal{O}_{1}^{\dagger },
\end{equation*}
with
\begin{eqnarray*}
\mathcal{O}_{1} &=&i\bar{f}M_{L}^{1}U\gamma ^{\mu }\left( D_{\mu }U\right)
^{\dagger }\frac{1-\gamma ^{5}}{2}f \\
&=&-if^{\dagger }\gamma ^{0}M_{L}^{1}\gamma ^{\mu }\left( D_{\mu }U\right)
U^{\dagger }\frac{1-\gamma ^{5}}{2}f,
\end{eqnarray*}
then we have
\begin{eqnarray*}
\mathcal{O}_{1}^{\dagger } &=&if^{\dagger }\frac{1-\gamma ^{5}}{2}\gamma
^{\mu \dagger }\gamma ^{0}M_{L}^{1\dagger }U\left( D_{\mu }U\right)
^{\dagger }f \\
&=&i\bar{f}\gamma ^{\mu }M_{L}^{1\dagger }U\left( D_{\mu }U\right) ^{\dagger
}\frac{1-\gamma ^{5}}{2}f \\
&=&i\bar{f}^{0}M_{L}^{1\dagger }U\gamma ^{\mu }\left( D_{\mu }U\right)
^{\dagger }\frac{1-\gamma ^{5}}{2}f,
\end{eqnarray*}
hence $M_{L}^{1}$ can be taken Hermitian without loss of generality. Under $%
CP$ we have
\begin{eqnarray*}
\mathcal{O}_{1} &\rightarrow &-if^{\intercal }\gamma ^{2}\gamma ^{0}\gamma
^{0}M_{L}^{1}\gamma ^{\mu }\left( D^{\mu }U\right) ^{\ast }U^{\intercal }%
\frac{1-\gamma ^{5}}{2}\gamma ^{2}\gamma ^{0}f^{\ast } \\
&=&-if^{\intercal }\frac{1-\gamma ^{5}}{2}\gamma ^{\mu \ast }M_{L}^{1}\left(
D^{\mu }U\right) ^{\ast }U^{\intercal }\gamma ^{0}f^{\ast } \\
&=&if^{\dagger }\gamma ^{0}M_{L}^{1T}U\left( D^{\mu }U\right) ^{\dagger
}\gamma ^{\mu \dagger }\frac{1-\gamma ^{5}}{2}f \\
&=&-if^{\dagger }\gamma ^{0}M_{L}^{1T}\gamma _{\mu }\left( D^{\mu }U\right)
U^{\dagger }\frac{1-\gamma ^{5}}{2}f.
\end{eqnarray*}
Again, in order to have $CP$ invariance of $\int d^{4}x\mathcal{L}_{L}^{1}$
we must have
\begin{equation*}
M_{L}^{1}=M_{L}^{1\ast },
\end{equation*}
so $M_{L}^{1}$ must be \emph{real symmetric}. Now taking
\begin{eqnarray*}
\mathcal{L}_{R}^{1} &=&\mathcal{O}_{2}+\mathcal{O}_{2}^{\dagger }, \\
\mathcal{O}_{2} &=&i\bar{f}M_{R}^{1}U^{\dagger }\gamma ^{\mu }\left( D_{\mu
}U\right) \frac{1+\gamma ^{5}}{2}f,
\end{eqnarray*}
and making a similar reasoning as in the case of $\mathcal{O}_{1}$ we obtain
that $M_{R}^{1}$ can be taken Hermitian without loss of generality. In order
to maintain $CP$ invariance it must be a \emph{real symmetric }matrix\emph{.
}Again taking
\begin{eqnarray*}
\mathcal{L}_{L}^{2} &=&\mathcal{O}_{3}+\mathcal{O}_{3}^{\dagger }, \\
\mathcal{O}_{3} &=&i\bar{f}M_{L}^{2}\gamma ^{\mu }\left( D_{\mu }U\right)
\tau ^{3}U^{\dagger }\frac{1-\gamma ^{5}}{2}f,
\end{eqnarray*}
we have
\begin{eqnarray*}
\mathcal{O}_{3}^{\dagger } &=&-if^{\dagger }M_{L}^{2\dagger }\frac{1-\gamma
^{5}}{2}U\tau ^{3}\gamma ^{\mu \dagger }\left( D_{\mu }U\right) ^{\dagger
}\gamma ^{0}f \\
&=&-i\bar{f}M_{L}^{2\dagger }U\tau ^{3}\gamma ^{\mu }\left( D_{\mu }U\right)
^{\dagger }\frac{1-\gamma ^{5}}{2}f,
\end{eqnarray*}
hence we have the Hermitian term
\begin{equation*}
\mathcal{L}_{L}^{2}=i\bar{f}\gamma ^{\mu }\left[ M_{L}^{2}\left( D_{\mu
}U\right) \tau ^{3}U^{\dagger }-M_{L}^{2\dagger }U\tau ^{3}\left( D_{\mu
}U\right) ^{\dagger }\right] \frac{1-\gamma ^{5}}{2}f.
\end{equation*}
so, under $CP$ we have
\begin{eqnarray*}
\mathcal{L}_{L}^{2} &\rightarrow &if^{\intercal }\gamma ^{2}\gamma
^{0}\gamma ^{0}\gamma ^{\mu }\left[ M_{L}^{2}\left( D^{\mu }U\right) ^{\ast
}\tau ^{3}U^{\intercal }-M_{L}^{2\dagger }U^{\ast }\tau ^{3}\left( D^{\mu
}U\right) ^{\intercal }\right] \frac{1-\gamma ^{5}}{2}\gamma ^{2}\gamma
^{0}f^{\ast } \\
&=&if^{\intercal }\frac{1-\gamma ^{5}}{2}\gamma ^{\mu \ast }\left[
M_{L}^{2}\left( D^{\mu }U\right) ^{\ast }\tau ^{3}U^{\intercal
}-M_{L}^{2\dagger }U^{\ast }\tau ^{3}\left( D^{\mu }U\right) ^{\intercal }%
\right] \gamma ^{0}f^{\ast } \\
&=&-i\bar{f}\gamma ^{\mu \dagger }\left[ M_{L}^{2T}U\tau ^{3}\left( D^{\mu
}U\right) ^{\dagger }-M_{L}^{2\ast }\left( D^{\mu }U\right) \tau
^{3}U^{\dagger }\right] \frac{1-\gamma ^{5}}{2}f.
\end{eqnarray*}
Hence, in order to have $CP$ invariance of $\int d^{4}x\mathcal{L}_{L}^{2}$
we must have
\begin{equation*}
M_{L}^{2}=M_{L}^{2\ast },
\end{equation*}
but the difference in this case is that we don't need the matrix $M_{L}^{2}$
to be Hermitian, so $M_{L}^{2}$ must be only \emph{real}. The same kind of
transformations can be done with the rest of dimension four operators. The
conclusion is that without any loss of generality we take the matrices in
family space $M_{L}^{1}$, $M_{R}^{1}$, $M_{L}^{3}$ and $M_{R}^{3}$
Hermitian, while $M_{L}^{2}$, $M_{R}^{2}$ and $M_{L}^{4}$ are completely
general. If we require those operators to be $CP$ conserving, the matrices $%
M_{LR}^{i}$ must be real.

\section{$CP$ violation in the effective couplings}

Generically the effective operators can be written as
\begin{equation}
\mathcal{L}_{L}=\mathrm{\bar{f}}\gamma _{\mu }S^{\mu }L\mathrm{f}+h.c.,
\label{eqa25}
\end{equation}
where
\begin{equation}
S^{\mu }\equiv \hat{M}_{L}\tau ^{u}O^{\mu }\tau ^{u}+\hat{M}_{L}K\tau
^{u}O^{\mu }\tau ^{d}+K^{\dagger }\hat{M}_{L}K\tau ^{d}O^{\mu }\tau
^{d}+K^{\dagger }\hat{M}_{L}\tau ^{d}O^{\mu }\tau ^{u}  \label{eqa26}
\end{equation}
then under $CP$ we have
\begin{equation}
\mathcal{L}_{L}\rightarrow \mathrm{\bar{f}}\gamma _{\mu }S^{\prime \mu }L%
\mathrm{f},  \label{eqa27}
\end{equation}
with
\begin{equation}
S^{\prime \mu }\equiv \hat{M}_{L}^{t}\tau ^{u}O^{\mu }\tau ^{u}+K^{t}\hat{M}%
_{L}^{t}\tau ^{d}O^{\mu }\tau ^{u}+K^{t}\hat{M}_{L}^{t}K^{\ast }\tau
^{d}O^{\mu }\tau ^{d}+\hat{M}_{L}^{t}K^{\ast }\tau ^{u}O^{\mu }\tau ^{d}
\label{eqa28}
\end{equation}
so in order to have $CP$ invariance we require
\begin{eqnarray}
\hat{M}_{L} &=&\hat{M}_{L}^{\ast },  \notag \\
\hat{M}_{L}K &=&\hat{M}_{L}K^{\ast },  \notag \\
K^{t}\hat{M}_{L}K^{\ast } &=&K^{\dagger }\hat{M}_{L}K,  \label{eqa29}
\end{eqnarray}
which can be fulfilled requiring
\begin{equation}
\hat{M}_{L}=\hat{M}_{L}^{\ast },\qquad K=K^{\ast },  \label{eqa30}
\end{equation}
Note that this last condition is sufficient but not necessary, however if we
ask for $CP$ invariance of the complete Lagrangian (as we should) the last
condition is both sufficient and necessary. Analogously, the right-handed
contribution, given by Eq. (\ref{l0fv}), is $CP$ invariant provided
\begin{equation}
\hat{M}_{R}^{p}=\hat{M}_{R}^{p\ast },  \label{eqa31}
\end{equation}

Eqs. (\ref{eqa24}), (\ref{Zcoupl1}) and (\ref{vertices1}) thus summarize the
contribution from dimension four operators to the observables. In addition
there will be contributions from other higher dimensional operators, such as
for instance dimension five ones (magnetic moment-type operators for
example). We expect these to be small in theories such as the ones we are
considering here. The reason is that we assume a large mass gap between the
energies at which our effective Lagrangian is going to be used and the scale
of new physics. This automatically suppresses the contribution of higher
dimensional operators. However, non-decoupling effects may be left in
dimension four operators, which may depend logarithmically in the scale of
the new physics. The clearest example of this is the Standard Model itself.
Since the Higgs is there an essential ingredient in proving the
renormalizability of the theory, removing it induces new divergences which
eventually manifest themselves as logarithms of the Higgs mass. This
enhances (for a relatively heavy Higgs) the importance of the $d=4$
coefficients, albeit in the Standard model they are small nonetheless since
the $\log M_{H}^{2}/M_{W}^{2}$ is preceded by a prefactor $y^{2}/16\pi ^{2}$%
, where $y$ is a Yukawa coupling (see \cite{nomix}).

Apart from the issue of wave-function renormalization, to which we shall
turn next, we have finished our theoretical analysis and we can start
drawing some conclusions.

One of the first things one observes is that there are no anomalous photon
or gluon couplings, diagonal or not in family. This excludes the appearance
of electromagnetic or strong penguin-like contributions from new physics to
the effective couplings and observables considered here. Since our analysis
is rather general this is an interesting observation.

Here it is worth to remember that the charged current sector cannot be
exactly described by a unitarity triangle because radiative corrections
spoil the relation between angles and sides of the observable vertex
couplings. In fact the $d=4$ operators we have analyzed do spoil that
relation too. To see this we need only to examine Eq. (\ref{vertices1}). The
total charged current vertex will be proportional to
\begin{equation}
U=K+\Delta K_{,}  \label{eqa32}
\end{equation}
where $\Delta K$ is a combination of the $\hat{M}_{L}$ matrices. Since $%
\Delta K$ is neither Hermitian nor anti-hermitian, $U$ is not unitary, not
even in a perturbative sense. The same happens when radiative corrections
are considered.

\section{Radiative corrections and renormalization}

\label{renorm}As we mentioned in the section \ref{effective}, the effective
couplings presented in (\ref{vertices1}) for the charged current vertices
are not the complete story because CKM and wave-function renormalization
gives a non-trivial contribution there. In this section we shall consider
the contribution to the observables due to wave-function renormalization and
the renormalization of the CKM matrix elements. The issue, as we shall see,
is far from trivial.

When we calculate cross sections in perturbation theory we have to take into
account the residues of the external leg propagators. The meaning of these
residues is clear when we do not have mixing. In this case, if we work in
the on-shell scheme, we can attempt to absorb these residues in the wave
function renormalization constants and forget about them. However the Ward
Identities force us to set up relations between the renormalization
constants that invalidate the naive on-shell scheme. The issue is resolved
in the following way: Take whatever renormalization scheme that respects
Ward Identities (such as minimal scheme) and use the corresponding
renormalization constants everywhere in except for the external legs
contributions. For the latter use the wfr. constants arising after the mass
pole and unit residue conditions are prescribed. This recipe is equivalent
to use the Ward identities-complying renormalization constants everywhere
and afterwards perform a finite renormalization of the external fields in
order to assure mass pole and residue one for the propagators. This is the
commonly used prescription \cite{Hollik} and, in the context of effective
theories was used in Chapter \ref{mattersectorchapter} and in \cite{nomix}
\cite{HE}.

Now let us now turn to the case where we have mixing. This was studied some
time ago by Aoki et al \cite{Aoki} and a on-shell scheme was proposed.
Unfortunately the issue is not settled. We have studied the problem with
some detail anew since, as already mentioned, the contribution from
wave-function renormalization is important in the present case. We have
found out that the set of conditions imposed by Aoki et al over-determine
the renormalization constants and is in fact incompatible with the analytic
structure of the theory. Moreover, even if this inconsistency is ``solved''
\cite{Denner} one has to be cautious to check that the resulting observable
quantity is gauge invariant. Since the whole issue is rather complex we will
analyze it separately in Chapter \ref{LSZchapter}. Here we will use the
results and conclusions of that chapter to proceed to the calculation of the
physical amplitudes showing explicitly the validity of Eqs. (\ref{eqa24}), (%
\ref{Zcoupl1}) and (\ref{vertices1}). The results of Chapter \ref{LSZchapter}
that are used here are given by Eqs. (\ref{zin}), (\ref{zout}), (\ref{zdiag}%
) and (\ref{wardf}).

\section{Contribution to wave-function renormalization}

\label{dk}The operator $\mathcal{L}_{L}^{4}$ is the only one contributing to
self-energies and, hence, to the wave-function renormalization constants. It
also gives a contribution (among others) to the neutral and charged current
vertices which (see Eq. (\ref{l7fv})). The bare contribution to the neutral
current is proportional to
\begin{eqnarray}
&&\left[ \left( \hat{M}_{L}^{4}+\hat{M}_{L}^{4\dagger }\right) \tau
^{u}-K^{\dagger }\left( \hat{M}_{L}^{4}+\hat{M}_{L}^{4\dagger }\right) K\tau
^{d}\right]  \notag \\
&&\times \left[ eQA_{\mu }+\frac{e}{c_{W}s_{W}}\left( \frac{\tau ^{3}}{2}%
-Qs_{W}^{2}\right) Z_{\mu }+g_{s}\frac{\mathbf{\lambda }}{2}\mathbf{\cdot G}%
_{\mu }\right] L,  \label{eqa43}
\end{eqnarray}
while its contribution to the charge current vertex is proportional to
\begin{equation}
\frac{e}{s_{W}}\left( \hat{M}_{L}^{4}-\hat{M}_{L}^{4\dagger }\right) \left( K%
\frac{\tau ^{-}}{2}W_{\mu }^{+}-K^{\dagger }\frac{\tau ^{+}}{2}W_{\mu
}^{-}\right) L.  \label{eq43b}
\end{equation}
The contribution from $\mathcal{L}_{L}^{4}$ to the bare self energies is
given by
\begin{eqnarray}
\Sigma ^{R\left( u,d\right) } &=&\Sigma ^{L\left( u,d\right) }=0,  \notag \\
\Sigma ^{\gamma Ru} &=&\Sigma ^{\gamma Rd}=0,  \notag \\
\Sigma ^{\gamma Ld} &=&K^{\dagger }\left( \hat{M}_{L}^{4}+\hat{M}%
_{L}^{4\dagger }\right) K,  \notag \\
\Sigma ^{\gamma Lu} &=&-\left( \hat{M}_{L}^{4}+\hat{M}_{L}^{4\dagger
}\right) ,  \label{eqa44}
\end{eqnarray}
hence using the on-shell wfr. constants given by Eqs. (\ref{zin}) and (\ref
{zout}) for $i\neq j$ we obtain
\begin{eqnarray}
\frac{1}{2}\delta Z_{ij}^{uL} &=&\frac{1}{2}\delta \bar{Z}_{ij}^{uL\dagger
}=-\frac{\left( \hat{M}_{L}^{4}+\hat{M}_{L}^{4\dagger }\right) _{ij}}{%
m_{j}^{2}-m_{i}^{2}}m_{j}^{2},  \notag \\
\frac{1}{2}\delta Z_{ij}^{uR} &=&\frac{1}{2}\delta \bar{Z}_{ij}^{uR\dagger
}=-m_{i}\frac{\left( \hat{M}_{L}^{4}+\hat{M}_{L}^{4\dagger }\right) _{ij}}{%
m_{j}^{2}-m_{i}^{2}}m_{j},  \notag \\
\frac{1}{2}\delta Z_{ij}^{dL} &=&\frac{1}{2}\delta \bar{Z}_{ij}^{dL\dagger }=%
\frac{\left( K^{\dagger }\left( \hat{M}_{L}^{4}+\hat{M}_{L}^{4\dagger
}\right) K\right) _{ij}}{m_{j}^{2}-m_{i}^{2}}m_{j}^{2},  \notag \\
\frac{1}{2}\delta Z_{ij}^{dR} &=&\frac{1}{2}\delta \bar{Z}_{ij}^{dR\dagger
}=m_{i}\frac{\left( K^{\dagger }\left( \hat{M}_{L}^{4}+\hat{M}_{L}^{4\dagger
}\right) K\right) _{ij}}{m_{j}^{2}-m_{i}^{2}}m_{j},  \label{offdiagm4}
\end{eqnarray}
and for $i=j$ using Eq. (\ref{zdiag}) we have
\begin{eqnarray}
\delta Z_{ii}^{uL} &=&\delta \bar{Z}_{ii}^{uL}=-\left( \hat{M}_{L}^{4}+\hat{M%
}_{L}^{4\dagger }\right) _{ii},  \notag \\
\delta Z_{ii}^{dL} &=&\delta \bar{Z}_{ii}^{dL}=\left( K^{\dagger }\left(
\hat{M}_{L}^{4}+\hat{M}_{L}^{4\dagger }\right) K\right) _{ii},  \notag \\
\delta \bar{Z}_{ii}^{uR} &=&\delta Z_{ii}^{uR}=\delta \bar{Z}%
_{ii}^{dR}=\delta Z_{ii}^{dR}=0,  \label{diagm4}
\end{eqnarray}
for the CKM counterterm we use the Ward identity (\ref{wardf}) taking
\begin{equation*}
\delta \hat{Z}^{dL}=\delta Z^{dL},
\end{equation*}
and using Ward identity (\ref{ward2inf})
\begin{eqnarray*}
\delta \hat{Z}^{uL} &=&\frac{1}{2}\left( \delta \hat{Z}^{uL}+\delta \hat{Z}%
^{uL\dagger }\right) +\frac{1}{2}\left( \delta \hat{Z}^{uL}-\delta \hat{Z}%
^{uL\dagger }\right) \\
&=&\frac{1}{2}K\left( \delta \hat{Z}^{dL\dagger }+\delta \hat{Z}^{dL}\right)
K^{\dagger }+\frac{1}{2}\left( \delta \hat{Z}^{uL}-\delta \hat{Z}^{uL\dagger
}\right) ,
\end{eqnarray*}
hence we still have freedom to prescribe $\delta \hat{Z}^{uL}-\delta \hat{Z}%
^{uL\dagger }.$ That is
\begin{equation*}
\delta K=\frac{1}{4}\left( \delta \hat{Z}^{uL}-\delta \hat{Z}^{uL\dagger
}\right) K-\frac{1}{4}K\left( \delta Z^{dL}-\delta Z^{dL\dagger }\right) .
\end{equation*}
The leading contribution of $\mathcal{L}_{L}^{4}$ to the charged vertex
including counterterms (see section \ref{wandtdecay} in Chapter \ref
{LSZchapter}) is proportional to
\begin{eqnarray}
\mathcal{A}_{\hat{M}_{L}^{4}}^{CC} &=&\left[ K+\delta K+\frac{1}{2}\delta
\bar{Z}^{Lu}K+\frac{1}{2}K\delta Z^{Ld}+\left( \hat{M}_{L}^{4}-\hat{M}%
_{L}^{4\dagger }\right) K\,\right] L  \notag \\
&=&\left[ K+\frac{1}{4}K\left( \delta Z^{dL}+\delta Z^{dL\dagger }\right) +%
\frac{1}{2}\delta \bar{Z}^{Lu}K\right.  \notag \\
&&+\left. \frac{1}{4}\left( \delta \hat{Z}^{uL}-\delta \hat{Z}^{uL\dagger
}\right) K+\left( \hat{M}_{L}^{4}-\hat{M}_{L}^{4\dagger }\right) K\right] L,
\label{m1pri}
\end{eqnarray}
where the last term in Eq. (\ref{m1pri}) corresponds to the direct
contribution of $\mathcal{L}_{L}^{4}$ (not through wfr. or CKM
counterterms). Taking
\begin{equation*}
\delta \hat{Z}^{uL}-\delta \hat{Z}^{uL\dagger }=\delta \bar{Z}^{Lu\dagger
}-\delta \bar{Z}^{Lu},
\end{equation*}
Eq. (\ref{m1pri}) becomes
\begin{equation*}
\mathcal{A}_{\hat{M}_{L}^{4}}^{CC}=\left[ K+\frac{1}{4}K\left( \delta
Z^{dL}+\delta Z^{dL\dagger }\right) +\frac{1}{4}\left( \delta \bar{Z}%
^{Lu}+\delta \bar{Z}^{Lu\dagger }\right) K+\left( \hat{M}_{L}^{4}-\hat{M}%
_{L}^{4\dagger }\right) K\right] L,
\end{equation*}
and from Eqs. (\ref{offdiagm4}) and (\ref{diagm4}) we finally obtain
\begin{eqnarray*}
\mathcal{A}_{\hat{M}_{L}^{4}}^{CC} &=&\left[ K+\frac{1}{2}\left( \hat{M}%
_{L}^{4}+\hat{M}_{L}^{4\dagger }\right) K-\frac{1}{2}\left( \hat{M}_{L}^{4}+%
\hat{M}_{L}^{4\dagger }\right) K+\left( \hat{M}_{L}^{4}-\hat{M}%
_{L}^{4\dagger }\right) K\right] L \\
&=&\left[ K+\left( \hat{M}_{L}^{4}-\hat{M}_{L}^{4\dagger }\right) K\right] L.
\end{eqnarray*}
Thus we observe that the total contribution of $\mathcal{L}_{kin}+\mathcal{L}%
_{L}^{4}$ is in fact equal to the contribution of $\mathcal{L}_{L}^{4}$
alone. The contributions coming from the wave function and CKM
renormalizations cancel out at tree level. Another point to note is that
this particular contribution preserves the perturbative unitarity of $K$, in
accordance with the equations-of-motion argument. For the neutral currents
we have
\begin{eqnarray}
\mathcal{A}_{\hat{M}_{L}^{4}}^{NC} &=&\left[ \left( \hat{M}_{L}^{4}+\hat{M}%
_{L}^{4\dagger }\right) \tau ^{u}-K^{\dagger }\left( \hat{M}_{L}^{4}+\hat{M}%
_{L}^{4\dagger }\right) K\tau ^{d}\right]  \notag \\
&&\times \left[ eQA_{\mu }+\frac{e}{c_{W}s_{W}}\left( \frac{\tau ^{3}}{2}%
-Qs_{W}^{2}\right) Z_{\mu }+g_{s}\frac{\mathbf{\lambda }}{2}\mathbf{\cdot G}%
_{\mu }\right] L  \notag \\
&&+\bar{Z}^{\frac{1}{2}}\left( eQA_{\mu }+\frac{e}{c_{W}s_{W}}\left[ \left(
\frac{\tau ^{3}}{2}-Qs_{W}^{2}\right) L-Qs_{W}^{2}R\right] Z_{\mu }+g_{s}%
\frac{\mathbf{\lambda }}{2}\cdot G_{\mu }\right) Z^{\frac{1}{2}}  \notag \\
&=&\mathcal{A}_{0}^{NC}+\left[ \left( \hat{M}_{L}^{4}+\hat{M}_{L}^{4\dagger
}\right) \tau ^{u}-K^{\dagger }\left( \hat{M}_{L}^{4}+\hat{M}_{L}^{4\dagger
}\right) K\tau ^{d}\right]  \notag \\
&&\times \left[ eQA_{\mu }+\frac{e}{c_{W}s_{W}}\left( \frac{\tau ^{3}}{2}%
-Qs_{W}^{2}\right) Z_{\mu }+g_{s}\frac{\mathbf{\lambda }}{2}\mathbf{\cdot G}%
_{\mu }\right] L  \notag \\
&&+\frac{1}{2}\left( eQA_{\mu }+g_{s}\frac{\mathbf{\lambda }}{2}\cdot G_{\mu
}\right) \left\{ \left[ \left( \delta \bar{Z}^{uL}+\delta Z^{uL}\right) \tau
^{u}+\left( \delta \bar{Z}^{dL}+\delta Z^{dL}\right) \tau ^{d}\right]
L\right.  \notag \\
&&+\left. \left[ \left( \delta \bar{Z}^{uR}+\delta Z^{uR}\right) \tau
^{u}+\left( \delta \bar{Z}^{dR}+\delta Z^{dR}\right) \tau ^{d}\right]
R\right\}  \notag \\
&&+\frac{1}{2}\frac{e}{c_{W}s_{W}}\left( \frac{\tau ^{3}}{2}%
-Qs_{W}^{2}\right) Z_{\mu }\left[ \left( \delta \bar{Z}^{uL}+\delta
Z^{uL}\right) \tau ^{u}+\left( \delta \bar{Z}^{dL}+\delta Z^{dL}\right) \tau
^{d}\right] L  \notag \\
&&-\frac{1}{2}\frac{e}{c_{W}s_{W}}Qs_{W}^{2}Z_{\mu }\left[ \left( \delta
\bar{Z}^{uR}+\delta Z^{uR}\right) \tau ^{u}+\left( \delta \bar{Z}%
^{dR}+\delta Z^{dR}\right) \tau ^{d}\right] R,  \label{ncm4l}
\end{eqnarray}
where we have defined the SM tree level contribution as
\begin{equation*}
\mathcal{A}_{0}^{NC}=QA_{\mu }+\frac{e}{c_{W}s_{W}}\left[ \left( \frac{\tau
^{3}}{2}-Qs_{W}^{2}\right) L-Qs_{W}^{2}R\right] Z_{\mu }+g_{s}\frac{\mathbf{%
\lambda }}{2}\cdot G_{\mu },
\end{equation*}
using Eqs. (\ref{offdiagm4}) and (\ref{diagm4}) we have that for all family
indices
\begin{eqnarray*}
\frac{1}{2}\left( \delta \bar{Z}^{uL}+\delta Z^{uL}\right) &=&-\hat{M}%
_{L}^{4}+\hat{M}_{L}^{4\dagger }, \\
\frac{1}{2}\left( \delta \bar{Z}^{dL}+\delta Z^{dL}\right) &=&K^{\dagger
}\left( \hat{M}_{L}^{4}+\hat{M}_{L}^{4\dagger }\right) K, \\
\frac{1}{2}\left( \delta \bar{Z}^{uR}+\delta Z^{uR}\right) &=&0, \\
\frac{1}{2}\left( \delta \bar{Z}^{dR}+\delta Z^{dR}\right) &=&0,
\end{eqnarray*}
and therefore replacing the above expressions in Eq. (\ref{ncm4l}) we obtain
\begin{equation*}
\mathcal{A}_{\hat{M}_{L}^{4}}^{NC}=\mathcal{A}_{0}^{NC}.
\end{equation*}
Hence, we observe that when renormalization constants are taken into account
the total contribution of $\mathcal{L}_{L}^{4}$ to the neutral current
vertices vanishes. This is a very non-trivial check of the whole procedure .
Of course nothing prevents the appearance of $\hat{M}_{L}^{4}$ at higher
orders when one, for instance, performs loops with the effective operators.
But this a purely academic question at this point.

This completes the theoretical analysis of the CKM and wave-function
renormalization.

\section{Some examples: a heavy doublet and a heavy Higgs}

\label{examples}Let us now try to get a feeling about the order of magnitude
of the coefficients of the effective Lagrangian. We shall consider two
examples: the effective theory induced by the integration of a heavy doublet
and the Standard Model itself in the limit of a heavy Higgs.

In the heavy doublet case we shall make use of some recent work by Del
Aguila and coworkers \cite{paco}. These authors have recently analyzed the
effect of integrating out heavy matter fields in different representations.
For illustration purposes we shall only consider the doublet case here. As
emphasized in \cite{paco} while additional chiral doublets are surely
excluded by the LEP data, vector multiplets are not.

Let us assume that the Standard Model is extended with a doublet of heavy
fermions $Q$ of mass $M$, with vector coupling to the gauge field. For the
time being we shall assume a light Higgs. In addition there will be an
extended Higgs-Yukawa term of the form
\begin{equation}
\lambda _{j}^{\left( u\right) }\bar{Q}\tilde{\phi}R\mathrm{u}_{j}+\lambda
_{j}^{\left( d\right) }\bar{Q}\phi R\mathrm{d}_{j},  \label{eqa50}
\end{equation}
where
\begin{equation}
\phi =\frac{1}{\sqrt{2}}\left(
\begin{array}{c}
\varphi _{1}+i\varphi _{2} \\
v+h+i\varphi _{3}
\end{array}
\right) ,\qquad \tilde{\phi}=i\mathbf{\tau }^{2}\phi ^{\ast },\qquad \mathrm{%
f}=\left(
\begin{array}{c}
\mathrm{u} \\
\mathrm{d}
\end{array}
\right) .  \label{eqb51}
\end{equation}

The heavy doublet can be exactly integrated. This procedure is described in
detail in \cite{paco}. After this operation we generate the following
effective couplings (all of them corresponding to operators of dimension
six)
\begin{eqnarray}
&&i\phi ^{\dagger }D_{\mu }\phi \mathrm{\bar{f}}\alpha _{\phi q}^{\left(
1\right) }\gamma ^{\mu }L\mathrm{f}+h.c.,  \notag \\
&&i\phi ^{\dagger }\mathbf{\tau }^{j}D_{\mu }\phi \mathrm{\bar{f}}\alpha
_{\phi q}^{\left( 3\right) }\gamma ^{\mu }\mathbf{\tau }^{j}L\mathrm{f}+h.c.,
\notag \\
&&i\phi ^{\dagger }D_{\mu }\phi \mathrm{\bar{f}}\alpha _{\phi u}\gamma ^{\mu
}\mathbf{\tau }^{u}R\mathrm{f}+h.c.,  \notag \\
&&i\phi ^{\dagger }D_{\mu }\phi \mathrm{\bar{f}}\alpha _{\phi d}\gamma ^{\mu
}\mathbf{\tau }^{d}R\mathrm{f}+h.c.,  \notag \\
&&\frac{1}{\sqrt{2}}\phi ^{t}\mathbf{\tau }^{2}D_{\mu }\phi \mathrm{\bar{f}}%
\alpha _{\phi \phi }\gamma ^{\mu }\mathbf{\tau }^{-}R\mathrm{f}+h.c.,  \notag
\\
&&-\phi ^{\dagger }\phi \mathrm{\bar{f}}\tilde{\phi}\alpha _{u\phi }R\mathrm{%
u}+h.c.,  \notag \\
&&-\phi ^{\dagger }\phi \mathrm{\bar{f}}\phi \alpha _{d\phi }R\mathrm{d}%
+h.c.,  \label{eqa51}
\end{eqnarray}
where
\begin{equation}
D_{\mu }\phi =\left( \partial _{\mu }+ig\frac{\mathbf{\tau }}{2}\cdot W_{\mu
}+i\frac{g^{\prime }}{2}B_{\mu }\right) \phi .  \label{eqa52}
\end{equation}
The coefficients appearing in (\ref{eqa51}) take the values
\begin{eqnarray}
\alpha _{\phi q}^{\left( 1\right) } &=&0,  \notag \\
\alpha _{\phi q}^{\left( 3\right) } &=&0,  \notag \\
\left( \alpha _{\phi u}\right) _{ij} &=&-\frac{1}{2}\lambda _{i}^{\left(
u\right) \dagger }\lambda _{j}^{\left( u\right) }\frac{1}{M^{2}},  \notag \\
\left( \alpha _{\phi d}\right) _{ij} &=&\frac{1}{2}\lambda _{i}^{\left(
d\right) \dagger }\lambda _{j}^{\left( d\right) }\frac{1}{M^{2}},  \notag \\
\left( \alpha _{\phi \phi }\right) _{ij} &=&\frac{1}{2}\lambda _{i}^{\left(
u\right) \dagger }\lambda _{j}^{\left( d\right) }\frac{1}{M^{2}},  \notag \\
\tilde{y}_{u} &\rightarrow &\tilde{y}_{u}\left( I-\alpha _{\phi
u}M^{2}\right) ,  \notag \\
\tilde{y}_{d} &\rightarrow &\tilde{y}_{d}\left( I+\alpha _{\phi
d}M^{2}\right) ,  \label{eqa53}
\end{eqnarray}

The above results are taken from \cite{paco} and have been derived in a
linear realization of the symmetry group, where the Higgs field, $h$, is
explicitly included, along with the Goldstone bosons. It is easy however to
recover the leading contribution to the coefficients of our effective
operators (\ref{effec}). The procedure would amount to integrating out the
Higgs field, of course. This would lead to two type of contributions:
tree-level and one loop. The latter are enhanced by logs of the Higgs mass,
but suppressed by the usual loop factor $1/16\pi ^{2}$. In addition there
are the multiplicative Yukawa couplings. It is not difficult to see though
that only the light fermion Yukawa couplings appear and hence the loop
contribution is small. To retain the tree-level contribution only we simply
replace $\phi $ by its vacuum expectation value.

Since $\alpha _{\phi q}^{(1)}$ and $\alpha _{\phi q}^{(3)}$ are zero there
is no net contribution to the left effective couplings. On the contrary, $%
\alpha _{\phi u},$ $\alpha _{\phi d},$ and $\alpha _{\phi \phi }$ contribute
to the effective operators containing right-handed fields
\begin{eqnarray}
\frac{M_{R}^{2\dagger }+M_{R}^{2\dagger }}{2} &=&-\frac{v^{2}}{8}\left(
\alpha _{\phi d}+\alpha _{\phi d}^{\dagger }+\alpha _{\phi u}+\alpha _{\phi
u}^{\dagger }\right) ,  \notag \\
\frac{M_{R}^{2}-M_{R}^{2\dagger }}{2} &=&\frac{v^{2}}{8}\left( \alpha _{\phi
\phi }-\alpha _{\phi \phi }^{\dagger }\right) ,  \notag \\
\frac{M_{R}^{1}+M_{R}^{1\dagger }}{2} &=&\frac{v^{2}}{16}\left( \alpha
_{\phi d}+\alpha _{\phi d}^{\dagger }-\alpha _{\phi u}-\alpha _{\phi
u}^{\dagger }+\alpha _{\phi \phi }+\alpha _{\phi \phi }^{\dagger }\right) ,
\notag \\
\frac{M_{R}^{3}+M_{R}^{3\dagger }}{2} &=&\frac{v^{2}}{16}\left( \alpha
_{\phi d}+\alpha _{\phi d}^{\dagger }-\alpha _{\phi u}-\alpha _{\phi
u}^{\dagger }-\alpha _{\phi \phi }-\alpha _{\phi \phi }^{\dagger }\right) ,
\label{eqa54}
\end{eqnarray}
In the process of integrating out the heavy fermions new mass terms have
been generated, so the mass matrix (of the light fermions) needs a further
re-diagonalization. This is quite standard and can be done by using the
formulae given in section \ref{physical}. After diagonalization we should
just replace $M_{R}^{i}\rightarrow \tilde{M}_{R}^{i}$ and this is the final
result in the physical basis. As we can see, the contribution to the
effective couplings, and hence to the observables, is always suppressed by a
power of $M^{-2}$, the scale of the new physics, as announced in the
introduction. The contribution from many other models involving heavy
fermions can be deduced from \cite{paco} in a similar way and general
patterns inferred.

The second example we would like to briefly discuss is the Standard Model
itself. Particularly, the Standard Model in the limit of a heavy Higgs. In
the case without mixing the effective coefficients were derived in \cite
{nomix}. The results in the general case where mixing is present are given
by
\begin{eqnarray}
\left( \tilde{M}_{R}^{2}-\tilde{M}_{R}^{2\dagger }\right) _{ij} &=&-\frac{1}{%
16\pi ^{2}}\frac{m_{i}^{u}K_{ij}m_{j}^{d}-m_{i}^{d}K_{ij}^{\dagger }m_{j}^{u}%
}{4v^{2}}(\frac{1}{\hat{\epsilon}}-\log \frac{M_{H}^{2}}{\mu ^{2}}+\frac{5}{2%
}),  \notag \\
\left( \tilde{M}_{R}^{2}+\tilde{M}_{R}^{2\dagger }\right) _{ij} &=&\frac{1}{%
16\pi ^{2}}\frac{m_{i}^{d2}-m_{i}^{u2}}{4v^{2}}\left( \frac{1}{\hat{\epsilon}%
}-\log \frac{M_{H}^{2}}{\mu ^{2}}+\frac{5}{2}\right) \delta _{ij},  \notag \\
\left( \tilde{M}_{R}^{1}+\tilde{M}_{R}^{1\dagger }\right) _{ij} &=&-\frac{1}{%
16\pi ^{2}}\frac{\left( m_{i}^{u2}+m_{i}^{d2}\right) \delta
_{ij}+m_{i}^{u}K_{ij}m_{j}^{d}+m_{i}^{d}K_{ij}^{\dagger }m_{j}^{u}}{8v^{2}}(%
\frac{1}{\hat{\epsilon}}-\log \frac{M_{H}^{2}}{\mu ^{2}}+\frac{5}{2}),
\notag \\
\left( \tilde{M}_{R}^{3}+\tilde{M}_{R}^{3\dagger }\right) _{ij} &=&-\frac{1}{%
16\pi ^{2}}\frac{\left( m_{i}^{u2}+m_{i}^{d2}\right) \delta
_{ij}-m_{i}^{u}K_{ij}m_{j}^{d}-m_{i}^{d}K_{ij}^{\dagger }m_{j}^{u}}{8v^{2}}(%
\frac{1}{\hat{\epsilon}}-\log \frac{M_{H}^{2}}{\mu ^{2}}+\frac{5}{2}),
\notag \\
\left( \hat{M}_{L}^{4}+\hat{M}_{L}^{4\dagger }\right) _{ij} &=&\frac{1}{%
16\pi ^{2}}\frac{m_{i}^{u2}\delta _{ij}-K_{ik}m_{k}^{d2}K_{kj}^{\dagger }}{%
4v^{2}}(\frac{1}{\hat{\epsilon}}-\log \frac{M_{H}^{2}}{\mu ^{2}}+\frac{1}{2}%
),  \notag \\
\left( \hat{M}_{L}^{2\dagger }+\hat{M}_{L}^{2}\right) _{ij} &=&\frac{1}{%
16\pi ^{2}}\frac{m_{i}^{u2}\delta _{ij}-K_{ik}m_{k}^{d2}K_{kj}^{\dagger }}{%
4v^{2}}\left( \frac{1}{\hat{\epsilon}}-\log \frac{M_{H}^{2}}{\mu ^{2}}+\frac{%
5}{2}\right) ,  \notag \\
\left( \hat{M}_{L}^{1}+\hat{M}_{L}^{1\dagger }\right) _{ij} &=&-\frac{1}{%
16\pi ^{2}}\frac{m_{i}^{u2}\delta _{ij}+K_{ik}m_{k}^{d2}K_{kj}^{\dagger }}{%
4v^{2}}(\frac{1}{\hat{\epsilon}}-\log \frac{M_{H}^{2}}{\mu ^{2}}+\frac{5}{2}%
),  \notag \\
\left( \hat{M}_{L}^{3}+\hat{M}_{L}^{3\dagger }\right) _{ij} &=&0,  \notag \\
\left( \hat{M}_{L}^{2}-\hat{M}_{L}^{2\dagger }\right) _{ij} &=&-\left( \hat{M%
}_{L}^{4}-\hat{M}_{L}^{4\dagger }\right) _{ij},  \label{effmix}
\end{eqnarray}
where we have used dimensional regularization with $d=4-2\epsilon $ and $%
\left\{ \gamma ^{5},\gamma ^{\mu }\right\} =0$; we have also defined $\frac{1%
}{\hat{\epsilon}}=\frac{1}{\epsilon }-\gamma +\log 4\pi $. Form Eqs. (\ref
{effmix}), (\ref{Zcoupl1}) and (\ref{vertices1}) we immediately obtain the
contribution to the $Z$ and $W$ current vertices

\begin{eqnarray}
g_{L}^{u} &=&\frac{1}{16\pi ^{2}}\frac{m_{i}^{u2}\delta _{ij}}{2v^{2}}\left(
\frac{1}{\hat{\epsilon}}-\log \frac{M_{H}^{2}}{\mu ^{2}}+\frac{5}{2}\right) ,
\notag \\
g_{L}^{d} &=&-\frac{1}{16\pi ^{2}}\frac{m_{i}^{d2}\delta _{ij}}{2v^{2}}(%
\frac{1}{\hat{\epsilon}}-\log \frac{M_{H}^{2}}{\mu ^{2}}+\frac{5}{2}),
\notag \\
g_{R}^{u} &=&-\frac{1}{16\pi ^{2}}\frac{m_{i}^{u2}\delta _{ij}}{2v^{2}}(%
\frac{1}{\hat{\epsilon}}-\log \frac{M_{H}^{2}}{\mu ^{2}}+\frac{5}{2}),
\notag \\
g_{R}^{d} &=&\frac{1}{16\pi ^{2}}\frac{m_{i}^{d2}\delta _{ij}}{2v^{2}}\left(
\frac{1}{\hat{\epsilon}}-\log \frac{M_{H}^{2}}{\mu ^{2}}+\frac{5}{2}\right) ,
\notag \\
h_{L} &=&\frac{1}{16\pi ^{2}}\frac{m_{i}^{u2}K_{ij}+K_{ij}m_{j}^{d2}}{4v^{2}}%
(\frac{1}{\hat{\epsilon}}-\log \frac{M_{H}^{2}}{\mu ^{2}}+\frac{5}{2}),
\notag \\
h_{R} &=&-\frac{1}{16\pi ^{2}}\frac{m_{i}^{u}K_{ij}m_{j}^{d}}{2v^{2}}(\frac{1%
}{\hat{\epsilon}}-\log \frac{M_{H}^{2}}{\mu ^{2}}+\frac{5}{2}).
\label{higgscoupl}
\end{eqnarray}
These coefficients summarize the non-decoupling effects of a heavy Higgs in
the Standard Model. Note that a heavy Higgs gives rise to radiative
corrections that do not contribute to flavor changing neutral currents, but
generates contributions to the charged currents that alter the unitarity of
the left mixing matrix $\mathcal{U}$ and produces a right mixing matrix
which is non-unitary and of course is not present at tree level.

The divergence of these coefficients just reflect that the Higgs is a
necessary ingredient for the Standard Model to be renormalizable. These
divergences cancel the singularities generated by radiative corrections in
the light sector. At the end of the day, this amounts to cancelling all $%
\frac{1}{\epsilon }$ and replacing $\mu \rightarrow M_{W}$.

Although, strictly speaking, the above results hold in the minimal Standard
Model, experience from a similar calculation (without mixing) in the
two-Higgs doublet model \cite{ciafaloni} leads us to conjecture that they
also hold for a large class of extended scalar sectors, provided that all
other scalar particles in the spectrum are made sufficiently heavy. Unless
some additional $CP$ violation is included in the two-doublet potential,
there is only one phase: the one of the Standard Model.

Thus we have seen how different type of theories lead to a very different
pattern for the coefficients of the effective theory and, eventually, to the
$CP$-violating observables. Theories with scalars are, generically,
non-decoupling, with large logs, which are nevertheless suppressed by the
usual loop factors. Theories with additional fermions are decoupling, but
provide contributions already at tree level. For heavy doublets only in the
right-handed sector, it turns out.

\section{Conclusions}

\label{conclusions}In this chapter we have performed a rather detailed
analysis of the issue of possible departures from the Standard Model in
effective vertices, with an special interest in the issue of possible new
sources of $CP$ violation and family mixing. The analysis we have performed
is rather general. We only assume that all ---so far--- undetected degrees
of freedom are heavy enough for an expansion in inverse powers of their mass
to be justified.

We have retained in all cases the leading contribution to the observables
from the effective Lagrangian. To be fully consistent one should, at the
same time, include the one-loop corrections from the Standard Model without
Higgs (universal). We have not done so, so our results are sensitive to the
contribution from the new physics ---encoded in the coefficients of the
effective Lagrangian--- inasmuch as this dominates over the Standard Model
radiative corrections. Anyhow, it is usually possible to treat radiative
corrections with the help of effective couplings, thus falling back again in
an effective Lagrangian treatment.

There are two main theoretical results presented in this chapter. First of
all, we have performed a complete study of all the possible new operators,
to leading order, and the way to implement the passage to the physical basis
when these additional interactions are included. Once this diagonalization
is performed we have found that new structures appear in the left effective
operators. In particular the CKM matrix shows up also in the neutral sector.

Secondly, we have included the contribution to physical amplitudes of the
wave function and CKM matrix element renormalization. Both need to be
included when the contribution from the effective operators to the different
observables is considered. In the next chapter we will analyze the
renormalization issue in detail providing the theoretical ground for the
wfr. and CKM counterterms used in this chapter.

Besides that, we have also computed the relevant coefficients in a number of
theories. Theories with extended matter sectors give, in principle,
relatively large contributions, since they contribute at the tree level.
When only heavy doublets are considered, the relevant left couplings are
left untouched. Observable effects should be sought after in the
right-handed sector. The contribution from the new physics is decoupling
(i.e. vanishes when the scale is sent to infinity). However the limits on
additional vector generations are weak, roughly one requires only their mass
to be heavier than the top one, so this may lead to large contributions. Of
course, there are mixing parameters $\lambda $, which can be bound from
flavor changing phenomenology. Measuring the right-handed couplings seems
the most promising way to test these possible effects. Stringent bounds
exist in this respect from $b\rightarrow s\gamma $, constraining the
couplings at the few per mille level \cite{cleo}. If one assumes some sort
of naturality argument for the scale of the coefficients in the effective
Lagrangian, this precludes observation unless at least the $1\%$ level of
accuracy is reached. Theories with extended scalar sectors are (unless fine
tuning of the potential is present such as in e.g. supersymmetric theories)
non-decoupling and in order to make its contribution larger than the
universal radiative corrections one requires a heavy Higgs (although their
contribution, with respect to universal radiative corrections is
nevertheless enhanced by the top Yukawa coupling).

In general, even if the physics responsible for the generation of the
additional effective operators is $CP$-conserving, phases which are present
in the Yukawa and kinetic couplings become observable. This should produce a
wealth of phases and new $CP$- violating effects. As we have seen,
contributions reaching the $1\%$ level are not easy to find, so it will be
extremely difficult to find any sizeable deviation with respect the Standard
Model in the ongoing experiments.

Moreover, since a good part of the radiative corrections in the Standard
Model itself can be incorporated in the $d=4$ effective operators (we have
seen that explicitly for the Higgs contribution) our results will be
relevant the day that experiments become accurate enough so that radiative
corrections are required. Finally, the effective Lagrangian approach
consists not only in writing down the Lagrangian itself, but it also comes
with a well defined set of counting rules. This set of counting rules allows
in the case of the CKM matrix elements a perturbative treatment of the
unitarity constraint. If one assumes that the contribution from new physics
and radiative corrections are comparable, then it is legitimate to use the
unitarity relations in all one-loop calculations. At tree-level, the
predictions should be modified to account for the presence of the
new-physics which introduces new phases. This procedure can be extended to
arbitrary order.

\chapter{Gauge invariance and wave-function renormalization}

\label{LSZchapter}In the previous chapters we have been concerned with the
contribution of effective operators to observable quantities. We have seen
also that low energy effects of genuine radiative corrections can also be
incorporated using this language. In particular electroweak radiative
corrections are known to be crucial in the neutral sector to bring theory
and experiment into agreement. Tree level results are incompatible with
experiment by many standard deviations \cite{Pich:1997ga}. Obviously we are
not there yet in the charged current sector, but in a few years electroweak
radiative corrections will be required in the studies analyzing the
``unitarity'' of the CKM matrix\footnote{%
The CKM matrix is certainly unitary, but the physical observables that at
tree level coincide with these matrix elements certainly do not necessarily
fulfil a unitarity constraint once quantum corrections are switched on (see
the previous chapter).}. It is hard to come with realistic models where new
physics gives contributions much larger than radiative corrections at low
energies, so it is crucial to have the latter under control.

These corrections are of several types. We need counter terms for the
electric charge, Weinberg angle and wave-function renormalization (wfr.) for
the $W$ gauge boson. We shall also require wfr. for the external fermions
and counter terms for the entries of the CKM matrix. The latter are in fact
related in a way that will be described below \cite{Balzereit:1999id}.
Finally one needs to include the 1PI vertex parts.

As explained in the introduction there has been some controversy in the
literature regarding the correct implementation of the on-shell scheme in
the presence of mixing. This chapter is dedicated to the analysis of this
problem showing that the source of conflict is located in the absorptive
parts of the fermionic self-energies. In the previous implementations of the
on-shell scheme such parts were dropped in the calculation of the wfr.
constants \cite{Denner}. We show that these parts are necessary for the
implementation of the no-mixing conditions on the fermionic propagator \cite
{Aoki} and furthermore to guarantee the gauge invariance of physical
amplitudes. In the following sections we shall compute the gauge dependence
of the absorptive parts in the self-energies and the vertex functions. We
shall see how the requirements of gauge invariance and proper on-shell
conditions (including exact diagonalization in flavor space) single out a
unique prescription for the wfr. We present the problem in detail in the
next section with the explicit expressions for the renormalization constants
given in sections \ref{offdiagonal} and \ref{diagonal}. Implementation for $%
W $ and top decay are shown in section \ref{wandtdecay}. A discussion to
extract the gauge dependence of all absorptive terms has been made in
section \ref{nielsen}. There extensive use of the Nielsen identities \cite
{Nielsen,Piguet,Sibold,Gambino} has been made. Previously in section \ref
{nielsenintro} we provide a brief introduction to such identities designed
for a quick understanding of their content. In section \ref{absorptive} and
\ref{cp} we return to $W$ and top decay to implement the previous results
and finally we conclude in section \ref{conc}.

\section{Statement of the problem and its solution}

\label{statement}We want to define an on-shell renormalization scheme that
guarantees the correct properties of the fermionic propagator in the $%
p^{2}\rightarrow m_{i}^{2}$ limit and at the same time renders the
observable quantities calculated in such a scheme gauge parameter
independent. In the first place up and down-type propagators have to be
family diagonal on-shell. The conditions necessary for that purpose were
first given by Aoki et. al. in \cite{Aoki}. Let us introduce some notation
in order to write them down. We renormalize the bare fermion fields $\Psi
_{0}$ and $\bar{\Psi}_{0}$ as
\begin{equation}
\Psi _{0}=Z^{\frac{1}{2}}\Psi \,,\qquad \bar{\Psi}_{0}=\bar{\Psi}\bar{Z}^{%
\frac{1}{2}}\,.  \label{fundamental}
\end{equation}
For reasons that will become clear along the discussion, we shall allow $Z$
and $\bar{Z}$ to be independent renormalization constants\footnote{%
This immediately raises some issues about hermiticity which we shall deal
with below.}. These renormalization constants contain flavor, family and
Dirac indices. We can decompose them into
\begin{equation}
Z^{\frac{1}{2}}=Z^{u\frac{1}{2}}\tau ^{u}+Z^{d\frac{1}{2}}\tau ^{d}\,,\qquad
\bar{Z}^{\frac{1}{2}}=\bar{Z}^{u\frac{1}{2}}\tau ^{u}+\bar{Z}^{d\frac{1}{2}%
}\tau ^{d}\,,  \label{ud}
\end{equation}
with $\tau ^{u}$ and $\tau ^{d}$ the up and down flavor projectors and
furthermore each piece in left and right chiral projectors, $L$ and $R$
respectively,
\begin{equation}
Z^{u\frac{1}{2}}=Z^{uL\frac{1}{2}}L+Z^{uR\frac{1}{2}}R\,,\qquad \bar{Z}^{u%
\frac{1}{2}}=\bar{Z}^{uL\frac{1}{2}}R+\bar{Z}^{uR\frac{1}{2}}L\,.  \label{LR}
\end{equation}
Analogous decompositions hold for $Z^{d\frac{1}{2}}$ and $\bar{Z}^{d\frac{1}{%
2}}$. Due to radiative corrections the propagator mixes fermion of different
family indices. Namely
\begin{equation*}
iS^{-1}\left( p\right) =\bar{Z}^{\frac{1}{2}}\left( \frac{{}}{{}}\not{p}%
-m-\delta m-\Sigma \left( p\right) \right) Z^{\frac{1}{2}}\,,
\end{equation*}
where the bare self-energy $\Sigma $ is non-diagonal and is given by $%
-i\Sigma =\sum $1PI. Within one-loop accuracy we can write $Z^{\frac{1}{2}%
}=1+\frac{1}{2}\delta Z$ etc. Introducing the family indices explicitly we
have
\begin{equation*}
iS_{ij}^{-1}\left( p\right) =\left( \not{p}-m_{i}\right) \delta _{ij}-\hat{%
\Sigma}_{ij}\left( p\right) \,,
\end{equation*}
where the one-loop renormalized self-energy is given by
\begin{equation}
\hat{\Sigma}_{ij}\left( p\right) =\Sigma _{ij}\left( p\right) -\frac{1}{2}%
\delta \bar{Z}_{ij}\left( \not{p}-m_{j}\right) -\frac{1}{2}\left( \not{p}%
-m_{i}\right) \delta Z_{ij}+\delta m_{i}\delta _{ij}\,.  \label{renself}
\end{equation}
Since we can project the above definition for up and down type-quarks,
flavor indices will be dropped in the sequel and only will be restored when
necessary. Recalling the following on-shell relations for Dirac spinors ($%
p^{2}\rightarrow m_{i}^{2}$)
\begin{eqnarray}
\left( \not{p}-m_{i}\right) u_{i}^{\left( s\right) }\left( p\right) &=&0\,,
\notag \\
\bar{u}_{i}^{\left( s\right) }\left( p\right) \left( \not{p}-m_{i}\right)
&=&0\,,  \notag \\
\left( \not{p}-m_{i}\right) v_{i}^{\left( s\right) }\left( -p\right) &=&0\,,
\notag \\
\bar{v}_{i}^{\left( s\right) }\left( -p\right) \left( \not{p}-m_{i}\right)
&=&0\,,  \label{onshellspinors}
\end{eqnarray}
the conditions \cite{Aoki} necessary to avoid mixing will be\footnote{%
Notice that, as a matter of fact, in \cite{Aoki} the conditions over
anti-fermions are not stated.}

\begin{eqnarray}
\hat{\Sigma}_{ij}\left( p\right) u_{j}^{\left( s\right) }\left( p\right)
&=&0\,,\qquad (p^{2}\rightarrow m_{j}^{2})\,,\quad \mathrm{(incoming}\text{ }%
\mathrm{particle)}  \label{inparticle} \\
\bar{v}_{i}^{\left( s\right) }\left( -p\right) \hat{\Sigma}_{ij}\left(
p\right) &=&0\,,\qquad (p^{2}\rightarrow m_{i}^{2})\,,\quad \mathrm{(incoming%
}\text{ }\mathrm{anti}\mathrm{-}\mathrm{particle)}  \label{inantiparticle} \\
\bar{u}_{i}^{\left( s\right) }\left( p\right) \hat{\Sigma}_{ij}\left(
p\right) &=&0\,,\qquad (p^{2}\rightarrow m_{i}^{2})\,,\quad \mathrm{(outgoing%
}\text{ }\mathrm{particle)}  \label{outparticle} \\
\hat{\Sigma}_{ij}\left( p\right) v_{j}^{\left( s\right) }\left( -p\right)
&=&0\,,\qquad (p^{2}\rightarrow m_{j}^{2})\,,\quad \mathrm{(outgoing}\text{ }%
\mathrm{anti}\mathrm{-}\mathrm{particle)}  \label{outantiparticle}
\end{eqnarray}
where no summation over repeated indices is assumed and $i\neq j.$ These
relations determine the non-diagonal parts of $Z$ and $\bar{Z}$ as will be
proven in the next section. Here, as a side remark, let us point out that
the need of different ``incoming'' and ``outgoing'' wfr. constants was
already recognized in \cite{EspMan}. Nevertheless, that paper was
unsuccessful in reconciling the on-shell prescription with the presence of
absorptive terms in the self-energies. However, since its results are
concerned with the leading contribution of an effective Lagrangian, no
absorptive terms are present and therefore conclusions still hold.

To obtain the diagonal parts $Z_{ii},$ $\bar{Z}_{ii}$, and $\delta m_{i}$
one imposes mass pole and unit residue conditions (to be discussed below).
Here it is worth to make one important comment regarding the above
conditions. First of all we note that in the literature the relation
\begin{equation}
\bar{Z}^{\frac{1}{2}}=\gamma ^{0}Z^{\frac{1}{2}\dagger }\gamma ^{0}\,,
\label{hermiticity}
\end{equation}
is taken for granted. This relation is tacitly assumed in \cite{Aoki} and
explicitly required in \cite{Denner}. Imposing Eq. (\ref{hermiticity}) would
guarantee hermiticity of the Lagrangian written in terms of the renormalized
physical fields. However, we are at this point concerned with external leg
renormalization, for which it is perfectly possible to use a different set
of renormalization constants (even ones that do not respect the requirement (%
\ref{hermiticity})), while keeping the Lagrangian hermitian. In fact, using
two sets of renormalization constants is a standard practice in the on-shell
scheme \cite{Hollik}, so one should not be concerned by this fact \textit{%
per se}. In case one is worried about the consistency of using a set of wfr.
constants not satisfying (\ref{hermiticity}) for the external legs while
keeping a hermitian Lagrangian, it should be pointed out that there is a
complete equivalence between the set of renormalization constants we shall
find out below and a treatment of the external legs where diagrams with
self-energies (including mass counter terms) are inserted instead of the
wfr. constants; provided, of course, that the mass counter term satisfy the
on-shell condition. Proceeding in this way gives results identical to ours
and different from those obtained using the wfr. proposed in \cite{Denner},
which do fulfil (\ref{hermiticity}). Further consistency checks are
presented in the following sections.

In any case, self-energies develop absorptive terms and this makes Eq. (\ref
{hermiticity}) incompatible with the diagonalizing conditions (\ref
{inparticle})-(\ref{outantiparticle}). Therefore in order to circumvent this
problem one can give up diagonalization conditions (\ref{inparticle})-(\ref
{outantiparticle}) or alternatively the hermiticity condition (\ref
{hermiticity}). The approach taken originally in \cite{Denner} and works
thereafter was the former alternative, while in this work we shall advocate
the second one. The approach of \cite{Denner} consists in dropping out
absorptive terms from conditions (\ref{inparticle})-(\ref{outantiparticle}).
That is for $i\neq j$
\begin{eqnarray}
\widetilde{Re}\left( \hat{\Sigma}_{ij}\left( p\right) \right) u_{j}^{\left(
s\right) }\left( p\right) &=&0\,,\qquad (p^{2}\rightarrow m_{j}^{2})\,,\quad
\mathrm{(incoming}\text{ }\mathrm{particle)}  \notag \\
\bar{v}_{i}^{\left( s\right) }\left( -p\right) \widetilde{Re}\left( \hat{%
\Sigma}_{ij}\left( p\right) \right) &=&0\,,\qquad (p^{2}\rightarrow
m_{i}^{2})\,,\quad \mathrm{(incoming}\text{ }\mathrm{anti}\mathrm{-}\mathrm{%
particle)}  \notag \\
\bar{u}_{i}^{\left( s\right) }\left( p\right) \widetilde{Re}\left( \hat{%
\Sigma}_{ij}\left( p\right) \right) &=&0\,,\qquad (p^{2}\rightarrow
m_{i}^{2})\,,\quad \mathrm{(outgoing}\text{ }\mathrm{particle)}  \notag \\
\widetilde{Re}\left( \hat{\Sigma}_{ij}\left( p\right) \right) v_{j}^{\left(
s\right) }\left( -p\right) &=&0\,,\qquad (p^{2}\rightarrow
m_{j}^{2})\,,\quad \mathrm{(outgoing}\text{ }\mathrm{anti}\mathrm{-}\mathrm{%
particle)}  \label{Denner1}
\end{eqnarray}
where $\widetilde{Re}$ includes the real part of the logarithms arising in
loop integrals appearing in the self-energies but not of the rest of
coupling factors of the Feynmann diagram. This approach is compatible with
the hermiticity condition (\ref{hermiticity}) but on the other hand have
several drawbacks. These drawbacks include

\begin{enumerate}
\item  Since only the $\widetilde{Re}$ part of the self-energies enters into
the diagonalizing conditions the on-shell propagator remains non-diagonal.

\item  The very definition of $\widetilde{Re}$ relies heavily on the
one-loop perturbative calculation where it is applied upon. In other words $%
\widetilde{Re}$ is not a proper function of its argument (in contrast to $Re$%
) and it is presumably cumbersome to implement in multi-loop calculations.

\item  As it will become clear in next sections, the on-shell scheme based
in the $\widetilde{Re}$ prescription leads to gauge parameter dependent
physical amplitudes. The reason for this unwanted dependence is the dropping
of absorptive gauge parameter dependent terms in the self-energies that are
necessary to cancel absorptive terms appearing in the vertices. As mentioned
in the introduction, in the SM, the gauge dependence drops in the modulus
squared of the amplitude, but not in the amplitude itself and it could be
eventually observable.
\end{enumerate}

Once stated the unwanted features of the $\widetilde{Re}$ approach let us
briefly state the consequences of dropping condition (\ref{hermiticity})

\begin{enumerate}
\item  Conditions (\ref{inparticle})-(\ref{outantiparticle}) readily
determine the off-diagonal $Z$ and $\bar{Z}$ wfr. which coincide with the
ones obtained using the $\widetilde{Re}$ prescription up to finite
absorptive gauge parameter dependent terms.

\item  The renormalized fermion propagator becomes exactly diagonal
on-shell, unlike in the $\widetilde{Re}$ scheme.

\item  Incoming and outgoing particles and anti-particles require different
renormalization constants when computing a physical amplitude. Annihilation
of particles and creation of anti-particles are accompanied by the
renormalization constant $Z$, while creation of particles and annihilation
of anti-particles are accompanied by the renormalization constant $\bar{Z}$.

\item  These constants $Z$ and $\bar{Z}$ are in what respects to their
dispersive parts identical to the ones in \cite{Denner}. They differ in
their absorptive parts. This might suggest to the alert reader there could
be problems with fundamental symmetries such as $CP$ or $CPT$. We shall
discuss this issue at the end of the chapter. Our conclusion is that
everything works out consistently in this respect.
\end{enumerate}

For explicit expressions for $Z$ and $\bar{Z}$ the reader should consult
formulae (\ref{zin}), (\ref{zout}) and (\ref{zdiag}) in the next two
sections. As an example how to implement them see section \ref{wandtdecay}.
The explicit dependence on the gauge parameter (for simplicity only the $W$
gauge parameter is considered) of the absorptive parts is given in section
\ref{absorptive}.

\section{Off-diagonal wave-function renormalization constants}

\label{offdiagonal}This section is devoted to a detailed derivation of the
off-diagonal renormalization constants deriving entirely from the on-shell
conditions (\ref{inparticle})-(\ref{outantiparticle}) and allowing for $\bar{%
Z}^{\frac{1}{2}}\neq \gamma ^{0}Z^{\frac{1}{2}\dagger }\gamma ^{0}$. First
of all we decompose the renormalized self-energy into all possible Dirac
structures

\begin{equation}
\hat{\Sigma}_{ij}\left( p\right) =\not{p}\left( \hat{\Sigma}_{ij}^{\gamma
R}\left( p^{2}\right) R+\hat{\Sigma}_{ij}^{\gamma L}\left( p^{2}\right)
L\right) +\hat{\Sigma}_{ij}^{R}\left( p^{2}\right) R+\hat{\Sigma}%
_{ij}^{L}\left( p^{2}\right) L\,,  \label{decomp}
\end{equation}
and use Eqs. (\ref{LR}), (\ref{renself}) and (\ref{decomp}) to obtain
\begin{eqnarray}
\hat{\Sigma}_{ij}\left( p\right) &=&\not{p}R\left( \Sigma _{ij}^{\gamma
R}\left( p^{2}\right) -\frac{1}{2}\delta \bar{Z}_{ij}^{R}-\frac{1}{2}\delta
Z_{ij}^{R}\right) +\not{p}L\left( \Sigma _{ij}^{\gamma L}\left( p^{2}\right)
-\frac{1}{2}\delta \bar{Z}_{ij}^{L}-\frac{1}{2}\delta Z_{ij}^{L}\right)
\notag  \label{renormself} \\
&&\hspace{-1.3cm}+R\left( \Sigma _{ij}^{R}\left( p^{2}\right) +\frac{1}{2}%
\left( \delta \bar{Z}_{ij}^{L}m_{j}+m_{i}\delta Z_{ij}^{R}\right) +\delta
_{ij}\delta m_{i}\right)  \notag \\
&&+L\left( \Sigma _{ij}^{L}\left( p^{2}\right) +\frac{1}{2}\left( \delta
\bar{Z}_{ij}^{R}m_{j}+m_{i}\delta Z_{ij}^{L}\right) +\delta _{ij}\delta
m_{i}\right) \,.
\end{eqnarray}
Repeated indices are not summed over. Hence from Eqs. (\ref{renormself}), (%
\ref{onshellspinors}) and (\ref{inparticle}) we obtain
\begin{eqnarray*}
\Sigma _{ij}^{\gamma R}\left( m_{j}^{2}\right) m_{j}-\frac{1}{2}\delta
Z_{ij}^{R}m_{j}+\Sigma _{ij}^{L}\left( m_{j}^{2}\right) +\frac{1}{2}%
m_{i}\delta Z_{ij}^{L} &=&0\,, \\
\Sigma _{ij}^{\gamma L}\left( m_{j}^{2}\right) m_{j}-\frac{1}{2}\delta
Z_{ij}^{L}m_{j}+\Sigma _{ij}^{R}\left( m_{j}^{2}\right) +\frac{1}{2}%
m_{i}\delta Z_{ij}^{R} &=&0\,.
\end{eqnarray*}
Exactly the same relations are obtained from Eqs. (\ref{renormself}), (\ref
{onshellspinors}) and Eq. (\ref{outantiparticle}). Analogously, Eqs. (\ref
{renormself}), (\ref{onshellspinors}) and Eq. (\ref{inantiparticle}) (or Eq.
(\ref{outparticle})) lead to
\begin{eqnarray*}
m_{i}\Sigma _{ij}^{\gamma R}\left( m_{i}^{2}\right) -\frac{1}{2}m_{i}\delta
\bar{Z}_{ij}^{R}+\Sigma _{ij}^{R}\left( m_{i}^{2}\right) +\frac{1}{2}\delta
\bar{Z}_{ij}^{L}m_{j} &=&0\,, \\
m_{i}\Sigma _{ij}^{\gamma L}\left( m_{i}^{2}\right) -\frac{1}{2}m_{i}\delta
\bar{Z}_{ij}^{L}+\Sigma _{ij}^{L}\left( m_{i}^{2}\right) +\frac{1}{2}\delta
\bar{Z}_{ij}^{R}m_{j} &=&0\,.
\end{eqnarray*}
Using the above expressions we immediately obtain
\begin{eqnarray}
\delta Z_{ij}^{L} &=&\frac{2}{m_{j}^{2}-m_{i}^{2}}\left[ \Sigma
_{ij}^{\gamma R}\left( m_{j}^{2}\right) m_{i}m_{j}+\Sigma _{ij}^{\gamma
L}\left( m_{j}^{2}\right) m_{j}^{2}+m_{i}\Sigma _{ij}^{L}\left(
m_{j}^{2}\right) +\Sigma _{ij}^{R}\left( m_{j}^{2}\right) m_{j}\right] \,,
\notag \\
\delta Z_{ij}^{R} &=&\frac{2}{m_{j}^{2}-m_{i}^{2}}\left[ \Sigma
_{ij}^{\gamma L}\left( m_{j}^{2}\right) m_{i}m_{j}+\Sigma _{ij}^{\gamma
R}\left( m_{j}^{2}\right) m_{j}^{2}+m_{i}\Sigma _{ij}^{R}\left(
m_{j}^{2}\right) +\Sigma _{ij}^{L}\left( m_{j}^{2}\right) m_{j}\right] \,,
\label{zin}
\end{eqnarray}
and
\begin{eqnarray}
\delta \bar{Z}_{ij}^{L} &=&\frac{2}{m_{i}^{2}-m_{j}^{2}}\left[ \Sigma
_{ij}^{\gamma R}\left( m_{i}^{2}\right) m_{i}m_{j}+\Sigma _{ij}^{\gamma
L}\left( m_{i}^{2}\right) m_{i}^{2}+m_{i}\Sigma _{ij}^{L}\left(
m_{i}^{2}\right) +\Sigma _{ij}^{R}\left( m_{i}^{2}\right) m_{j}\right] \,,
\notag \\
\delta \bar{Z}_{ij}^{R} &=&\frac{2}{m_{i}^{2}-m_{j}^{2}}\left[ \Sigma
_{ij}^{\gamma L}\left( m_{i}^{2}\right) m_{i}m_{j}+\Sigma _{ij}^{\gamma
R}\left( m_{i}^{2}\right) m_{i}^{2}+m_{i}\Sigma _{ij}^{R}\left(
m_{i}^{2}\right) +\Sigma _{ij}^{L}\left( m_{i}^{2}\right) m_{j}\right] \,.
\label{zout}
\end{eqnarray}
At the one-loop level in the SM we can define
\begin{equation*}
\Sigma _{ij}^{R}\left( p^{2}\right) \equiv \Sigma _{ij}^{S}\left(
p^{2}\right) m_{j}\,,\qquad \Sigma _{ij}^{L}\left( p^{2}\right) \equiv
m_{i}\Sigma _{ij}^{S}\left( p^{2}\right) \,,
\end{equation*}
and therefore
\begin{eqnarray*}
\delta \bar{Z}_{ij}^{L}-\delta Z_{ij}^{L\dagger } &=&\frac{2}{%
m_{i}^{2}-m_{j}^{2}}\left\{ \frac{{}}{{}}\left( \Sigma _{ij}^{\gamma
R}\left( m_{i}^{2}\right) -\Sigma _{ji}^{\gamma R\ast }\left(
m_{i}^{2}\right) \right) m_{i}m_{j}\right. +\left( \Sigma _{ij}^{\gamma
L}\left( m_{i}^{2}\right) -\Sigma _{ji}^{\gamma L\ast }\left(
m_{i}^{2}\right) \right) m_{i}^{2} \\
&&\left. +\left( m_{i}^{2}+m_{j}^{2}\right) \left( \frac{{}}{{}}\Sigma
_{ij}^{S}\left( m_{i}^{2}\right) -\Sigma _{ji}^{S\ast }\left(
m_{i}^{2}\right) \frac{{}}{{}}\right) \frac{{}}{{}}\right\} \neq 0\,,
\end{eqnarray*}
and a similar relation holds for $\delta \bar{Z}_{ij}^{R}-\delta
Z_{ij}^{R\dagger }.$ The above non-vanishing difference is due to the
presence of branch cuts in the self-energies that invalidate the
pseudo-hermiticity relation
\begin{equation}
\Sigma _{ij}\left( p\right) \neq \gamma ^{0}\Sigma _{ij}^{\dagger }\left(
p\right) \gamma ^{0}\,.  \label{pseudo}
\end{equation}
Eq. (\ref{pseudo}) is assumed in \cite{Aoki} and if we, temporally, ignore
those branch cut contributions our results reduces to the ones depicted in
\cite{DennerSack} or \cite{Denner}. In the SM these branch cuts are
generically gauge dependent as a cursory look to the appropriate integrals
shows at once.

\section{Diagonal wave-function renormalization constants}

\label{diagonal} Once the off-diagonal wfr. are obtained we focus our
attention in the diagonal sector. Near the on-shell limit we can neglect the
off-diagonal parts of the inverse propagator and write
\begin{equation}
iS_{ij}^{-1}\left( p\right) =\left( \not{p}-m_{i}-\hat{\Sigma}_{ii}\left(
p\right) \right) \delta _{ij}=\left( \frac{{}}{{}}\not{p}\left( aL+bR\right)
+cL+dR\frac{{}}{{}}\right) \delta _{ij}\,,
\end{equation}
and therefore after some algebra
\begin{equation*}
-iS_{ij}\left( p\right) =\frac{\not{p}\left( aL+bR\right) -dL-cR}{p^{2}ab-cd}%
\delta _{ij}\,,
\end{equation*}
in our case we have
\begin{eqnarray}
a &=&1-\Sigma _{ii}^{\gamma L}\left( p^{2}\right) +\frac{1}{2}\delta \bar{Z}%
_{ii}^{L}+\frac{1}{2}\delta Z_{ii}^{L}\,,  \notag \\
b &=&1-\Sigma _{ii}^{\gamma R}\left( p^{2}\right) +\frac{1}{2}\delta \bar{Z}%
_{ii}^{R}+\frac{1}{2}\delta Z_{ii}^{R}\,,  \notag \\
c &=&-\Sigma _{ii}^{L}\left( p^{2}\right) -\left( 1+\frac{1}{2}\delta \bar{Z}%
_{ii}^{R}+\frac{1}{2}\delta Z_{ii}^{L}\right) m_{i}-\delta m_{i}\,,  \notag
\\
d &=&-\Sigma _{ii}^{R}\left( p^{2}\right) -\left( 1+\frac{1}{2}\delta \bar{Z}%
_{ii}^{L}+\frac{1}{2}\delta Z_{ii}^{R}\right) m_{i}-\delta m_{i}\,.
\label{abcd}
\end{eqnarray}
In the limit $p^{2}\rightarrow m_{i}^{2}$ the chiral structures in the
numerator has to cancel ($a\rightarrow b$ and $c\rightarrow d$), this
requirement leads to
\begin{eqnarray}
\delta \bar{Z}_{ii}^{R}-\delta \bar{Z}_{ii}^{L} &=&\Sigma _{ii}^{\gamma
R}\left( m_{i}^{2}\right) -\Sigma _{ii}^{\gamma L}\left( m_{i}^{2}\right) +%
\frac{\Sigma _{ii}^{R}\left( m_{i}^{2}\right) -\Sigma _{ii}^{L}\left(
m_{i}^{2}\right) }{m_{i}}\,,  \notag \\
\delta Z_{ii}^{R}-\delta Z_{ii}^{L} &=&\Sigma _{ii}^{\gamma R}\left(
m_{i}^{2}\right) -\Sigma _{ii}^{\gamma L}\left( m_{i}^{2}\right) -\frac{%
\Sigma _{ii}^{R}\left( m_{i}^{2}\right) -\Sigma _{ii}^{L}\left(
m_{i}^{2}\right) }{m_{i}}\,.  \label{minus}
\end{eqnarray}
and we also have that
\begin{eqnarray*}
&&p^{2}b-cda^{-1} \\
&=&p^{2}\left( 1-\Sigma _{ii}^{\gamma R}\left( p_{i}^{2}\right) +\frac{1}{2}%
\delta \bar{Z}_{ii}^{R}+\frac{1}{2}\delta Z_{ii}^{R}\right) -m_{i}^{2}\left(
1+\Sigma _{ii}^{\gamma L}\left( p_{i}^{2}\right) -\frac{1}{2}\delta \bar{Z}%
_{ii}^{L}-\frac{1}{2}\delta Z_{ii}^{L}\right) \\
&&-m_{i}\left( \Sigma _{ii}^{R}\left( p_{i}^{2}\right) +\Sigma
_{ii}^{L}\left( p_{i}^{2}\right) +\left( \frac{1}{2}\delta \bar{Z}_{ii}^{L}+%
\frac{1}{2}\delta Z_{ii}^{R}+\frac{1}{2}\delta \bar{Z}_{ii}^{R}+\frac{1}{2}%
\delta Z_{ii}^{L}\right) m_{i}+2\delta m_{i}\right)
\end{eqnarray*}
since in the limit $p^{2}\rightarrow m_{i}^{2}$ we want to have a zero in
the real part of the inverse of the propagator we impose
\begin{eqnarray*}
0 &=&\lim_{p^{2}\rightarrow m_{i}^{2}}Re\left( p^{2}b-cda^{-1}\right) \\
&=&Re\left\{ m_{i}^{2}\left( -\Sigma _{ii}^{\gamma R}\left( m_{i}^{2}\right)
-\Sigma _{ii}^{\gamma L}\left( m_{i}^{2}\right) \right) \right. \\
&&-\left. \left( \Sigma _{ii}^{R}\left( m_{i}^{2}\right) +\Sigma
_{ii}^{L}\left( m_{i}^{2}\right) +2\delta m_{i}\right) m_{i}\right\}
\end{eqnarray*}
from where $\delta m_{i}$ is obtained
\begin{equation}
\delta m_{i}=-\frac{1}{2}Re\left\{ m_{i}\Sigma _{ii}^{\gamma L}\left(
m_{i}^{2}\right) +m_{i}\Sigma _{ii}^{\gamma R}+\Sigma _{ii}^{L}\left(
m_{i}^{2}\right) +\Sigma _{ii}^{R}\left( m_{i}^{2}\right) \right\} \,.
\label{deltam}
\end{equation}
This condition defines a mass and a width that agrees at the one-loop level
with the ones given in \cite{Sirlin:1991fd}, \cite{Sirlin:1998dz}, \cite
{Grassi:2001dz} and \cite{Grassi:2001bz}. Mass and width are defined as the
real an imaginary parts of the propagator pole in the complex plane
respectively. Note also that from Eqs. (\ref{abcd}) (\ref{minus}) and (\ref
{deltam}) we have
\begin{equation}
\lim_{p^{2}\rightarrow m_{i}^{2}}\left( -ca^{-1}\right) =m_{i}+\frac{i}{2}%
Im\left( \Sigma _{ii}^{\gamma R}\left( m_{i}^{2}\right) m_{i}+\Sigma
_{ii}^{\gamma L}\left( m_{i}^{2}\right) m_{i}+\Sigma _{ii}^{R}\left(
m_{i}^{2}\right) +\Sigma _{ii}^{L}\left( m_{i}^{2}\right) \right) \,,
\end{equation}
and therefore
\begin{equation*}
\lim_{p^{2}\rightarrow m_{i}^{2}}\frac{\not{p}\left( aL+bR\right) -dL-cR}{%
p^{2}ab-cd}=\frac{\not{p}+m_{i}-i\Gamma /2}{im_{i}\Gamma }\,,
\end{equation*}
where the width is defined as
\begin{equation*}
\Gamma \equiv -Im\left( \Sigma _{ii}^{\gamma R}\left( m_{i}^{2}\right)
m_{i}+\Sigma _{ii}^{\gamma L}\left( m_{i}^{2}\right) m_{i}+\Sigma
_{ii}^{R}\left( m_{i}^{2}\right) +\Sigma _{ii}^{L}\left( m_{i}^{2}\right)
\right) \,.
\end{equation*}
This quantity is ultraviolet finite. In order to find the residue in the
complex plane we expand the propagator around the physical mass obtaining
for $p^{2}\sim m_{i}^{2}$
\begin{equation}
S_{ij}\left( p\right) =\frac{i\left[ \not{p}+m_{i}-i\Gamma /2+\mathcal{O}%
\left( p^{2}-m_{i}^{2}\right) \right] }{im_{i}\Gamma +\left(
p^{2}-m_{i}^{2}\right) a^{-1}\left[ ab+m_{i}^{2}\left( a^{\prime
}b+ab^{\prime }\right) -\left( c^{\prime }d+cd^{\prime }\right) \right] }+%
\mathcal{O}\left( \left( p^{2}-m_{i}^{2}\right) ^{2}\right) \,,
\end{equation}
where $a=b$ and $c=d$ are evaluated at $p^{2}=m_{i}^{2}$. Hereafter primed
quantities denote derivatives with respect to $p^{2}$. $\mathcal{O}\left(
\left( p^{2}-m_{i}^{2}\right) ^{n}\right) $ stands for non-essential
corrections of order $(p^{2}-m_{i}^{2})^{n}$. Note that the $\mathcal{O}%
\left( p^{2}-m_{i}^{2}\right) $ corrections in the numerator do not mix with
the ones of the same order in the denominator since the first ones are of
order $\Gamma ^{-1}$ and the second ones are of order $\Gamma ^{-2}.$ Taking
into account these comments the unit residue condition amounts to requiring
\begin{eqnarray*}
1 &=&\frac{a+b}{2}+m_{i}^{2}\left( a^{\prime }+b^{\prime }\right) -\left(
c^{\prime }d+cd^{\prime }\right) a^{-1} \\
&=&\frac{a+b}{2}+m_{i}^{2}\left( a^{\prime }+b^{\prime }\right) +\left(
m_{i}-i\Gamma /2\right) \left( c^{\prime }+d^{\prime }\right) \,,
\end{eqnarray*}
from where we obtain
\begin{eqnarray}
\frac{1}{2}\left( \delta \bar{Z}_{ii}^{L}+\delta \bar{Z}_{ii}^{R}\right)
&=&\Sigma _{ii}^{\gamma L}\left( m_{i}^{2}\right) +\Sigma _{ii}^{\gamma
R}\left( m_{i}^{2}\right) -\frac{1}{2}\left( \delta Z_{ii}^{L}+\delta
Z_{ii}^{R}\right)  \notag \\
&&+2m_{i}^{2}\left( \Sigma _{ii}^{\gamma L\prime }\left( m_{i}^{2}\right)
+\Sigma _{ii}^{\gamma R\prime }\left( m_{i}^{2}\right) \right)  \notag \\
&&+2m_{i}\left( \Sigma _{ii}^{L\prime }\left( m_{i}^{2}\right) +\Sigma
_{ii}^{R\prime }\left( m_{i}^{2}\right) \right)
\end{eqnarray}
from where
\begin{eqnarray}
\frac{1}{2}\left( \delta \bar{Z}_{ii}^{L}+\delta \bar{Z}_{ii}^{R}\right)
&=&\Sigma _{ii}^{\gamma L}\left( m_{i}^{2}\right) +\Sigma _{ii}^{\gamma
R}\left( m_{i}^{2}\right) -\frac{1}{2}\left( \frac{{}}{{}}\delta
Z_{ii}^{L}+\delta Z_{ii}^{R}\frac{{}}{{}}\right) +2m_{i}^{2}\left( \Sigma
_{ii}^{\gamma L\prime }\left( m_{i}^{2}\right) +\Sigma _{ii}^{\gamma R\prime
}\left( m_{i}^{2}\right) \right)  \notag \\
&&+2m_{i}\left( \frac{{}}{{}}\Sigma _{ii}^{L\prime }\left( m_{i}^{2}\right)
+\Sigma _{ii}^{R\prime }\left( m_{i}^{2}\right) \frac{{}}{{}}\right) \,.
\label{residue}
\end{eqnarray}
We have already required all the necessary conditions to fix the correct
properties of the on-shell propagator but still there is some freedom left
in the definition of the diagonal $Z$'s. This freedom can be expressed in
terms of a set of finite coefficients $\alpha _{i}$ given by
\begin{equation*}
\frac{1}{2}\left( \delta Z_{ii}^{L}+\delta Z_{ii}^{R}\right) =\frac{1}{2}%
\left( \delta \bar{Z}_{ii}^{L}+\delta \bar{Z}_{ii}^{R}\right) +\alpha _{i}\,.
\end{equation*}
Bearing in mind that ambiguity and using Eqs. (\ref{minus}) and (\ref
{residue}) we obtain
\begin{eqnarray}
\delta \bar{Z}_{ii}^{L} &=&\Sigma _{ii}^{\gamma L}\left( m_{i}^{2}\right) -X-%
\frac{\alpha _{i}}{2}+D\,,  \notag \\
\delta \bar{Z}_{ii}^{R} &=&\Sigma _{ii}^{\gamma R}\left( m_{i}^{2}\right) +X-%
\frac{\alpha _{i}}{2}+D\,,  \notag \\
\delta Z_{ii}^{L} &=&\Sigma _{ii}^{\gamma L}\left( m_{i}^{2}\right) +X+\frac{%
\alpha _{i}}{2}+D\,,  \notag \\
\delta Z_{ii}^{R} &=&\Sigma _{ii}^{\gamma R}\left( m_{i}^{2}\right) -X+\frac{%
\alpha _{i}}{2}+D\,,  \label{zdiag}
\end{eqnarray}
where
\begin{eqnarray*}
X &=&\frac{1}{2}\frac{\Sigma _{ii}^{R}\left( m_{i}^{2}\right) -\Sigma
_{ii}^{L}\left( m_{i}^{2}\right) }{m_{i}}\,, \\
D &=&m_{i}^{2}\left( \Sigma _{ii}^{\gamma L\prime }\left( m_{i}^{2}\right)
+\Sigma _{ii}^{\gamma R\prime }\left( m_{i}^{2}\right) \right) +m_{i}\left(
\frac{{}}{{}}\Sigma _{ii}^{L\prime }\left( m_{i}^{2}\right) +\Sigma
_{ii}^{R\prime }\left( m_{i}^{2}\right) \frac{{}}{{}}\right) \,.
\end{eqnarray*}
Note that since $X=0$ at the one-loop level and choosing $\alpha _{i}=0$ we
obtain $\delta \bar{Z}_{ii}^{L}=\delta Z_{ii}^{L}$ and $\delta \bar{Z}%
_{ii}^{R}=\delta Z_{ii}^{R}.$ However we have the freedom to choose $\alpha
_{i}\neq 0$.$.$Note that the presence of $\alpha _{i}$ does not affect mass
terms since they renormalized as
\begin{equation*}
\left( \delta \bar{Z}_{ii}^{L}+\delta Z_{ii}^{R}\right) R+\left( \delta \bar{%
Z}_{ii}^{R}+\delta Z_{ii}^{L}\right) L,
\end{equation*}
which is $\alpha _{i}$ independent. Moreover all neutral currents
renormalized as
\begin{equation*}
g_{R}\left( \delta \bar{Z}_{ii}^{R}+\delta Z_{ii}^{R}\right) R+g_{L}\left(
\delta \bar{Z}_{ii}^{L}+\delta Z_{ii}^{L}\right) L,
\end{equation*}
which also $\alpha _{i}$ independent. However charged currents renormalize
as
\begin{eqnarray*}
&&\left( \delta \bar{Z}_{ik}^{uL}K_{kj}+K_{ik}\delta Z_{kj}^{dL}\right) \tau
^{-}+\left( \delta \bar{Z}_{ik}^{dL}K_{kj}^{\dagger }+K_{ik}^{\dagger
}\delta Z_{kj}^{uL}\right) \tau ^{+} \\
&=&\left( -\frac{\alpha _{i}^{u}}{2}K_{ij}+K_{ij}\frac{\alpha _{j}^{d}}{2}%
\right) \tau ^{-}+\left( -\frac{\alpha _{i}^{d}}{2}K_{ij}^{\dagger
}+K_{ij}^{\dagger }\frac{\alpha _{j}^{u}}{2}\right) \tau ^{+}
\end{eqnarray*}
hence if we take the tree level plus the above renormalization contribution
and multiply this by its Hermitian conjugate we obtain
\begin{eqnarray}
&&2\left[ \left( K-\frac{\alpha ^{u}}{2}K+K\frac{\alpha ^{d}}{2}\right)
\left( K-\frac{\alpha ^{u}}{2}K+K\frac{\alpha ^{d}}{2}\right) ^{\dagger }%
\right] _{ij}\tau ^{d}  \notag \\
&&+2\left[ \left( K^{\dagger }-\frac{\alpha _{i}^{d}}{2}K_{ij}^{\dagger
}+K_{ij}^{\dagger }\frac{\alpha _{j}^{u}}{2}\right) \left( K^{\dagger }-%
\frac{\alpha _{i}^{d}}{2}K_{ij}^{\dagger }+K_{ij}^{\dagger }\frac{\alpha
_{j}^{u}}{2}\right) ^{\dagger }\right] _{ij}\tau ^{u}  \notag \\
&=&2\left[ \delta _{ij}-\frac{\alpha _{i}^{u}+\alpha _{i}^{u\ast }}{2}\delta
_{ij}+K_{ik}\frac{\alpha _{k}^{d}+\alpha _{k}^{d\ast }}{2}K_{kj}^{\dagger }%
\right] \tau ^{d}  \notag \\
&&+2\left[ \delta _{ij}-\frac{\alpha _{i}^{d}+\alpha _{i}^{d\ast }}{2}\delta
_{ij}+K_{ik}^{\dagger }\frac{\alpha _{k}^{u}+\alpha _{k}^{u\ast }}{2}K_{kj}%
\right] \tau ^{u}  \label{unit}
\end{eqnarray}
Note that when $\alpha _{i}$ are pure imaginary quantities we can interpret
this freedom as the one we have to add phases to the CKM matrix. That
freedom does not alter the unitarity of that matrix as can be immediately
seen from (\ref{unit}). However when $\alpha _{i}$ are real this freedom
alters such unitarity. Hereafter we will set $\alpha _{i}=0.$ This does not
affect the mass terms or neutral current couplings, but changes the charged
coupling currents by multiplying the CKM matrix $K$ by diagonal matrices.
Except for this last freedom, the on-shell conditions determine one unique
solution, the one presented here, with $\bar{Z}^{\frac{1}{2}}\neq \gamma
^{0}Z^{\frac{1}{2}\dagger }\gamma ^{0}$.

\section{The role of Ward Identities}

\label{wardidentities}Let us obtain the Ward Identities that relate internal
wfr. between themselves and to the CKM counterterm. The non-physical basis
belongs to an $\emph{irreducible}$ \emph{representation} of $SU_{L}\left(
2\right) $ (weak doublet) and we if we ask the renormalization group to
respect this representation we have
\begin{eqnarray}
\mathrm{u}_{L}^{0} &=&Z_{w}^{L\frac{1}{2}}\mathrm{u}_{L},  \notag \\
\mathrm{d}_{L}^{0} &=&Z_{w}^{L\frac{1}{2}}\mathrm{d}_{L},
\end{eqnarray}
and
\begin{eqnarray}
\mathrm{\bar{u}}_{L}^{0}\gamma ^{\mu } &=&\mathrm{\bar{u}}_{L}\gamma ^{\mu }%
\bar{Z}_{w}^{L\frac{1}{2}},  \notag \\
\mathrm{\bar{d}}_{L}^{0}\gamma ^{\mu } &=&\mathrm{\bar{d}}_{L}\gamma ^{\mu }%
\bar{Z}_{w}^{L\frac{1}{2}},
\end{eqnarray}
where the wfr. $Z_{w}^{L\frac{1}{2}}$and $\bar{Z}_{w}^{L\frac{1}{2}}$ are
the \emph{same} for the two components of the $SU_{L}\left( 2\right) $ weak
doublet. The non-physical basis is related to the basis diagonalizing the
mass matrix in the Lagrangian via a bi-unitary transformation given by
\begin{eqnarray}
\mathrm{u}_{L}^{0} &=&V_{Lu}^{0}u_{L}^{0},\quad \mathrm{u}_{L}=V_{Lu}u_{L},
\notag \\
\mathrm{d}_{L}^{0} &=&V_{Ld}^{0}d_{L}^{0},\quad \mathrm{d}_{L}=V_{Ld}d_{L},
\label{eqa42}
\end{eqnarray}
and
\begin{eqnarray}
\mathrm{\bar{u}}_{L}^{0} &=&\bar{u}_{L}^{0}V_{Lu}^{0\dagger },\quad \mathrm{%
\bar{u}}_{L}=\bar{u}_{L}V_{Lu}^{\dagger },  \notag \\
\mathrm{\bar{d}}_{L}^{0} &=&\bar{d}_{L}^{0}V_{Ld}^{0\dagger },\quad \mathrm{%
\bar{d}}_{L}=\bar{d}_{L}V_{Ld}^{\dagger },  \label{eqa42a}
\end{eqnarray}
so we obtain
\begin{eqnarray}
u_{L}^{0} &=&V_{Lu}^{0\dagger }Z_{w}^{L\frac{1}{2}}V_{Lu}u_{L}\equiv \hat{Z}%
^{uL\frac{1}{2}}u_{L},  \notag \\
d_{L}^{0} &=&V_{Ld}^{0\dagger }Z_{w}^{L\frac{1}{2}}V_{Ld}d_{L}\equiv \hat{Z}%
^{dL\frac{1}{2}}d_{L},  \label{renormrel}
\end{eqnarray}
and
\begin{eqnarray}
\bar{u}_{L}^{0}\gamma ^{\mu } &=&\bar{u}_{L}\gamma ^{\mu }V_{Lu}^{\dagger }%
\bar{Z}_{w}^{L\frac{1}{2}}V_{Lu}^{0}\equiv \bar{u}_{L}\gamma ^{\mu }\widehat{%
\bar{Z}}^{uL\frac{1}{2}},  \notag \\
\bar{d}_{L}^{0}\gamma ^{\mu } &=&\bar{d}_{L}\gamma ^{\mu }V_{Ld}^{\dagger }%
\bar{Z}_{w}^{L\frac{1}{2}}V_{Ld}^{0}\equiv \bar{d}_{L}\gamma ^{\mu }\widehat{%
\bar{Z}}^{dL\frac{1}{2}}.  \label{renormrela}
\end{eqnarray}
Note that $u_{L}$, $\bar{u}_{L}$, $d_{L}$ and $\bar{d}_{L}$ are not the
physical fields since a diagonal mass matrix in the renormalized Lagrangian
does not guarantee the diagonalization of the physical propagator containing
radiative corrections. The physical fields can be obtained from these ones
by a supplementary finite renormalization. From Eqs. (\ref{renormrel}-\ref
{renormrela}) we immediately obtain
\begin{eqnarray}
K^{0} &=&V_{Lu}^{0\dagger }V_{Ld}^{0}=\hat{Z}^{uL\frac{1}{2}}V_{Lu}^{\dagger
}V_{Ld}\hat{Z}^{dL\frac{-1}{2}}  \notag \\
&=&\widehat{\bar{Z}}^{uL\frac{1}{2}}K\hat{Z}^{dL\frac{-1}{2}}=\widehat{\bar{Z%
}}^{uL\frac{-1}{2}}K\widehat{\bar{Z}}^{dL\frac{1}{2}}  \label{ward1}
\end{eqnarray}
and
\begin{eqnarray}
\hat{Z}^{uL\dagger \frac{1}{2}}\hat{Z}^{uL\frac{1}{2}} &=&V_{Lu}^{\dagger
}Z_{w}^{L\dagger \frac{1}{2}}Z_{w}^{L\frac{1}{2}}V_{Lu}  \notag \\
&=&V_{Lu}^{\dagger }V_{Ld}\hat{Z}^{dL\dagger \frac{1}{2}}\hat{Z}^{dL\frac{1}{%
2}}V_{Ld}^{\dagger }V_{Lu}  \notag \\
&=&K\hat{Z}^{dL\dagger \frac{1}{2}}\hat{Z}^{dL\frac{1}{2}}K^{\dagger },
\label{ward2}
\end{eqnarray}
together with
\begin{eqnarray}
\widehat{\bar{Z}}^{uL\frac{1}{2}}\widehat{\bar{Z}}^{uL\dagger \frac{1}{2}}
&=&V_{Lu}^{\dagger }\bar{Z}_{w}^{L\frac{1}{2}}\bar{Z}_{w}^{L\dagger \frac{1}{%
2}}V_{Lu}  \notag \\
&=&V_{Lu}^{\dagger }V_{Ld}\widehat{\bar{Z}}^{dL\frac{1}{2}}\widehat{\bar{Z}}%
^{dL\dagger \frac{1}{2}}V_{Ld}^{\dagger }V_{Lu}  \notag \\
&=&K\widehat{\bar{Z}}^{dL\frac{1}{2}}\widehat{\bar{Z}}^{dL\dagger \frac{1}{2}%
}K^{\dagger },  \label{ward2a}
\end{eqnarray}
If we define the CKM renormalization constant as $K^{0}=K+\delta K$ we can
rewrite Eqs. (\ref{ward1}) and (\ref{ward2}-\ref{ward2a}) in the
perturbative way as
\begin{eqnarray}
\delta K &=&\frac{1}{2}\left( \delta \hat{Z}^{uL}K-K\delta \hat{Z}%
^{dL}\right) =\frac{1}{2}\left( \delta \widehat{\bar{Z}}^{dL}K-K\delta
\widehat{\bar{Z}}^{uL}\right) ,  \label{ward1inf} \\
\delta \hat{Z}^{uL\dagger }+\delta \hat{Z}^{uL} &=&K\left( \delta \hat{Z}%
^{dL\dagger }+\delta \hat{Z}^{dL}\right) K^{\dagger },  \label{ward2inf} \\
\delta \widehat{\bar{Z}}^{uL\dagger }+\delta \widehat{\bar{Z}}^{uL}
&=&K\left( \delta \widehat{\bar{Z}}^{dL\dagger }+\delta \widehat{\bar{Z}}%
^{dL}\right) K^{\dagger }.  \label{ward3inf}
\end{eqnarray}
Using that equations we can rewrite $\delta K$ as
\begin{eqnarray}
\delta K &=&\frac{1}{4}\left( \delta \hat{Z}^{uL}-\delta \hat{Z}^{uL\dagger
}\right) K-\frac{1}{4}K\left( \delta \hat{Z}^{dL}-\delta \hat{Z}^{dL\dagger
}\right)  \notag \\
&=&\frac{1}{4}\left( \delta \widehat{\bar{Z}}^{dL}-\delta \widehat{\bar{Z}}%
^{dL\dagger }\right) K-\frac{1}{4}K\left( \delta \widehat{\bar{Z}}%
^{uL}-\delta \widehat{\bar{Z}}^{uL\dagger }\right) ,  \label{wardf}
\end{eqnarray}

Obviously these identities constrain the $\delta K$ counterterm be such that
$K+\delta K$ is a unitary matrix. Here it is worth remembering that the $%
\hat{Z}$'s and $\widehat{\bar{Z}}$'s are not the renormalization constants
that allow us to obtain an up (down) propagator with the desired properties
listed in the on-shell scheme, this properties must be attained performing
an additional finite renormalization on the\emph{\ }external up (down)
fermions. This point is illustrated in section \ref{dk} of Chapter \ref
{cpviolationandmixing} where we have calculated the contribution to the
vertices of the effective operators including the renormalization of the CKM
matrix given by Eq. (\ref{wardf}) and the contribution of the operator $%
\mathcal{L}_{L}^{4}$ via the wfr.

\section{W$^{+}$ and top decay}

\label{wandtdecay}Let us now apply the above mechanism to $W^{+}$ and top
decay. We write
\begin{eqnarray}
W^{+}\left( q\right) &\rightarrow &f_{i}\left( p_{1}\right) \bar{f}%
_{j}\left( p_{2}\right) \,,  \label{wdecay} \\
f_{i}\left( p_{1}\right) &\rightarrow &W^{+}\left( q\right) f_{j}\left(
p_{2}\right) \,,  \label{tdecay}
\end{eqnarray}
where $f$ indicates particle and $\bar{f}$ anti-particle. The Latin indices
are reserved for family indices. Leptonic and quark channels can be
considered with the same notation, and confusion should not arise. For the
process (\ref{wdecay}) there are at next-to-leading order two different type
of Lorentz structures
\begin{eqnarray}
M_{L}^{\left( 1\right) } &=&\bar{u}_{i}\left( p_{1}\right) \not{\varepsilon}%
\left( q\right) Lv_{j}\left( p_{2}\right) \,,\qquad \left( L\leftrightarrow
R\right) \,,  \notag \\
M_{L}^{\left( 2\right) } &=&\bar{u}_{i}\left( p_{1}\right) Lv_{j}\left(
p_{2}\right) p_{1}\cdot \varepsilon \left( q\right) \,,\qquad \left(
L\leftrightarrow R\right) \,,  \label{wdtree}
\end{eqnarray}
where $\varepsilon $ stands for the vector polarization of the $W^{+}$.
Equivalently for the process (\ref{tdecay}) we shall use
\begin{eqnarray}
M_{L}^{\left( 1\right) } &=&\bar{u}_{j}\left( p_{2}\right) \not{\varepsilon}%
^{\ast }\left( q\right) Lu_{i}\left( p_{1}\right) \,,\qquad \left(
L\leftrightarrow R\right) \,,  \notag \\
M_{L}^{\left( 2\right) } &=&\bar{u}_{j}\left( p_{2}\right) Lu_{i}\left(
p_{1}\right) p_{1}\cdot \varepsilon ^{\ast }\left( q\right) \,,\qquad \left(
L\leftrightarrow R\right) \,.  \label{tdtree}
\end{eqnarray}
The transition amplitude at tree level for the processes (\ref{wdecay}) and (%
\ref{tdecay}) is given by
\begin{equation*}
\mathcal{M}_{0}=-\frac{eK_{ij}}{2s_{W}}M_{L}^{\left( 1\right) }\,,
\end{equation*}
where Eq. (\ref{wdtree}) is used for $M_{L}^{\left( 1\right) }$ in $W^{+}$
decay and Eq. (\ref{tdtree}) instead for $M_{L}^{\left( 1\right) }$ in $t$
decay. The one-loop corrected transition amplitude can be written as
\begin{eqnarray}
\mathcal{M}_{1} &=&-\frac{e}{2s_{W}}M_{L}^{\left( 1\right) }\left[
K_{ij}\left( 1+\frac{\delta e}{e}-\frac{\delta s_{W}}{s_{W}}+\frac{1}{2}%
\delta Z_{W}\right) +\delta K_{ij}+\frac{1}{2}\sum_{r}\left( \delta \bar{Z}%
_{ir}^{Lu}K_{rj}+K_{ir}\delta Z_{rj}^{Ld}\right) \right]  \notag \\
&&-\frac{e}{2s_{W}}\left( \delta F_{L}^{\left( 1\right) }M_{L}^{\left(
1\right) }+M_{L}^{\left( 2\right) }\delta F_{L}^{\left( 2\right)
}+M_{R}^{\left( 1\right) }\delta F_{R}^{\left( 1\right) }+M_{R}^{\left(
2\right) }\delta F_{R}^{\left( 2\right) }\right) \,.  \label{vertex}
\end{eqnarray}
In this expression $\delta F_{L,R}^{\left( 1,2\right) }$ are the electroweak
form factors coming from one-loop vertex diagrams. The renormalization
constants are given by
\begin{eqnarray*}
\frac{\delta e}{e} &=&-\frac{1}{2}\left[ \left( \delta Z_{2}^{A}-\delta
Z_{1}^{A}\right) +\delta Z_{2}^{A}\right] =-\frac{s_{W}}{c_{W}M_{Z}^{2}}\Pi
^{ZA}\left( 0\right) +\frac{1}{2}\frac{\partial \Pi ^{AA}}{\partial k^{2}}%
\left( 0\right) \,, \\
\frac{\delta s_{W}}{s_{W}} &=&-\frac{c_{W}^{2}}{2s_{W}^{2}}\left( \frac{%
\delta M_{W}^{2}}{M_{W}^{2}}-\frac{\delta M_{Z}^{2}}{M_{Z}^{2}}\right) =-%
\frac{c_{W}^{2}}{2s_{W}^{2}}Re\left( \frac{\Pi ^{WW}\left( M_{W}^{2}\right)
}{M_{W}^{2}}-\frac{\Pi ^{ZZ}\left( M_{Z}^{2}\right) }{M_{Z}^{2}}\right) \,,
\\
\delta Z_{W} &=&-\frac{\partial \Pi ^{WW}}{\partial k^{2}}\left(
M_{W}^{2}\right) \,,
\end{eqnarray*}
and the fermionic wfr. constants are depicted in Eqs. (\ref{zin}), (\ref
{zout}) and (\ref{zdiag}) where the indices $u$ or $d$ must be restored in
the masses. The index $A$ refers to the photon field.

As for the $\delta K_{ij}$ renormalization constants, a SU(2) Ward identity (%
\ref{wardf}) \cite{Grassi} fixes these counter terms to be
\begin{equation}
\delta K_{jk}=\frac{1}{4}\left[ \left( \delta \hat{Z}^{uL}-\delta \hat{Z}%
^{uL\dagger }\right) K-K\left( \delta \hat{Z}^{dL}-\delta \hat{Z}^{dL\dagger
}\right) \right] _{jk}\,,  \label{deltaK}
\end{equation}
where the hat in $\hat{Z}$ means that the wfr. constants appearing in the
above expression are not the same ones used to renormalize and guarantee the
proper on-shell residue for the external legs as already has been emphasized
(see section \ref{wardidentities}). One may, for instance, use minimal
subtraction $Z$'s for the former.

We know \cite{Marciano} that the combination $\frac{\delta e}{e}-\frac{%
\delta s_{W}}{s_{W}}$ is gauge parameter independent. All the other vertex
functions and renormalization constants are gauge dependent. For the reasons
stated in the introduction we want the amplitude (\ref{vertex}) to be
exactly gauge independent ---not just its modulus--- so the gauge dependence
must cancel between all the remaining terms.

In section \ref{nielsen} we shall make use of the Nielsen identities \cite
{Gambino,Nielsen,Piguet,Sibold} to determine that three of the form factors
appearing in the vertex (\ref{vertex}) are by themselves gauge independent,
namely
\begin{equation*}
\partial _{\xi }\delta F_{L}^{\left( 2\right) }=\partial _{\xi }\delta
F_{R}^{\left( 1\right) }=\partial _{\xi }\delta F_{R}^{\left( 2\right) }=0\,.
\end{equation*}
$\xi $ is the gauge-fixing parameter. We shall also see that the gauge
dependence in the remaining form factor $\delta F_{L}^{\left( 1\right) }$
cancels exactly with the one contained in $\delta Z_{W}$ and in $\delta Z$
and $\delta \bar{Z}$. Therefore to guarantee a gauge-fixing parameter
independent amplitude $\delta K$ must be gauge independent as well.

The difficulties related to a proper definition of $\delta K$ were first
pointed out in \cite{Grassi,Gambino}, where it was realized that using the
on-shell $Z$'s of \cite{DennerSack} in Eq.~(\ref{deltaK}) led to a gauge
dependent $K$ and amplitude. They suggested a modification of the on-shell
scheme based on a subtraction at $p^{2}=0$ for all flavors that ensured
gauge independence. We want to stress that the choice for $\delta K$ is not
unique and different choices may differ by gauge independent finite parts
\cite{Kniehl}. Note that the gauge independence of $\delta K$ is in
contradistinction with the conclusions of \cite{Barroso} and in addition
these authors have a non-unitary bare CKM matrix which does not respect the
Ward identity.

As we shall see, if instead of using our prescription for $\delta Z$ and $%
\delta \bar{Z}$ one makes use of the wfr. constants of \cite{Denner} to
renormalize the external fermion legs, it turns out that the gauge
cancellation dictated by the Nielsen identities does not actually take place
in the amplitude. The culprits are of course the absorptive parts. These
absorptive parts of the self-energies are absent in \cite{Denner} due to the
use of the $\widetilde{Re}$ prescription, which throws them away. Notice,
though, that the vertex contribution has gauge dependent absorptive parts
(calculated in the next section) and they remain in the final result.

One might think of absorbing these additional terms in the counter term for $%
\delta K$. This does not work. Indeed one can see from explicit calculations
that wfr. constants decompose as
\begin{equation}
\delta Z^{Lu}=A^{uL}+iB^{uL}\,,\qquad \delta \bar{Z}^{Lu}=A^{uL\dagger
}+iB^{uL\dagger }\,,\qquad \left( L\leftrightarrow R\text{, }%
u\leftrightarrow d\right) \,,  \label{zdecomp}
\end{equation}
where the matrices $A$'s or $B$'s contain the dispersive and absorptive
parts of the self-energies, respectively. Moreover if one substitutes back
Eq. (\ref{zdecomp}) into Eq. (\ref{vertex}) one immediately sees that a
necessary requirement allowing the $A^{u}$ and $A^{d}$ (respectively $B^{u}$
and $B^{d}$) contribution to be absorbed into a CKM matrix counter term of
the form given in Eq. (\ref{deltaK}) is that $A^{u}$ and $A^{d}$
(respectively $B^{u}$ and $B^{d}$) were anti-hermitian (respectively
hermitian) matrices. By direct inspection one can conclude that all $A$'s or
$B$'s are neither hermitian nor anti-hermitian matrices and therefore any of
such redefinitions are impossible unless one is willing to give up the
unitarity of the bare $K$. A problem somewhat similar to that was
encountered in \cite{Barroso} (but different, they did not consider
absorptive parts at all, the inconsistency showed up already with the
dispersive parts of the on-shell scheme of \cite{DennerSack}).

It turns out that in the SM these gauge dependent absorptive parts, leading
to a gauge dependent amplitude if they are dropped, do actually cancel, at
least at the one-loop level, in the modulus of the $S$-matrix element. Thus
at this level the use of $\widetilde{Re}$ is irrelevant. It is also shown in
section \ref{absorptive} that gauge independent absorptive parts do survive
even in the modulus of the amplitude for top or anti-top decay (and only in
these cases). Therefore we have to conclude that the difference between
using $\widetilde{Re}$, as advocated in \cite{Denner}, or not, as we do, is
not just a semantic one. As we have seen such difference cannot be
attributed to a finite renormalization of $K$, provided the bare $K$ remains
unitary as required by the Ward identity (\ref{deltaK}).

\section{Introduction to the Nielsen Identities.}

\label{nielsenintro}

This section is aimed to provide a basic introduction to the so-called
Nielsen Identities. The literature dealing with this subject is rather
extensive and we refer the interested reader to \cite
{Nielsen,Piguet,Sibold,Gambino} for details. Let us start by defining the
complete Lagrangian necessary to work with. This Lagrangian includes the
standard classical term $\mathcal{L}_{\varphi }$ plus the gauge fixing and
Fadeev-Popov terms
\begin{equation}
\mathcal{L}_{GF}+\mathcal{L}_{FP}=\sum_{i}\alpha _{i}\left( \mathcal{L}%
_{GF}^{\left( i\right) }+\mathcal{L}_{FP}^{\left( i\right) }\right) ,%
\mathcal{\qquad L}_{GF}^{\left( i\right) }+\mathcal{L}_{FP}^{\left( i\right)
}=s\mathcal{\tilde{L}}_{i},  \label{FPGF}
\end{equation}
where $\alpha _{i}$ are gauge fixing parameters and $s$ is the BRST operator
\cite{Becchi:1975md,Becchi:1976nq}. An essential ingredient to obtain
Nielsen Identities is a source term $\mathcal{L}_{\chi }$ given by
\begin{equation*}
\mathcal{L}_{\chi }=-\sum_{i}\chi _{i}\mathcal{\tilde{L}}_{i},
\end{equation*}
where $\chi _{i}$ are grassman source terms and the $\mathcal{\tilde{L}}_{i}$
factors are given by Eq. (\ref{FPGF}). The existence of $\mathcal{\tilde{L}}%
_{i}$ means that $\mathcal{L}_{GF}+\mathcal{L}_{FP}$ is a trivial term in
the BRST cohomology generated by $s.$ Besides this source term we need the
standard source term
\begin{equation*}
\mathcal{L}_{J}=\sum_{\varphi }\mathcal{L}_{J_{\varphi }},\mathcal{\qquad L}%
_{J_{\varphi }}=J_{\varphi }\varphi ,
\end{equation*}
where $\varphi $ represent matter and gauge fields and $J_{\varphi }$ are
their corresponding sources. And finally a $\mathcal{L}_{\eta }$ term with
sources $\eta _{\varphi }$ coupled to the BRST variations of the fields
\begin{equation*}
\mathcal{L}_{\eta }=\sum_{\varphi }\mathcal{L}_{\eta _{\varphi }},\mathcal{%
\qquad L}_{\eta _{\varphi }}=\eta _{\varphi }s\varphi ,
\end{equation*}
Recapitulating, we have the complete Lagrangian $\mathcal{L}$ given by
\begin{equation*}
\mathcal{L}=\mathcal{L}_{\varphi }+\mathcal{L}_{GF}+\mathcal{L}_{FP}+%
\mathcal{L}_{J}+\mathcal{L}_{\eta _{\varphi }}+\mathcal{L}_{\chi }.
\end{equation*}
Then we introduce the partition function
\begin{equation*}
Z\left[ J,\eta ,\chi ,\alpha \right] =\int D\varphi \exp \left( i\mathcal{L}%
\right) ,
\end{equation*}
and the generator of connected Green functions $W$ given by
\begin{equation*}
e^{iW}=Z\left[ J,\eta ,\chi ,\alpha \right] .
\end{equation*}
Finally the effective action $\Gamma $ is given by the Legendre
transformation of $W$ with respect to $J_{\varphi }$ only, that is
\begin{equation*}
\Gamma \left[ \varphi ^{cl},\eta ,\chi ,\alpha \right] =W\left[ J,\eta ,\chi
,\alpha \right] -\sum_{\varphi }\varphi ^{cl}J_{\varphi },
\end{equation*}
with
\begin{equation}
J_{\varphi }=-\frac{\delta \Gamma }{\delta \varphi ^{cl}},\mathcal{\qquad }%
\varphi ^{cl}=\frac{\delta W}{\delta J_{\varphi }}.  \label{Legendre}
\end{equation}
We are now ready to derive Nielsen Identities much in the same way as when
deriving Ward Identities. Namely we will set the variation of $Z$ with
respect to any change of variables to zero. We choose this change of
variable as a BRST variation
\begin{equation*}
\varphi =\tilde{\varphi}+\delta \tilde{\varphi}=\tilde{\varphi}+\left( s%
\tilde{\varphi}\right) \lambda ,
\end{equation*}
which is a super-change of variables that is a symmetry of $\mathcal{L}%
_{\varphi }+\mathcal{L}_{GF}+\mathcal{L}_{FP}$ and has Berezinian equal to
1. Therefore
\begin{eqnarray*}
0 &=&\delta Z=\int D\tilde{\varphi}\left( Ber\left( \frac{\delta \varphi }{%
\delta \tilde{\varphi}}\right) \exp \left( i\mathcal{\tilde{L}}\left( \tilde{%
\varphi}\right) \right) -\exp \left( i\mathcal{L}\left( \tilde{\varphi}%
\right) \right) \right) \\
&=&i\int D\varphi \exp \left( i\mathcal{L}\right) \delta \mathcal{L} \\
&=&i\int D\varphi \exp \left( i\mathcal{L}\right) \left( \sum_{\varphi
}J_{\varphi }\delta \varphi -\sum_{i}\chi _{i}\delta \mathcal{\tilde{L}}%
_{i}\right) \\
&=&i\int D\varphi \exp \left( i\mathcal{L}\right) \left( \sum_{\varphi
}J_{\varphi }s\varphi -\sum_{i}\chi _{i}\left( \mathcal{L}_{GF}^{\left(
i\right) }+\mathcal{L}_{FP}^{\left( i\right) }\right) \right) \lambda \\
&=&i\int D\varphi \exp \left( i\mathcal{L}\right) \left( \sum_{\varphi
}J_{\varphi }\frac{\delta \mathcal{L}}{\delta \eta _{\varphi }}-\sum_{i}\chi
_{i}\frac{\delta \mathcal{L}}{\delta \alpha _{i}}\right) \lambda \\
&=&\left( \sum_{\varphi }J_{\varphi }\frac{\delta Z}{\delta \eta _{\varphi }}%
-\sum_{i}\chi _{i}\frac{\delta Z}{\delta \alpha _{i}}\right) \lambda .
\end{eqnarray*}
Hence
\begin{eqnarray*}
0 &=&\sum_{\varphi }J_{\varphi }\frac{\delta Z}{\delta \eta _{\varphi }}%
+\sum_{i}\chi _{i}\frac{\delta Z}{\delta \alpha _{i}} \\
&=&\sum_{\varphi }J_{\varphi }\frac{\delta W}{\delta \eta _{\varphi }}%
+\sum_{i}\chi _{i}\frac{\delta W}{\delta \alpha _{i}},
\end{eqnarray*}
but using Eq. (\ref{Legendre}) and the fact that all sources but $J_{\varphi
}$ were not Legendre transformed we obtain
\begin{equation*}
0=-\sum_{\varphi }\frac{\delta \Gamma }{\delta \varphi ^{cl}}\frac{\delta
\Gamma }{\delta \eta _{\varphi }}-\sum_{i}\chi _{i}\frac{\delta \Gamma }{%
\delta \alpha _{i}},
\end{equation*}
or deriving with respect to $\chi _{i}$ and evaluating at $\chi _{i}=0$ we
finally obtain
\begin{equation}
\left. \frac{\delta \Gamma }{\delta \alpha _{i}}\right| _{\chi
_{i}=0}=-\sum_{\varphi }\left[ \frac{\delta ^{2}\Gamma }{\delta \chi
_{i}\delta \varphi ^{cl}}\frac{\delta \Gamma }{\delta \eta _{\varphi }}+%
\frac{\delta \Gamma }{\delta \varphi ^{cl}}\frac{\delta ^{2}\Gamma }{\delta
\eta _{\varphi }\delta \chi _{i}}\right] _{\chi _{i}=0},  \label{Nielsen}
\end{equation}
where all derivatives are right derivatives and care in the ordering must be
noticed. Eq. (\ref{Nielsen}) is the generator of all Nielsen identities that
are obtained taking derivatives with respect to the $\varphi ^{cl}$
different sources.

\section{Nielsen Identities in $W^{+}$ and top decay}

\label{nielsen}

In this section we derive in detail the gauge dependence of the vertex
three-point function. It is therefore rather technical and it can be omitted
by readers just interested in the physical conclusions. In order to have
control on gauge dependence, a useful tool is provided by the Nielsen
identities discussed in the previous section. For such purpose besides the
``classical'' Lagrangian $\mathcal{L}_{\mathrm{SM}}$ we have to take into
account the gauge fixing term $\mathcal{L}_{\mathrm{GF}}$, the Fadeev-Popov
term $\mathcal{L}_{\mathrm{FP}}$ and source terms. Such source terms are the
ones given by BRST variations of matter ($\bar{\eta}^{u},\eta ^{u},\ldots $)
and gauge fields together with Goldstone and ghost fields (not including
anti-ghosts). We refer the reader to \cite{Hollik}, \cite{Gambino} for
notation and further explanations (we have absorbed a factor $i$ in the
definition of the charged goldstone bosons $G^{\pm }$ with respect to the
conventions in \cite{Hollik}). We also include source terms ($\chi $) for
the composite operators whose BRST variation generate $\mathcal{L}_{\mathrm{%
GF}}+\mathcal{L}_{\mathrm{FP}}.$ Schematically
\begin{eqnarray*}
\mathcal{L} &=&\mathcal{L}_{\mathrm{SM}}+\mathcal{L}_{\mathrm{GF}}+\mathcal{L%
}_{\mathrm{FP}}-\frac{1}{2\xi }\chi \left( \frac{{}}{{}}\left( \partial
^{\mu }W_{\mu }^{-}+\xi M_{W}G^{-}\right) \bar{c}^{+}+\left( \partial ^{\mu
}W_{\mu }^{+}+\xi M_{W}G^{+}\right) \bar{c}^{-}\frac{{}}{{}}\right) \\
&&+\frac{ig}{\sqrt{2}}\bar{\eta}_{i}^{u}K_{ir}Ld_{r}-\frac{ig}{\sqrt{2}}\bar{%
c}^{+}\bar{d}_{r}K_{rj}^{\dagger }R\eta _{j}^{u}+\bar{s}_{i}^{u}u_{i}+\bar{u}%
_{j}s_{j}^{u}+\bar{s}_{i}^{d}d_{i}+\bar{d}_{j}s_{j}^{d}+\ldots \,,
\end{eqnarray*}
where the ellipsis stands for the remaining source terms. The effective
action, $\Gamma $, is introduced in the standard manner
\begin{equation}
\Gamma \left[ \chi ,\bar{\eta}^{u},\eta ^{u},\bar{u}^{cl},u^{cl},\ldots %
\right] =W\left[ \chi ,\bar{\eta}^{u},\eta ^{u},\bar{s}^{u},s^{u},\ldots %
\right] -\left( \bar{s}_{i}^{u}u_{i}^{cl}+\bar{u}_{j}^{cl}s_{j}^{u}+\bar{s}%
_{i}^{d}d_{i}^{cl}+\bar{d}_{j}^{cl}s_{j}^{d}+\ldots \right) \,,
\label{gamma}
\end{equation}
with
\begin{equation}
e^{iW}=Z\left[ \chi ,\bar{\eta}^{u},\eta ^{u},\bar{s}^{u},s^{u},\ldots %
\right] \equiv \int D\Phi \exp \left( i\mathcal{L}\right) \,.  \label{omega}
\end{equation}
From the above expressions and using BRST transformations we can extract the
Nielsen identities for the three-point functions (see \cite{Nielsen} for
details)
\begin{eqnarray}
\partial _{\xi }\Gamma _{W_{\mu }^{+}\bar{u}_{i}d_{j}} &=&-\Gamma _{\chi
W_{\mu }^{+}\gamma _{W_{\alpha }}^{-}}\Gamma _{W_{\alpha }^{+}\bar{u}%
_{i}d_{j}}-\Gamma _{\chi \bar{u}_{i}\eta _{r}^{u}}\Gamma _{W_{\mu }^{+}\bar{u%
}_{r}d_{j}}  \notag \\
&&-\Gamma _{W_{\mu }^{+}\bar{u}_{i}d_{r}}\Gamma _{\bar{\eta}%
_{r}^{d}d_{j}\chi }-\Gamma _{\chi W_{\mu }^{+}\gamma _{G_{\alpha
}}^{-}}\Gamma _{G_{\alpha }^{+}\bar{u}_{i}d_{j}}  \notag \\
&&-\Gamma _{\chi \gamma _{G_{\alpha }}^{+}\bar{u}_{i}d_{j}}\Gamma
_{G_{\alpha }^{-}W_{\mu }^{+}}-\Gamma _{\chi \gamma _{W_{\alpha }}^{+}\bar{u}%
_{i}d_{j}}\Gamma _{W_{\alpha }^{-}W_{\mu }^{+}}  \notag \\
&&-\Gamma _{\chi W_{\mu }^{+}\bar{u}_{i}\eta _{r}^{d}}\Gamma _{\bar{d}%
_{r}d_{j}}-\Gamma _{\bar{u}_{i}u_{r}}\Gamma _{\chi W_{\mu }^{+}\bar{\eta}%
_{r}^{u}d_{j}}\,,  \label{nielsen00}
\end{eqnarray}
where we have omitted the momentum dependence and defined
\begin{equation*}
\Gamma _{\chi \bar{u}_{i}\eta _{j}^{u}}\equiv \frac{\vec{\delta}}{\delta
\chi }\frac{\vec{\delta}}{\delta \bar{u}_{i}^{cl}\left( p\right) }\frac{%
\delta }{\delta \eta _{j}^{u}\left( p\right) }\Gamma \,,\quad \Gamma _{\bar{%
\eta}_{i}^{u}u_{j}\chi }\equiv \frac{\delta }{\delta \bar{\eta}%
_{i}^{u}\left( p\right) }\frac{\vec{\delta}}{\delta u_{j}^{cl}\left(
p\right) }\frac{\vec{\delta}}{\delta \chi }\Gamma \,.
\end{equation*}
In the rest of this section we shall evaluate the on-shell contributions to
Eq.~(\ref{nielsen00}). Analogously we can also derive Nielsen identities for
two-point functions
\begin{eqnarray}
\partial _{\xi }\Gamma _{W_{\mu }^{+}W_{\beta }^{-}}^{\left( 1\right) }
&=&-2\left( \Gamma _{\chi W_{\mu }^{+}\gamma _{W_{\alpha }}^{-}}^{\left(
1\right) }\Gamma _{W_{\alpha }^{+}W_{\beta }^{-}}+\Gamma _{\chi W_{\mu
}^{+}\gamma _{G_{\alpha }}^{-}}^{\left( 1\right) }\Gamma _{G_{\alpha
}^{+}W_{\beta }^{-}}\right) \,,  \label{bose1} \\
\partial _{\xi }\Gamma _{W_{\mu }^{+}G_{\beta }^{-}}^{\left( 1\right) }
&=&-2\left( \Gamma _{\chi W_{\mu }^{+}\gamma _{W_{\alpha }}^{-}}^{\left(
1\right) }\Gamma _{W_{\alpha }^{+}G_{\beta }^{-}}+\Gamma _{\chi W_{\mu
}^{+}\gamma _{G_{\alpha }}^{-}}^{\left( 1\right) }\Gamma _{G_{\alpha
}^{+}G_{\beta }^{-}}\right) \,.  \label{bose2}
\end{eqnarray}
On-shell these reduce to
\begin{equation}
\Gamma _{\chi W^{+}\gamma _{W}^{-}}^{T\left( 1\right) }\left(
M_{W}^{2}\right) =-\frac{1}{2}\partial _{\xi }\left. \frac{\partial \Gamma
_{W^{+}W^{-}}^{T\left( 1\right) }}{\partial q^{2}}\left( q^{2}\right)
\right| _{q^{2}=M_{W}^{2}}=\frac{1}{2}\partial _{\xi }\delta Z_{W}\,,\quad
\Gamma _{\chi W^{+}\gamma _{G}^{-}}^{T\left( 1\right) }\left( q\right) =0\,,
\label{nielsenp}
\end{equation}
where the superscript $T$ refers to the transverse part and the superscript $%
(1)$ makes reference to the one-loop order correction.

Using these two sets of results and restricting Eq.~(\ref{nielsen00}) to the
1PI function appropriate for (on-shell) top-decay
\begin{eqnarray}
&&\bar{u}_{u}\left( p_{i}\right) \epsilon ^{\mu }\left( q\right) \partial
_{\xi }\Gamma _{W_{\mu }^{+}\bar{u}_{i}d_{j}}^{\left( 1\right) }v_{d}\left(
-p_{j}\right)  \notag \\
&=&\frac{g}{\sqrt{2}}\bar{u}_{u}\left( p_{i}\right) \left\{ \Gamma _{\chi
\bar{u}_{i}\eta _{r}^{u}}K_{rj}\not{\epsilon}L+K_{ir}\not{\epsilon}L\Gamma _{%
\bar{\eta}_{r}^{d}d_{j}\chi }+\frac{1}{2}\partial _{\xi }\delta Z_{W}K_{ij}%
\not{\epsilon}L\right\} v_{d}\left( -p_{j}\right) \,.  \label{nielsen0}
\end{eqnarray}

At the one-loop level we also have the Nielsen identity
\begin{equation}
\partial _{\xi }\Sigma _{ij}^{u}\left( p\right) =\Gamma _{\chi \bar{u}%
_{i}\eta _{j}^{u}}^{\left( 1\right) }\left( p\right) \left( \not{p}%
-m_{j}^{u}\right) +\left( \not{p}-m_{i}^{u}\right) \Gamma _{\bar{\eta}%
_{i}^{u}u_{j}\chi }^{\left( 1\right) }\left( p\right) \,,  \label{nielsen2}
\end{equation}
which is the fermionic counterpart of Eqs. (\ref{bose1}) and (\ref{bose2}).
Similar relation holds interchanging $u\leftrightarrow d$. With the use of
Eq.~(\ref{nielsen2}) and an analogous decomposition to Eq. (\ref{decomp})
for $\Gamma $,
\begin{eqnarray}
\Gamma _{\chi \bar{u}_{i}\eta _{j}^{u}}^{\left( 1\right) }\left( p\right) &=&%
\not{p}\left( \Gamma _{\chi \bar{u}_{i}\eta _{j}^{u}}^{\gamma R\left(
1\right) }\left( p^{2}\right) R+\Gamma _{\chi \bar{u}_{i}\eta
_{j}^{u}}^{\gamma L\left( 1\right) }\left( p^{2}\right) L\right) +\Gamma
_{\chi \bar{u}_{i}\eta _{j}^{u}}^{R\left( 1\right) }\left( p^{2}\right)
R+\Gamma _{\chi \bar{u}_{i}\eta _{j}^{u}}^{L\left( 1\right) }\left(
p^{2}\right) L\,,  \notag \\
\Gamma _{\bar{\eta}_{i}^{u}u_{j}\chi }^{\left( 1\right) }\left( p\right) &=&%
\not{p}\left( \Gamma _{\bar{\eta}_{i}^{u}u_{j}\chi }^{\gamma R\left(
1\right) }\left( p^{2}\right) R+\Gamma _{\bar{\eta}_{i}^{u}u_{j}\chi
}^{\gamma L\left( 1\right) }\left( p^{2}\right) L\right) +\Gamma _{\bar{\eta}%
_{i}^{u}u_{j}\chi }^{R\left( 1\right) }\left( p^{2}\right) R+\Gamma _{\bar{%
\eta}_{i}^{u}u_{j}\chi }^{L\left( 1\right) }\left( p^{2}\right) L\,,
\label{decomp2}
\end{eqnarray}
we obtain after equating Dirac structures
\begin{eqnarray}
\partial _{\xi }\Sigma _{ij}^{u\gamma R}\left( p^{2}\right) &=&\Gamma _{\chi
\bar{u}_{i}\eta _{j}^{u}}^{L\left( 1\right) }\left( p^{2}\right)
-m_{j}\Gamma _{\chi \bar{u}_{i}\eta _{j}^{u}}^{\gamma R\left( 1\right)
}\left( p^{2}\right) +\Gamma _{\bar{\eta}_{i}^{u}u_{j}\chi }^{R\left(
1\right) }\left( p^{2}\right) -m_{i}\Gamma _{\bar{\eta}_{i}^{u}u_{j}\chi
}^{\gamma R\left( 1\right) }\left( p^{2}\right) \,,  \notag \\
\partial _{\xi }\Sigma _{ij}^{uR}\left( p^{2}\right) &=&p^{2}\Gamma _{\chi
\bar{u}_{i}\eta _{j}^{u}}^{\gamma L\left( 1\right) }\left( p^{2}\right)
-m_{j}\Gamma _{\chi \bar{u}_{i}\eta _{j}^{u}}^{R\left( 1\right) }\left(
p^{2}\right) +p^{2}\Gamma _{\bar{\eta}_{i}^{u}u_{j}\chi }^{\gamma R\left(
1\right) }\left( p^{2}\right) -m_{i}\Gamma _{\bar{\eta}_{i}^{u}u_{j}\chi
}^{R\left( 1\right) }\left( p^{2}\right) \,,  \label{rels}
\end{eqnarray}
and analogous expressions exchanging $L\leftrightarrow R$ and $%
u\leftrightarrow d$. Moreover from Eqs. (\ref{nielsen0}) and (\ref{decomp2})
we obtain
\begin{eqnarray}
&&\hspace{-0.8cm}\bar{u}_{u}\left( p_{i}\right) \epsilon ^{\mu }\left(
q\right) \partial _{\xi }\Gamma _{W_{\mu }^{+}\bar{u}_{i}d_{j}}^{\left(
1\right) }v_{d}\left( -p_{j}\right) =  \notag \\
&&\frac{g}{\sqrt{2}}\left\{ \frac{{}}{{}}\bar{u}_{u}\left( p_{i}\right)
\left( m_{i}^{u}\Gamma _{\chi \bar{u}_{i}\eta _{r}^{u}}^{\gamma R\left(
1\right) }\left( m_{i}^{u2}\right) +\Gamma _{\chi \bar{u}_{i}\eta
_{r}^{u}}^{R\left( 1\right) }\left( m_{i}^{u2}\right) \right) K_{rj}\not{%
\epsilon}Lv_{d}\left( -p_{j}\right) \right.  \notag \\
&&+\hspace{-0.9cm}\bar{u}_{u}\left( p_{i}\right) K_{ir}\not{\epsilon}L\left(
m_{j}^{d}\Gamma _{\bar{\eta}_{r}^{d}d_{j}\chi }^{\gamma R\left( 1\right)
}\left( m_{j}^{d2}\right) +\Gamma _{\bar{\eta}_{r}^{d}d_{j}\chi }^{L\left(
1\right) }\left( m_{j}^{d2}\right) \right) v_{d}\left( -p_{j}\right)  \notag
\\
&&+\left. \frac{1}{2}\partial _{\xi }\delta Z_{W}\bar{u}_{u}\left(
p_{i}\right) K_{ij}\not{\epsilon}Lv_{d}\left( -p_{j}\right) \frac{{}}{{}}%
\right\} \,.  \label{final}
\end{eqnarray}
Using Eqs.~(\ref{zin}), (\ref{zout}) and (\ref{rels}) one arrives at
\begin{eqnarray}
m_{j}^{u}\Gamma _{\bar{\eta}_{i}^{u}u_{j}\chi }^{\gamma R\left( 1\right)
}\left( m_{j}^{u2}\right) +\Gamma _{\bar{\eta}_{i}^{u}u_{j}\chi }^{L\left(
1\right) }\left( m_{j}^{u2}\right) &=&\frac{1}{2}\partial _{\xi }\delta
Z_{ij}^{uL}\,,\quad \left( i\neq j\right) \,,  \label{gammalin} \\
m_{i}^{u}\Gamma _{\chi \bar{u}_{i}\eta _{j}^{u}}^{\gamma R\left( 1\right)
}\left( m_{i}^{u2}\right) +\Gamma _{\chi \bar{u}_{i}\eta _{j}^{u}}^{R\left(
1\right) }\left( m_{i}^{u2}\right) &=&\frac{1}{2}\partial _{\xi }\delta \bar{%
Z}_{ij}^{uL}\,,\quad \left( i\neq j\right) \,,  \label{gammalout}
\end{eqnarray}
and once more similar relations hold exchanging $L\leftrightarrow R$ and $%
u\leftrightarrow d$. Notice that absorptive parts are present in the 1PI
Green functions and hence in $\delta Z$ and $\delta \bar{Z}$ too. If we
forget about such absorptive parts we would have pseudo-hermiticity. Namely
\begin{equation*}
\Gamma _{\chi \bar{u}_{i}\eta _{j}^{u}}^{\left( 1\right) }=\gamma ^{0}\Gamma
_{\bar{\eta}_{i}^{u}u_{j}\chi }^{\left( 1\right) \dagger }\gamma ^{0}\,,
\end{equation*}
where $\Gamma _{\bar{\eta}_{i}^{u}u_{j}\chi }^{\dagger }$ means complex
conjugating $\Gamma _{\bar{\eta}_{i}^{u}u_{j}\chi }$ and interchanging \emph{%
both} Dirac and family indices. However the imaginary branch cuts terms
prevent the above relation to hold and then Eq. (\ref{hermiticity}) does not
hold.

At this point one might be tempted to plug expressions (\ref{gammalin}), (%
\ref{gammalout}) in Eq. (\ref{final}). However such relations are obtained
only in the restricted case $i\neq j$. For $i=j$ Eqs. (\ref{rels}) are
insufficient to determine the combinations appearing in the l.h.s. of Eqs. (%
\ref{gammalin}), (\ref{gammalout}) and further information is required. That
is also necessary even in the actual case where the r.h.s. of Eqs. (\ref
{gammalin}), (\ref{gammalout}) are not singular at $m_{i}\rightarrow m_{j}$
\cite{Yamada}. In the rest of this section we shall proceed to calculate
such diagonal combinations and as by product we shall also cross-check the
results already obtained for the off-diagonal contributions and in addition
produce some new ones.

By direct computation one generically finds
\begin{eqnarray}
\Gamma _{\chi \bar{u}_{i}\eta _{j}^{u}}^{\left( 1\right) } &=&\left( \frac{{}%
}{{}}\not{p}m_{i}^{u}B_{ij}^{u}\left( p^{2}\right) +C_{ij}^{u}\left(
p^{2}\right) +A_{ij}^{u}\left( p^{2}\right) \frac{{}}{{}}\right) R\,,  \notag
\\
\Gamma _{\bar{\eta}_{i}^{u}u_{j}\chi }^{\left( 1\right) } &=&L\left( \frac{{}%
}{{}}\not{p}B_{ij}^{u}\left( p^{2}\right) m_{j}^{u}+C_{ij}^{u}\left(
p^{2}\right) +A_{ij}^{u}\left( p^{2}\right) \frac{{}}{{}}\right) \,,
\label{integrals}
\end{eqnarray}
and analogous relations interchanging $u\leftrightarrow d.$ The $A$ function
comes from the diagram containing a charged gauge boson propagator and $B$
and $C$ from the diagram containing a charged Goldstone boson propagator.
From Eqs. (\ref{nielsen2}) and (\ref{integrals}) we obtain
\begin{eqnarray}
\partial _{\xi }\Sigma _{ij}^{\gamma R}\left( p^{2}\right)
&=&-2m_{i}B_{ij}\left( p^{2}\right) m_{j}\,,  \notag \\
\partial _{\xi }\Sigma _{ij}^{\gamma L}\left( p^{2}\right) &=&2\left( \frac{%
{}}{{}}A_{ij}\left( p^{2}\right) +C_{ij}\left( p^{2}\right) \frac{{}}{{}}%
\right) \,,  \notag \\
\partial _{\xi }\Sigma _{ij}^{R}\left( p^{2}\right) &=&\left( \frac{{}}{{}}%
p^{2}B_{ij}\left( p^{2}\right) -C_{ij}\left( p^{2}\right) -A_{ij}\left(
p^{2}\right) \frac{{}}{{}}\right) m_{j}\,,  \notag \\
\partial _{\xi }\Sigma _{ij}^{L}\left( p^{2}\right) &=&m_{i}\left( \frac{{}}{%
{}}p^{2}B_{ij}\left( p^{2}\right) -C_{ij}\left( p^{2}\right) -A_{ij}\left(
p^{2}\right) \frac{{}}{{}}\right) \,.  \label{sigs}
\end{eqnarray}
The above system of equations is overdetermined and therefore some
consistency identities between bare self-energies arise, namely
\begin{equation}
\partial _{\xi }\left( \frac{{}}{{}}m_{i}\Sigma _{ij}^{R}\left( p^{2}\right)
-\Sigma _{ij}^{L}\left( p^{2}\right) m_{j}\frac{{}}{{}}\right) =0\,,
\label{const0}
\end{equation}
and
\begin{equation}
\partial _{\xi }\left( p^{2}\Sigma _{ij}^{\gamma R}\left( p^{2}\right)
+\Sigma _{ij}^{\gamma L}\left( p^{2}\right) m_{i}m_{j}+m_{i}\Sigma
_{ij}^{R}\left( p^{2}\right) +\Sigma _{ij}^{L}\left( p^{2}\right)
m_{j}\right) =0\,.  \label{const2}
\end{equation}
These constrains must hold independently of any renormalization scheme and
we have checked them by direct computation. Actually the former trivially
holds since, at least at the one-loop level in the SM,
\begin{equation}
m_{i}\Sigma _{ij}^{R}\left( p^{2}\right) -\Sigma _{ij}^{L}\left(
p^{2}\right) m_{j}=0\,.  \label{const1}
\end{equation}

Finally, projecting Eq. (\ref{integrals}) over spinors we also have
\begin{eqnarray}
\bar{u}_{u}\left( p_{i}\right) \Gamma _{\chi \bar{u}_{i}\eta
_{j}^{u}}^{\left( 1\right) } &=&\bar{u}_{u}\left( p_{i}\right) \left( \frac{%
{}}{{}}m_{i}^{u2}B_{ij}^{u}\left( m_{i}^{u2}\right) +C_{ij}^{u}\left(
m_{i}^{u2}\right) +A_{ij}^{u}\left( m_{i}^{u2}\right) \frac{{}}{{}}\right)
R\,,  \notag \\
\Gamma _{\bar{\eta}_{i}^{d}d_{j}\chi }^{\left( 1\right) }v_{d}\left(
-p_{j}\right) &=&L\left( \frac{{}}{{}}B_{ij}^{d}\left( m_{j}^{d2}\right)
m_{j}^{d2}+C_{ij}^{d}\left( m_{j}^{d2}\right) +A_{ij}^{d}\left(
m_{j}^{d2}\right) \right) v_{d}\left( -p_{j}\right) \,.  \label{calc0}
\end{eqnarray}
The r.h.s. of the previous expressions can be evaluated in terms of the wfr.
via the use of Eqs. (\ref{sigs})
\begin{eqnarray}
&\partial _{\xi }&\hspace{-0.3cm}\left( m_{j}^{u}m_{i}^{u}\Sigma
_{ij}^{u\gamma R}\left( p^{2}\right) +p^{2}\Sigma _{ij}^{u\gamma L}\left(
p^{2}\right) +m_{j}^{u}\Sigma _{ij}^{uR}\left( p^{2}\right) +m_{i}^{u}\Sigma
_{ij}^{uL}\left( p^{2}\right) \right) =  \notag \\
&&B_{ij}^{u}\left( p^{2}\right) \left( \frac{{}}{{}}p^{2}\left(
m_{j}^{u2}+m_{i}^{u2}\right) -2m_{j}^{u2}m_{i}^{u2}\frac{{}}{{}}\right)
\notag \\
&&+\left( 2p^{2}-m_{j}^{u2}-m_{i}^{u2}\right) \left( A_{ij}^{u}\left(
p^{2}\right) +C_{ij}^{u}\left( p^{2}\right) \frac{{}}{{}}\right) \,,
\label{calcu} \\
&\partial _{\xi }&\hspace{-0.3cm}\left( \Sigma _{ij}^{d\gamma R}\left(
p^{2}\right) m_{i}^{d}m_{j}^{d}+\Sigma _{ij}^{d\gamma L}\left( p^{2}\right)
p^{2}+m_{i}^{d}\Sigma _{ij}^{dL}\left( p^{2}\right) +\Sigma _{ij}^{dR}\left(
p^{2}\right) m_{j}^{d}\right) =  \notag \\
&&B_{ij}^{d}\left( p^{2}\right) \left( \frac{{}}{{}}p^{2}\left(
m_{i}^{d2}+m_{j}^{d2}\right) -2m_{i}^{d2}m_{j}^{d2}\right)  \notag \\
&&+\left( 2p^{2}-m_{i}^{d2}-m_{j}^{d2}\right) \left( \frac{{}}{{}}%
A_{ij}^{d}\left( p^{2}\right) +C_{ij}^{d}\left( p^{2}\right) \right) \,.
\label{calcd}
\end{eqnarray}
Hence using the off-diagonal wfr. expressions (\ref{zin}), (\ref{zout}) we
re-obtain
\begin{equation}
\bar{u}_{u}\left( p_{i}\right) \frac{1}{2}\partial _{\xi }\delta \bar{Z}%
_{ij}^{uL}R=\bar{u}\left( p_{i}\right) \Gamma _{\chi \bar{u}_{i}\eta
_{j}^{u}}^{\left( 1\right) }\,,\qquad L\frac{1}{2}\partial _{\xi }\delta
Z_{ij}^{dL}v_{d}\left( -p_{j}\right) =\Gamma _{\bar{\eta}_{i}^{d}d_{j}\chi
}^{\left( 1\right) }v_{d}\left( -p_{j}\right) \,.  \label{basicrel}
\end{equation}
For the diagonal wfr. we use Eqs. (\ref{zdiag}) together with (\ref{sigs})
and (\ref{calc0}) obtaining \emph{exactly} the same result as in Eq. (\ref
{basicrel}) with $i=j$ therein. Note however that since in Eq. (\ref{calc0})
we have no derivatives with respect to $p^{2}$ obtaining Eq. (\ref{basicrel}%
) involves a subtle cancellation between the $p^{2}$ derivatives of the bare
self-energies appearing in the definition of the diagonal wfr; for instance
\begin{eqnarray*}
&&\bar{u}\left( p_{i}\right) \frac{1}{2}\partial _{\xi _{W}}\delta \bar{Z}%
_{ii}^{uL}R \\
&=&\frac{1}{2}\bar{u}\left( p_{i}\right) \partial _{\xi _{W}}\left\{ 2\left(
A_{ii}\left( m_{i}^{u2}\right) +C_{ii}\left( m_{i}^{u2}\right) \right)
\right. \\
&&+2m_{i}^{u2}\left( -B_{ii}^{\prime }\left( m_{i}^{u2}\right)
m_{i}^{u2}+A_{ii}^{\prime }\left( m_{i}^{u2}\right) +C_{ii}^{\prime }\left(
m_{i}^{u2}\right) \right) \\
&&+\left. 2m_{i}^{u2}\left( B_{ii}\left( m_{i}^{u2}\right)
+m_{i}^{u2}B_{ii}^{\prime }\left( m_{i}^{u2}\right) -C_{ii}^{\prime }\left(
m_{i}^{u2}\right) -A_{ii}^{\prime }\left( m_{i}^{u2}\right) \right) \right\}
R \\
&=&\bar{u}\left( p_{i}\right) \Gamma _{\chi \bar{u}_{i}\eta
_{j}^{u}}^{\left( 1\right) }\,.
\end{eqnarray*}

Before proceeding let us make a side remark concerning the regularity
properties of the gauge derivative in Eqs. (\ref{calcu}) and (\ref{calcu})
in the limit $m_{i}\rightarrow m_{j}$. Note that evaluating Eq. (\ref{calcu}%
) at $p^{2}=m_{i}^{u2}$ and Eq. (\ref{calcd}) at $p^{2}=m_{j}^{d2},$ a
global factor $\left( m_{i}^{u2}-m_{j}^{u2}\right) $ appears in the first
equation and $\left( m_{j}^{d2}-m_{i}^{d2}\right) $ in the second one.
Therefore it can be immediately seen that Nielsen identities together with
the information provided by Eq. (\ref{integrals}) assures the regularity of
the gauge derivative for the off-diagonal wfr. constants when $%
m_{i}\rightarrow m_{j}$. Moreover we have seen that such limit is not only
regular but also equal to the expression obtained from the diagonal wfr.
which is not \emph{a priori} obvious \cite{Grassi}, \cite{Yamada}.

\begin{figure}[!hbp]
\begin{center}
\includegraphics[width=\figwidth]{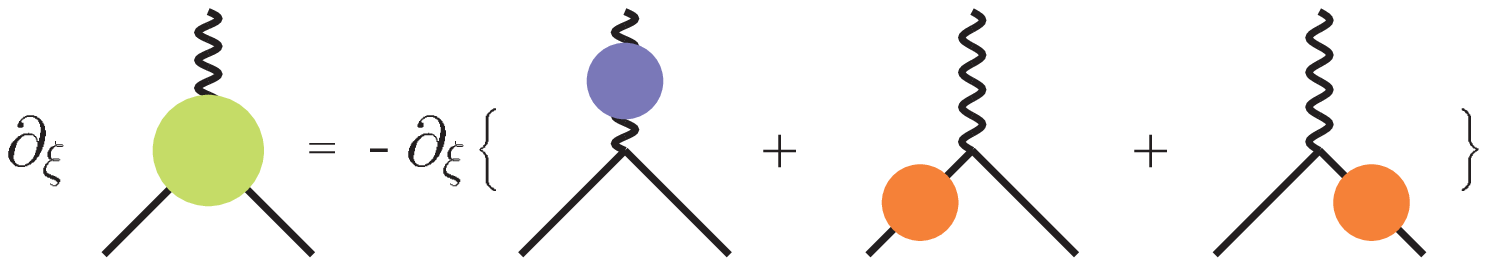}
\end{center}
\caption{Pictorial representation of the on-shell Nielsen identity given by
Eq.(\ref{vertexgauge}). The blobs in the lhs. represent bare one-loop
contributions to the on-shell vertex and the blobs in the rhs. wfr. counter
terms.}
\label{FigLSZ1}
\end{figure}
Replacing Eq. (\ref{basicrel}) in Eq. (\ref{nielsen0}) we obtain
\begin{eqnarray}
&&\partial _{\xi }\left( \bar{u}_{u}\left( p_{i}\right) \epsilon ^{\mu
}\left( q\right) \Gamma _{W_{\mu }^{+}\bar{u}_{i}d_{j}}^{\left( 1\right)
}v_{d}\left( -p_{j}\right) \right)  \notag \\
&=&\frac{e}{2s_{W}}M_{L}^{\left( 1\right) }\partial _{\xi }\left( \delta
\bar{Z}_{ir}^{uL}K_{rj}+K_{ir}\delta Z_{rj}^{dL}+\delta Z_{W}K_{ij}\right)
\notag \\
&=&-\frac{e}{2s_{W}}\partial _{\xi }\left( M_{L}^{\left( 1\right) }\delta
F_{L}^{\left( 1\right) }+M_{L}^{\left( 2\right) }\delta F_{L}^{\left(
2\right) }+M_{R}^{\left( 1\right) }\delta F_{R}^{\left( 1\right)
}+M_{R}^{\left( 2\right) }\delta F_{R}^{\left( 2\right) }\right) \,,
\label{vertexgauge}
\end{eqnarray}
where Eq.~(\ref{vertex}) and the gauge independence of the electric charge
and Weinberg angle has been used in the last equality. In the previous
expression $M_{L,R}^{\left( i\right) }$ are understood with the physical
momenta $p_{1}$ and $p_{2}$ of Eq. (\ref{wdtree}) replaced by the
diagrammatic momenta $p_{i}$ and $-p_{j}$ respectively. Note that Eq. (\ref
{vertexgauge}) states that the gauge dependence of the on-shell bare
one-loop vertex function cancels out the renormalization counter terms
appearing in Eq. (\ref{vertex}) (see Fig. \ref{FigLSZ1}). This is one of the
crucial results and special care should be taken not to ignore any of the
absorptive parts ---including those in the wfr. constants. As a consequence
\begin{equation*}
\partial _{\xi }\mathcal{M}_{1}=-\frac{e}{2s_{W}}M_{L}^{\left( 1\right)
}\partial _{\xi }\delta K_{ij}\,,
\end{equation*}
and asking for a gauge independent amplitude the counter term for $K_{ij}$
must be separately gauge independent, as originally derived in \cite{Grassi}.

Finally, since each structure $M_{L,R}^{\left( i\right) }$ must cancel
separately we have that the Nielsen identities enforce
\begin{equation*}
\partial _{\xi }\delta F_{L}^{\left( 2\right) }=\partial _{\xi }\delta
F_{R}^{\left( 1\right) }=\partial _{\xi }\delta F_{R}^{\left( 2\right) }=0\,.
\end{equation*}

\section{Absorptive parts}

\label{absorptive} Having determined in the previous section, thanks to an
extensive use of the Nielsen identities, the gauge dependence of the
different quantities appearing in top or $W$ decay in terms of the
self-energies, we shall now proceed to list the absorptive parts of the wfr.
constants, with special attention to their gauge dependence. The aim of this
section is to state the differences between the wfr. constants given in our
scheme and the ones in \cite{Denner}. Recall that at one-loop such
difference reduces to the absorptive ($\widetilde{Im}$) contribution to the $%
\delta Z$'s. In what concerns the gauge dependent part (with $\xi \geq 0$)
the absorptive contribution ($\widetilde{Im}_{\xi }$) in the fermionic $%
\delta Z$'s amounts to

\begin{eqnarray}
i\widetilde{Im}_{\xi }\left( \delta Z_{ij}^{uL}\right) &=&\sum_{h}\frac{%
iK_{ih}K_{hj}^{\dagger }}{8\pi v^{2}m_{j}^{u2}}\theta \left(
m_{j}^{u}-m_{h}^{d}-\sqrt{\xi }M_{W}\right) \left( m_{j}^{u2}-m_{h}^{d2}-\xi
M_{W}^{2}\right)  \notag \\
&&\times \sqrt{\left( \left( m_{j}^{u}-m_{h}^{d}\right) ^{2}-\xi
M_{W}^{2}\right) \left( \left( m_{j}^{u}+m_{h}^{d}\right) ^{2}-\xi
M_{W}^{2}\right) }\,,  \notag \\
i\widetilde{Im}_{\xi }\left( \delta \bar{Z}_{ij}^{uL}\right) &=&\sum_{h}%
\frac{iK_{ih}K_{hj}^{\dagger }}{8\pi v^{2}m_{i}^{u2}}\theta \left(
m_{i}^{u}-m_{h}^{d}-\sqrt{\xi }M_{W}\right) \left( m_{i}^{u2}-m_{h}^{d2}-\xi
M_{W}^{2}\right)  \notag \\
&&\times \sqrt{\left( \left( m_{i}^{u}-m_{h}^{d}\right) ^{2}-\xi
M_{W}^{2}\right) \left( \left( m_{i}^{u}+m_{h}^{d}\right) ^{2}-\xi
M_{W}^{2}\right) }\,,  \notag \\
\widetilde{Im}_{\xi }\left( \delta Z_{ij}^{uR}\right) &=&\widetilde{Im}_{\xi
}\left( \delta \bar{Z}_{ij}^{uR}\right) =0\,,  \label{absorptive1}
\end{eqnarray}
where $\theta $ is the Heaviside function and $v$ is the Higgs vacuum
expectation value (see appendix \ref{selfenergiesapp}). For the down $\delta
Z$ we have the same formulae replacing $u\leftrightarrow d$ and $%
K\leftrightarrow K^{\dagger }.$ Note that the vanishing of $\widetilde{Im}%
_{\xi }\left( \delta Z_{ij}^{uR}\right) $ and $\widetilde{Im}_{\xi }\left(
\delta \bar{Z}_{ij}^{uR}\right) $ could be anticipated from constraint (\ref
{const2}) derived from Nielsen identities. Using these results we can write
\begin{eqnarray}
&&i\partial _{\xi }\widetilde{Im}\left[ \sum_{r}\left( \delta \bar{Z}%
_{ir}^{uL}K_{rj}+K_{ir}\delta Z_{rj}^{dL}\right) +\delta Z_{W}K_{ij}\right]
\notag \\
&=&K_{ij}\partial _{\xi }\left\{ \frac{i}{8\pi v^{2}}\left[ \frac{1}{%
m_{i}^{u2}}\theta \left( m_{i}^{u}-m_{j}^{d}-\sqrt{\xi }M_{W}\right) \left(
m_{i}^{u2}-m_{j}^{d2}-\xi M_{W}^{2}\right) \right. \right.  \notag \\
&&+\left. \frac{1}{m_{j}^{d2}}\theta \left( m_{j}^{d}-m_{i}^{u}-\sqrt{\xi }%
M_{W}\right) \left( m_{j}^{d2}-m_{i}^{u2}-\xi M_{W}^{2}\right) \frac{{}}{{}}%
\right]  \notag \\
&&\times \left. \sqrt{\left( \left( m_{j}^{d}-m_{i}^{u}\right) ^{2}-\xi
M_{W}^{2}\right) \left( \left( m_{j}^{d}+m_{i}^{u}\right) ^{2}-\xi
M_{W}^{2}\right) }+i\widetilde{Im}_{\xi }\left( \delta Z_{W}\right) \frac{{}%
}{{}}\right\} \,.  \label{gauge0}
\end{eqnarray}
In the case $\left| m_{i}^{u}-m_{j}^{d}\right| \leq \sqrt{\xi }M_{W}$ the
above expression reduces to
\begin{equation}
\partial _{\xi }\sum_{r}\widetilde{Im}\left( \delta \bar{Z}%
_{ir}^{uL}K_{rj}+K_{ir}\delta Z_{rj}^{dL}\right) =0\,,  \label{gauge1}
\end{equation}
while for $\left| m_{i}^{u}-m_{j}^{d}\right| \geq \sqrt{\xi }M_{W}$ we have
\begin{eqnarray}
&&i\partial _{\xi }\sum_{r}\widetilde{Im}\left( \delta \bar{Z}%
_{ir}^{uL}K_{rj}+K_{ir}\delta Z_{rj}^{dL}\right)  \notag \\
&=&K_{ij}\partial _{\xi }\left\{ \frac{i}{4\pi v^{2}}\frac{\left|
m_{i}^{u2}-m_{j}^{d2}\right| -\xi M_{W}^{2}}{m_{i}^{u2}+m_{j}^{d2}+\left|
m_{i}^{u2}-m_{j}^{d2}\right| }\right.  \notag \\
&&\times \left. \sqrt{\left( \left( m_{j}^{d}-m_{i}^{u}\right) ^{2}-\xi
M_{W}^{2}\right) \left( \left( m_{j}^{d}+m_{i}^{u}\right) ^{2}-\xi
M_{W}^{2}\right) }\right\} \,.  \label{gauge2}
\end{eqnarray}
Moreover the $\xi $-dependent absorptive contribution to $\delta Z_{W}$ ($%
\widetilde{Im}_{\xi }\left( \delta Z_{W}\right) $) has no dependence in
quark masses since the diagram with a fermion loop is gauge independent.
Because of that we can conclude that the derivative in Eq. (\ref{gauge0})
does not vanish. Defining $\Delta _{ij}$ as the difference between the
vertex observable calculated in our scheme and the same in the scheme using $%
\widetilde{Re}$ we have
\begin{equation*}
\Delta _{ij}\sim \left| K_{ij}\right| ^{2}\mathrm{Re}\left( i\widetilde{Im}%
\delta Z_{W}\right) +\mathrm{Re}\left\{ iK_{ij}^{\ast }\sum_{r}\left[
\widetilde{Im}\left( \delta \bar{Z}_{ir}^{uL}\right) K_{rj}+K_{ir}\widetilde{%
Im}\left( \delta Z_{rj}^{dL}\right) \right] \right\} \,.
\end{equation*}
In the case of $\delta Z_{W}$ one can easily check that $\widetilde{Im}%
\left( \delta Z_{W}\right) =\mathrm{Im}\left( \delta Z_{W}\right) $
obtaining
\begin{equation}
\Delta _{ij}\sim \mathrm{Re}\left\{ iK_{ij}^{\ast }\sum_{r}\left[ \widetilde{%
Im}\left( \delta \bar{Z}_{ir}^{uL}\right) K_{rj}+K_{ir}\widetilde{Im}\left(
\delta Z_{rj}^{dL}\right) \right] \right\} \,.  \label{gauge3}
\end{equation}
Thus from Eqs. (\ref{gauge1}), (\ref{gauge2}) and (\ref{gauge3}) we
immediately obtain
\begin{equation}
\partial _{\xi }\Delta _{ij}\sim \mathrm{Re}\left\{ iK_{ij}^{\ast }\sum_{r}%
\left[ \partial _{\xi }\widetilde{Im}\left( \delta \bar{Z}_{ir}^{uL}\right)
K_{rj}+K_{ir}\partial _{\xi }\widetilde{Im}\left( \delta Z_{rj}^{dL}\right) %
\right] \right\} =0\,.  \label{res1}
\end{equation}
However gauge independent absorptive parts, included if our prescription is
used but not if one uses the one of \cite{Denner} which makes use of the $%
\widetilde{Re}$, do contribute to Eq. (\ref{gauge3}). In order to see that
we can take $\xi =1$ obtaining for the physical values of the masses
\begin{eqnarray}
\widetilde{Im}_{\xi =1}\left( \delta Z_{rj}^{dL}\right) &=&0\,,  \notag \\
\widetilde{Im}_{\xi =1}\left( \delta \bar{Z}_{ir}^{uL}\right) &=&\sum_{h}%
\frac{K_{ih}K_{hr}^{\dagger }}{8\pi v^{2}m_{i}^{u2}}\frac{\theta \left(
m_{i}^{u}-m_{h}^{d}-M_{W}\right) }{m_{i}^{u2}-m_{r}^{u2}}  \notag \\
&&\times \sqrt{\left( m_{i}^{u2}-\left( M_{W}-m_{h}^{d}\right) ^{2}\right)
\left( m_{i}^{u2}-\left( M_{W}+m_{h}^{d}\right) ^{2}\right) }  \notag \\
&&\times \left( \frac{1}{2}\left( m_{r}^{u2}+m_{h}^{d2}+2M_{W}^{2}\right)
\left( m_{i}^{u2}+m_{h}^{2d}-M_{W}^{2}\right) -\left(
m_{i}^{u2}+m_{r}^{u2}\right) m_{h}^{d2}\right) \,,  \notag \\
&&  \label{zulb}
\end{eqnarray}
where only the results for $i\neq j$ have been presented. Note that $%
\widetilde{Im}_{\xi =1}\left( \delta \bar{Z}_{ir}^{uL}\right) \neq 0$ only
when $i=3$, that is when the renormalized up-particle is a top. In addition,
since the $m_{r}^{u2}$ dependence in Eq. (\ref{zulb}) does not vanish, CKM
phases do not disappear from Eq. (\ref{gauge3}) and therefore
\begin{equation}
\Delta _{3j}\sim \mathrm{Re}\left\{ iK_{3j}^{\ast }\sum_{r}\left[ \widetilde{%
Im}\left( \delta \bar{Z}_{3r}^{uL}\right) K_{rj}+K_{3r}\widetilde{Im}\left(
\delta Z_{rj}^{dL}\right) \right] \right\} \neq 0\,.  \label{res2}
\end{equation}
Eqs. (\ref{res1}) and (\ref{res2}) show that even though the difference $%
\Delta _{3j}$ is gauge independent, does not actually vanish. There are
genuine gauge independent pieces that contribute not only to the amplitude,
but also to the observable. As discussed these additional pieces cannot be
absorbed by a redefinition of $K_{ij}$. Numerically such gauge independent
corrections amounts roughly to $\Delta _{3j}\simeq 5\times 10^{-3}O_{\mathrm{%
tree}}$ where $O_{\mathrm{tree}}$ is the observable quantity calculated at
leading order.

\section{$CP$ violation and $CPT$ invariance}

\label{cp}In this section we want to show that using wfr. constants that do
not verify a pseudo-hermiticity condition does not lead to any unwanted
pathologies. In particular: (a) No new sources of $CP$ violation appear
besides the ones already present in the SM. (b) The total width of particles
and anti-particles coincide, thus verifying the $CPT$ theorem. Let us start
with the latter, which is not completely obvious since not all external
particles and anti-particles are renormalized with the same constant due to
the different absorptive parts.

The optical theorem asserts that
\begin{eqnarray}
\Gamma _{t}\sim \sum_{f}\int d\Pi _{f}\left| M\left( t^{\left( \hat{n}%
\right) }\left( p\right) \rightarrow f\right) \right| ^{2} &=&2\mathrm{Im}%
\left[ M\left( t^{\left( \hat{n}\right) }\left( p\right) \rightarrow
t^{\left( \hat{n}\right) }\left( p\right) \right) \right] \,,
\label{topdecay} \\
\Gamma _{\bar{t}}\sim \sum_{f}\int d\Pi _{f}\left| M\left( \bar{t}^{\left(
\hat{n}\right) }\left( p\right) \rightarrow f\right) \right| ^{2} &=&2%
\mathrm{Im}\left[ M\left( \bar{t}^{\left( \hat{n}\right) }\left( p\right)
\rightarrow \bar{t}^{\left( \hat{n}\right) }\left( p\right) \right) \right]
\,,  \label{antitopdecay}
\end{eqnarray}
where we have consider, just as an example, top ($t^{\left( \hat{n}\right)
}\left( p\right) $) and anti-top ($\bar{t}^{\left( \hat{n}\right) }\left(
p\right) $) decay, with $p$ and $\hat{n}$ being their momentum and
polarization. Recalling that the incoming fermion and outgoing anti-fermion
spinors are renormalized with a common constant (see Eq. (\ref{fundamental}%
)) as are the outgoing fermion and incoming anti-fermion ones, it is
immediate to see that
\begin{eqnarray*}
M\left( t^{\left( \hat{n}\right) }\left( p\right) \rightarrow t^{\left( \hat{%
n}\right) }\left( p\right) \right) &=&\bar{u}^{\left( \hat{n}\right) }\left(
p\right) A_{33}\left( p\right) u^{\left( \hat{n}\right) }\left( p\right) \,,
\\
M\left( \bar{t}^{\left( \hat{n}\right) }\left( p\right) \rightarrow \bar{t}%
^{\left( \hat{n}\right) }\left( p\right) \right) &=&-\bar{v}^{\left( \hat{n}%
\right) }\left( p\right) A_{33}\left( -p\right) v^{\left( \hat{n}\right)
}\left( p\right) \,,
\end{eqnarray*}
where the minus sign comes from an interchange of two fermion operators and
where the subscripts in $A$ indicate family indices. Using the fact that
\begin{equation*}
u^{\left( \hat{n}\right) }\left( p\right) \otimes \bar{u}^{\left( \hat{n}%
\right) }\left( p\right) =\frac{\not{p}+m}{2m}\frac{1+\gamma ^{5}\not{n}}{2}%
\,,\qquad -v^{\left( \hat{n}\right) }\left( p\right) \otimes \bar{v}^{\left(
\hat{n}\right) }\left( p\right) =\frac{-\not{p}+m}{2m}\frac{1+\gamma ^{5}%
\not{n}}{2}\,,
\end{equation*}
with $n=\frac{1}{\sqrt{\left( p^{0}\right) ^{2}-\left( \vec{p}\cdot \hat{n}%
\right) ^{2}}}\left( \vec{p}\cdot \hat{n},p^{0}\hat{n}\right) $ being the
polarization four-vector and performing some elementary manipulations we
obtain
\begin{eqnarray*}
&&\bar{u}^{\left( \hat{n}\right) }\left( p\right) A_{33}\left( p\right)
u^{\left( \hat{n}\right) }\left( p\right) \\
&=&Tr\left[ \left( \frac{\not{p}+m}{2m}\frac{1+\gamma ^{5}\not{n}}{2}\right)
\left( a\left( p^{2}\right) \not{p}L+b\left( p^{2}\right) \not{p}R+c\left(
p^{2}\right) L+d\left( p^{2}\right) R\right) \right] \\
&=&\frac{1}{4}Tr\left\{ \frac{\not{p}+m}{2m}\left[ \left( a\left(
p^{2}\right) +b\left( p^{2}\right) \right) \not{p}+c\left( p^{2}\right)
+d\left( p^{2}\right) \right] \right\} \\
&=&\frac{1}{4}Tr\left\{ \frac{-\not{p}+m}{2m}\left[ -\left( a\left(
p^{2}\right) +b\left( p^{2}\right) \right) \not{p}+c\left( p^{2}\right)
+d\left( p^{2}\right) \right] \right\} \\
&=&Tr\left[ \frac{-\not{p}+m}{2m}\frac{1+\gamma ^{5}\not{n}}{2}\left(
-a\left( p^{2}\right) \not{p}L-b\left( p^{2}\right) \not{p}R+c\left(
p^{2}\right) L+d\left( p^{2}\right) R\right) \right] \\
&=&-\bar{v}^{\left( \hat{n}\right) }\left( p\right) A_{33}\left( -p\right)
v^{\left( \hat{n}\right) }\left( p\right) \,,
\end{eqnarray*}
where we have decomposed $A_{33}\left( p\right) $ into its most general
Dirac structure. We thus conclude the equality between Eqs. (\ref{topdecay})
and (\ref{antitopdecay}) verifying that the lifetimes of top and anti-top
are identical. The detailed form of the wfr. constants, or whether they have
absorptive parts or not, does not play any role.

Even thought total decay widths for top and anti-top are identical the
partial ones need not to if $CP$ violation is present and some compensation
between different processes must take place. Here we shall show that when $%
K=K^{\ast }$ the $CP$ invariance of the Lagrangian manifests itself in a
zero asymmetry between the partial differential decay rate of top and its $%
CP $ conjugate process. The fact that the external renormalization constants
have dispersive parts does not alter this conclusion. This is of course
expected on rather general grounds, so the following discussion has to be
taken really as a verification that no unexpected difficulties arise.

To illustrate this point let us consider the top decay channel $t\left(
p_{1}\right) \rightarrow W^{+}\left( p_{1}-p_{2}\right) +b\left(
p_{2}\right) $ and its $CP$ conjugate process $\bar{t}\left( \tilde{p}%
_{1}\right) \rightarrow W^{-}\left( \tilde{p}_{1}-\tilde{p}_{2}\right)
+b\left( \tilde{p}_{2}\right) .$ Let us note the respective amplitudes by $%
\mathcal{A}$ and $\mathcal{B}$ which are given as
\begin{eqnarray*}
\mathcal{A} &=&\varepsilon ^{\mu }\bar{u}^{\left( s_{2}\right) }\left(
p_{2}\right) A_{\mu }u^{\left( s_{1}\right) }\left( p_{1}\right) \,, \\
\mathcal{B} &=&-\tilde{\varepsilon}^{\mu }\bar{v}^{\left( s_{1}\right)
}\left( \tilde{p}_{1}\right) B_{\mu }v^{\left( s_{2}\right) }\left( \tilde{p}%
_{2}\right) \,,
\end{eqnarray*}
where $\tilde{a}^{\mu }=a_{\mu }=\left( a^{0},-a^{i}\right) $ for any
four-vector. Considering contributions up to including next-to-leading
corrections we have
\begin{eqnarray*}
A_{\mu } &=&-i\frac{e}{\sqrt{2}s_{W}}\left[ \left( \bar{Z}^{\frac{1}{2}%
bL}K^{\dagger }Z^{\frac{1}{2}tL}+K^{\dagger }\delta _{V}+\delta K^{\dagger
}\right) \gamma _{\mu }L+\delta F_{\mu }\right] \,, \\
B_{\mu } &=&-i\frac{e}{\sqrt{2}s_{W}}\left[ \left( \bar{Z}^{\frac{1}{2}%
tL}KZ^{\frac{1}{2}bL}+K\delta _{V}+\delta K\right) \gamma _{\mu }L+\delta
G_{\mu }\right] \,,
\end{eqnarray*}
with $\delta _{V}$ $=\frac{\delta e}{e}-\frac{\delta s_{W}}{s_{W}}+\frac{1}{2%
}\delta Z_{W}$ and $\delta F_{\mu }$ and $\delta G_{\mu }$ are given by the
one-loop diagrams. From a direct computation it can be seen that if $%
K=K^{\ast }$ this implies

\begin{equation}
\bar{Z}^{\frac{1}{2}L}=\left( Z^{\frac{1}{2}L}\right) ^{T}\,,\quad \bar{Z}^{%
\frac{1}{2}R}=\left( Z^{\frac{1}{2}R}\right) ^{T}\,,\quad \tilde{\varepsilon}%
^{\mu }\delta G_{\mu }=\varepsilon ^{\mu }\gamma ^{2}\delta F_{\mu
}^{T}\gamma ^{2}\,,  \label{CPfacts}
\end{equation}
where the superscript $T$ means transposition with respect to all indices
(family indices in the case of $Z^{\frac{1}{2}L}$ and $Z^{\frac{1}{2}R}$ and
Dirac indices in the case of $\delta F_{\mu }$ ). Using
\begin{equation*}
i\gamma ^{2}\bar{u}^{\left( s\right) T}\left( p\right) =sv^{\left( s\right)
}\left( \tilde{p}\right) \,,\qquad u^{\left( s\right) T}\left( p\right)
i\gamma ^{2}=-s\bar{v}^{\left( s\right) }\left( \tilde{p}\right) \,,
\end{equation*}
where $s=\pm 1$, depending on the spin direction in the $\hat{z}$ axis, we
obtain
\begin{eqnarray*}
\mathcal{A} &=&\frac{-ie}{\sqrt{2}s_{W}}\varepsilon ^{\mu }\bar{u}^{\left(
s_{2}\right) }\left( p_{2}\right) \left[ \left( \bar{Z}^{\frac{1}{2}%
bL}K^{\dagger }Z^{\frac{1}{2}tL}+K^{\dagger }\delta _{V}+\delta K^{\dagger
}\right) \gamma _{\mu }L+\delta F_{\mu }\right] u^{\left( s_{1}\right)
}\left( p_{1}\right) \\
&=&\frac{-ie}{\sqrt{2}s_{W}}\varepsilon ^{\mu }u^{\left( s_{1}\right)
T}\left( p_{1}\right) \left[ L\left( \left( Z^{\frac{1}{2}tL}\right)
^{T}K^{\ast }\left( \bar{Z}^{\frac{1}{2}bL}\right) ^{T}+K^{\ast }\delta
_{V}+\delta K^{\ast }\right) \gamma _{\mu }^{T}+\delta F_{\mu }^{T}\right]
\bar{u}^{\left( s_{2}\right) T}\left( p_{2}\right) \\
&=&\frac{-s_{1}s_{2}ie}{\sqrt{2}s_{W}}\varepsilon ^{\mu }\bar{v}^{\left(
s_{1}\right) }\left( \tilde{p}_{1}\right) \gamma ^{2}\left[ L\left( \left(
Z^{\frac{1}{2}tL}\right) ^{T}K^{\ast }\left( \bar{Z}^{\frac{1}{2}bL}\right)
^{T}+K^{\ast }\delta _{V}+\delta K^{\ast }\right) \gamma _{\mu }^{T}+\delta
F_{\mu }^{T}\right] \gamma ^{2}v^{\left( s_{2}\right) }\left( \tilde{p}%
_{2}\right) \\
&=&\frac{-s_{1}s_{2}ie}{\sqrt{2}s_{W}}\varepsilon ^{\mu }\bar{v}^{\left(
s_{1}\right) }\left( \tilde{p}_{1}\right) \left[ \left( \left( Z^{\frac{1}{2}%
tL}\right) ^{T}K^{\ast }\left( \bar{Z}^{\frac{1}{2}bL}\right) ^{T}+K^{\ast
}\delta _{V}+\delta K^{\ast }\right) \gamma _{\mu }^{\dagger }L+\gamma
^{2}\delta F_{\mu }^{T}\gamma ^{2}\right] v^{\left( s_{2}\right) }\left(
\tilde{p}_{2}\right) ,
\end{eqnarray*}
now using Eq. (\ref{CPfacts}) we see that if no $CP$ violating phases are
present in the CKM matrix $K$ (and therefore neither in $\delta K,$ Eq. (\ref
{deltaK})) we obtain that $\mathcal{A}=-s_{1}s_{2}\mathcal{B}$ and thus
\begin{equation*}
\left| \mathcal{A}\right| ^{2}=\left| \mathcal{B}\right| ^{2}\,.
\end{equation*}

Note again that when $CP$ violating phases are present we can expect in
general non-vanishing phase-space dependent asymmetries for the different
channels. Once we sum over all channels and integrate over the final state
phase space a compensation must take place as we have seen guaranteed by
unitarity and $CPT$ invariance. Using a set of wfr. constants with
absorptive parts as advocated here (and required by gauge invariance) leads
to different results than using the prescription originally advocated in
\cite{Denner}, in particular using Eq. (\ref{res2}) for $K\neq K^{\ast }$ we
expect $\Delta _{3j}^{\left( t~decay\right) }-\Delta _{3j}^{\left( \bar{t}%
~decay\right) }\neq 0$.

\section{Conclusions}

\label{conc}

Let us recapitulate the main results of this chapter. We hope, first of all,
to have convinced the reader that \emph{there is} a problem with what
appears to be the commonly accepted prescription for dealing with wave
function renormalization when mixing is present. The situation is even
further complicated by the appearance of $CP$ violating phases. The problem
has a twofold aspect. On the one hand the prescription of \cite{Denner} does
not diagonalize the propagator matrix in family space in what respects to
the absorptive parts. On the other hand it yields gauge \emph{dependent}
amplitudes, albeit gauge \emph{independent} modulus squared amplitudes. This
is not satisfactory: interference with e.g. strong phases may reveal an
unacceptable gauge dependence.

The only solution is to accept wfr. constants that do not satisfy a
pseudo-hermiticity condition due to the presence of the absorptive parts,
which are neglected in \cite{Denner}. This immediately brings about some
gauge \emph{independent} absorptive parts which appear even in the modulus
squared amplitude and which are neglected in the treatment of \cite{Denner}.
Furthermore, these parts (and the gauge dependent ones) cannot be absorbed
in unitary redefinitions of the CKM matrix which are the only ones allowed
by Ward identities. We have checked that ---although unconventional--- the
presence of the absorptive parts in the wfr. constants is perfectly
compatible with basic tenets of field theory and the Standard Model.
Numerically we have found the differences to be important, at the order of
the half per cent. Small, but relevant in the future. This information will
be relevant to extract the experimental values of the CKM mixing matrix.

Traditionally, wave function renormalization seems to have been the ``poor
relative'' in the Standard Model renormalization program. We have seen here
that it is important on two counts. First because it is related to the
counter terms for the CKM mixing matrix, although the on-shell values for
wave function constants cannot be directly used there. Second because they
are crucial to obtain gauge independent $S$ matrix elements and observables.
While using our wfr. constants (but not the ones in \cite{Denner}) for the
external legs is strictly equivalent to considering reducible diagrams (with
on-shell mass counter terms) the former procedure is considerably more
practical.

\chapter{Probing LHC phenomenology: single top production}

\label{LHCphenomenology}With the current limit on the Higgs mass already
placed at 113.5 GeV \cite{LEPC} and no clear evidence for the existence of
an elementary scalar (despite much controversy regarding the results of the
last days of LEP) it makes sense to envisage an alternative to the minimal
Standard Model described by an effective theory without any physical light
scalar fields. This in spite of the seemingly good agreement between
experiment and radiative corrections computed in the framework of the
minimal Standard Model (see \cite{chanowitz} however).

The four dimensional operators contributing to his effective theory were
already analyzed in previous chapters. Here we plan to investigate some
features that physics encoded in these operators introduce in the production
of top (or anti-top) quarks at the LHC. In Chapter \ref{mattersectorchapter}
we have discussed, among other things, some phenomenological consequences of
this effective Lagrangian in the neutral current sector. Here we have chosen
single top production because we are interested in probing the charged
current sector.

\begin{figure}[!hbp]
\begin{center}
\includegraphics[width=8cm]{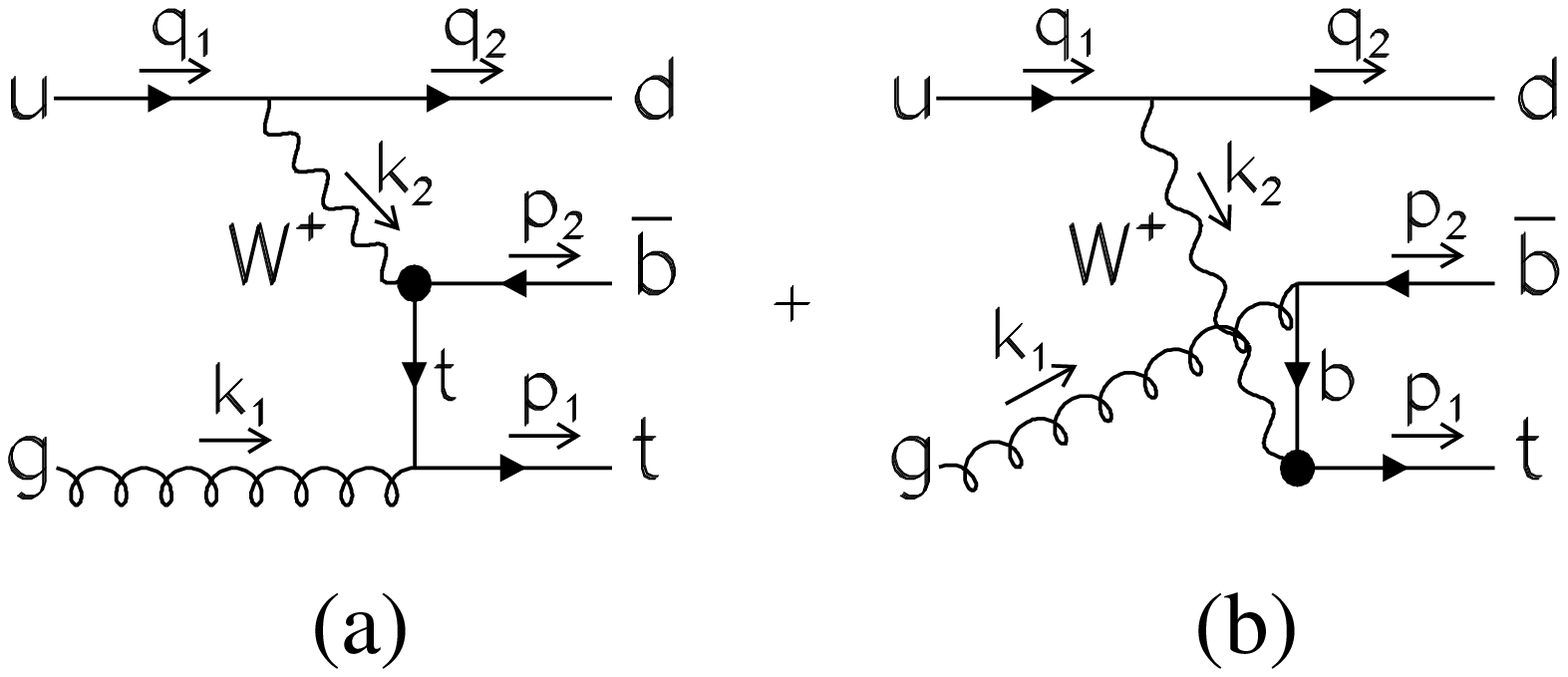}
\end{center}
\caption{Feynman diagrams contributing single top production subprocess. In
this case we have a $d$ as spectator quark}
\label{u+gt+b-d+tot}
\end{figure}

\begin{figure}[!hbp]
\begin{center}
\includegraphics[width=7.5cm]{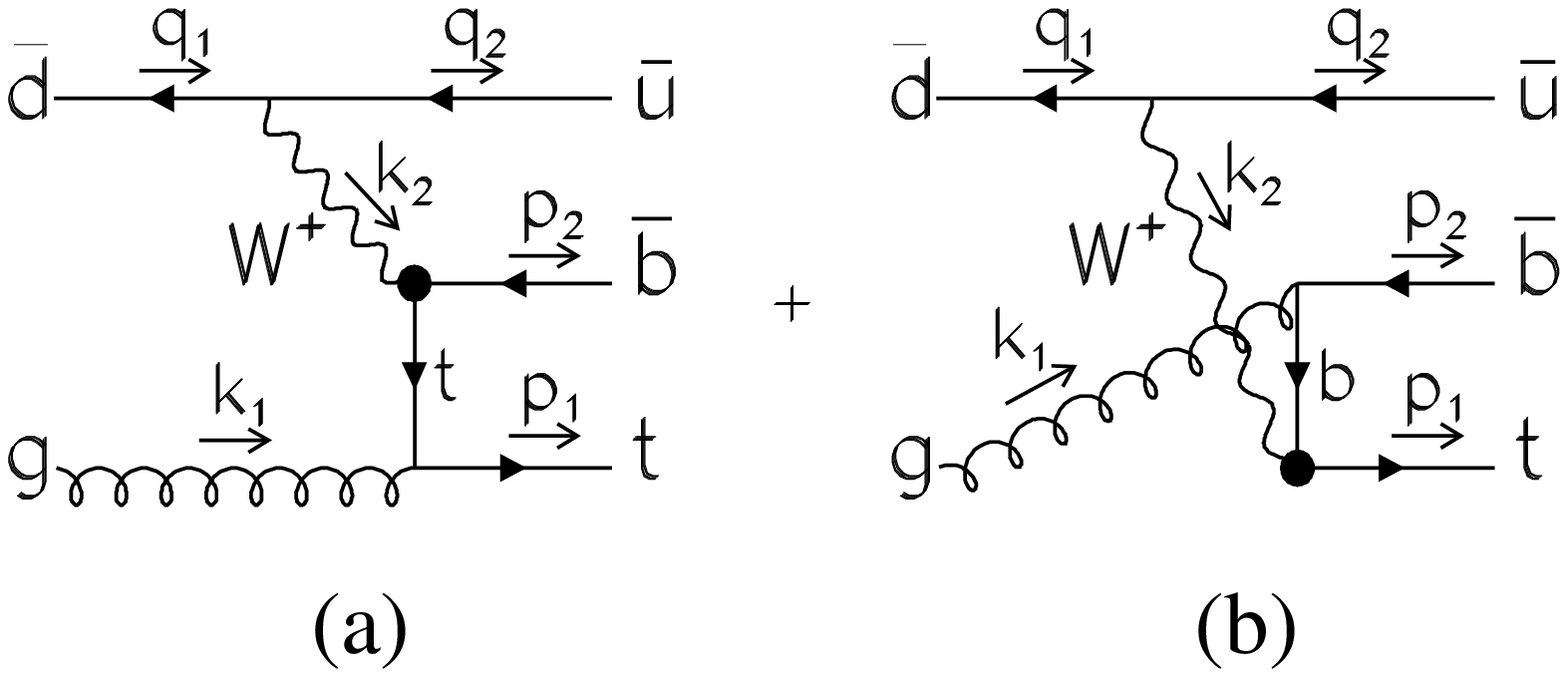}
\end{center}
\caption{Feynman diagrams contributing single top production subprocess. In
this case we have a $\bar{u}$ as spectator quark}
\label{d-gt+b-u-tot}
\end{figure}

In the electroweak sector tree level contribution to neutral and charged
currents can be written as
\begin{equation}
-\frac{e}{4c_{W}s_{W}}\bar{f}\gamma ^{\mu }\left( \kappa _{L}^{NC}L+\kappa
_{R}^{NC}R\right) Z_{\mu }f-\frac{e}{s_{W}}\bar{f}\gamma ^{\mu }\left(
\kappa _{L}^{CC}L+\kappa _{R}^{CC}R\right) \frac{\tau ^{-}}{2}W_{\mu
}^{+}f+h.c.
\end{equation}
The dominant process at LHC energies that tests $\kappa _{L}^{CC}$ and $%
\kappa _{R}^{CC}$ in a direct way (i.e. not through top decay) is single top
production in the so-called $W-$gluon fusion channel. The electroweak
subprocesses corresponding to this channel are depicted in Figs. (\ref
{u+gt+b-d+tot}) and (\ref{d-gt+b-u-tot}), where light $u$-type quarks or $%
\bar{d}$-type antiquarks are extracted from the protons, respectively.
Besides this dominant channel (250 pb at LHC \cite{SSW}) single tops are
also produced through the process where the $W^{+}$ boson interacts with a $%
b $-quark extracted from the sea of the proton (50 pb) and in the
quark-quark fusion process (10 pb). This last process (\textit{s}-channel)
will be analyzed in the last chapter with top decay taken also into account.

In a proton-proton collision a bottom-anti-top pair is also produced through
analogous subprocesses. The analysis of such anti-top production processes
is similar to the top ones and the corresponding cross sections can be
easily derived doing the appropriate changes (see appendix \ref{TchannelappA}%
).

In this chapter we will analyze the sensitivity of different LHC observables
to the magnitude of charged current couplings $\kappa _{L}^{CC}$ and $\kappa
_{R}^{CC}$ through single top production in the $W$-gluon fusion channel. In
section \ref{firstlook} we show how the measurement of top spin plays a
central role in the isolation of observables sensitive to left and right
coupling variations. In this Chapter we do not analyze in detail top decay
but we perform a theoretical approach at the issue of measuring top spin
from its decay products. In this regard we show in section \ref
{differentialc} that the presence of effective right-handed couplings
implies that the top is not in a pure spin state, which is a fact that was
overlooked in earlier works in the literature.

Moreover, in section \ref{decay} we show that there is a unique spin basis
allowing the calculation of top production an decay convoluting the top
decay products angular distribution with the polarized top differential
cross section. In the next chapter we show explicitly this basis both for
the \textit{t}- and \textit{s}- channels.

\section{Effective couplings and observables}

Including family mixing and, possibly, $CP$ violation, the complete set of
dimension four effective operators which may contribute to the top effective
couplings and are relevant for the present discussion is the set given by
Eq. (\ref{effec}) \cite{ABCH,nomix,EspMan}. In addition, as we have seen in
Chapter \ref{cpviolationandmixing} we have the `universal' terms given by
Eq. (\ref{lm}) which are present in the Standard Model. In Eq. (\ref{lm}) we
allow for general couplings $X_{L}$, $X_{R}^{(u,d)}$; in the Standard Model
these couplings can be renormalized away via a change of basis, but as we
have seen Chapter \ref{cpviolationandmixing} in more general theories they
leave traces in other operators not present in the Standard Model \cite
{EspMan}.

In Chapter \ref{cpviolationandmixing} we have also seen that when we
diagonalize the mass matrix present in Eq. (\ref{lm}) via a redefinition of
the matter fields $\left( \mathrm{f}\rightarrow f\right) $ we change also
the structure of operators (\ref{effec}). Taking that into account, the
contribution to the different gauge boson-fermion-fermion vertices is as
follows
\begin{eqnarray}
\mathcal{L}_{bff} &=&-g_{s}\bar{f}\gamma ^{\mu }\left( a_{L}L+a_{R}R\right)
\frac{\mathbf{\lambda }}{2}\mathbf{\cdot G}_{\mu }f,  \notag \\
&&-e\bar{f}\gamma ^{\mu }\left( b_{L}L+b_{R}R\right) A_{\mu }f,  \notag \\
&&-\frac{e}{2c_{W}s_{W}}\bar{f}\gamma ^{\mu }\left[ \left( c_{L}^{u}\tau
^{u}+c_{L}^{d}\tau ^{d}\right) L+\left( c_{R}^{u}\tau ^{u}+c_{R}^{d}\tau
^{d}\right) R\right] Z_{\mu }f  \notag \\
&&-\frac{e}{s_{W}}\bar{f}\gamma ^{\mu }\left[ \left( d_{L}L+d_{R}R\right)
\frac{\tau ^{-}}{2}W_{\mu }^{+}+\left( d_{L}^{\dagger }L+d_{R}^{\dagger
}R\right) \frac{\tau ^{+}}{2}W_{\mu }^{-}\right] f,  \label{vert2}
\end{eqnarray}
where $\tau ^{u}$ and $\tau ^{d}$ are the up and down projectors and $f$
represents the matter fields in the physical, diagonal basis. It was shown
in Chapter \ref{cpviolationandmixing} that once the all the renormalization
(vertex, CKM elements, wave-function) counterterms are taken into account we
obtain $a_{L,R}=1$, $b_{L,R}=Q$; i.e. we have no contribution from the
effective operators to the vertices of the gluon and photon. For the $Z$
couplings we get instead
\begin{eqnarray}
c_{L}^{u} &=&1-2Qs_{W}^{2}-\hat{M}_{L}^{1}-\hat{M}_{L}^{1\dagger }+\hat{M}%
_{L}^{2\dagger }+\hat{M}_{L}^{2}+\hat{M}_{L}^{3}+\hat{M}_{L}^{3\dagger },
\notag \\
c_{L}^{d} &=&-1-2Qs_{W}^{2}+K^{\dagger }\left( \hat{M}_{L}^{1}+\hat{M}%
_{L}^{1\dagger }+\hat{M}_{L}^{2\dagger }+\hat{M}_{L}^{2}-\hat{M}_{L}^{3}-%
\hat{M}_{L}^{3\dagger }\right) K,  \notag \\
c_{R}^{u} &=&-2s_{W}^{2}Q+\hat{M}_{R}^{1}+\hat{M}_{R}^{1\dagger }+\hat{M}%
_{R}^{2}+\hat{M}_{R}^{2\dagger }+\hat{M}_{R}^{3}+\hat{M}_{R}^{3\dagger },
\notag \\
c_{R}^{d} &=&-2s_{W}^{2}Q+\hat{M}_{R}^{1}+\hat{M}_{R}^{1\dagger }-\hat{M}%
_{R}^{2}-\hat{M}_{R}^{2\dagger }-\hat{M}_{R}^{3}-\hat{M}_{R}^{3\dagger },
\label{Zcoupl2}
\end{eqnarray}
where $K$ is the CKM matrix, and the matrices $\hat{M}_{L}$'s and $\hat{M}%
_{R}$'s are redefined matrices according to the results of Chapter \ref
{cpviolationandmixing} (the exact relation of these matrices to the $%
M_{L,R}^{i}$ of Eqs. (\ref{effec}) has no relevance for the present
discussion). Finally for the charged couplings we have
\begin{eqnarray}
d_{L} &=&K+\left( -\hat{M}_{L}^{1}-\hat{M}_{L}^{1\dagger }+\hat{M}_{L}^{2}-%
\hat{M}_{L}^{2\dagger }-\hat{M}_{L}^{3}-\hat{M}_{L}^{3\dagger }+\hat{M}%
_{L}^{4}-\hat{M}_{L}^{4\dagger }\right) K,  \notag \\
d_{R} &=&\hat{M}_{R}^{1}+\hat{M}_{R}^{1\dagger }+\hat{M}_{R}^{2}-\hat{M}%
_{R}^{2\dagger }-\hat{M}_{R}^{3}-\hat{M}_{R}^{3\dagger }.  \label{vertices2}
\end{eqnarray}

Since the set of operators (\ref{effec}) is the most general one allowed by
general requirements of gauge invariance, locality and hermiticity; it is
clear that radiative corrections, when expanded in powers of $p^{2}$, can be
incorporated into them. In fact, such an approach has proven to be very
fruitful in the past. Once everything is included we are allowed to identify
the couplings $d_{L,R}$ with $\kappa _{LR}^{CC}$. In this work we shall be
concerned with the bounds that the LHC experiments will be able to set on
the couplings $\kappa _{LR}^{CC}$, more specifically on the entries $tj$ of
these matrices (those involving the top). In the rest of the chapter we do
not consider mixing and we consider non-tree level and new physics
contributions only on the $tb$ effective couplings, therefore in the
numerical simulations we have taken
\begin{eqnarray*}
d_{L} &=&diag\ (K_{ud},K_{cs},g_{L}), \\
d_{R} &=&diag\ (0,0,g_{R}).
\end{eqnarray*}
When we talk along this chapter about the results for the Standard Model at
tree level we mean $g_{L}=1$, and $g_{R}=0$. However, even though numerical
results are presented considering only the $tb$ entry ($g_{L}$ and $g_{R}$),
since flavor indices and masses are kept all along in the analytical
expressions (see appendix \ref{TchannelappA}), the appropriate changes to
include other entries are immediate.

As we have seen in Chapter \ref{mattersectorchapter} the effective couplings
of the neutral sector (\ref{Zcoupl2}) can be determined from the $%
Z\rightarrow f~\bar{f}$ vertex\footnote{%
A 3 $\sigma $ discrepancy with respect to the Standard Model results, mostly
due to the right-handed coupling, remains in the $Z$ couplings of the $b$
quark to this date.} \cite{nomix}, but at present not much is known from the
$tb$ effective coupling. This is perhaps best evidenced by the fact that the
current experimental results for the (left-handed) $K_{tb}$ matrix element
give \cite{PDG}
\begin{equation}
\frac{|K_{tb}|^{2}}{|K_{td}|^{2}+|K_{ts}|^{2}+|K_{tb}|^{2}}=0.99\pm 0.29.
\end{equation}
In the Standard Model this matrix element is expected to be close to 1. It
should be emphasized that these are the `measured' or `effective' values of
the CKM matrix elements, and that they do not necessarily correspond, even
in the Standard Model, to the entries of a unitary matrix on account of the
presence of radiative corrections. These deviations with respect to unitary
are expected to be small ---at the few per cent level at most--- unless new
physics is present. At the Tevatron the left-handed couplings are expected
to be eventually measured with a 5\% accuracy \cite{TEVA}. The present work
is a contribution to such an analysis in the case of the LHC experiments.

As far as experimental bounds for the right handed effective couplings is
concerned, the more stringent ones come at present from the measurements on
the $b\rightarrow s\gamma $ decay at CLEO \cite{cleo}. Due to a $m_{t}/m_{b}$
enhancement of the chirality flipping contribution, a particular combination
of mixing angles and $\kappa _{R}^{CC}$ can be found. The authors of \cite
{LPY} reach the conclusion that $|\mathrm{Re}(\kappa _{R}^{CC})|\leq
0.4\times 10^{-2}$. However, considering $\kappa _{R}^{CC}$ as a matrix in
generation space, this bound only constraints the $tb$ element. Other
effective couplings involving the top remain virtually unrestricted from the
data. The previous bound on the right-handed coupling is a very stringent
one. It is pretty obvious that the LHC will not be able to compete with such
a bound. Yet, the measurement will be a direct one, not through loop
corrections. Equally important is that it will yield information on the $td$
and $ts$ elements too, by just replacing the $\bar{b}$ quark in Figs. (\ref
{u+gt+b-d+tot}) and (\ref{d-gt+b-u-tot}) by a $\bar{d}$ or a $\bar{s}$
respectively.

Now we shall proceed to analyze the bounds that single top production at the
LHC can set on the effective couplings. This combined with the data from $Z$
physics will allow an estimation of the six effective couplings (\ref
{Zcoupl2}-\ref{vertices2}) in the matter sector of the effective electroweak
Lagrangian. We will, in the present work limit ourselves to the
consideration of the cross-sections for production of polarized top quarks.
We shall not consider at this stage the potential of measuring top decays
angular distributions in order to establish relevant bounds on the effective
electroweak couplings. This issue merits a more detailed analysis, including
the possibility of detecting $CP$ violation \cite{DambESp}.

\section{The cross section in the \textit{t}-channel}

\label{cross}In order to calculate the cross section $\sigma $ of the
process $pp\rightarrow t\bar{b}$ we have used the CTEQ4 set of structure
functions \cite{CTEQ4} to determine the probability of extracting a parton
with a given fraction of momenta from the proton. Hence we write
schematically
\begin{equation}
\sigma =\sum_{q}\int_{0}^{1}\int_{0}^{1}f_{g}(y)f_{q}(x)\hat{\sigma}%
(xP_{1},yP_{2})dxdy,  \label{pdfs}
\end{equation}
where $f_{q}$ denote the parton distribution function (PDF) corresponding to
the partonic quarks and antiquarks and $f_{g}$ indicate the PDF
corresponding to the gluon. In Eq.(\ref{pdfs}) we have set the light quark
and gluon momenta to $xP_{1}$ and $yP_{1}$, respectively. ($P_{1}$ and $%
P_{2} $ are the four-momenta of the two colliding protons.) The
approximation thus involves neglecting the transverse momenta of the
incoming partons; the transverse fluctuations are integrated over by doing
the appropriate integrals over $k_{T}$. We have then proceeded as follows.
We have multiplied the parton distribution function of a gluon of a given
momenta from the first proton by the sum of parton distribution functions
for obtaining a $u$ type quark from the second proton. This result is then
multiplied by the cross sections of the subprocesses of Fig. (\ref
{u+gt+b-d+tot}). We perform also the analogous process with the $\bar{d}$
type anti-quarks of Fig. (\ref{d-gt+b-u-tot}). At the end, these two partial
results are add up to obtain the total $pp\rightarrow t\bar{b}$ cross
section.

Typically the top quark decays weakly well before strong interactions become
relevant, we can in principle measure its polarization state with virtually
no contamination of strong interactions (see e.g. \cite{parke} for
discussions this point and section \ref{decay}). For this reason we have
considered polarized cross sections and provide general formulas for the
production of polarized tops or anti-tops. To this end one needs to
introduce the spin projector
\begin{equation*}
\left( \frac{1+\gamma _{5}\not{n}}{2}\right) ,
\end{equation*}
with
\begin{eqnarray}
n^{\mu } &=&\frac{1}{\sqrt{\left( p_{1}^{0}\right) ^{2}-\left( \vec{p}%
_{1}\cdot \hat{n}\right) ^{2}}}\left( \vec{p}_{1}\cdot \hat{n},p_{1}^{0}\hat{%
n}\right) ,  \label{spinfour} \\
\hat{n}^{2} &=&1,\qquad n^{2}=-1,  \notag
\end{eqnarray}
as the polarization projector for a particle or anti-particle of momentum $%
p_{1}$ with spin in the $\hat{n}$ direction. The calculation of the
subprocesses cross sections have been performed for tops and anti-tops
polarized in an arbitrary direction $\hat{n}$. Later we have analyzed
numerically different spin frames defined as follows

\begin{itemize}
\item  Lab helicity frame: the polarization vector is taken in the direction
of the three momentum of the top or anti-top (right helicity) or in the
opposite direction (left helicity).

\item  Lab spectator frame: the polarization vector is taken in the
direction of the three momentum of the spectator quark jet or in the
opposite direction. The spectator quark is the $d$-type quark in Fig. (\ref
{u+gt+b-d+tot}) or the $\bar{u}$-type quark in Fig. (\ref{d-gt+b-u-tot}).

\item  Rest spectator frame: like in the Lab spectator frame we choose the
spectator jet to define the polarization of the top or anti-top. Here,
however, we define $\hat{n}$ as $\pm $ the direction of the three momentum
of the spectator quark in the top or anti-top rest frame (given by a pure
boost transformation $\Lambda $ of the lab frame). Then we have $%
n_{r}=\left( 0,\hat{n}\right) $ in that frame and $n=\Lambda ^{-1}n_{r}$
back to the lab frame.
\end{itemize}

The calculation of the subprocess polarized cross-section we present is
completely analytical from beginning to end and the results are given in
appendix.\ref{TchannelappA} Both the kinematics and the polarization vector
of the top (or anti-top) are completely general. Since the calculation is of
a certain complexity a number of checks have been done to ensure that no
mistakes have been made. The integrated cross section agrees well with the
results in \cite{SSW} when the same cuts, scale, etc. are used. The mass of
the top is obviously kept, but so is the bottom mass. The latter in fact
turns out to be more relevant than expected as we shall see in a moment. As
we have already discussed, the production of flavors other than ${\bar{b}}$
in association with the top can be easily derived from our results.

In single top production a distinction is often made between $2\rightarrow 2$
and $2\rightarrow 3$ processes. The latter corresponds, in fact, to the
processes we have been discussing, the ones represented in Fig. (\ref
{u+gt+b-d+tot}), in which a gluon from the sea splits into a $b$ $\bar{b}$
pair. In the $2\rightarrow 2$ process the $b$ quark is assumed to be
extracted from the sea of the proton, and both $b$ and $\bar{b}$ are
collinear. Of course since the proton has no net $b$ content, a $\bar{b}$
quark must be present somewhere in the final state and the distinction
between the two processes is purely kinematical. As is well known, when
calculating the total cross section for single top production a logarithmic
mass singularity \cite{SSW} appears in the total cross section due to the
collinear regime where the $b$ quark (and the $\bar{b}$) quark have $%
k_{T}\rightarrow 0$. This kinematic singularity is actually regulated by the
mass of the bottom; it appears to all orders in perturbation theory and a
proper treatment of this singularity requires the use of the
Altarelli-Parisi equations and its resumation into a $b$ parton distribution
function. While the evolution of the parton distribution functions is
governed by perturbation theory, their initial values are not and some
assumptions are unavoidable. Clearly an appropriate cut in $p_{T}$ should
allow us to retain the perturbative regime of the $2\rightarrow 3$ process,
while suppressing the $2\rightarrow 2$ one.

Two experimental approaches can be used at this point. One ---advocated by
Willenbrock and coworkers \cite{SSW} is to focus on the low $p_{T}$ regime.
The idea is to minimize the contribution of the $t,\bar{t}$ background,
whose characteristic angular distributions are more central. Then one is
actually interested in processes where one does not see the $\bar{b}$ (resp.
$b$) quark which is produced in association with the $t$ quark (resp. $\bar{t%
}$), and accordingly sets an upper cut on the $p_{T}$ of the $\bar{b}$.
Clearly one then has to take into account the $2\rightarrow 2$ process and,
in particular, one must pay attention not to double count the low $p_{T}$
region (for the $\bar{b}$ (or $b$) quark) of the $2\rightarrow 3$ process,
which is already included via a $b$ PDF and has to be subtracted. This
strategy has some risks. First of all, the separation between the $%
2\rightarrow 3$ and $2\rightarrow 2$ is not a clear cut one. The separation
takes place in a region where the cross section is rapidly varying so the
results do depend to some extend on the way the separation is done. Also as
we just said relies on some initial condition for the $b$ PDF at some
initial scale (for instance at $\mu =m_{b}$. Moreover, this strategy does
not completely avoid the background originated in $\bar{t}t$ production
either; for instance when in the decaying $\bar{t}\rightarrow W^{-}\bar{b}%
\rightarrow \bar{u}d\bar{b}$ the $\bar{b}$ is missed along with the $\bar{u}$%
-type anti-quark in which case the $d$-type quark is taken as the spectator
or when the $\bar{b}$ is missed along with the $d$-type quark in which case
the $\bar{u}$-type anti-quark is taken as the spectator.

On the other hand, measuring the $\bar{b}$ (or $b$ for anti-top production)
momenta will allow a better kinematic reconstruction of the individual
processes. This should allow for a separation from the dominant mechanism of
top production through gluon fusion. Setting a sufficiently high upper cut
for the jet energy and a good jet separation might be sufficient to avoid
contamination from $t,\bar{t}$ when one hadronic jet is missed. Finally, the
spin structure of the top is completely different in both cases due to the
chiral couplings in electroweak production. Therefore, according to this
philosophy we have implemented a lower cut of 30 GeV in the transversal
momentum of the $\bar{b}$ (resp. $b$) in top (resp. anti-top) production.

We do not really want to make strong claims as to which strategy should
prove more efficient eventually. Many different ingredients have to be taken
into account. Just to mention one more: the results of our analysis show
that the sensitivity to the right handed effective coupling is not very big
and that the (subdominant) \textit{s}-channel process may actually be more
adequate for this purpose. Yet, this is again more central, so one will need
to consider the \textit{t}-channel process for largish values of $p_{T}$
anyway.

\section{A first look at the results}

\label{firstlook}We shall now present the results of our analysis. To
calculate the total event production corresponding to different observables
we have used the integrating Monte Carlo program VEGAS \cite{vegas}. We
present results after one year (defined as 10$^{7}$ seg.) run at full
luminosity in one detector (100 $\mathrm{fb}^{-1}$ at LHC).

The total contribution to the electroweak vertices $g_{L}$, $g_{R}$ has two
sources: the effective operators parametrizing new physics, and the
contribution from the universal radiative corrections. In the standard
model, neglecting mixing, for example, we have a tree level contribution to
the $\bar{t}W_{\mu }^{+}b$ vertex given by $-\frac{i}{\sqrt{2}}\gamma _{\mu
}gK_{tb}L$. Radiative corrections (universal and $M_{H}$ dependent) modify $%
g_{L}$ and generate a non zero $g_{R}$. These radiative corrections depend
weakly on the energy of the process and thus in a first approximation we can
take them as constant. Our purpose is to estimate the dependence of
different LHC observables on these total effective couplings and how the
experimental results can be used to set bounds on them. Assuming that the
radiative corrections are known, this implies in turn a bound on the
coefficients of the effective electroweak Lagrangian.

\begin{figure}[!hbp]
\begin{center}
\includegraphics[width=\figwidth]{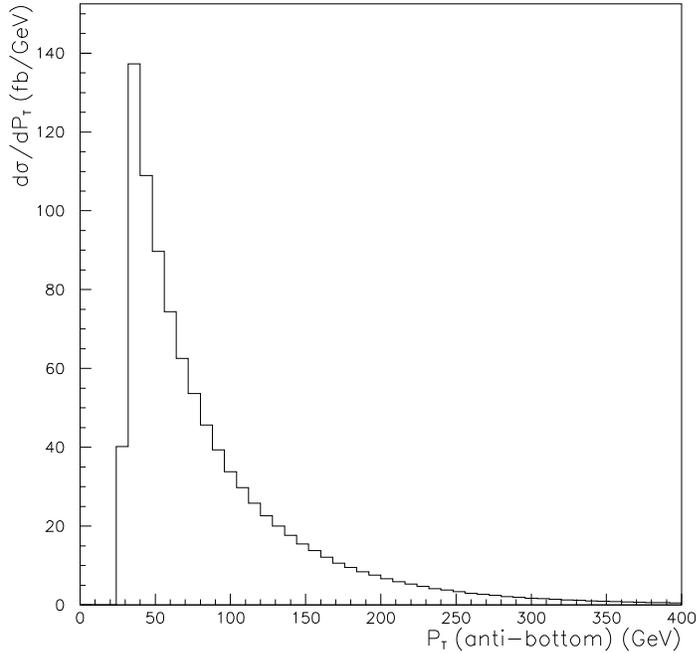}
\end{center}
\par
\caption{Anti-bottom transversal momentum distribution corresponding to
unpolarized single top production at the LHC. The calculation was performed
at the tree level in the Standard Model. Note the 30 GeV. cut implemented to
avoid large logs due to the massless singularity in the total cross section.
In this plot $\protect\mu ^{2}=\hat{s}=\left( q_{1}+q_{2}\right) ^{2}$ too.}
\label{pbottrans_lab_unpol_gl=1_gr=0}
\end{figure}

Let us start by discussing the experimental cuts. Due to geometrical
detector constraints we cut off very low angles for the outgoing particles.
The top, anti-bottom, and spectator quark have to come out with an angle in
between 10 and 170 degrees. These angular cuts correspond to a cut in
pseudorapidity $\left| \eta \right| <2.44$. In order to be able to detect
the three jets corresponding to the outgoing particles we implements
isolation cuts of 20 degrees between each other.

As already discussed we use a lower cut of 30 GeV in the $\bar{b}$ jet. This
reduces the cross section to less than one third of its total value, since
typically the $\bar{b}$ quark comes out in the same direction as the
incoming gluon and a large fraction of them do not pass the cut (see Fig. (%
\ref{pbottrans_lab_unpol_gl=1_gr=0})). Similarly, $p_{T}>20$ GeV cuts are
set for the top and spectator quark jets. These cuts guarantee the validity
of perturbation theory and will serve to separate from the overwhelming
background of low $p_{T}$ physics. These values come as a compromise to
preserve a good signal, while suppressing unwanted contributions. They are
similar, but not identical to the ones used in \cite{SSW} and \cite{mandp}.
To summarize the allowed regions are
\begin{eqnarray}
\mathrm{detector~geometry~cuts} &:&10^{o}\leq \theta _{i}\leq 170^{o},\quad
i=t,\bar{b},q_{s},  \notag \\
\mathrm{isolation~cuts} &:&20^{o}\leq \theta _{ij},\quad i,j=t,\bar{b},q_{s},
\notag  \label{cuts} \\
\mathrm{theoretical~cuts} &:&20~\mathrm{GeV}\leq p_{1}^{T},\quad 20~\mathrm{%
GeV}\leq q_{2}^{T},\quad 30~\mathrm{GeV}\leq p_{2}^{T},
\end{eqnarray}
where $\theta _{t}$, $\theta _{\bar{b}},$ $\theta _{q_{s}}$ are the polar
angles with respect to the beam line of the top, anti-bottom and spectator
quark respectively; $\theta _{t\bar{b}}$, $\theta _{tq_{s}},$ $\theta _{\bar{%
b}q_{s}}$ are the angles between top and anti-bottom, top and spectator, and
anti-bottom and spectator, respectively. The momenta conventions are given
in Figs. (\ref{u+gt+b-d+tot}) and (\ref{d-gt+b-u-tot}).
\begin{figure}[!hbp]
\begin{center}
\includegraphics[width=\figwidth]{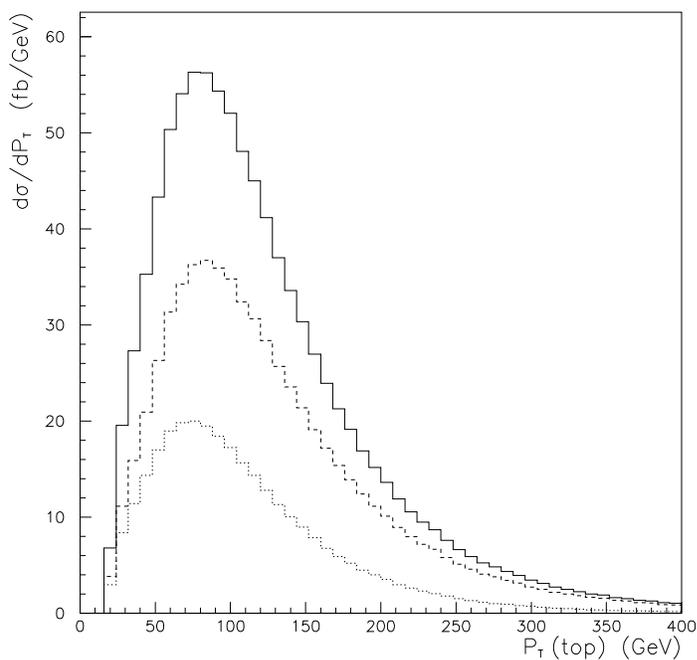}
\end{center}
\par
\caption{Top transversal momentum distribution corresponding to polarized
single top production at the LHC in the LAB system. The solid line
corresponds to unpolarized top production and the dashed (dotted) line
corresponds to tops of negative (positive) helicity. The subprocesses
contributing to these histograms have been calculated at tree level in the
electroweak theory. The cuts are described in the text. The degree of
polarization in this spin basis and reference frame is only 69\% . The QCD
scale is taken to be $\protect\mu ^{2}=\hat{s}=\left( q_{1}+q_{2}\right)
^{2} $. }
\label{ptoptrans_lab_lr_gl=1_gr=0}
\end{figure}

Numerically, the dominant contribution to the process comes from the diagram
where a $b$ quark is exchanged in the $t$ channel, but a large amount of
cancellation takes place with the crossed interference term with the diagram
with a top quark in the $t$ channel. The smallest contribution (but
obviously non-negligible) corresponds to this last diagram. It is then easy
to see, given the relative smallness of the $b$ mass, why the process is so
much forward.

Undoubtedly the largest theoretical uncertainty in the whole calculation is
the choice of a scale for $\alpha _{s}$ and the PDF's. We perform a leading
order calculation in QCD and the scale dependence is large. We have made two
different choices. We present some results with the scale $p_{T}^{cut}$ used
in $\alpha _{s}$ and the gluon PDF, while the virtuality of the $W$ boson is
used as scale for the PDF of the light quarks in the proton. When we use
these scales and compute, for instance, the total cross section above a cut
of $p_{T}=20$ GeV in the $\bar{b}$ momentum, we get an excellent agreement
with the calculations in \cite{SSW}. Most of our results are however
presented with a common scale $\mu ^{2}=\hat{s}$, $\hat{s}$ being the
center-of-mass energy squared of the $qg$ subprocess. The total cross
section above the cut is then roughly speaking two thirds of the previous
one, but no substantial change in the distributions takes place. It remains
to be seen which one is the correct choice.

From our Monte Carlo simulation for single top production at the LHC after 1
year of full luminosity and with the cuts given above we obtain the total
number of events. This number depends on the value of the effective
couplings and on the top polarization vector $n$ given in the frames defined
in section \ref{cross} . If we call $N\left( g_{L},g_{R},\hat{n}%
,(frame)\right) $ to this quantity, we obtain the following results
\begin{eqnarray}
N\left( g_{L},g_{R},\hat{n}=\pm \frac{\vec{p}_{1}}{\left| \vec{p}_{1}\right|
},(lab)\right) &=&g_{L}^{2}\times \left( 3.73\mp 1.31\right) \times
10^{5}+g_{R}^{2}\times \left( 3.54\pm .97\right) \times 10^{5}  \notag \\
&&+g_{L}g_{R}\times \left( -.237\mp .0283\right) \times 10^{5},  \notag \\
N\left( g_{L},g_{R},\hat{n}=\pm \frac{\vec{q}_{2}}{\left| \vec{q}_{2}\right|
},(lab)\right) &=&g_{L}^{2}\times \left( 3.73\pm 2.22\right) \times
10^{5}+g_{R}^{2}\times \left( 3.54\mp 2.12\right) \times 10^{5}  \notag \\
&&+g_{L}g_{R}\times \left( -.237\mp 0.001\right) \times 10^{5},  \notag \\
N\left( g_{L},g_{R},\hat{n}=\pm \frac{\vec{q}_{2}}{\left| \vec{q}_{2}\right|
},(rest)\right) &=&g_{L}^{2}\times \left( 3.73\pm 2.49\right) \times
10^{5}+g_{R}^{2}\times \left( 3.54\mp 2.15\right) \times 10^{5}  \notag \\
&&+g_{L}g_{R}\times \left( -.237\mp .0180\right) \times 10^{5},
\label{results}
\end{eqnarray}
where we have omitted the $O\left( \sqrt{N}\right) $ statistical errors and
we have neglected possible $CP$ phases ($g_{L}$ and $g_{R}$ real). One can
observe from the simulations that the production of negative helicity (left)
tops represents the 69\% of the total single top production (see Fig. (\ref
{ptoptrans_lab_lr_gl=1_gr=0})), this predominance of left tops in the tree
level electroweak approximation is expected due to the suppression at high
energies of right-handed tops because of the zero right coupling in the
charged current sector. In fact the production of right-handed tops would be
zero were it not for the chirality flip, due to the top mass, in the \textit{%
t}-channel. Of course the name `left' and `right' are a bit misleading; we
really mean negative and positive helicity states.

\begin{figure}[!hbp]
\begin{center}
\includegraphics[width=12cm]{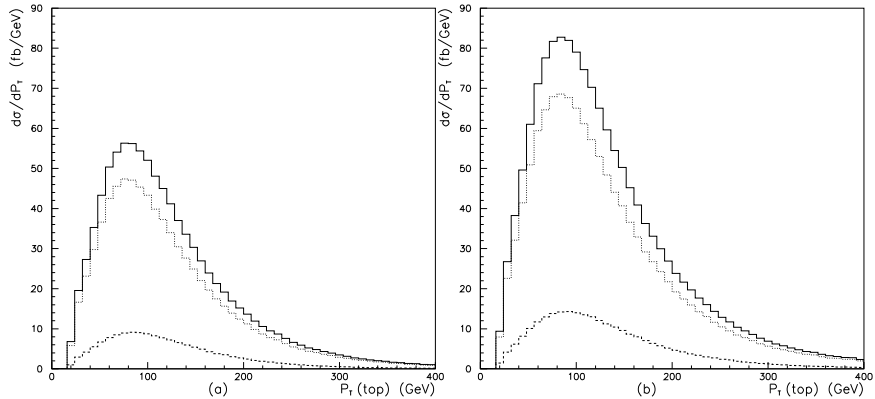}
\end{center}
\par
\caption{Top transversal momentum distribution corresponding to polarized
single top production at the LHC. The solid line corresponds to unpolarized
top production and the dashed (dotted) line corresponds to tops polarized in
the spectator jet negative (positive) direction in the top rest frame. In
(a) the QCD scale is taken $\protect\mu ^{2}=\hat{s}=\left(
q_{1}+q_{2}\right) ^{2}$ and in (b) $\protect\mu =p_{cut}^{T(bot)}=30$ GeV.
The subprocesses contributing to these histograms have been calculated at
tree level in the electroweak theory. With our set of cuts, the polarization
is in both cases 84 \% }
\label{ptoptrans_rep_+-q2_gl=1_gr=0}
\end{figure}

Chirality states cannot be used, because the production is peaked in the 200
to 400 GeV region for the energy of the top and the mass cannot be
neglected. The results for the production of tops polarized in the spectator
jet direction in the top rest frame can be summarized in Fig. (\ref
{ptoptrans_rep_+-q2_gl=1_gr=0}).
\begin{figure}[!hbp]
\begin{center}
\includegraphics[width=\figwidth]{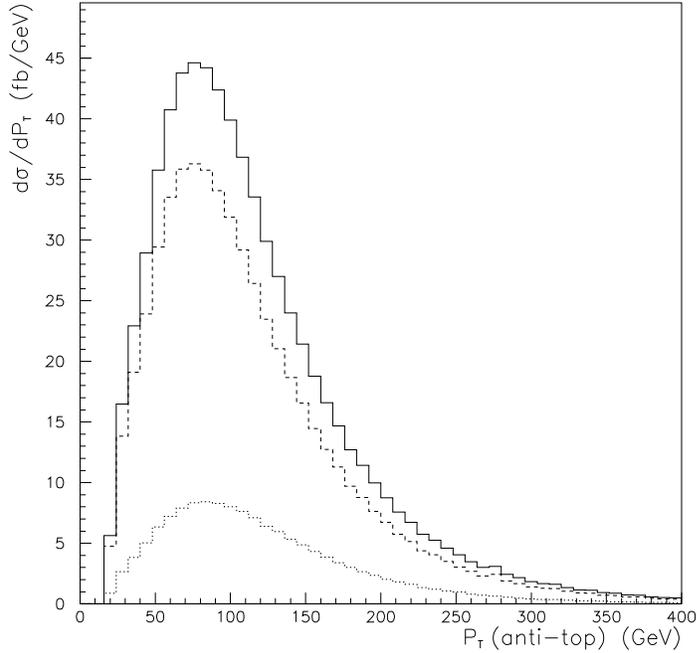}
\end{center}
\par
\caption{Anti-top transversal momentum distribution corresponding to
polarized single anti-top production at the LHC. The solid line corresponds
to unpolarized anti-top production and the dashed (dotted) line corresponds
to anti-tops polarized in the spectator jet negative (positive) direction in
the top rest frame. The subprocesses contributing to these histograms have
been calculated at tree level in the electroweak theory, using the same cuts
and conventions as in the previous figures.}
\label{patoptrans_rep_+-q2_gl=1_gr=0}
\end{figure}

We have also calculated single anti-top production obtaining a pattern
similar to that of single top production but suppressed by an approximately
75\% factor. This can be observed for example in Fig. (\ref
{patoptrans_rep_+-q2_gl=1_gr=0}). This suppression is generated by the
parton distribution functions corresponding to negatively charged quarks
that are smaller than the ones corresponding to positively charged quarks.
Because of that the conclusions for anti-top production are practically the
same as the ones for top production taking into account such suppression and
that, because of the transformations (\ref{change}) (see appendix \ref
{TchannelappA}), passing from top to anti-top is equivalent to changing the
spin direction.
\begin{figure}[!hbp]
\begin{center}
\includegraphics[width=\figwidth]{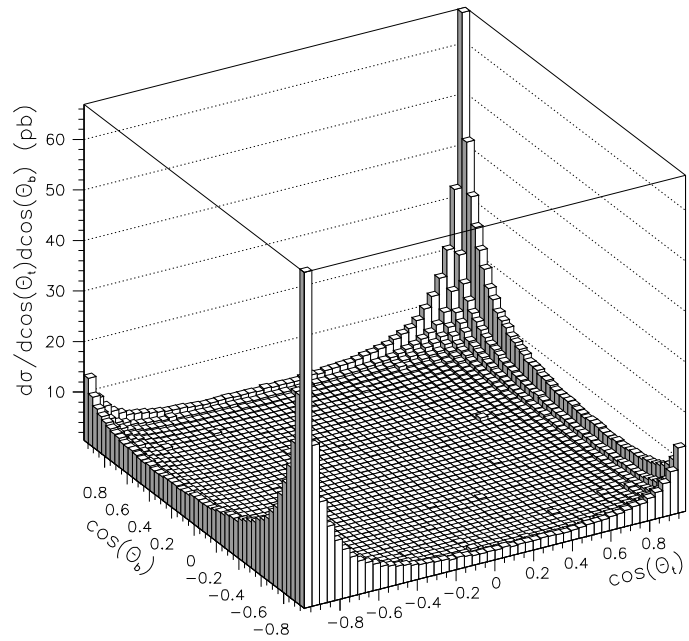}
\end{center}
\par
\caption{Distribution of the cosines of the polar angles of the top and
anti-bottom with respect to the beam line. The plot corresponds to
unpolarized single top production at the LHC. The calculation was performed
at the tree level in Standard Model with $\protect\mu ^{2}=\hat{s}=\left(
q_{1}+q_{2}\right) ^{2}$.}
\label{costcosb_lab_unpol_gl=1_gr=0}
\end{figure}
\begin{figure}[!hbp]
\begin{center}
\includegraphics[width=\figwidth]{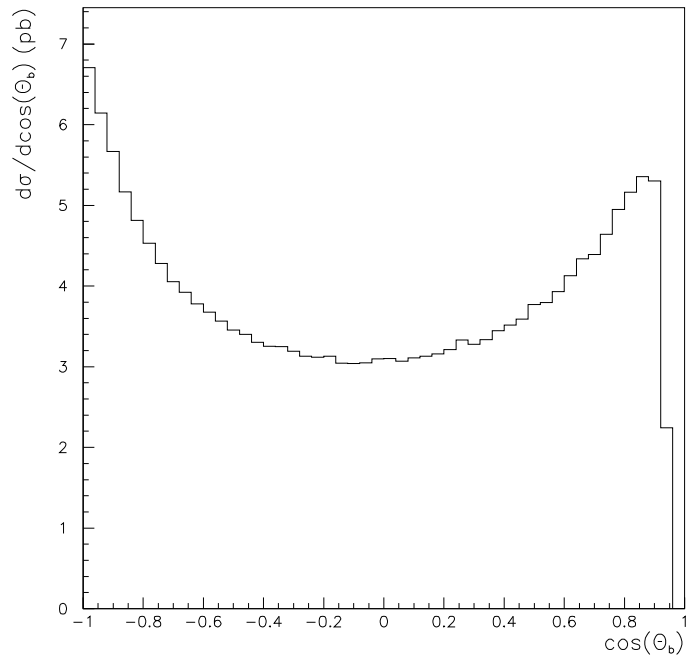}
\end{center}
\par
\caption{Distribution of the cosine of the angle between top and anti-bottom
corresponding to unpolarized single top production at the LHC. The
calculation was performed at the tree level in Standard Model with $\protect%
\mu ^{2}=\hat{s}=\left( q_{1}+q_{2}\right) ^{2}$. The abrupt fall near 1 is
due to the 20 degrees isolation cut.}
\label{costopbot_lab_unpol_gl=1_gr=0}
\end{figure}
\begin{figure}[!hbp]
\begin{center}
\includegraphics[width=\figwidth]{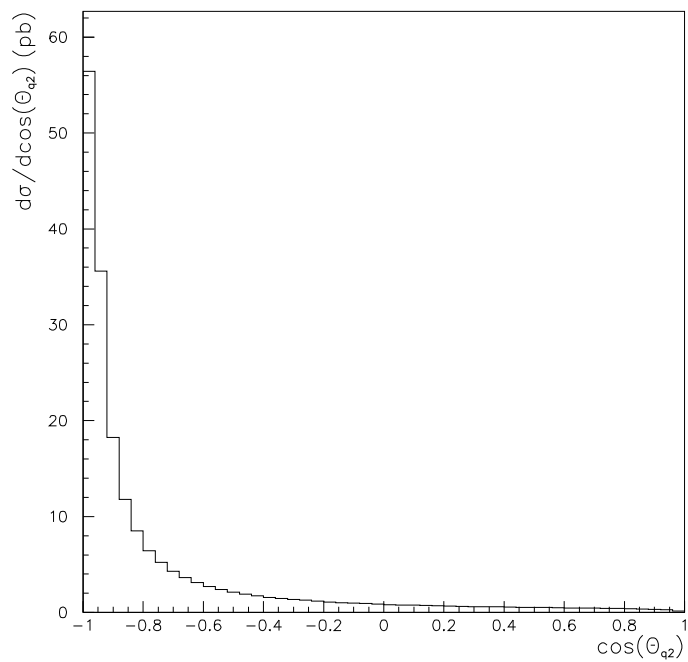}
\end{center}
\par
\caption{Distribution of the cosine of the angle between the spectator quark
and the gluon corresponding to unpolarized single top production at the LHC.
The momentum of the gluon is in the beam line direction but its sense is not
observable so to obtain an observable distribution we have to symmetrize the
above one. The calculation was performed at the tree level in Standard Model
with $\protect\mu ^{2}=\hat{s}=\left( q_{1}+q_{2}\right) ^{2}$.}
\label{cosq2_lab_unpol_gl=1_gr=0}
\end{figure}

In Fig. (\ref{costcosb_lab_unpol_gl=1_gr=0}) we plot the cross section
distribution of the polar angles of the top and anti-bottom with respect to
the beam line for unpolarized single top production at the LHC. In Fig. (\ref
{costopbot_lab_unpol_gl=1_gr=0}) we plot the distribution of the cosine of
the angle between the top and the anti-bottom for unpolarized single top
production at the LHC. Everything is calculated in the (tree-level) Standard
Model in the LAB frame. In both figures the above cuts are implemented, in
particular the isolation cut of 20 degrees in the angle between the top and
the anti-bottom is clearly visible in Fig. (\ref
{costopbot_lab_unpol_gl=1_gr=0}). In Fig. (\ref{cosq2_lab_unpol_gl=1_gr=0})
we also present the distribution of the cosine of the angle between the
spectator quark and the gluon. From inspection of these figures two facts
emerge: a) the top-bottom distribution is strongly peaked in the beam
direction as expected. b) Even with the presence of the isolation cut, near
the beam axis configurations with top and anti-bottom almost parallel are
flavored with respect to back-to-back configurations. Therefore this is an
indication that almost back-to-back configurations are distributed more
uniformly in space than parallel configurations favoring the beam line
direction.
\begin{figure}[!hbp]
\begin{center}
\includegraphics[width=12cm]{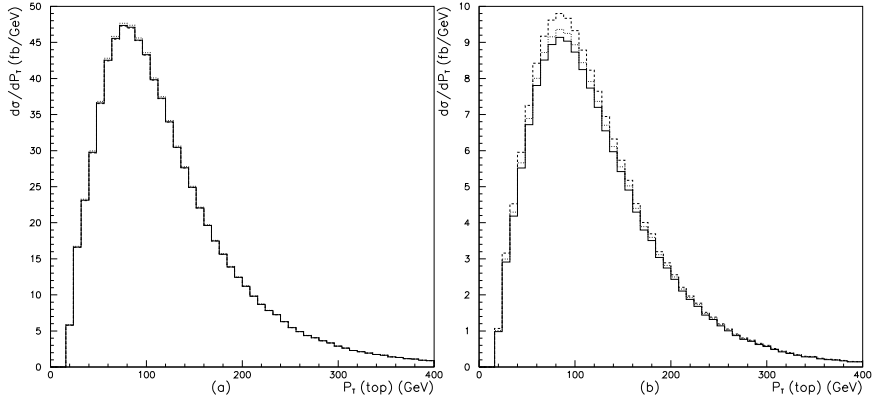}
\end{center}
\par
\caption{Top transversal momentum distribution corresponding to polarized
single top production at the LHC. plots (a), (b) correspond to tops
polarized in the spectator jet \emph{positive}, \emph{negative} direction
respectively in the \emph{top rest frame}. The subprocesses contributing to
the solid line histogram have been calculated at tree level in the SM ($%
g_{L}=1$, $g_{R}=0$). The dashed (dotted) line histogram have been
calculated at tree level with $g_{L}=1$, and $g_{R}=0.1$ ($g_{L}=1$, and $%
g_{R}=-0.1$). Note in (a) that the variation in the cross section due to the
variation of the right coupling around its SM tree level value is
practically inappreciable.}
\label{ptoptrans_rep_+-q2_gl=1_gr=+-.1}
\end{figure}

Let us now depart from the tree-level Standard Model and consider non-zero
values for $\delta g_{L}$ and $\delta g_{R}$. In what concerns the
dependence on the right effective coupling, our results are summarized in
Fig. (\ref{ptoptrans_rep_+-q2_gl=1_gr=+-.1}). From that figure it is quite
apparent that negatively polarized tops (in the top rest frame, as
previously described) are more sensitive to the value of the right coupling.

Taking into account the results of Eq. (\ref{results}) we can establish the
intervals where the effective couplings are indistinguishable from their
tree level Standard Model values taking a 1 sigma deviation as a rough
statistical criterion. Evidently we do not pretend to make here a serious
experimental analysis since we are not taking into account the full set of
experimental and theoretical uncertainties. Our aim is just to present an
order of magnitude estimate of the sensitivity of the different spin basis
to the value of the effective coupling around their tree level Standard
Model value. The results are given in Table \ref{sensitivity}, where we
indicate also the polarization vector chosen in each case. Of course those
sensitivities (which, as said, are merely indicative) are calculated with
the assumption that one could perfectly measure the top polarization in any
of the above basis. As it is well known the top polarization is only
measurable in an indirect way through the angular distribution of its decay
products. In section \ref{decay} we outline the procedure to use our results
to obtain a final angular distribution for the polarized top decay products
(we believe that some confusion exists on this point). Obtaining that
angular distribution involves a convolution of the single top production
cross section with the decay products angular distribution and because of
that we expect the true sensitivity to be worse than the ones given in Table
\ref{sensitivity}. Obviously such distribution is an observable quantity and
therefore must be independent of the spin basis one uses at an intermediate
step calculation (in other words, the results must be independent of the
basis in which the top spin density matrix is written). Because of that the
discussion as to which is the ``best'' basis for the top polarization is
somewhat academic in our view (see Chapter \ref{LSZchapter}). Any basis will
do; if any, the natural basis is that one where the density matrix becomes
diagonal, where production and decay factorize. This basis corresponds to
none of the above. However it may still be useful to know that some basis
are more sensitive to the effective couplings than others if one \emph{%
assumes} (at least as a gedanken experiment) that the polarization of the
top could be measured directly.

It is worth mentioning that the bottom mass, which appears in the cross
section in crossed left-right terms, such as $m_{b}g_{L}g_{R}$, plays a
crucial role in the actual determination of $g_{R}$. This is because from
the $|\mathrm{Re}(\kappa _{R}^{CC})|\leq 0.4\times 10^{-2}$ bound \cite{LPY}
we expect $g_{L}g_{R}m_{b}>g_{R}^{2}m_{t}.$ Evidently for the $ts$ or $td$
couplings these terms are not expected to be so relevant.

\begin{table}[tbh]
\centering
\begin{tabular}{|c|c|c|}
\hline
polarization, frame & $g_{L}$ & $g_{R}$ \\ \hline
$\hat{n}=\pm \frac{\vec{p}_{1}}{\left| \vec{p}_{1}\right| },~lab$ &
\multicolumn{1}{|l|}{$\left[ 0.\,9986,1.\,0014\right] \quad ~~\left(
-\right) $} & \multicolumn{1}{|l|}{$\left[ -0.26,0.85\right] \qquad
~~~~\left( +\right) $} \\ \hline
$\hat{n}=\pm \frac{\vec{q}_{2}}{\left| \vec{q}_{2}\right| },~lab$ &
\multicolumn{1}{|l|}{$\left[ 0.9987,1.0013\right] \qquad \left( +\right) $}
& \multicolumn{1}{|l|}{$\left[ -0.013,0.063\right] \qquad \left( -\right) $}
\\ \hline
$\hat{n}=\pm \frac{\vec{q}_{2}}{\left| \vec{q}_{2}\right| },~rest$ &
\multicolumn{1}{|l|}{$\left[ 0.9987,1.0013\right] \qquad \left( +\right) $}
& \multicolumn{1}{|l|}{$\left[ -0.021,0.059\right] \qquad \left( -\right) $}
\\ \hline
\end{tabular}
\caption{Sensitivity of the polarized single top production to variations of
the effective couplings. To calculate the intervals we have taken 2 sigma
statistical deviations (95.5\% confidence level) from tree level values as
an order of magnitude criterion. Of course, given the uncertainties in the
QCD scale, the overall normalization is dubious and the actual precision on $%
g_{L}$ a lot less. The purpose of these figures is to illustrate the
relative accuracy. Between parenthesis we indicate the spin direction taken
to calculate each interval.}
\label{sensitivity}
\end{table}

\section{The differential cross section for polarized tops}

\label{differentialc}We define the matrix elements of the subprocess of
Figs. (\ref{u+gt+b-d+tot}) and (\ref{d-gt+b-u-tot}) as $M_{+}^{d}$ and $%
M_{+}^{\bar{u}}$, respectively. We also define the matrix elements
corresponding to the processes producing anti-tops as $M_{-}^{u}$, and $%
M_{-}^{\bar{d}}$. With these definitions the differential cross section for
polarized tops $d\sigma $ can be written schematically as
\begin{equation*}
d\sigma =\beta \left( f_{u}\left| M_{+}^{d}\right| ^{2}+f_{\bar{d}}\left|
M_{+}^{\bar{u}}\right| ^{2}\right) ,
\end{equation*}
where $f_{u}$ and $f_{\bar{d}}$ denote the parton distribution functions
corresponding to extracting a $u$-type quark and a $\bar{d}$-type quark
respectively and $\beta $ is a proportionality factor incorporating the
kinematics$.$ Now using our analytical results for the matrix elements given
in appendix \ref{TchannelappA} along with Eq. (\ref{change}) and symmetries (%
\ref{sym}) we obtain
\begin{eqnarray}
d\sigma &=&\beta f_{u}\left[ \left| g_{L}\right| ^{2}\left( a+a_{n}\right)
+\left| g_{R}\right| ^{2}\left( b+b_{n}\right) +\frac{g_{R}^{\ast
}g_{L}+g_{R}g_{L}^{\ast }}{2}\left( c+c_{n}\right) +i\frac{g_{L}^{\ast
}g_{R}-g_{R}^{\ast }g_{L}}{2}d_{n}\right]  \notag \\
&&+\beta f_{\bar{d}}\left[ \left| g_{R}\right| ^{2}\left( a-a_{n}\right)
+\left| g_{L}\right| ^{2}\left( b-b_{n}\right) +\frac{g_{R}^{\ast
}g_{L}+g_{R}g_{L}^{\ast }}{2}\left( c-c_{n}\right) -i\frac{g_{L}^{\ast
}g_{R}-g_{R}^{\ast }g_{L}}{2}d_{n}\right] \notag  \\
&=&\left(
\begin{array}{cc}
g_{L}^{\ast } & g_{R}^{\ast }
\end{array}
\right) A\left(
\begin{array}{c}
g_{L} \\
g_{R}
\end{array}
\right) ,   \label{t-channeldecomp}
\end{eqnarray}
where
\begin{equation}
A=\beta \left(
\begin{array}{cc}
f_{u}\left( a+a_{n}\right) +f_{\bar{d}}\left( b-b_{n}\right) & \frac{1}{2}%
f_{u}\left( c+c_{n}+id_{n}\right) +\frac{1}{2}f_{\bar{d}}\left(
c-c_{n}-id_{n}\right) \\
\frac{1}{2}f_{u}\left( c+c_{n}-id_{n}\right) +\frac{1}{2}f_{\bar{d}}\left(
c-c_{n}+id_{n}\right) & f_{u}\left( b+b_{n}\right) +f_{\bar{d}}\left(
a-a_{n}\right)
\end{array}
\right) ,  \label{new}
\end{equation}
and where $a,$ $b$, $c$, $a_{n}$, $b_{n}$, $c_{n}$ and $d_{n}$ are
independent of the effective couplings $g_{R}$ and $g_{L}$ and the
subscripts $n$ indicate linear dependence on the top spin four-vector $n.$.
From Eq. (\ref{new}) we observe that $A$ is an Hermitian matrix and
therefore it is diagonalizable with real eigenvalues. Moreover, from the
positivity of $d\sigma $ we immediately arrive at the constraints
\begin{eqnarray}
\det A &\geq &0,  \label{constr1} \\
TrA &\geq &0,  \label{constr2}
\end{eqnarray}
that is
\begin{eqnarray}
&&\left( f_{u}\left( a+a_{n}\right) +f_{\bar{d}}\left( b-b_{n}\right)
\right) \left( f_{u}\left( b+b_{n}\right) +f_{\bar{d}}\left( a-a_{n}\right)
\right)  \notag \\
&\geq &\frac{1}{4}\left( c^{2}\left( f_{u}+f_{\bar{d}}\right) ^{2}+\left(
c_{n}^{2}+d_{n}^{2}\right) \left( f_{u}-f_{\bar{d}}\right)
^{2}+2cc_{n}\left( f_{u}^{2}-f_{\bar{d}}^{2}\right) \right) ,  \label{co1}
\end{eqnarray}
and
\begin{equation}
\left( f_{u}+f_{\bar{d}}\right) \left( a+b\right) +\left( f_{u}-f_{\bar{d}%
}\right) \left( a_{n}+b_{n}\right) \geq 0.  \label{co2}
\end{equation}
Note that it is not possible to saturate both constraints for the same
configuration because this would imply a vanishing $A$ which in turn would
imply relations such as
\begin{equation*}
\frac{a+b}{a_{n}+b_{n}}=\frac{f_{\bar{d}}-f_{u}}{f_{\bar{d}}+f_{u}}=\frac{%
a_{n}-b_{n}}{a-b},
\end{equation*}
which evidently do not hold. Moreover, since constraints (\ref{co1}) and (%
\ref{co2}) must be satisfied for any set of positive PDF's we immediately
obtain the bounds
\begin{eqnarray*}
ab+a_{n}b_{n}-\frac{1}{4}\left( c^{2}+c_{n}^{2}+d_{n}^{2}\right) &\geq
&\left| a_{n}b+ab_{n}-\frac{1}{2}cc_{n}\right| \\
b^{2}+a^{2}-\left( b_{n}^{2}+a_{n}^{2}\right) &\geq &\frac{1}{2}\left(
c^{2}-\left( c_{n}^{2}+d_{n}^{2}\right) \right) .
\end{eqnarray*}
In order to have a 100\% polarized top we need a spin four-vector $n$ that
saturates the constraint (\ref{constr1}) (that is Eq.(\ref{co1})) for each
kinematical situation, that is we need $A\left( n\right) $ to have a zero
eigenvalue which is equivalent to have a unitary matrix $C$ satisfying
\begin{equation*}
C^{\dagger }AC=\mathrm{diag}\left( \lambda ,0\right) ,
\end{equation*}
for some positive eigenvalue $\lambda $. In general such $n$ need not exist
and, should it exist, is in any case independent of the effective couplings $%
g_{R}$ and $g_{L}$. Moreover, provided this $n$ exists there is only one
solution (up to a global complex normalization factor $\alpha $) for the
pair $\left( g_{R},g_{L}\right) $ to the equation $d\sigma =0,$ This
solution is just
\begin{eqnarray}
g_{L} &=&\alpha C_{12},  \notag \\
g_{R} &=&\alpha C_{22}.  \label{100}
\end{eqnarray}
Note that if one of the effective couplings vanishes we can take the other
constant and arbitrary. However if both effective couplings are
non-vanishing we would have a quotient $g_{R}/g_{L}$ that would depend in
general on the kinematics. This is not possible so we can conclude that for
a non-vanishing $g_{R}$ ( $g_{L}$ is evidently non-vanishing) it is not
possible to have a pure spin state (or, else, only for fine tuned $g_{R}$ a
100\% polarization is possible).

To illustrate these considerations let us give an example: in the unphysical
situation where $m_{t}\rightarrow 0$ it can be shown that there exists two
solutions to the saturated constraint (\ref{constr1}), namely
\begin{equation}
m_{t}n^{\mu }\rightarrow \pm \left( \left| \vec{p}_{1}\right| ,p_{1}^{0}%
\frac{\vec{p}_{1}}{\left| \vec{p}_{1}\right| }\right) ,
\end{equation}
once we have found this result we plug it in the expression (\ref{100}) and
we find the solutions $\left( 0,g_{L}\right) $ with $g_{L}$ arbitrary for
the $+$ sign and $\left( g_{R},0\right) $ with $g_{R}$ arbitrary for the $-$
sign. That is, physically we have zero probability of producing a right
handed top when we have only a left handed coupling and viceversa when we
have only a right handed coupling. Note that in this case it is clear that
having both effective couplings non-vanishing would imply the absence of 100
\% polarization in any spin basis. This can be understood in general
remembering that the top particle forms in general an entangled state with
the other particles of the process. Since we are tracing over the unknown
spin degrees of freedom and over the flavors of the spectator quark we do
not expect in general to end up with a top in a pure polarized state;
although this is not impossible as it is shown the in the last example.

In the physical situation where $m_{t}\neq 0$ (we use $m_{t}=175.6$ GeV and $%
m_{b}=5$ GeV in this work) we have found that a spin basis with relatively
high polarization is the one with the spin $\hat{n}$ taken in the direction
of the spectator quark in the top rest frame. This is in accordance to the
results in \cite{mandp}. In general the degree of polarization ($\frac{N_{+%
\hat{n}}}{N_{+\hat{n}}+N_{-\hat{n}}}$) depends not only on the spin frame
but also on the particular cuts chosen. We have found that the lower cut for
the transverse momentum of the bottom worsens the polarization degree but,
in spite of that, from Eq.(\ref{results}) we see that we have a 84\% of
polarization in the Standard Model ($g_{L}=1,$ $g_{R}=0$) that is much
bigger than the 69\% obtained with the helicity frame. The above results
follow the general trend of those presented by Mahlon and Parke \cite{mandp}%
, but in general, their degree of polarization is higher. We understand that
this is due to the different cuts (in particular for the transversal
momentum of the bottom) along with the different set of PDF's used in our
simulations.

\section{Measuring the top polarization from its decay products}

\label{decay}A well know result in the tree level SM regarding the measure
of the top polarization from its decay products is the formula that states
the following: Given a top polarized in the $\hat{n}$ direction in its rest
frame, the lepton $l^{+}$ produced in the decay of the top via the process
\begin{equation}
t\rightarrow b\left( W^{+}\rightarrow l^{+}\nu _{l}\right) ,
\end{equation}
presents an angular distribution \cite{cos}

\begin{equation}
\sigma _{l}=\alpha \left( 1+\cos \theta \right) ,  \label{pol1}
\end{equation}
where $\alpha $ is a normalization factor and $\theta $ is the axial angle
measured from the direction of $\hat{n}.$ What can we do when the top is in
a mixed state with no 100\% polarization in any direction? The first naive
answer would be: With any axis $\hat{n}$ in the top rest frame the top will
have a polarization $p_{+}$ (with $0\leq p_{+}\leq 1)$ in that direction and
a polarization $p_{-}=1-p_{+}$ in the opposite direction so the angular
distribution for the lepton is
\begin{eqnarray}
\sigma _{l} &=&\alpha \left( p_{+}\left( 1+\cos \theta \right) +p_{-}\left(
1-\cos \theta \right) \right)  \notag \\
&=&\alpha \left( 1+\left( p_{+}-p_{-}\right) \cos \theta \right)  \notag \\
&=&\alpha \left( 1+\left( 2p_{+}-1\right) \cos \theta \right) .  \label{pol}
\end{eqnarray}
The problem with formula (\ref{pol}) is that the angular distribution for
the lepton depends on the arbitrary chosen axis $\hat{n}$ and this cannot be
correct. The correct answer can be obtained by noting the following facts:

\begin{itemize}
\item  Given an arbitrary chosen axis $\hat{n}$ in the rest frame and the
associated spin basis to it $\left\{ \left| +\hat{n}\right\rangle ,\left| -%
\hat{n}\right\rangle \right\} $ the top spin state in given by a $2\times 2$
density matrix $\rho $
\begin{equation}
\rho =\rho _{+}\left| +\hat{n}\right\rangle \left\langle +\hat{n}\right|
+\rho _{-}\left| -\hat{n}\right\rangle \left\langle -\hat{n}\right| +b\left|
+\hat{n}\right\rangle \left\langle -\hat{n}\right| +b^{\ast }\left| -\hat{n}%
\right\rangle \left\langle +\hat{n}\right| ,
\end{equation}
which is in general not diagonal ($b\neq 0$) and whose coefficients depend
on the rest of kinematical variables determining the differential cross
section.

\item  From the calculation of the polarized cross section \emph{we only know%
} the diagonal elements $\rho _{\pm }=p_{\pm }=\left| M\right| _{\pm \hat{n}%
}^{2}/\left( \left| M\right| _{+\hat{n}}^{2}+\left| M\right| _{-\hat{n}%
}^{2}\right) .$

\item  Given $\rho $ in any orthogonal basis determined (up to phases) by $%
\hat{n}$ we can change to another basis that diagonalizes $\rho .$ Since the
top is a spin $1/2$ particle, this basis will correspond to another
direction $\hat{n}_{d}$.

\item  Once we have $\rho $ diagonalized then Eq.(\ref{pol}) is trivially
correct with $p_{\pm }=\rho _{\pm }$ and now $\theta $ is unambiguously
measured from the direction of $\hat{n}_{d}.$
\end{itemize}

From the above facts the first question that comes to our minds is if there
exists a way to determine $\hat{n}_{d}$ without knowing the off-diagonal
matrix elements of $\rho .$ The answer is yes. It is an easy exercise of
elementary quantum mechanics that given a $2\times 2$ Hermitian matrix $\rho
$ the eigenvector with largest (lowest) eigenvalue correspond to the unitary
vector that maximizes (minimizes) the bilinear form $\left\langle v\right|
\rho \left| v\right\rangle $ constrained to $\left\{ \left| v\right\rangle
,~\left\langle v|v\right\rangle =1\right\} $. Since an arbitrary normalized $%
\left| v\right\rangle $ can be written (up to phases) as $\left| +\hat{n}%
\right\rangle $ and in that case $\rho _{+}=p_{+}$ then the correct $\hat{n}%
_{d}$ entering in Eq.(\ref{pol}) is the one that maximizes the differential
cross section $\left| M\right| _{\hat{n}}^{2}$ for each kinematical
configuration. At the end, the correct angular distribution for the leptons
is given by the cross section for polarized tops\emph{\ in this basis} ($%
\hat{n}_{d}$) convoluted with formula (\ref{pol1}) (or improvements of it
\cite{Mauser}).

The above analysis was carried out in the Standard Model ($g_{R}=0$) but it
is correct also for $g_{R}\neq 0$ using the complete formula for this case
\begin{equation}
\sigma _{l}=\alpha \left( 1+\left( p_{+}-p_{-}\right) \cos \theta \left( 1-%
\frac{1}{4}\left| g_{R}\right| ^{2}h\left( \frac{M_{W}^{2}}{m_{t}^{2}}%
\right) \right) \right) ,  \label{final2}
\end{equation}
where $h\left( \frac{M_{W}^{2}}{m_{t}^{2}}\right) \simeq 0.566$ \cite{cos}.
Formula (\ref{final2}) deserves some comments:

\begin{itemize}
\item  First of all we remember that $\theta $ is the angle (in the top rest
frame) between the $\hat{n}$ that maximizes the difference $\left(
p_{+}-p_{-}\right) $ and the three momentum of the lepton.

\item  Taking into account the above comment and that $\left(
p_{+}-p_{-}\right) $ depends on $g_{L}$ and $g_{R}$ we see that also $\theta
$ depends on $g_{L}$ and $g_{R}$.

\item  From the computational point of view, formula (\ref{final2}) is not
an explicit formula because involves a process of maximization for each
kinematical configuration.

\item  In some works in the literature \cite{mandp} formula (\ref{final2})
is presented for an arbitrary choice of the spin basis $\left\{ \left| \pm
\hat{n}\right\rangle \right\} $ in the top rest frame. This is incorrect
because it does not take into account that, in general, the top spin density
matrix is not diagonal.

\item  In a recent work \cite{Mauser} $O\left( \alpha _{s}\right) $
corrections are incorporated to the polarized top decay angular analysis. In
this work the density matrix for the top spin is properly taken into
account. To connect this work with ours we have to replace their
polarization vector $\vec{P}$ by $P\hat{n}_{d}$ where the magnitude of the
top polarization $P$ is just the spin asymmetry $\left( \left| M\right| _{+%
\hat{n}_{d}}^{2}-\left| M\right| _{-\hat{n}_{d}}^{2}\right) /\left( \left|
M\right| _{+\hat{n}_{d}}^{2}+\left| M\right| _{-\hat{n}_{d}}^{2}\right) $ in
our language. This taken into account, the density matrix the authors of
\cite{Mauser} quote is in the basis $\left\{ \left| +\hat{n}%
_{W^{+}}\right\rangle ,\left| -\hat{n}_{W^{+}}\right\rangle \right\} $ where
$\hat{n}_{W^{+}}$ is a normal vector in the direction of the three momentum
of the $W^{+}$ (in the top rest frame).
\end{itemize}

\section{Conclusions}

We have done a complete calculation of the subprocess cross sections for
polarized tops or anti-tops including the right effective coupling and
bottom mass corrections. We have used a $p_{T}>30$ GeV cut in the transverse
momentum of the produced $\bar{b}$ quark and, accordingly we have retained
only the so called $2\rightarrow 3$ process, for the reasons described in
the text.

Our analysis here is completely general. No approximation is made. We use
the most general set of couplings and, since our approach is completely
analytical, we can describe the contribution from other intermediate quarks
in the $t$ channel, mixing, etc. Masses and mixing angles are retained. On
the contrary, the analysis has to be considered only preliminary from an
experimental point of view. No detailed study of the backgrounds has been
made, except for the dominant $gg\rightarrow t\bar{t}$ process which has
been considered to some extent (although again without quantitative
evaluation).

Given the (presumed) smallness of the right handed couplings, the bottom
mass plays a role which is more important than anticipated, as the mixed
crossed $g_{L}g_{R}$ term, which actually is the most sensitive one to $%
g_{R} $ is accompanied by a $b$ quark mass. The statistical sensitivity to
different values of this coupling is given in the text.

We present a variety of $p_{T}$ and angular distributions both for the $t$
and the $\bar{b}$ quarks. Obviously, the top decays shortly after
production, but we have not made detailed simulations of this part. In fact,
the interest of this decay is obvious: one can measure the spin of the top
through the angular distribution of the leptons produced in this decay. In
the Standard Model, single top production gives a high degree of
polarization (84 \% in the optimal basis, with the present set of cuts).
This is a high degree of polarization, but well below the 90+ claimed by
Mahlon and Parke in \cite{mandp}. We understand this being due to the
presence of the 30 GeV cut. In fact, if we remove this cut completely we get
a 91 \% polarization. Still below the result of \cite{mandp} but in rough
agreement (note that we do not include the $2\rightarrow 2$ process).
Inasmuch as they can be compared our results are in good agreement with
those presented in \cite{SSW} in what concerns the total cross-section. Two
different choices for the strong scale $\mu ^{2}$ are presented.

In addition, it turns out that when $g_{R}\neq 0$ the top can never be 100\%
polarized. In other words, it is in a mixed state. In this case we show that
a unique spin basis is singled out which allows one to connect top decay
products angular distribution with the polarized top differential cross
section.

Finally it should be mentioned that a previous study for this process in the
present context was performed in \cite{effW} using the effective W
approximation \cite{dawson}, in which the $W$ is treated as a parton of the
proton. While this is certainly not an exact treatment, it was expected to
be sufficiently good for our purposes. In the course of this work we have
found, however, a number of differences.

\chapter{Single top production in the \textit{s}-channel and top decay}

\label{topdecaychapter} In the previous chapter we have analyzed in detail
the dominant \textit{t}-channel mechanism of single top production and the
sensitivity to values of the effective couplings $g_{L}$ and $g_{R}$
departing from their SM tree-level values.

As it has been discussed in Chapter \ref{LHCphenomenology}, to be able to
tell the corrections due to a right effective coupling from those due to a
left one needs to measure the polarization of the top. Of course the top
decays shortly after its production and all one can hope to see are the
decay products. Since there is a correlation between the angular momentum of
the top and the angular distribution of those products, one might hope to be
able to `measure' the top polarization and thus separate left from right
contributions. Obviously, since the correlation is not a delta function
(prohibited by quantum mechanics), some information must be lost along the
top decay process.

Although the calculations in Chapter \ref{LHCphenomenology} were presented
for the production of polarized tops, with respect to an arbitrary axis, and
something was said there about the subsequent top decay, the issue was not
discussed in great detail. In this chapter we would like to analyze this
point more deeply. We shall not do so, however, in the \textit{t}-channel
process, but rather in the much simpler \textit{s}-channel production
mechanism. Although this mechanism is subleading (see the discussion in the
previous chapter concerning the different contributions to the cross section
for single top production) it is not negligible at all. Furthermore, the
results obtained here can be carried over to the \textit{t}-channel process
without much difficulty.

\begin{figure}[!hbp]
\begin{center}
\includegraphics[width=6cm]{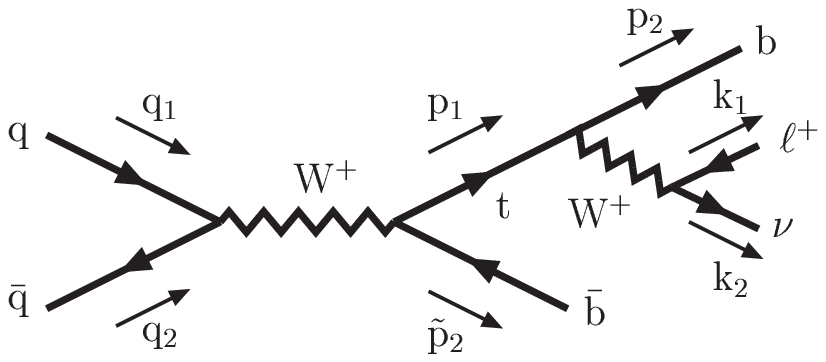}
\end{center}
\par
\caption{Feynman diagram contributing to single top production and decay
process in the s-channel.}
\label{singletopschannelanddecay}
\end{figure}

In this chapter we will thus complete some of the more subtle aspects of
single top production that were not taken into account in last chapter. The
process analyzed in this chapter is given by Fig. (\ref
{singletopschannelanddecay}).

Here, exactly as we did in Chapter 5, we shall assume that the produced top
is on-shell, namely we study the production and subsequent decay of real
tops. This way of proceeding goes under the name of narrow width
approximation and it is a standard practice to analyze complicated processes
consisting on the production of an unstable particle followed by its decay.

\section{Cross sections for top production and decay}

Using the momenta conventions of Fig. (\ref{singletopschannelanddecay}) and
averaging over colors and spins of the initial fermions and summing over
colors and spins of the final fermions (remember that we have included a
spin projector for the top) the squared amplitude for top production is
given by
\begin{eqnarray}
\left| M_{n}\right| ^{2} &=&\frac{e^{4}N_{c}}{s_{W}^{4}}\left( \frac{1}{%
k^{2}-M_{W}^{2}}\right) ^{2}  \notag \\
&&\times \left\{ \left| \tilde{g}_{L}\right| ^{2}\left[ \left| g_{R}\right|
^{2}\left( q_{1}\cdot \frac{p_{1}+m_{t}n}{2}\right) \left( q_{2}\cdot \tilde{%
p}_{2}\right) +\left| g_{L}\right| ^{2}\left( q_{2}\cdot \frac{p_{1}-m_{t}n}{%
2}\right) \left( q_{1}\cdot \tilde{p}_{2}\right) \right. \right.  \notag \\
&&+m_{b}\frac{g_{L}g_{R}^{\ast }+g_{R}g_{L}^{\ast }}{4}\left[ m_{t}\left(
q_{1}\cdot q_{2}\right) +\left( q_{2}\cdot p_{1}\right) \left( q_{1}\cdot
n\right) -\left( q_{2}\cdot n\right) \left( q_{1}\cdot p_{1}\right) \right]
\notag \\
&&+\left. \left. im_{b}\frac{g_{L}g_{R}^{\ast }-g_{R}g_{L}^{\ast }}{4}%
\varepsilon _{\mu \alpha \rho \sigma }n^{\mu }p_{1}^{\alpha }q_{1}^{\rho
}q_{2}^{\sigma }\right] \right\}  \notag \\
&&+\left| \tilde{g}_{R}\right| ^{2}\left[ \left| g_{R}\right| ^{2}\left(
q_{2}\cdot \frac{p_{1}+m_{t}n}{2}\right) \left( q_{1}\cdot \tilde{p}%
_{2}\right) +\left| g_{L}\right| ^{2}\left( q_{1}\cdot \frac{p_{1}-m_{t}n}{2}%
\right) \left( q_{2}\cdot \tilde{p}_{2}\right) \right.  \notag \\
&&+m_{b}\frac{g_{L}g_{R}^{\ast }+g_{R}g_{L}^{\ast }}{4}\left[ m_{t}\left(
q_{1}\cdot q_{2}\right) +\left( q_{1}\cdot p_{1}\right) \left( q_{2}\cdot
n\right) -\left( q_{1}\cdot n\right) \left( q_{2}\cdot p_{1}\right) \right]
\notag \\
&&+\left. \left. im_{b}\frac{g_{L}g_{R}^{\ast }-g_{R}g_{L}^{\ast }}{4}%
\varepsilon _{\mu \alpha \rho \sigma }n^{\mu }p_{1}^{\alpha }q_{2}^{\rho
}q_{1}^{\sigma }\right] \right\} ,  \label{schanproduction}
\end{eqnarray}
where $\tilde{g}_{L},$ and $\tilde{g}_{R}$ are left and right couplings to
the light quarks and $g_{L}$ and $g_{R}$ the coupling to the heavy up bottom
system. In the simulations we have taken $\tilde{g}_{L}=1,$ $\tilde{g}%
_{R}=0. $ Hence the differential cross section for producing polarized tops
is given by
\begin{eqnarray*}
d\sigma _{\hat{n}} &=&f\left( \tilde{x}_{1},\tilde{x}_{2},\left(
q_{1}+q_{2}\right) ^{2},\Lambda _{QCD}\right) d\tilde{x}_{1}d\tilde{x}_{2}%
\frac{1}{4\left| q_{2}^{0}\overrightarrow{q_{1}}-\overrightarrow{q_{2}}%
q_{1}^{0}\right| } \\
&&\times \frac{d^{3}p_{1}}{\left( 2\pi \right) ^{3}2p_{1}^{0}}\frac{d^{3}%
\tilde{p}_{2}}{\left( 2\pi \right) ^{3}2\tilde{p}_{2}^{0}}\left|
M_{n}\right| ^{2}\left( 2\pi \right) ^{4}\delta ^{4}\left(
q_{1}+q_{2}-p_{1}-p_{2}\right)
\end{eqnarray*}
where $f\left( \tilde{x}_{1},\tilde{x}_{2},\left( q_{1}+q_{2}\right)
^{2},\Lambda _{QCD}\right) d\tilde{x}_{1}d\tilde{x}_{2}$ accounts for the
PDF contribution. For the top the total decay rate is given by
\begin{eqnarray*}
\Gamma &=&\frac{e^{2}}{s_{W}^{2}}\left\{ \left( \left| g_{L}\right|
^{2}+\left| g_{R}\right| ^{2}\right) \left( m_{t}^{2}+m_{b}^{2}-2M_{W}^{2}+%
\frac{\left( m_{t}^{2}-m_{b}^{2}\right) ^{2}}{M_{W}^{2}}\right) \right. \\
&&\left. -12m_{t}m_{b}\frac{g_{L}g_{R}^{\ast }+g_{R}g_{L}^{\ast }}{2}%
\right\} \frac{\sqrt{\left( m_{t}^{2}+m_{b}^{2}-M_{W}^{2}\right)
^{2}-4m_{t}^{2}m_{b}^{2}}}{64\pi m_{t}^{2}p_{1}^{0}}.
\end{eqnarray*}
The squared amplitude corresponding to the decay rate in the channel
depicted in Fig. (\ref{singletopschannelanddecay}) summing over the top
polarizations (with a spin projector inserted), averaging over its colors
and summing over colors and polarizations of decay products is given by
\begin{equation*}
\left| M_{n}^{D}\right| ^{2}=\frac{-4}{N_{c}}\left| M_{n}\right| ^{2}\left(
q_{1}\rightarrow k_{2},\ q_{2}\rightarrow k_{1},\ \tilde{p}_{2}\rightarrow
-p_{2}\right) ,
\end{equation*}
where $\left| M_{n}\right| ^{2}\left( q_{1}\rightarrow k_{2},\
q_{2}\rightarrow k_{1},\ \tilde{p}_{2}\rightarrow -p_{2}\right) $ is just
expression (\ref{schanproduction}) with the indicated changes in momenta Now
$\tilde{g}_{L},$ and $\tilde{g}_{R}$ are left and right couplings
corresponding to the lepton-neutrino vertex and we have taken again $\tilde{g%
}_{L}=1,$ $\tilde{g}_{R}=0.$ Hence the decay rate differential cross section
for this channel is given by
\begin{equation*}
d\Gamma _{n}=\frac{\left| M_{n}^{D}\right| ^{2}}{2p_{1}^{0}}\frac{d^{3}k_{1}%
}{\left( 2\pi \right) ^{3}2k_{2}^{0}}\frac{d^{3}k_{2}}{\left( 2\pi \right)
^{3}2k_{1}^{0}}\frac{d^{3}p_{2}}{\left( 2\pi \right) ^{3}2p_{2}^{0}}\left(
2\pi \right) ^{4}\delta ^{4}\left( k_{1}+k_{2}+p_{2}-p_{1}\right) .
\end{equation*}
Using the narrow-width approximation we have that the differential cross
section $d\sigma $ corresponding to Fig. (\ref{singletopschannelanddecay})
is given by
\begin{equation}
d\sigma =\sum_{\pm n}d\sigma _{n}\times \frac{d\Gamma _{n}}{\Gamma }.
\label{finalschannel}
\end{equation}

\section{The role of spin in the narrow-width approximation}

Within the narrow-width approximation we decompose the process depicted in
Fig. (\ref{singletopschannelanddecay}) in two consecutive processes, the top
production and its consecutive decay. In that set up we denote the single
top production amplitude as $A_{p,\pm \hat{n}\left( p\right) }$ and the top
decay amplitude as $B_{p,\pm \hat{n}\left( p\right) }.$ In the polar
representation we write
\begin{eqnarray*}
A_{p,\pm \hat{n}\left( p\right) } &=&\left| A_{p,\pm \hat{n}\left( p\right)
}\right| e^{i\varphi _{\pm }\left( p\right) }, \\
B_{p,\pm \hat{n}\left( p\right) } &=&\left| B_{p,\pm \hat{n}\left( p\right)
}\right| e^{i\omega _{\pm }\left( p\right) },
\end{eqnarray*}
where $p$ indicate external momenta and $\hat{n}\left( p\right) $ a given
spin basis for the top (see section \ref{cross} in Chapter \ref
{LHCphenomenology}). The differential cross section for the whole process $%
\mathcal{M}$ is schematically given by
\begin{equation}
d\mathcal{\sigma }=\int \left| A_{p,+\hat{n}\left( p\right) }B_{p,+\hat{n}%
\left( p\right) }+A_{p,-\hat{n}\left( p\right) }B_{p,-\hat{n}\left( p\right)
}\right| ^{2}dp,  \label{exact}
\end{equation}
where the integration over momenta is taken outside the modulus squared
because these are \emph{unseen external} momenta (like neutrino momenta, or
angular and longitudinal momenta variables in $p_{T}$ histograms, etc.).
Since there are still some kinematical variables still pending for
integration (for example $p_{T}$ in $p_{T}$ distributions) we keep the ``$d$%
''$\mathcal{\ }$in front of $\mathcal{\sigma }$. Hence
\begin{eqnarray}
d\mathcal{\sigma } &=&\int \left| A_{p,+\hat{n}\left( p\right) }\right|
^{2}\left| B_{p,+\hat{n}\left( p\right) }\right| ^{2}dp+\int \left| A_{p,-%
\hat{n}\left( p\right) }\right| ^{2}\left| B_{p,-\hat{n}\left( p\right)
}\right| ^{2}dp  \notag \\
&&+2\int \left| A_{p,+\hat{n}\left( p\right) }\right| \left| B_{p,+\hat{n}%
\left( p\right) }\right| \left| A_{p,-\hat{n}\left( p\right) }\right| \left|
B_{p,-\hat{n}\left( p\right) }\right|  \notag \\
&&\times \cos \left( \varphi _{+}\left( p\right) -\varphi _{-}\left(
p\right) +\omega _{+}\left( p\right) -\omega _{-}\left( p\right) \right) dp
\notag \\
&\simeq &\int \left| A_{p,+\hat{n}\left( p\right) }\right| ^{2}\left| B_{p,+%
\hat{n}\left( p\right) }\right| ^{2}dp+\int \left| A_{p,-\hat{n}\left(
p\right) }\right| ^{2}\left| B_{p,-\hat{n}\left( p\right) }\right| ^{2}dp.
\label{approx}
\end{eqnarray}
Since the axis with respect to which the spin basis is defined is completely
arbitrary $d\mathcal{\sigma }$ is independent on this choice of basis.
However within the narrow width approximation one does not compute $d%
\mathcal{\sigma }$ following formula (\ref{exact}). The practical procedure
relies in computing the \emph{probability } of producing a polarized top and
then multiplying this probability by the \emph{probability} of a given decay
channel (see Eq. (\ref{finalschannel})). This procedure is equivalent to the
neglection of the interference term in formula (\ref{approx}) as indicated
there.

Let us see whether this approximation can justified. Clearly, the
integration over momenta enhances the positive-definite terms in front of
the interference oscillating one. If in addition we make a choice for $\hat{n%
}\left( p\right) $ that diagonalizes the top spin density matrix (see
Chapter \ref{LHCphenomenology}) and thus maximizes $\left| A_{p,+\hat{n}%
\left( p\right) }\right| $ and minimizes $\left| A_{p,-\hat{n}\left(
p\right) }\right| $, then we expect the interference term to be negligible
when compared to $\int \left| A_{p,+\hat{n}\left( p\right) }\right|
^{2}\left| B_{p,+\hat{n}\left( p\right) }\right| ^{2}dp$ even for small
amount of phase space integration. In the \textit{s}-channel we will see in
the next section that the limit of $g_{R}\rightarrow 0$ there exists a spin
basis $\hat{n}\left( p\right) $ where $\left| A_{p,-\hat{n}\left( p\right)
}\right| $ is strictly zero. This basis is given by
\begin{equation*}
n=\frac{1}{m_{t}}\left( \frac{m_{t}^{2}}{\left( q_{2}\cdot p_{1}\right) }%
q_{2}-p_{1}\right) .
\end{equation*}
From this it follows that for small $g_{R}$ if we use that basis the
interference integrand is already negligible with respect to the dominant
term $\int \left| A_{p,+\hat{n}\left( p\right) }\right| ^{2}\left| B_{p,+%
\hat{n}\left( p\right) }\right| ^{2}dp$. For $g_{R}\neq 0$ one can still
find a basis that maximizes $\left| A_{p,+\hat{n}\left( p\right) }\right| $
(and minimizes $\left| A_{p,-\hat{n}\left( p\right) }\right| $) and
therefore diagonalizes the top density matrix $\rho $ (see section \ref
{decay} in Chapter \ref{LHCphenomenology}) In the next section we will show
how to obtain such a basis that will be the one used in our numerical
integration. In these simulations we have checked numerically that this
basis is the one that maximizes $d\mathcal{\sigma }$ and therefore, on the
same grounds, the one that minimizes the interference term.

Given that the observables are strictly independent of the choice of spin
basis \emph{only} if the interference term is included, we can easily assess
the importance of the latter by checking to what extent a residual spin
basis dependence is present. We have checked numerically this point by
changing the definition of the spin basis $\hat{n}\left( p\right) $ and
noting that our results are weakly dependent on the choice of $\hat{n}\left(
p\right) $. A $3.8\%$ maximum variation was found between our diagonal basis
and another orthogonal to the beam axis (that is, almost orthogonal to all
momenta). Moreover we have checked that if spin is ignored altogether the
same amount of variation is observed. Thus we conclude that even though the
dependence on the choice of spin basis is not dramatic, its consideration is
a must for a precise description using the narrow-width approximation.

\section{The diagonal basis}

As stated in the previous section in order to calculate the top decay we
have to find the basis where the polarized single top production cross
section is maximal. The can do this maximizing in the $4$-dimensional space
generated by the components of $n$ constrained by
\begin{equation}
n\cdot p_{1}=0,\qquad n^{2}=-1,  \label{spinconstraints}
\end{equation}
where $p_{1}$ is the top four-moment, that is
\begin{eqnarray*}
n^{0} &=&\frac{n^{1}p_{1}^{1}+n^{2}p_{1}^{2}+n^{2}p_{1}^{2}}{p_{1}^{0}}, \\
\left( p_{1}^{0}\right) ^{2} &=&\left( p_{1}^{0}\right) ^{2}\left\| \vec{n}%
\right\| ^{2}-\left( n^{1}p_{1}^{1}+n^{2}p_{1}^{2}+n^{2}p_{1}^{2}\right)
^{2},
\end{eqnarray*}
where $\left\| \vec{n}\right\| =\sqrt{\left( n^{1}\right) ^{2}+\left(
n^{2}\right) ^{2}+\left( n^{3}\right) ^{2}}$, that is $n^{i}=\left\| \vec{n}%
\right\| \hat{n}^{i}$ with $\hat{n}$ the normalized spin three-vector. From
above equations we obtain
\begin{eqnarray*}
\left\| \vec{n}\right\| &=&\frac{p_{1}^{0}}{\sqrt{\left( p_{1}^{0}\right)
^{2}-\left( \hat{n}^{1}p_{1}^{1}+\hat{n}^{2}p_{1}^{2}+\hat{n}%
^{2}p_{1}^{2}\right) ^{2}}}, \\
n^{0} &=&\left\| \vec{n}\right\| \frac{\hat{n}^{1}p_{1}^{1}+\hat{n}%
^{2}p_{1}^{2}+\hat{n}^{2}p_{1}^{2}}{p_{1}^{0}},
\end{eqnarray*}
from which Eq. (\ref{spinfour}) follows immediately. Let us now find the
polarization vector that maximizes and minimizes the differential cross
section of single top production.

\subsection{The \textit{t}-channel}

We will begin with the \textit{t}-channel the was analyzed in the previous
chapter. Using Eq. (\ref{t-channeldecomp}) we define
\begin{eqnarray}
a_{n} &=&n\cdot a,\qquad b_{n}=n\cdot b,  \notag \\
c_{n} &=&n\cdot c,\qquad d_{n}=n\cdot d,  \label{ndecomp}
\end{eqnarray}
and using Lagrange multipliers $\lambda _{1}$ and $\lambda _{2}$ for
constraints (\ref{spinconstraints}) we maximize
\begin{equation*}
\sigma +\lambda _{1}\left( n^{2}+1\right) +\lambda _{2}n\cdot p_{1},
\end{equation*}
obtaining the equations
\begin{eqnarray}
n &=&-\frac{\beta }{2\lambda _{1}}f_{u}\left[ \left| g_{L}\right|
^{2}a+\left| g_{R}\right| ^{2}b+\frac{g_{R}^{\ast }g_{L}+g_{R}g_{L}^{\ast }}{%
2}c+i\frac{g_{L}^{\ast }g_{R}-g_{R}^{\ast }g_{L}}{2}d\right]  \notag \\
&&+\frac{\beta }{2\lambda _{1}}f_{\bar{d}}\left[ \left| g_{R}\right|
^{2}a+\left| g_{L}\right| ^{2}b+\frac{g_{R}^{\ast }g_{L}+g_{R}g_{L}^{\ast }}{%
2}c+i\frac{g_{L}^{\ast }g_{R}-g_{R}^{\ast }g_{L}}{2}d\right] -\frac{\lambda
_{2}}{2\lambda _{1}}p_{1},  \label{seq1} \\
0 &=&n^{2}+1,  \label{seq2} \\
0 &=&n\cdot p_{1},  \label{seq3}
\end{eqnarray}
and thus using Eqs. (\ref{seq1}) and (\ref{seq3})
\begin{eqnarray}
\lambda _{2} &=&-\frac{\beta }{m_{t}^{2}}f_{u}\left[ \left| g_{L}\right|
^{2}a\cdot p_{1}+\left| g_{R}\right| ^{2}b\cdot p_{1}+\frac{g_{R}^{\ast
}g_{L}+g_{R}g_{L}^{\ast }}{2}c\cdot p_{1}+i\frac{g_{L}^{\ast
}g_{R}-g_{R}^{\ast }g_{L}}{2}d\cdot p_{1}\right]  \notag \\
&&+\frac{\beta }{m_{t}^{2}}f_{\bar{d}}\left[ \left| g_{R}\right| ^{2}a\cdot
p_{1}+\left| g_{L}\right| ^{2}b\cdot p_{1}+\frac{g_{R}^{\ast
}g_{L}+g_{R}g_{L}^{\ast }}{2}c\cdot p_{1}+i\frac{g_{L}^{\ast
}g_{R}-g_{R}^{\ast }g_{L}}{2}d\cdot p_{1}\right] ,  \notag
\end{eqnarray}
and therefore
\begin{eqnarray*}
n &=&\frac{\beta }{2\lambda _{1}}\left\{ \left( f_{u}\left| g_{L}\right|
^{2}-f_{\bar{d}}\left| g_{R}\right| ^{2}\right) \left( \frac{a\cdot p_{1}}{%
m_{t}^{2}}p_{1}-a\right) +\left( f_{u}\left| g_{R}\right| ^{2}-f_{\bar{d}%
}\left| g_{L}\right| ^{2}\right) \left( \frac{b\cdot p_{1}}{m_{t}^{2}}%
p_{1}-b\right) \right. \\
&&+\left. \frac{g_{R}^{\ast }g_{L}+g_{R}g_{L}^{\ast }}{2}\left( f_{u}-f_{%
\bar{d}}\right) \left( \frac{c\cdot p_{1}}{m_{t}^{2}}p_{1}-c\right) +i\frac{%
g_{L}^{\ast }g_{R}-g_{R}^{\ast }g_{L}}{2}\left( f_{u}-f_{\bar{d}}\right)
\left( \frac{d\cdot p_{1}}{m_{t}^{2}}p_{1}-d\right) \right\} ,
\end{eqnarray*}
with the normalization factor $\lambda _{1}$ given by Eq. (\ref{seq2}). Note
that in the idealized case $f_{u}=f_{\bar{d}}=f$ we obtain
\begin{equation*}
n=\alpha \left\{ \frac{\left( a-b\right) \cdot p_{1}}{m_{t}^{2}}p_{1}-\left(
a-b\right) \right\} ,
\end{equation*}
where $\alpha $ is the normalization constant that does not depend on $f$ or
the effective couplings. In the SM ($g_{R}=0$) we obtain
\begin{equation*}
n=\alpha \left( f_{u}\left( \frac{a\cdot p_{1}}{m_{t}^{2}}p_{1}-a\right) -f_{%
\bar{d}}\left( \frac{b\cdot p_{1}}{m_{t}^{2}}p_{1}-b\right) \right) ,
\end{equation*}
where $\alpha $ is a normalizing factor.

\subsection{The \textit{s}-channel}

The \textit{s}-channel differential cross section has the form
\begin{eqnarray*}
d\sigma &=&\beta \left( f_{u}f_{\bar{d}}+f_{c}f_{\bar{s}}\right) \left[
\left| g_{L}\right| ^{2}\left( a_{s}+a_{n}\right) +\left| g_{R}\right|
^{2}\left( b_{s}+b_{n}\right) \right. \\
&&+\left. \frac{g_{R}^{\ast }g_{L}+g_{R}g_{L}^{\ast }}{2}\left(
c_{s}+c_{n}\right) +i\frac{g_{L}^{\ast }g_{R}-g_{R}^{\ast }g_{L}}{2}d_{n}%
\right] ,
\end{eqnarray*}
where again $\beta $ is a proportionality incorporating the kinematics, and
where $f_{u,c}$ and $f_{\bar{d},\bar{s}}$ denote the parton distribution
functions corresponding to extracting a $u,c$-type quarks and a $\bar{d},%
\bar{s}$-type quarks respectively. Using again the decomposition (\ref
{ndecomp}) and proceeding analogously to the \textit{t}-channel calculation
we obtain
\begin{eqnarray}
n &=&\alpha \left\{ \left| g_{L}\right| ^{2}\left( \frac{a\cdot p_{1}}{%
m_{t}^{2}}p_{1}-a\right) +\left| g_{R}\right| ^{2}\left( \frac{b\cdot p_{1}}{%
m_{t}^{2}}p_{1}-b\right) \right.  \notag \\
&&+\left. \frac{g_{R}^{\ast }g_{L}+g_{R}g_{L}^{\ast }}{2}\left( \frac{c\cdot
p_{1}}{m_{t}^{2}}p_{1}-c\right) +i\frac{g_{L}^{\ast }g_{R}-g_{R}^{\ast }g_{L}%
}{2}\left( \frac{d\cdot p_{1}}{m_{t}^{2}}p_{1}-d\right) \right\} ,
\label{nschan}
\end{eqnarray}
where $\alpha $ is the normalizing factor that in this case (compared to the
\textit{t}-channel result) does not depend on the PDF's. From Eq. (\ref
{schanproduction}) we obtain
\begin{eqnarray*}
a^{\mu } &=&-m_{t}q_{2}^{\mu }\left( q_{1}\cdot \tilde{p}_{2}\right) , \\
b^{\mu } &=&+m_{t}q_{1}^{\mu }\left( q_{2}\cdot \tilde{p}_{2}\right) , \\
c^{\mu } &=&+m_{b}\left( q_{1}^{\mu }\left( q_{2}\cdot p_{1}\right)
-q_{2}^{\mu }\left( q_{1}\cdot p_{1}\right) \right) , \\
d^{\mu } &=&-m_{b}\varepsilon _{~\alpha \rho \sigma }^{\mu }p_{1}^{\alpha
}q_{1}^{\rho }q_{2}^{\sigma },
\end{eqnarray*}
hence replacing in Eq. (\ref{nschan}) we arrive at
\begin{eqnarray}
n^{\mu } &=&\alpha \left\{ \left| g_{L}\right| ^{2}\left( \left( q_{1}\cdot
\tilde{p}_{2}\right) \left( q_{2}\cdot p_{1}\right) p_{1}^{\mu }-\left(
q_{1}\cdot \tilde{p}_{2}\right) m_{t}^{2}q_{2}^{\mu }\right) \right.  \notag
\\
&&+\left| g_{R}\right| ^{2}\left( \left( q_{2}\cdot \tilde{p}_{2}\right)
\left( q_{1}\cdot p_{1}\right) p_{1}^{\mu }-\left( q_{2}\cdot \tilde{p}%
_{2}\right) m_{t}^{2}q_{1}^{\mu }\right)  \notag \\
&&+m_{b}m_{t}\frac{g_{R}^{\ast }g_{L}+g_{R}g_{L}^{\ast }}{2}\left(
q_{1}^{\mu }\left( q_{2}\cdot p_{1}\right) -q_{2}^{\mu }\left( q_{1}\cdot
p_{1}\right) \right)  \notag \\
&&+\left. i\frac{g_{R}^{\ast }g_{L}-g_{L}^{\ast }g_{R}}{2}%
m_{b}m_{t}\varepsilon _{~~\alpha \rho \sigma }^{\mu }p_{1}^{\alpha
}q_{1}^{\rho }q_{2}^{\sigma }\right\} ,  \label{finalbasis}
\end{eqnarray}
which is the basis we use in our numerical simulations. If we neglect $g_{R}$
we obtain
\begin{equation*}
n^{\mu }=\pm \frac{\left( q_{1}\cdot \tilde{p}_{2}\right) \left( q_{2}\cdot
p_{1}\right) p_{1}^{\mu }-\left( q_{1}\cdot \tilde{p}_{2}\right)
m_{t}^{2}q_{2}^{\mu }}{\sqrt{\left( q_{1}\cdot \tilde{p}_{2}\right)
^{2}\left( q_{2}\cdot p_{1}\right) ^{2}m_{t}^{2}-\left( q_{1}\cdot \tilde{p}%
_{2}\right) ^{2}m_{t}^{4}q_{2}^{2}}},
\end{equation*}
where we have included the normalization factor and since $q_{2}^{2}=0$ the
above reduces to
\begin{equation*}
m_{t}n=\pm \left( \frac{m_{t}^{2}}{\left( q_{2}\cdot p_{1}\right) }%
q_{2}-p_{1}\right) ,
\end{equation*}
which is the result we have quoted in the previous section coinciding with
\cite{mandp}

\section{Numerical results}

To calculate the differential cross section corresponding to the \textit{s}%
-channel we employ a set of cuts that are compatible with the ones used in
the \textit{t}-channel. Since in the previous chapter top decay was not
considered, the equivalence is only approximate and a more detailed
phenomenological analysis will be required in due course. The present study
should however suffice to identify the most promising observables. The
allowed kinematical regions we shall employ are

\begin{eqnarray}
\mathrm{detector~geometry~cuts} &:&10^{o}\leq \theta _{i}\leq 170^{o},\quad
i=b,\bar{b},l,  \notag \\
\mathrm{isolation~cuts} &:&20^{o}\leq \theta _{ij},\quad i,j=b,\bar{b},l,
\notag \\
\mathrm{theoretical~cuts} &:&20~\mathrm{GeV}\leq p_{b}^{T},\quad 20~\mathrm{%
GeV}\leq p_{\bar{b}}^{T},
\end{eqnarray}

The details concerning luminosity, parton distribution functions, $Q^{2}$
dependence and so on have already been presented in Chapter \ref
{LHCphenomenology}.

The more salient results of the present analysis for the \textit{s}-channel
top production can be seen in Figs. (\ref{cosptnpl_gl=1_gr=+-5.d-2_ss}-\ref
{cosptnpb_gl=1_gr=+-5.d-2_ss}). We present two types of graphs. The first
type involves the anti-lepton plus bottom invariant mass. In the hadronic
decays of the top a full reconstruction of the top mass would be feasible.

\begin{figure}[!hbp]
\begin{center}
\includegraphics[width=12cm]{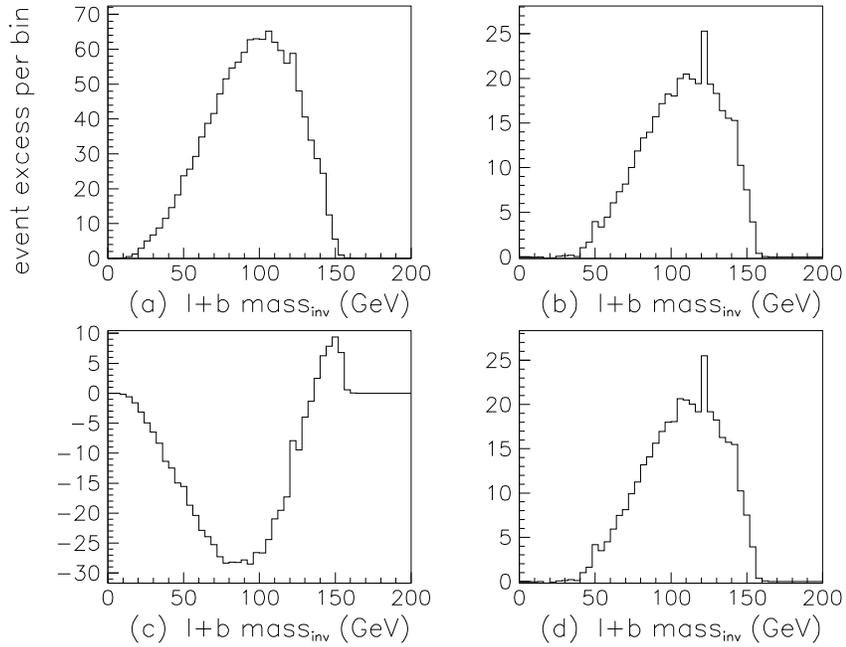}
\end{center}
\par
\caption{Event production difference between non-vanishing $g_{R}$ coupling
caculations and the tree level SM ones ($g_{R}=0$). Differences are plotted
versus the invariant mass of the lepton (electron or muon) plus bottom
system arising in top decay from single top production at the LHC. We have
taken $g_{R}=+5\times 10^{-2}$, $+i5\times 10^{-2},$ $-5\times 10^{-2}$ and $%
-i5\times 10^{-2}$ in plots (a), (b), (c) and (d) respectively. Calculation
are performed at $\protect\mu ^{2}=\hat{s}=\left( q_{1}+q_{2}\right) ^{2}$.}
\label{plpbmvv_gl=1_gr=+-5.d-2_dif}
\end{figure}

We have found that the anti-lepton plus bottom invariant mass distribution
is sensitive to $g_{R}$. Figs. (\ref{plpbmvv_gl=1_gr=+-5.d-2_dif}) and (\ref
{plpbmvv_gl=1_gr=+-5.d-2_ss}) reflect this sensitivity with the second
figure showing the statistical significance per bin. The other set of graphs
corresponds to various $p_{T}$ and angular distributions of the final
particles. The sensitivity to departures from the tree level SM is shown in
Figs. (\ref{pabtrans_gl=1_gr=+-5.d-2_ss}), (\ref{pbtrans_gl=1_gr=+-5.d-2_ss}%
) and (\ref{pltrans_gl=1_gr=+-5.d-2_ss}). We also include the statistical
significance per bin for the signal vs $\cos \left( \theta _{tl}\right) $ in
Fig. (\ref{cosptnpl_gl=1_gr=+-5.d-2_ss}) and vs $\cos \left( \theta
_{tb}\right) $ in Fig. (\ref{cosptnpb_gl=1_gr=+-5.d-2_ss}). $\cos \left(
\theta _{tl}\right) $ and $\cos \left( \theta _{tb}\right) $ are the cosines
of the angle between the best reconstruction of top momentum and the momenta
of anti-lepton and bottom, respectively. In these figures we can clearly see
that low angles corresponds to bigger sensitivities. This is in qualitative
accordance with Eq. (\ref{pol1}) which tells us that anti-leptons are
predominantly produced in the direction of the top spin and therefore most
of those produced predominantly in the top direction come from a top mainly
polarized in a positive helicity state. Thus the quantity of those
anti-leptons is more sensitive to variations in $g_{R}.$ Even though this
argument applies in the top rest frame, the fact that most of the kinematics
lies in the beam direction makes it valid at least for this kinematics. With
the cuts considered here, the SM prediction at tree level for the total
number of events at LHC with one year full luminosity (100 $\mathrm{fb}^{-1}$%
) is $180700$ (with a 20\% error due to theoretical uncertainties). Using
the values $g_{L}=1,$ $g_{R}=+5\times 10^{-2}$ leads to an excess of $1220$
events which corresponds to a $2.87$ standard deviations signal. The $%
g_{L}=1,$ $g_{R}=-5\times 10^{-2}$ model has a deficit of $480$ events which
corresponds to a $1.13$ standard deviations signal. Finally the $g_{L}=1,$ $%
g_{R}=\pm i5\times 10^{-2}$ model has an excess of $367$ events which
corresponds to a $0.86$ standard deviations. We see that there is a large
dependence on the phase of $g_{R}.$

The implementation of careful selected cuts or an accurate $\chi ^{2}$ test
can improve those statistical significances but since here we are interested
in an order of magnitude estimate we will not enter into such analysis here.
Moreover since backgrounds are bound to worsen the sensitivity the above
results must be taken as order of magnitude estimates only. A more detailed
analysis goes beyond the scope of this chapter.

\newpage

\begin{figure}[!htp]
\begin{center}
\includegraphics[width=8cm]{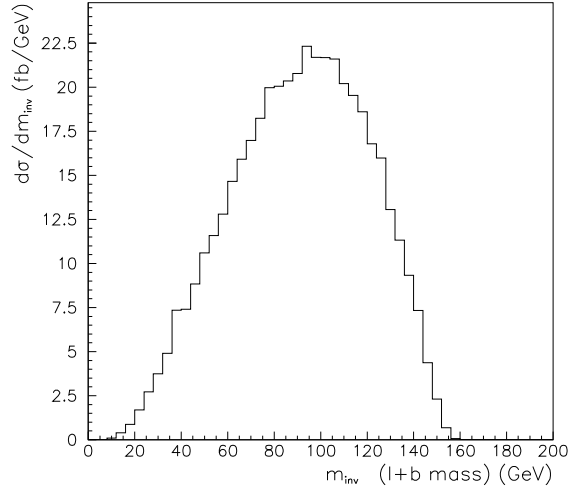}
\end{center}
\par
\caption{Distribution of the invariant mass of the lepton (electron or muon)
plus bottom system arising in top decay from single top production at the
LHC. The calculation was performed at the tree level in Standard Model with $%
\protect\mu ^{2}=\hat{s}=\left( q_{1}+q_{2}\right) ^{2}$.}
\label{plpbmvv_gl=1_gr=0}
\end{figure}

\begin{figure}[!hbp]
\begin{center}
\includegraphics[width=\figwidth]{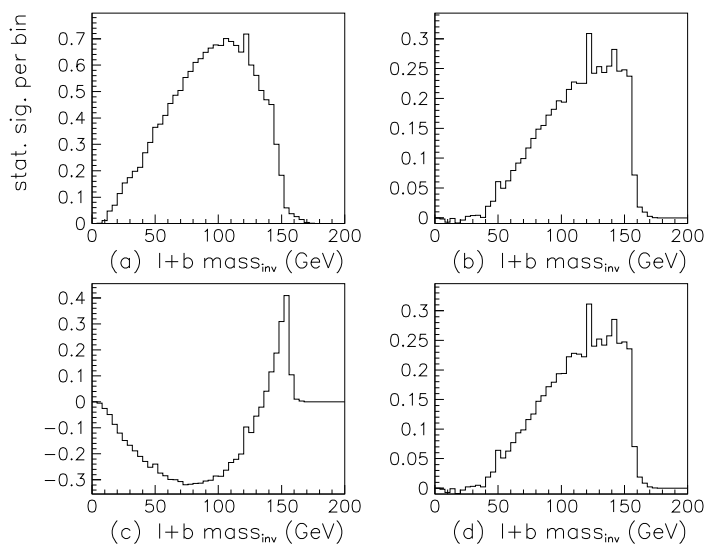}
\end{center}
\par
\caption{Plots corresponding to differences (a), (b) (c) and (d) of Fig. (%
\ref{plpbmvv_gl=1_gr=+-5.d-2_dif}) divided by the square root of the event
production per bin at LHC. The square of the quotient denominator can be
obtained from Fig. (\ref{plpbmvv_gl=1_gr=0}) multiplying $d\protect\sigma
/dm_{inv}$ by the LHC 1-year full luminosity (100 $\mathrm{fb}^{-1}$) and by
the width of each bin (4 GeV. in Fig. (\ref{plpbmvv_gl=1_gr=0})). Taking the
modulus of the above plots we obtain the statistical significance of the
corresponding signals per bin. Note that statistical significance has a
strong and non-linear dependence both on the invariant mass and the right
coupling $g_{R.}$ However purely imaginary couplings are almost insensible
to their sign.}
\label{plpbmvv_gl=1_gr=+-5.d-2_ss}
\end{figure}
\begin{figure}[!hbp]
\begin{center}
\includegraphics[width=\figwidth]{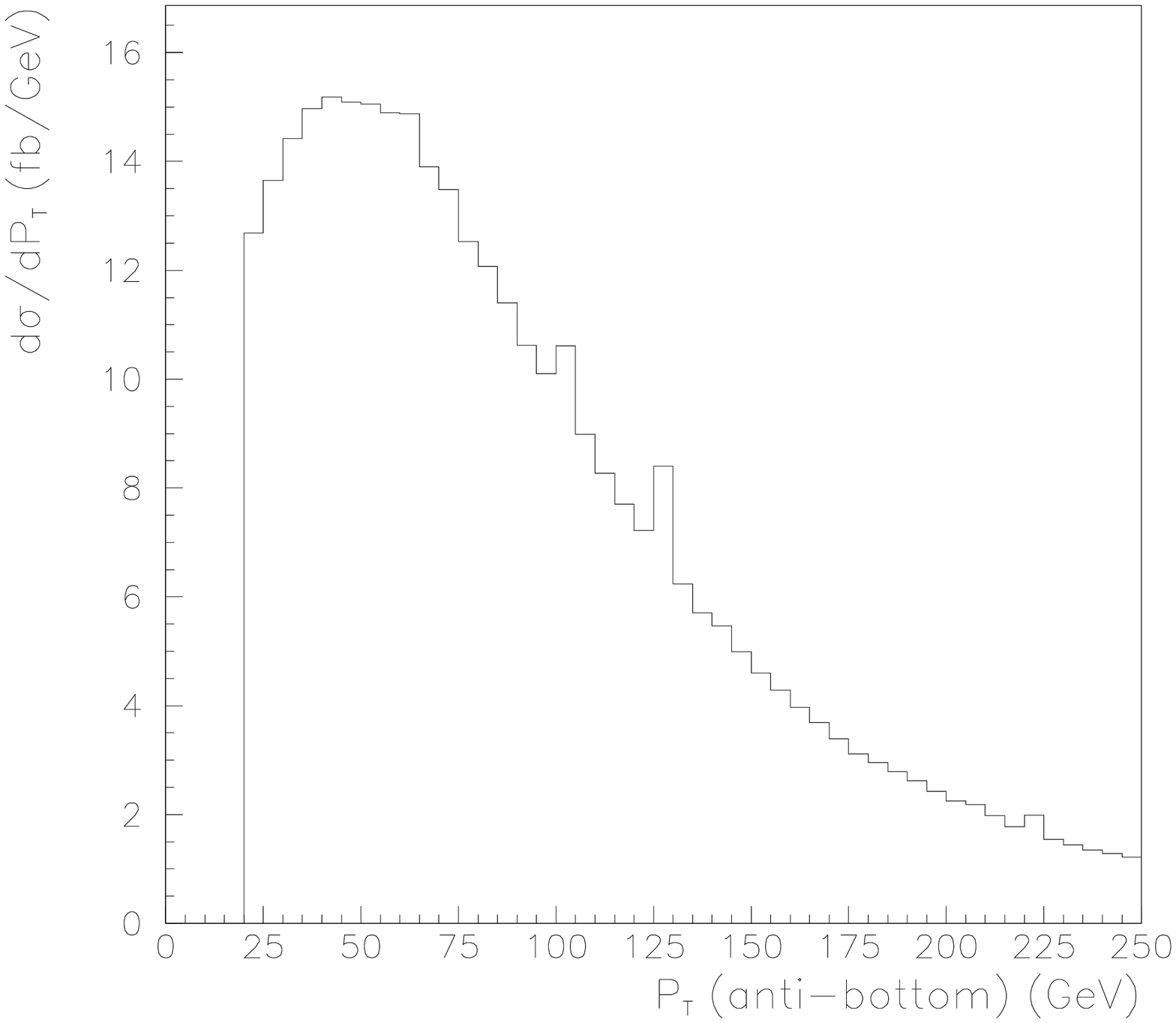}
\end{center}
\par
\caption{Anti-bottom transversal momentum distribution corresponding to
single top production at the LHC. The calculation has been performed at tree
level in the SM ($g_{L}=1$, $g_{R}=0$). }
\label{pabtrans_gl=1_gr=0}
\end{figure}
\begin{figure}[!hbp]
\begin{center}
\includegraphics[width=\figwidth]{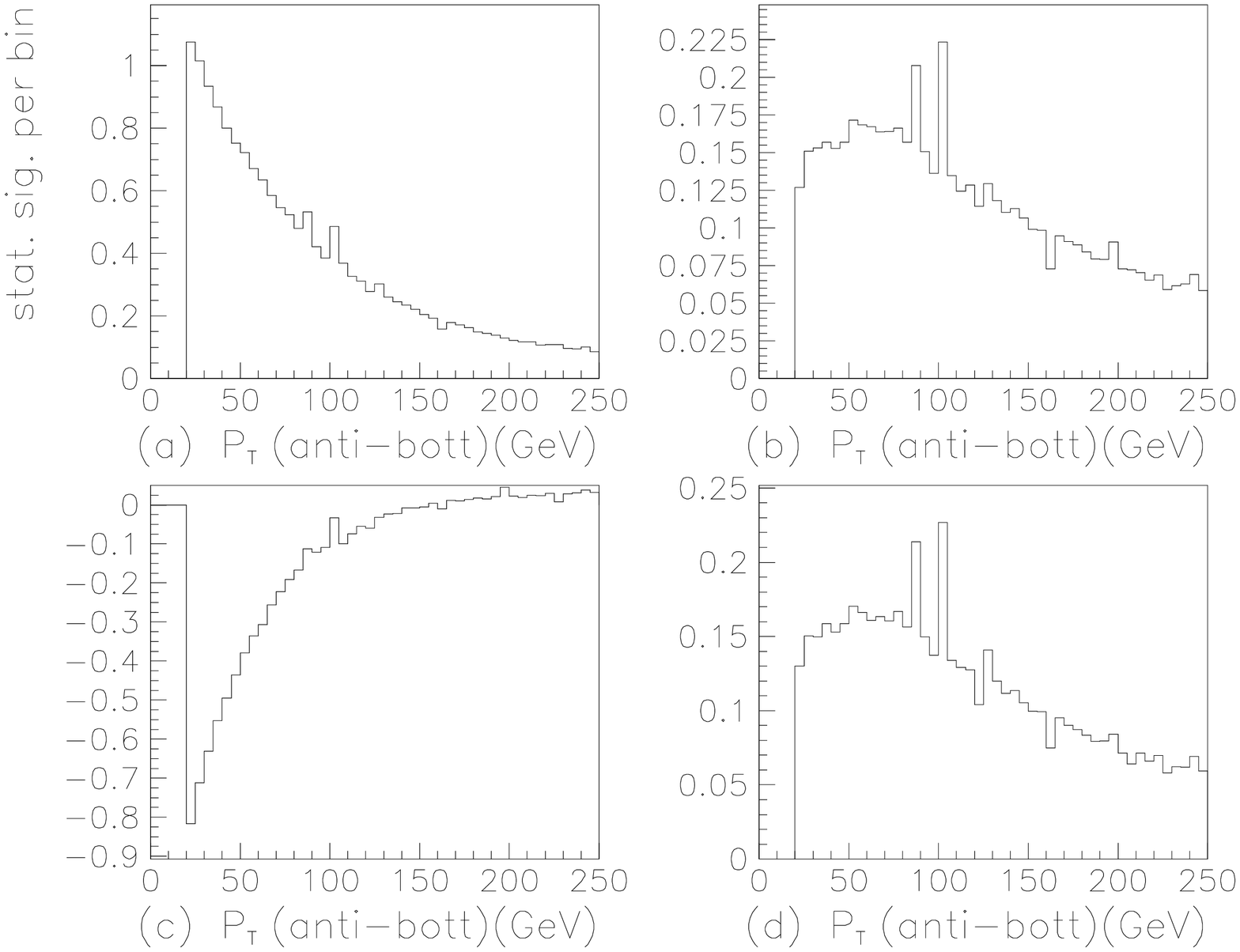}
\end{center}
\par
\caption{Taking the modulus of the above plots we obtain the statistical
significance of the corresponding signals per bin with respect to
anti-bottom transversal momentum. Like in Fig. (\ref
{plpbmvv_gl=1_gr=+-5.d-2_ss}) we have taken $g_{R}=+5\times 10^{-2}$, $%
+i5\times 10^{-2},$ $-5\times 10^{-2}$ and $-i5\times 10^{-2}$ in plots (a),
(b), (c) and (d) respectively. Note that here statistical significance has a
strong dependence on the anti-bottom transversal momentum but is almost
linear on $\emph{Re}\left( g_{R}\right) $ and almost insensible to the sign
of $\emph{Im}\left( g_{R}\right) $. }
\label{pabtrans_gl=1_gr=+-5.d-2_ss}
\end{figure}
\begin{figure}[!hbp]
\begin{center}
\includegraphics[width=\figwidth]{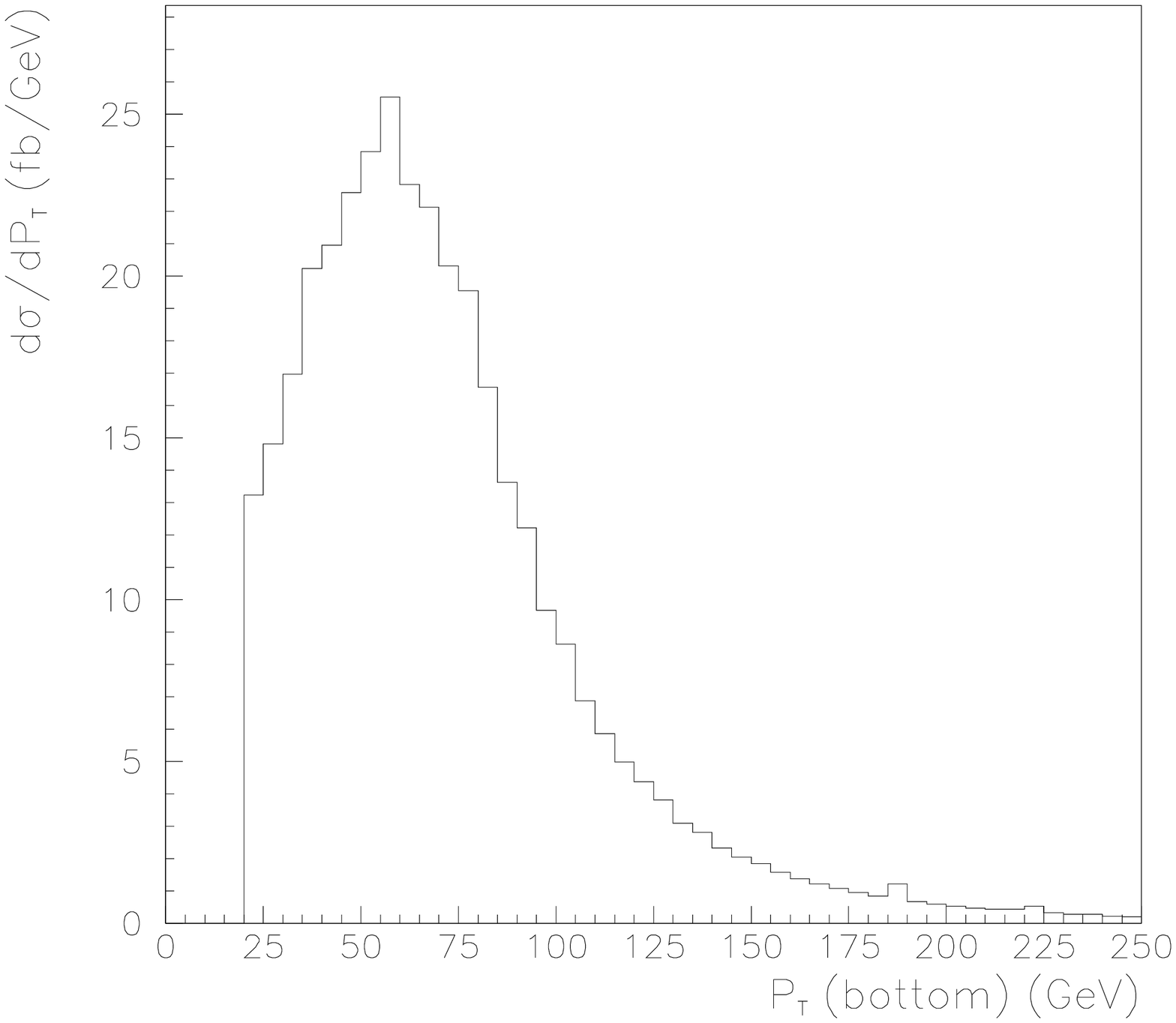}
\end{center}
\par
\caption{Bottom transversal momentum distribution corresponding to single
top production at the LHC. The calculation has been performed at tree level
in the SM ($g_{L}=1$, $g_{R}=0$). }
\label{pbtrans_gl=1_gr=0}
\end{figure}
\begin{figure}[!hbp]
\begin{center}
\includegraphics[width=\figwidth]{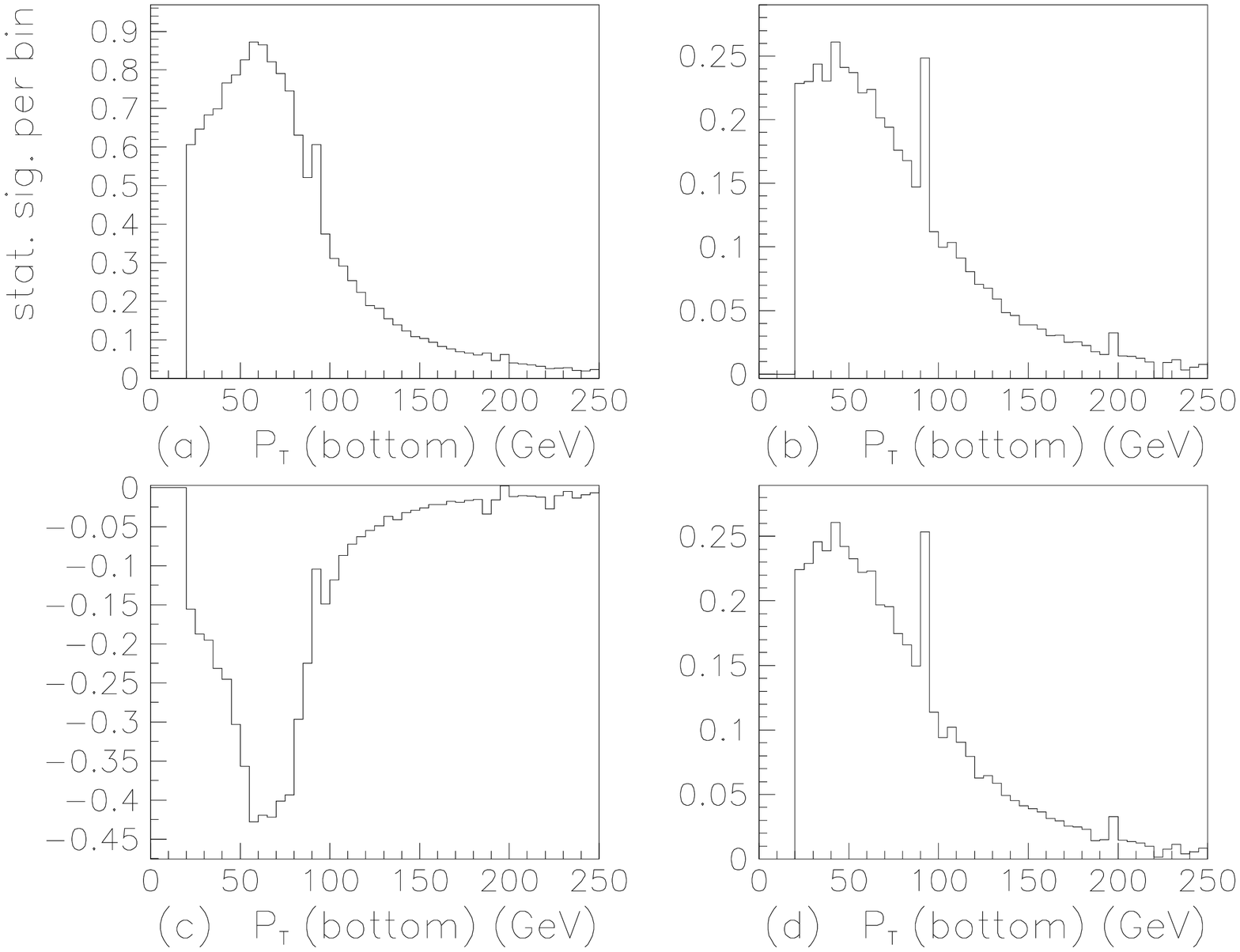}
\end{center}
\par
\caption{Taking the modulus of the above plots we obtain the statistical
significance of the corresponding signals per bin with respect to bottom
transversal momentum. Like in Fig. (\ref{plpbmvv_gl=1_gr=+-5.d-2_ss}) we
have taken $g_{R}=+5\times 10^{-2}$, $+i5\times 10^{-2},$ $-5\times 10^{-2}$
and $-i5\times 10^{-2}$ in plots (a), (b), (c) and (d) respectively. Note
that here statistical significance has a strong dependence on the bottom
transversal momentum and clearly favors positive values of $\emph{Re}\left(
g_{R}\right) $ and again is insensible to the sign of $\emph{Im}\left(
g_{R}\right) $. }
\label{pbtrans_gl=1_gr=+-5.d-2_ss}
\end{figure}
\begin{figure}[!hbp]
\begin{center}
\includegraphics[width=\figwidth]{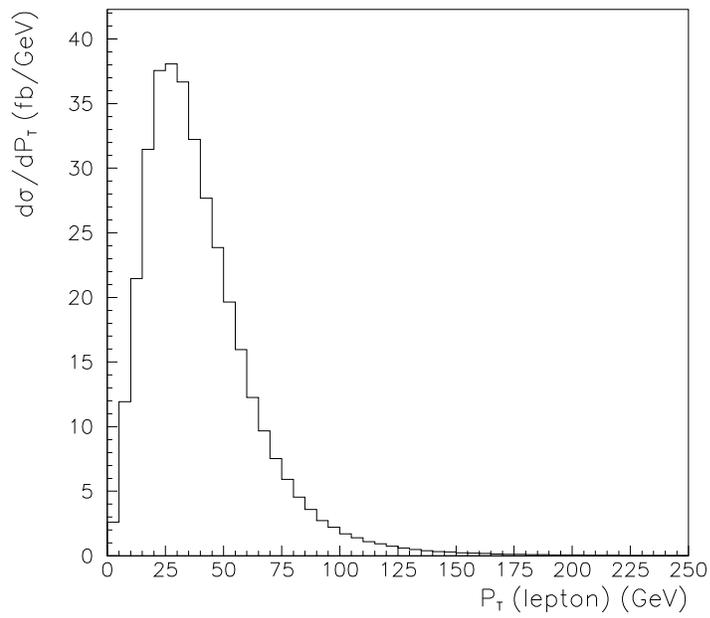}
\end{center}
\par
\caption{Lepton (electron or muon) transversal momentum distribution
corresponding to single top production at the LHC. The calculation has been
performed at tree level in the SM ($g_{L}=1$, $g_{R}=0$). }
\label{pltrans_gl=1_gr=0}
\end{figure}
\begin{figure}[!hbp]
\begin{center}
\includegraphics[width=\figwidth]{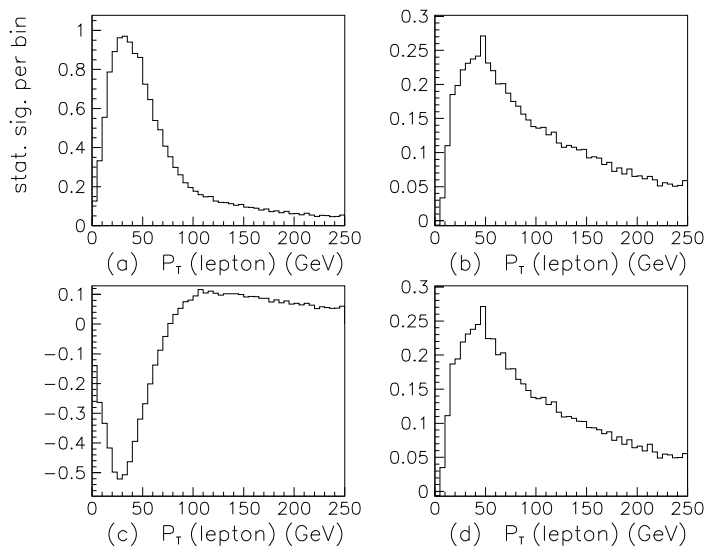}
\end{center}
\par
\caption{Taking the modulus of the above plots we obtain the statistical
significance of the corresponding signals per bin with respect to lepton
(electron or muon) transversal momentum. Like in Fig. (\ref
{plpbmvv_gl=1_gr=+-5.d-2_ss}) we have taken $g_{R}=+5\times 10^{-2}$, $%
+i5\times 10^{-2},$ $-5\times 10^{-2}$ and $-i5\times 10^{-2}$ in plots (a),
(b), (c) and (d) respectively. Note that again statistical significance has
a strong dependence on the lepton transversal momentum and clearly favors
positive values of $\emph{Re}\left( g_{R}\right) .$ The sign of $\emph{Im}%
\left( g_{R}\right) $ cannot be distinguished.}
\label{pltrans_gl=1_gr=+-5.d-2_ss}
\end{figure}
\begin{figure}[!hbp]
\begin{center}
\includegraphics[width=\figwidth]{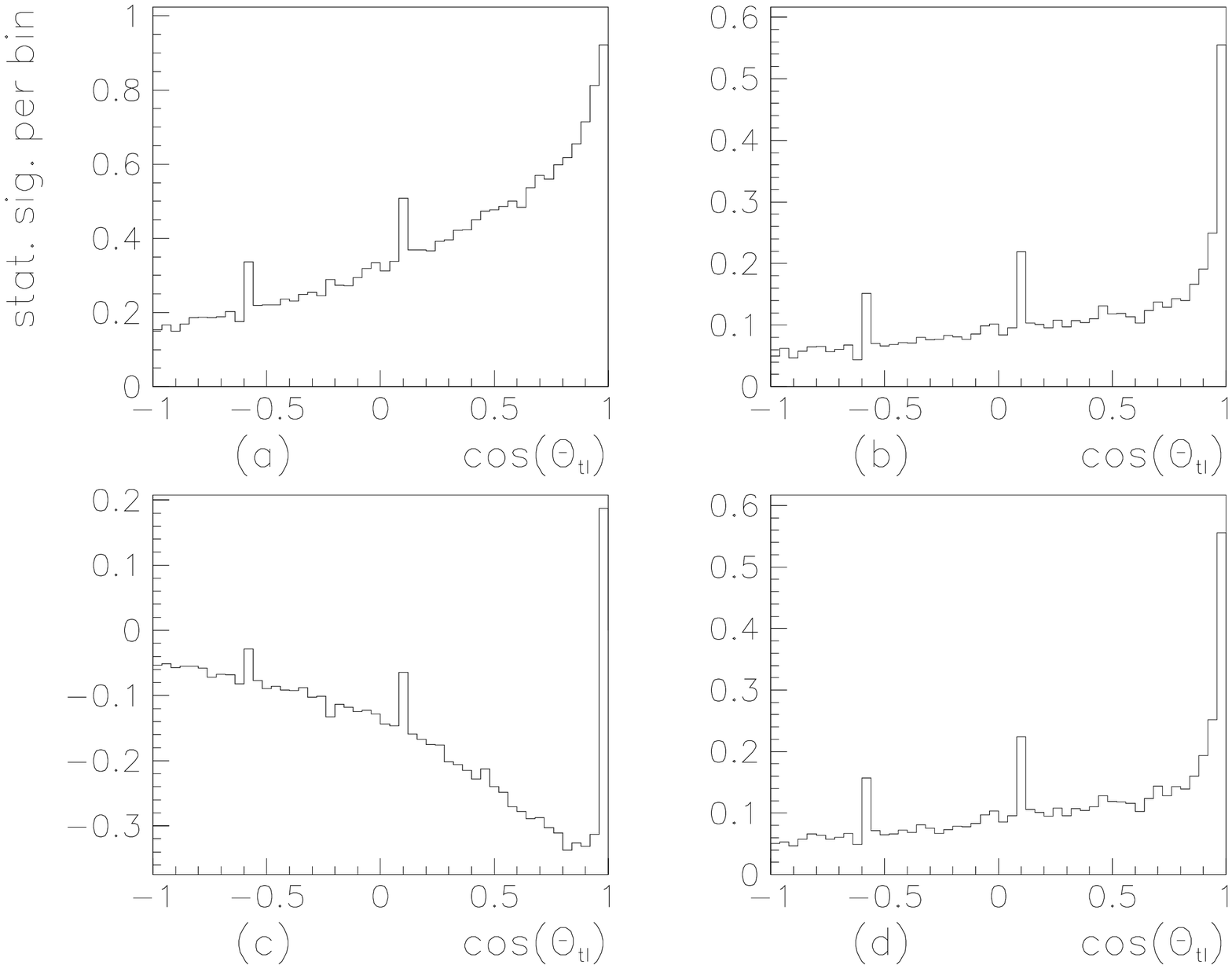}
\end{center}
\par
\caption{Taking the modulus of the above plots we obtain the statistical
significance of the corresponding signals per bin with respect to $\cos
\left( \protect\theta _{tl}\right) =$ $\vec{p}_{l}\cdot \left( \vec{p}_{l}+%
\vec{p}_{b}\right) $ $/$ $\left| \vec{p}_{l}\right| \left| \vec{p}_{l}+\vec{p%
}_{b}\right| $ where $\vec{p}_{l}$ and $\vec{p}_{b}$ are respectively the
tree momenta of the lepton (positron or anti-muon) and bottom. The
combination $\vec{p}_{l}+\vec{p}_{b}$ is the best experimental
reconstruction of the top momemtum provided the neutrino information is
lost. Like in Fig. (\ref{plpbmvv_gl=1_gr=+-5.d-2_ss}) we have taken $%
g_{R}=+5\times 10^{-2}$, $+i5\times 10^{-2},$ $-5\times 10^{-2}$ and $%
-i5\times 10^{-2}$ in plots (a), (b), (c) and (d) respectively. Note that
again statistical significance has a strong dependence on $\cos \left(
\protect\theta _{tl}\right) $.}
\label{cosptnpl_gl=1_gr=+-5.d-2_ss}
\end{figure}
\begin{figure}[!hbp]
\begin{center}
\includegraphics[width=\figwidth]{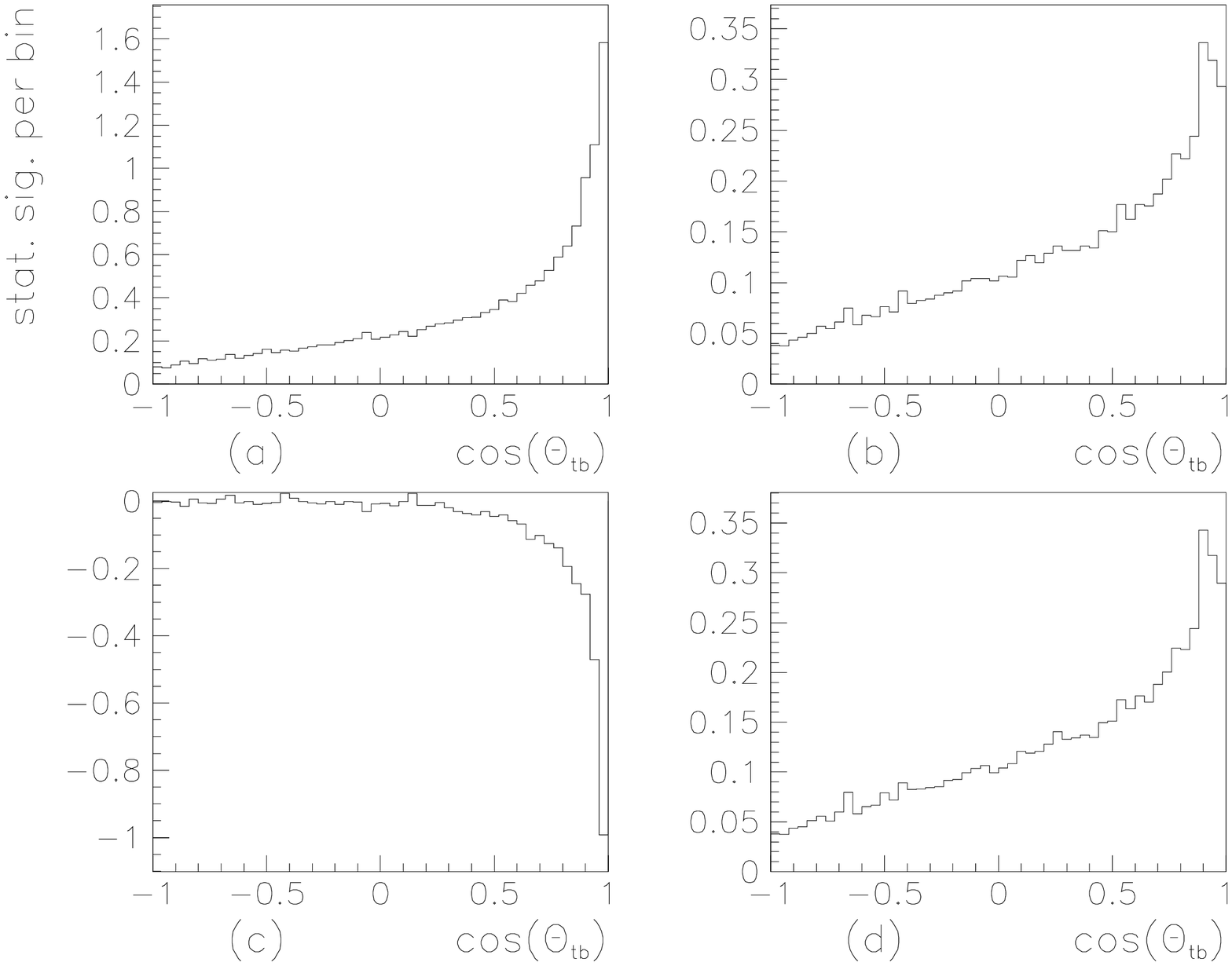}
\end{center}
\par
\caption{Taking the modulus of the above plots we obtain the statistical
significance of the corresponding signals per bin with respect to $\cos
\left( \protect\theta _{tb}\right) =$ $\vec{p}_{b}\cdot \left( \vec{p}_{l}+%
\vec{p}_{b}\right) $ $/$ $\left| \vec{p}_{l}\right| \left| \vec{p}_{l}+\vec{p%
}_{b}\right| $ where $\vec{p}_{l}$ and $\vec{p}_{b}$ are respectively the
tree momenta of the lepton (positron or anti-muon) and bottom. The
combination $\vec{p}_{l}+\vec{p}_{b}$ is the best experimental
reconstruction of the top momemtum provided the neutrino information is
lost. Like in Fig. (\ref{plpbmvv_gl=1_gr=+-5.d-2_ss}) we have taken $%
g_{R}=+5\times 10^{-2}$, $+i5\times 10^{-2},$ $-5\times 10^{-2}$ and $%
-i5\times 10^{-2}$ in plots (a), (b), (c) and (d) respectively. Note that
again statistical significance has a strong dependence on $\cos \left(
\protect\theta _{tb}\right) $.}
\label{cosptnpb_gl=1_gr=+-5.d-2_ss}
\end{figure}

\begin{figure}[!hbp]
\begin{center}
\includegraphics[width=\figwidth]{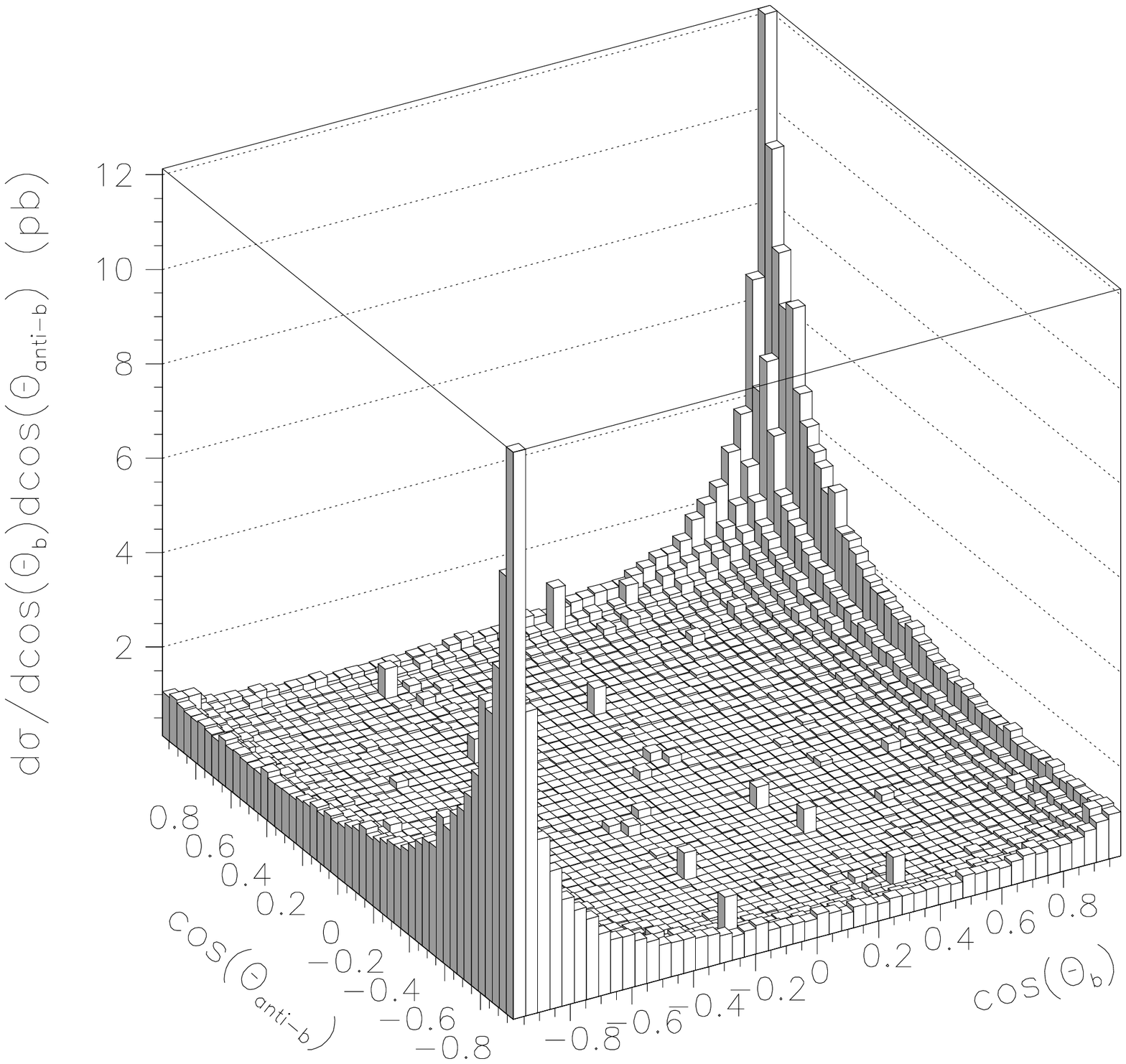}
\end{center}
\par
\caption{Distribution of the cosines of the polar angles of the botom and
anti-bottom with respect to the beam line. The plot corresponds to single
top production at the LHC with top decay included. The calculation was
performed at the tree level in Standard Model with $\protect\mu ^{2}=\hat{s}%
=\left( q_{1}+q_{2}\right) ^{2}$.}
\label{cosbcosab_gl=1_gr=0}
\end{figure}

\section{Conclusions}

In this chapter we have performed a full analysis of the sensitivity of
single top production in the \textit{s}-channel to the presence of anomalous
couplings in the effective electroweak theory. The analysis has been done in
the context of the LHC experiments.

Unlike in the discussion concerning the single top production through the
dominant \textit{t}-channel, top decay has been considered. The only
approximation involved is to consider the top as a real particle (narrow
width approximation).

We have paid careful attention to the issue of the top polarization. We have
argued, first of all, why it is not unjustified to neglect the interference
term and to proceed as if the top spin was determined at an intermediate
stage. We have provided a spin basis where the interference term is
minimized. A similar analysis applies to the \textit{t}-channel process. We
present here and explicit basis for this case too. We get a sensitivity to $%
g_{R}$ in the same ballpark as the one obtained in the \textit{t}-channel
(where decay was not considered). Finally we have obtained that observables
most sensible to $g_{R}$ are those where anti-lepton and bottom momenta are
cut to be almost collinear.

\chapter{Results and Conclusions}

Here we present a summary of the main results obtained in this thesis.

\begin{itemize}
\item  In Chapter \ref{mattersectorchapter}:

\begin{itemize}
\item  We present a complete classification of four-fermion operators giving
mass to physical fermions and gauge vector bosons in models of dynamical
symmetry breaking. This is done when new particles appear in the usual
representations of the $SU(2)_{L}\times SU(3)_{c}$ group, and a partial
classification it is done in the general case also. Only a single family is
considered and therefore the problem of mixing is not addressed here.

\item  We investigate the phenomenological consequences for the electroweak
neutral sector of such class of models. This is done matching the
four-fermion description to a lower energy theory that only contain all
degrees of freedom of the SM (but the Higgs). The coefficients of such low
energy effective Lagrangian for dynamical symmetry breaking models are then
compared to those of models with elementary scalars (such as the minimal
Standard Model).

\item  We determine the value of the $Zb\bar{b}$ effective coupling in
models of dynamical symmetry breaking and verify that the contribution is
large, but its sign is not defined, contrary to some claims. The current
value of this coupling is off the SM value by nearly a 3 $\sigma $ effect.
We estimate the effects for light fermions too, where they are not
observable at present. Some general observations concerning the mechanism of
dynamical symmetry breaking are presented.
\end{itemize}

\item  In Chapter \ref{cpviolationandmixing}:

\begin{itemize}
\item  We analyze the structure of the four-dimensional effective operators
in the electroweak matter sector when $CP$ violations and family mixing is
allowed.

\item  We perform the diagonalization of the mass and kinetic terms showing
that, besides the presence of the CKM matrix in the SM charged vertex, new
structures show up in the effective operators constructed with left handed
fermions. In particular the CKM matrix is also present in the neutral sector.

\item  We calculate also the contribution to the effective operators in the
minimal SM with a heavy Higgs and in the SM supplemented with an additional
heavy fermion doublet.

\item  In general, even if the physics responsible for the generation of the
additional effective operators is $CP$-conserving, phases which are present
in the Yukawa and kinetic couplings become observable in the effective
operators after diagonalization.
\end{itemize}

\item  In Chapter \ref{LSZchapter}:

\begin{itemize}
\item  We present and solve the issue of defining a 1-loop set of wfr.
constants consistent with the on-shell requirements and the gauge invariance
of physical amplitudes. We demonstrate using Nielsen identities that our set
of wfr. constants together with a gauge independent CKM renormalization
yields gauge independent physical amplitudes for top and W decays.

\item  We show that the previous on-shell prescription given in \cite{Denner}
does not diagonalize the propagator in family space and yields gauge
dependent amplitudes for the charged electroweak vertex, albeit gauge
independent modulus. This is not satisfactory since interference with e.g.
strong phases may reveal an unacceptable gauge dependence. In the case of
top decay we find that the numerical difference in the squared amplitude
between our result and the one using the prescription in \cite{Denner}
amounts to a half per cent. This difference will be relevant to future
experiments testing the $tb$ vertex.

\item  We check the consistency of our scheme with the $CPT$ theorem. This
is done showing that although our wfr. constants do not verify the
pseudo-hermiticity condition ($\bar{Z}\neq \gamma ^{0}Z^{\dagger }\gamma
^{0} $) the total width of particles and anti-particles coincide.
\end{itemize}

\item  In Chapter \ref{LHCphenomenology}:

\begin{itemize}
\item  We present a complete calculation of the \textit{t}-channel cross
sections for polarized tops or anti-tops including right effective couplings
and bottom mass contributions.

\item  We perform a Monte Carlo simulation of the production of single
polarized tops at LHC presenting a variety of $p_{T}$ and angular
distributions both for the $t$ and the $\bar{b}$ quarks. We show, without
considering backgrounds or top decay, that we can expect a 2 $\sigma $
sensitivity to $g_{R}$ variations of the order of $5\times 10^{-2}.$

\item  We show based on general theoretical grounds that the top cannot be
produced in a pure spin state. Moreover we indicate which is the adequate
spin basis to correctly fold top production cross section with top decay.
This is necessary in order to calculate the whole process in the framework
of the narrow width approximation.
\end{itemize}

\item  In Chapter \ref{topdecaychapter}:

\begin{itemize}
\item  We present a complete calculation of the \textit{s}-channel cross
section for single top production including top decay. Calculations include
right effective couplings and bottom mass contributions.

\item  We perform a Monte Carlo simulation of the production and decay of
single polarized tops at LHC in the \textit{s}-channel. We plot several $%
p_{T}$, invariant mass and angular distributions constructed with observable
anti-lepton momentum and bottom and anti-bottom jets momenta. We find that
variations of $g_{R}$ of the order of $5\times 10^{-2}$ are visible with
signals ranging from $3$ to $1$ standard deviations depending on the phase
of $g_{R}$ and the observables selected.

\item  We present explicit expressions both for the \textit{t}- and \textit{s%
}- channels of the top spin basis that diagonalizes the top density matrix.
We check numerically for the \textit{s}-channel that such basis minimizes
the interference terms not taken into account in the narrow width
approximation.
\end{itemize}
\end{itemize}

\appendix

\chapter{Conventions and useful formulae.}

\label{appendixA}We use the metric $\eta _{\mu \nu }=\left(
1,-1,-1,-1\right) $ and the Dirac representation with
\begin{eqnarray*}
\gamma ^{0} &=&\left(
\begin{array}{cc}
I & 0 \\
0 & -I
\end{array}
\right) ,\quad \gamma ^{i}=\left(
\begin{array}{cc}
0 & \tau ^{i} \\
-\tau ^{i} & 0
\end{array}
\right) ,\quad \gamma ^{5}=i\gamma ^{0}\gamma ^{1}\gamma ^{2}\gamma
^{3}=\left(
\begin{array}{cc}
0 & I \\
I & 0
\end{array}
\right) , \\
\tau ^{1} &=&\left(
\begin{array}{ll}
0 & 1 \\
1 & 0
\end{array}
\right) ,\quad \tau ^{2}=\left(
\begin{array}{ll}
0 & -i \\
i & 0
\end{array}
\right) ,\quad \tau ^{3}=\left(
\begin{array}{ll}
1 & 0 \\
0 & -1
\end{array}
\right) , \\
\tau ^{+} &=&\frac{\tau ^{1}-i\tau ^{2}}{\sqrt{2}}=\left(
\begin{array}{ll}
0 & 0 \\
\sqrt{2} & 0
\end{array}
\right) ,\quad \tau ^{-}=\frac{\tau ^{1}+i\tau ^{2}}{\sqrt{2}}=\left(
\begin{array}{ll}
0 & \sqrt{2} \\
0 & 0
\end{array}
\right) .
\end{eqnarray*}
we also define the projectors
\begin{equation*}
P_{\pm }=\frac{1\pm \gamma ^{5}}{2},\tau ^{u}=\frac{I+\tau ^{3}}{2},\tau
^{d}=\frac{I-\tau ^{3}}{2},
\end{equation*}
where $P_{+}$ is the right projector ($R$), $P_{-}$ the left projector ($L$%
), $\tau ^{u}$ is the up projector and $\tau ^{d}$ the down projector
satisfying
\begin{eqnarray*}
\left( P_{\pm }\right) ^{2} &=&P_{\pm }, \\
P_{+}P_{-} &=&P_{-}P_{+}=0, \\
P_{+}+P_{-} &=&I, \\
\left( \tau ^{u}\right) ^{2} &=&\tau ^{u}, \\
\left( \tau ^{d}\right) ^{2} &=&\tau ^{d}, \\
\tau ^{u}\tau ^{d} &=&\tau ^{d}\tau ^{u}=0, \\
\tau ^{u}+\tau ^{d} &=&I.
\end{eqnarray*}
Let us write the matrices
\begin{equation*}
G_{L}=e^{i\mathbf{\theta \cdot }\frac{\mathbf{\tau }}{2}},\qquad
G_{R}=e^{i\beta \frac{\tau ^{3}}{2}},\qquad G_{z}=e^{i\beta z},
\end{equation*}
where $\alpha ,$ $\mathbf{\theta }$ and $\beta $ parametrize a
representation of $SU\left( 3\right) _{c}\times SU\left( 2\right) _{L}\times
U\left( 1\right) _{Y}$, with $\mathbf{\tau }$ the Pauli matrices, $\mathbf{%
\lambda }$ the Gell-Mann matrices and $z$ a real parameter which takes the
value $\frac{1}{6}$ for quarks and $\frac{-1}{2}$ for leptons. Then under $%
SU\left( 3\right) _{c}\times SU\left( 2\right) _{L}\times U\left( 1\right)
_{Y}$ we have that SM matter and gauge fields transform as
\begin{eqnarray}
q_{L} &\rightarrow &G_{c}G_{\frac{1}{6}}G_{L}q_{L},  \notag \\
l_{L} &\rightarrow &G_{c}G_{\frac{-1}{2}}G_{L}l_{L},  \notag \\
q_{R} &\rightarrow &G_{c}G_{\frac{1}{6}}G_{R}q_{R},  \notag \\
l_{R} &\rightarrow &G_{c}G_{\frac{-1}{2}}G_{R}l_{R},  \notag \\
U &\rightarrow &G_{L}UG_{R}^{\dagger },  \notag \\
\frac{\mathbf{\tau }}{2}\mathbf{\cdot W}_{\mu } &\rightarrow &G_{L}\left(
\frac{\mathbf{\tau }}{2}\mathbf{\cdot W}_{\mu }-\frac{i}{g}\partial _{\mu
}\right) G_{L}^{\dagger }  \notag \\
\frac{\tau ^{3}}{2}B_{\mu } &\rightarrow &\frac{\tau ^{3}}{2}B_{\mu }-\frac{i%
}{g^{\prime }}G_{R}\partial _{\mu }G_{R}^{\dagger },  \notag \\
\frac{\mathbf{\lambda }}{2}\mathbf{\cdot G}_{\mu } &\rightarrow &G_{c}\left(
\frac{\mathbf{\lambda }}{2}\mathbf{\cdot G}_{\mu }-\frac{i}{g_{s}}\partial
_{\mu }\right) G_{c}^{\dagger }.  \label{a1}
\end{eqnarray}
These transformations allow for the covariant derivatives
\begin{eqnarray*}
D_{\mu }U &=&\partial _{\mu }U+ig\frac{\mathbf{\tau }}{2}\mathbf{\cdot W}%
_{\mu }U-ig^{\prime }U\frac{\tau ^{3}}{2}B_{\mu }, \\
D_{\mu }^{L}f_{L} &=&\partial _{\mu }f_{L}+ig\frac{\mathbf{\tau }}{2}\mathbf{%
\cdot W}_{\mu }f_{L}+ig^{\prime }\left( \frac{\tau ^{3}}{2}-Q\right) B_{\mu
}f_{L}+ig_{s}\frac{\mathbf{\lambda }}{2}\mathbf{\cdot G}_{\mu }f_{L}, \\
D_{\mu }^{R}f_{R} &=&\partial _{\mu }f_{R}+ig^{\prime }QB_{\mu }f_{R}+ig_{s}%
\frac{\mathbf{\lambda }}{2}\mathbf{\cdot G}_{\mu }f_{R},
\end{eqnarray*}
where the charge $Q$ and hypercharge $Y$ are given by
\begin{equation*}
Q=\frac{\tau ^{3}}{2}+z,\qquad Y=\left\{
\begin{tabular}{ll}
$z$ & $\mathrm{for\ lefts}$ \\
$\frac{\tau ^{3}}{2}+z$ & $\mathrm{for\ rights}$%
\end{tabular}
\right. ,
\end{equation*}
It is useful also to introduce the notation
\begin{equation*}
\tau ^{\pm }\equiv \frac{\tau ^{1}\mp i\tau ^{2}}{\sqrt{2}},\qquad W_{\mu
}^{\pm }\equiv \frac{W_{\mu }^{1}\mp iW_{\mu }^{2}}{\sqrt{2}}.
\end{equation*}
when diagonalizing the mass matrices we perform
\begin{eqnarray*}
W_{\mu }^{3} &=&s_{W}A_{\mu }+c_{W}Z_{\mu }, \\
B_{\mu } &=&c_{W}A_{\mu }-s_{W}Z_{\mu }, \\
s_{W} &\equiv &\sin \theta _{W}\equiv \frac{g^{\prime }}{\sqrt{%
g^{2}+g^{\prime \,2}}}, \\
c_{W} &\equiv &\cos \theta _{W}\equiv \frac{g}{\sqrt{g^{2}+g^{\prime \,2}}},
\\
e &\equiv &gs_{W}=g^{\prime }c_{W},
\end{eqnarray*}
obtaining the SM kinetic term given by Eq.(\ref{SMkin})

\section{Some facts}

We have
\begin{eqnarray*}
\left\{ \gamma ^{\mu },\gamma ^{\nu }\right\} &=&2\eta ^{\mu \nu }, \\
\gamma ^{0}\gamma ^{\mu }\gamma ^{0} &=&\gamma ^{\mu \dagger }=\gamma _{\mu
}, \\
\gamma ^{2}\gamma ^{\mu }\gamma ^{2} &=&\gamma ^{\mu \ast },
\end{eqnarray*}
we also have
\begin{eqnarray*}
\left( D_{\mu }U\right) ^{\dagger }U &=&\left( \partial _{\mu }U^{\dagger
}-igU^{\dagger }\frac{\mathbf{\tau }}{2}\mathbf{\cdot W}_{\mu }+ig^{\prime }%
\frac{\tau ^{3}}{2}U^{\dagger }B_{\mu }\right) \\
&=&-U^{\dagger }\partial _{\mu }U-igU^{\dagger }\frac{\mathbf{\tau }}{2}%
\mathbf{\cdot W}_{\mu }U+ig^{\prime }U^{\dagger }U\frac{\tau ^{3}}{2}B_{\mu }
\\
&=&-U^{\dagger }D_{\mu }U,
\end{eqnarray*}
and
\begin{eqnarray*}
U^{\intercal }\tau ^{2} &=&\tau ^{2}U^{\dagger }, \\
\left( D_{\mu }U\right) ^{\intercal }\tau ^{2} &=&\tau ^{2}\left( D_{\mu
}U\right) ^{\dagger }.
\end{eqnarray*}
and for a $2\times 2$ matrix $A$ we have
\begin{equation*}
\det \left( A\right) =\frac{\epsilon ^{lm}\epsilon ^{ij}}{2}A_{il}A_{jm}=-%
\frac{1}{2}Tr\left( \epsilon A\epsilon A^{\intercal }\right) =\frac{1}{2}%
Tr\left( \tau ^{2}A\tau ^{2}A^{\intercal }\right) ,
\end{equation*}
Other useful properties are
\begin{eqnarray*}
\left[ \tau ^{i},\tau ^{j}\right] &=&i2\epsilon ^{ijk}\tau ^{k}, \\
\left\{ \tau ^{i},\tau ^{j}\right\} &=&2\delta ^{ij},
\end{eqnarray*}
implying
\begin{eqnarray*}
e^{-i\eta _{i}\frac{\tau ^{i}}{2}}\tau ^{p}e^{i\eta _{j}\frac{\tau ^{j}}{2}}
&=&\tau ^{p}+\left[ \tau ^{p},i\eta _{j}\frac{\tau ^{j}}{2}\right] +\frac{1}{%
2!}\left[ \left[ \tau ^{p},i\eta _{j}\frac{\tau ^{j}}{2}\right] ,i\eta _{k}%
\frac{\tau ^{k}}{2}\right] +\cdots \\
&=&\tau ^{p}-\eta _{j}\epsilon ^{pjk}\tau ^{k}+\frac{\left( -1\right) ^{2}}{%
2!}\eta _{j}\eta _{k}\epsilon ^{pjl}\epsilon ^{lkm}\tau ^{m}+\cdots \\
&=&\left( e^{A}\right) _{pk}\tau ^{k},
\end{eqnarray*}
where
\begin{equation*}
A_{ij}\equiv \epsilon ^{ijk}\eta _{k}.
\end{equation*}
and finally it will be useful to keep the following set of algebraic
relations
\begin{eqnarray*}
\left\{ \tau ^{3},\tau ^{\pm }\right\} &=&0, \\
\tau ^{3}\tau ^{\pm } &=&\mp \tau ^{\pm }, \\
\tau ^{\pm }\tau ^{3} &=&\pm \tau ^{\pm }, \\
\tau ^{+}\tau ^{d} &=&\tau ^{d}\tau ^{-}=\tau ^{-}\tau ^{u}=\tau ^{u}\tau
^{+}=0, \\
\tau ^{-}\tau ^{d} &=&\tau ^{u}\tau ^{-}=\tau ^{-}, \\
\tau ^{+}\tau ^{u} &=&\tau ^{d}\tau ^{+}=\tau ^{+}, \\
\left( \tau ^{+}\right) ^{2} &=&\left( \tau ^{-}\right) ^{2}=0, \\
\left( \tau ^{u}\right) ^{2} &=&\tau ^{u}, \\
\left( \tau ^{d}\right) ^{2} &=&\tau ^{d}, \\
\tau ^{u}\tau ^{d} &=&\tau ^{d}\tau ^{u}=0, \\
\tau ^{+}\tau ^{-} &=&2\tau ^{u}, \\
\tau ^{-}\tau ^{+} &=&2\tau ^{d}.
\end{eqnarray*}
The Dirac spinors used for calculations are given by
\begin{eqnarray*}
u^{\left( +\right) }\left( p\right) &=&\frac{\not{p}+m}{\sqrt{2m\left(
m+p^{0}\right) }}\left(
\begin{array}{c}
1 \\
0 \\
0 \\
0
\end{array}
\right) ,\qquad u^{\left( -\right) }\left( p\right) =\frac{\not{p}+m}{\sqrt{%
2m\left( m+p^{0}\right) }}\left(
\begin{array}{c}
0 \\
1 \\
0 \\
0
\end{array}
\right) , \\
v^{\left( +\right) }\left( p\right) &=&\frac{-\not{p}+m}{\sqrt{2m\left(
m+p^{0}\right) }}\left(
\begin{array}{c}
0 \\
0 \\
0 \\
1
\end{array}
\right) ,\qquad v^{\left( -\right) }\left( p\right) =\frac{-\not{p}+m}{\sqrt{%
2m\left( m+p^{0}\right) }}\left(
\begin{array}{c}
0 \\
0 \\
1 \\
0
\end{array}
\right) , \\
\bar{u}^{\left( s\right) }\left( p\right) &=&u^{\left( s\right) \dagger
}\left( p\right) \gamma ^{0}, \\
\bar{v}^{\left( s\right) }\left( p\right) &=&v^{\left( s\right) \dagger
}\left( p\right) \gamma ^{0},
\end{eqnarray*}
hence
\begin{eqnarray*}
\left( i\gamma ^{0}\gamma ^{2}\right) \bar{u}^{\left( s\right) T}\left(
p\right) &=&-i\gamma ^{2}u^{\left( s\right) \ast }\left( p\right) \\
&=&-i\gamma ^{2}\frac{\not{p}^{\ast }+m}{\sqrt{2m\left( m+p^{0}\right) }}%
u^{\left( s\right) }\left( \left( m,0\right) \right) \\
&=&\frac{-\not{p}+m}{\sqrt{2m\left( m+p^{0}\right) }}\left( -i\gamma
^{2}\right) u^{\left( s\right) }\left( \left( m,0\right) \right) \\
&=&\frac{-\not{p}+m}{\sqrt{2m\left( m+p^{0}\right) }}\left(
\begin{array}{cccc}
0 & 0 & 0 & -1 \\
0 & 0 & 1 & 0 \\
0 & 1 & 0 & 0 \\
-1 & 0 & 0 & 0
\end{array}
\right) u^{\left( s\right) }\left( \left( m,0\right) \right) \\
&=&-sv^{\left( s\right) }\left( p\right) ,
\end{eqnarray*}
and
\begin{eqnarray*}
i\gamma ^{2}\bar{u}^{\left( s\right) T}\left( p\right) &=&-i\gamma
^{0}\gamma ^{2}u^{\left( s\right) \ast }\left( p\right) \\
&=&-i\gamma ^{0}\gamma ^{2}\frac{\not{p}^{\ast }+m}{\sqrt{2m\left(
m+p^{0}\right) }}u^{\left( s\right) }\left( \left( m,0\right) \right) \\
&=&\frac{-\tilde{\not{p}}+m}{\sqrt{2m\left( m+p^{0}\right) }}\left( -i\gamma
^{0}\gamma ^{2}\right) u^{\left( s\right) }\left( \left( m,0\right) \right)
\\
&=&\frac{-\tilde{\not{p}}+m}{\sqrt{2m\left( m+p^{0}\right) }}\left(
\begin{array}{cccc}
0 & 0 & 0 & -1 \\
0 & 0 & 1 & 0 \\
0 & -1 & 0 & 0 \\
1 & 0 & 0 & 0
\end{array}
\right) u^{\left( s\right) }\left( \left( m,0\right) \right) \\
&=&sv^{\left( s\right) }\left( \tilde{p}\right) ,
\end{eqnarray*}
with
\begin{equation*}
\tilde{p}^{\mu }=p_{\mu }=\left( p^{0},-\vec{p}\right) ,
\end{equation*}
and
\begin{eqnarray*}
u^{\left( s\right) T}\left( p\right) i\gamma ^{2} &=&u^{\left( s\right)
T}\left( \left( m,0\right) \right) \frac{\not{p}^{T}+m}{\sqrt{2m\left(
m+p^{0}\right) }}i\gamma ^{2} \\
&=&u^{\left( s\right) T}\left( \left( m,0\right) \right) i\gamma ^{2}\frac{-%
\tilde{\not{p}}+m}{\sqrt{2m\left( m+p^{0}\right) }} \\
&=&u^{\left( s\right) T}\left( \left( m,0\right) \right) \left(
\begin{array}{cccc}
0 & 0 & 0 & 1 \\
0 & 0 & -1 & 0 \\
0 & -1 & 0 & 0 \\
1 & 0 & 0 & 0
\end{array}
\right) \frac{-\tilde{\not{p}}+m}{\sqrt{2m\left( m+p^{0}\right) }} \\
&=&sv^{\left( s\right) T}\left( \left( m,0\right) \right) \frac{-\tilde{\not{%
p}}+m}{\sqrt{2m\left( m+p^{0}\right) }}\gamma ^{0}\gamma ^{0} \\
&=&-sv^{\left( s\right) T}\left( \left( m,0\right) \right) \frac{-\tilde{%
\not{p}}^{\dagger }+m}{\sqrt{2m\left( m+p^{0}\right) }}\gamma ^{0} \\
&=&-s\bar{v}^{\left( s\right) }\left( \tilde{p}\right) ,
\end{eqnarray*}
and summing up we have
\begin{eqnarray}
i\gamma ^{2}\bar{u}^{\left( s\right) T}\left( p\right) &=&sv^{\left(
s\right) }\left( \tilde{p}\right) ,  \notag \\
u^{\left( s\right) T}\left( p\right) i\gamma ^{2} &=&-s\bar{v}^{\left(
s\right) }\left( \tilde{p}\right) .  \label{trans}
\end{eqnarray}

\chapter{Matter sector appendices}

\label{Msectorapp}

\section{{$d=4$} operators}

\label{MsectorappA}The procedure we have followed to obtain operators (\ref
{2.6b}--\ref{2.6d}) is very simple. We have to look for operators of the
form $\bar{\psi}\Gamma \psi $, where $\psi =q_{L},\;q_{R}$ and $\Gamma $
contains a covariant derivative, $D_{\mu }$, and an arbitrary number of $U$
matrices. These operators must be gauge invariant so not any form of $\Gamma
$ is possible. Moreover, we can drop total derivatives and, since $U$ is
unitary, we have the following relation
\begin{equation}
D_{\mu }U=-U(D_{\mu }U)^{\dagger }U.
\end{equation}
Apart from the obvious structure $D_{\mu }U$ which transform as $U$ does, we
immediately realize that the particular form of $G_{R}$ implies the
following simple transformations for the combinations $U\tau ^{3}U^{\dagger
} $ and $(D_{\mu }U)\tau ^{3}U^{\dagger }$
\begin{eqnarray}
U\tau ^{3}U^{\dagger } &\mapsto &G_{L}\;U\tau ^{3}U^{\dagger
}\;G_{L}^{\dagger } \\
(D_{\mu }U)\tau ^{3}U^{\dagger } &\mapsto &G_{L}\;(D_{\mu }U)\tau
^{3}U^{\dagger }\;G_{L}^{\dagger }
\end{eqnarray}
Keeping all these relations in mind, we simply write down all the
possibilities for $\bar{\psi}\Gamma \psi $ and find the list of operators (%
\ref{2.6b}--\ref{2.6d}). It is worth mentioning that there appears to be
another family of four operators in which the $U$ matrices also occur within
a trace: $\bar{\psi}\Gamma \psi \;Tr\Gamma ^{\prime }$. One can check,
however, that these are not independent. More precisely, using the
remarkable identities
\begin{eqnarray*}
i\left( D_{\mu }U\right) \tau ^{3}U^{\dagger }+h.c. &=&i\tau ^{3}U^{\dagger
}\left( D_{\mu }U\right) +h.c \\
&=&iTr\left( \left( D_{\mu }U\right) \tau ^{3}U^{\dagger }\right) ,
\end{eqnarray*}
we have
\begin{eqnarray}
i\bar{q}_{L}\gamma ^{\mu }q_{L}\;Tr\left[ (D_{\mu }U)\tau ^{3}U^{\dagger }%
\right] &=&\frac{1}{M_{L}^{2}}\mathcal{L}_{L}^{2}, \\
i\bar{q}_{L}\gamma ^{\mu }U\tau ^{3}U^{\dagger }q_{L}\;Tr\left[ (D_{\mu
}U)\tau ^{3}U^{\dagger }\right] &=&\frac{1}{2M_{L}^{3}}\mathcal{L}_{L}^{3}-%
\frac{1}{2M_{L}^{1}}\mathcal{L}_{L}^{1}, \\
i\bar{q}_{R}\gamma ^{\mu }q_{R}\;Tr\left[ (D_{\mu }U)\tau ^{3}U^{\dagger }%
\right] &=&\frac{1}{M_{R}^{2}}\mathcal{L}_{R}^{2}, \\
i\bar{q}_{R}\gamma ^{\mu }\tau ^{3}q_{R}\;Tr\left[ (D_{\mu }U)\tau
^{3}U^{\dagger }\right] &=&\frac{1}{2M_{R}^{1}}\mathcal{L}_{R}^{1}+\frac{1}{%
2M_{R}^{3}}\mathcal{L}_{R}^{3},
\end{eqnarray}
Note that $\mathcal{L}_{L}^{4}$ (as well as $\mathcal{L}_{R}^{\prime }$
discussed above) can be reduced by equations of motion to operators of lower
dimension which do not contribute to the physical processes we are
interested in. We have checked that its contribution indeed drops from the
relevant $S$-matrix elements.

\section{Feynman rules}

\label{MsectorappB}We write the effective $d=4$ Lagrangian as
\begin{equation*}
\mathcal{L}_{c}{}_{\mathrm{eff}}=\sum_{k=1}^{3}\left( \mathcal{L}_{L}^{k}+%
\mathcal{L}_{R}^{k}\right) +\mathcal{L}_{L}^{4}+\mathcal{L}_{R}^{\prime },
\end{equation*}
where the real coefficients $M_{L,R}^{i}$ appearing in the definitions (\ref
{2.6b}-\ref{2.6d}) are to be determined through the matching. We need to
match the effective theory described by $\mathcal{L}_{c}{}_{\mathrm{eff}}$
to both, the MSM and the underlying theory parametrized by the four-fermion
operators. It has proven more convenient to work with the physical fields $%
W^{\pm }$, $Z$ and $\gamma $ in the former case whereas the use of the
Lagrangian fields $W^{1}$, $W^{2}$, $W^{3}$ and $B$ is clearly more
straightforward for the latter. Thus, we give the Feynman rules in terms of
both the physical and unphysical basis.\setlength{\unitlength}{3mm}
\begin{eqnarray}
\begin{picture}(15,7)(-2,0) \put(0,1.5){\makebox(0,0)[c]{$d $}}
\put(11,1.5){\makebox(0,0)[c]{$\bar d$}} \put(0,0){\line(1,0){10}}
\put(2.5,0){\vector(1,0){0.1}} \put(7.5,0){\vector(1,0){0.1}}
\put(6,5){\makebox(0,0)[l]{$Z_\mu$}} \multiput(5,0.25)(0,1){5}{\zvs}
\end{picture}\kern-0.5cm &=&{\frac{ie}{2s_{W}c_{W}}}\gamma _{\mu }\left\{
\frac{1}{2}\left( -2M_{L}^{1}+2M_{R}^{1}+2M_{L}^{3}+2M_{R}^{3}\right)
-M_{L}^{2}-M_{R}^{2}\right.  \notag \\
&&-\left. \left( 1-{\frac{2}{3}}s_{W}^{2}\right) M_{L}^{4}+{\frac{1}{3}}%
s_{W}^{2}\;M_{R}^{\prime }\right\}  \notag \\
&+&{\frac{ie}{2s_{W}c_{W}}}\gamma _{\mu }\gamma _{5}\left\{ \frac{1}{2}%
\left( 2M_{L}^{1}+2M_{R}^{1}-2M_{L}^{3}+2M_{R}^{3}\right)
+M_{L}^{2}-M_{R}^{2}\right.  \notag \\
&&\left. +\left( 1-{\frac{2}{3}}s_{W}^{2}\right) M_{L}^{4}+{\frac{1}{3}}%
s_{W}^{2}\;M_{R}^{\prime }\right\} ,  \label{newN1d1}
\end{eqnarray}
\begin{eqnarray}
\begin{picture}(15,7)(-2,0) \put(0,1.5){\makebox(0,0)[c]{$u $}}
\put(11,1.5){\makebox(0,0)[c]{$\bar u$}} \put(0,0){\line(1,0){10}}
\put(2.5,0){\vector(1,0){0.1}} \put(7.5,0){\vector(1,0){0.1}}
\put(6,5){\makebox(0,0)[l]{$Z_\mu$}} \multiput(5,0.25)(0,1){5}{\zvs}
\end{picture}\kern-0.5cm &=&{\frac{ie}{2s_{W}c_{W}}}\gamma _{\mu }\left\{
\frac{1}{2}\left( 2M_{L}^{1}-2M_{R}^{1}-2M_{L}^{3}-2M_{R}^{3}\right)
-M_{L}^{2}-M_{R}^{2}\right.  \notag \\
&&-\left. \left( 1-{\frac{4}{3}}s_{W}^{2}\right) M_{L}^{4}+{\frac{2}{3}}%
s_{W}^{2}\;M_{R}^{\prime }\right\}  \notag \\
&+&{\frac{ie}{2s_{W}c_{W}}}\gamma _{\mu }\gamma _{5}\left\{ \frac{1}{2}%
\left( -2M_{L}^{1}-2M_{R}^{1}+2M_{L}^{3}-2M_{R}^{3}\right)
+M_{L}^{2}-M_{R}^{2}\right.  \notag \\
&&+\left. \left( 1-{\frac{4}{3}}s_{W}^{2}\right) M_{L}^{4}+{\frac{2}{3}}%
s_{W}^{2}\;M_{R}^{\prime }\right\} ,  \label{newN1d2}
\end{eqnarray}
\begin{eqnarray}
\begin{picture}(15,7)(-2,0) \put(0,1.5){\makebox(0,0)[c]{$d $}}
\put(11,1.5){\makebox(0,0)[c]{$\bar d$}} \put(0,0){\line(1,0){10}}
\put(2.5,0){\vector(1,0){0.1}} \put(7.5,0){\vector(1,0){0.1}}
\put(6,5){\makebox(0,0)[l]{$A_\mu$}} \multiput(5,0.25)(0,1){5}{\zvs}
\end{picture}\kern-0.5cm &=&-ie{\frac{1}{3}}\gamma _{\mu }\left( M_{L}^{4}+%
\frac{1}{2}M_{R}^{\prime }\right) +ie{\frac{1}{3}}\gamma _{\mu }\gamma
_{5}\left( M_{L}^{4}-\frac{1}{2}M_{R}^{\prime }\right) ,  \label{newN1d3} \\
\begin{picture}(15,7)(-2,0) \put(0,1.5){\makebox(0,0)[c]{$u $}}
\put(11,1.5){\makebox(0,0)[c]{$\bar u$}} \put(0,0){\line(1,0){10}}
\put(2.5,0){\vector(1,0){0.1}} \put(7.5,0){\vector(1,0){0.1}}
\put(6,5){\makebox(0,0)[l]{$A_\mu$}} \multiput(5,0.25)(0,1){5}{\zvs}
\end{picture}\kern-0.5cm &=&-ie{\frac{2}{3}}\gamma _{\mu }\left( M_{L}^{4}+%
\frac{1}{2}M_{R}^{\prime }\right) +ie{\frac{2}{3}}\gamma _{\mu }\gamma
_{5}\left( M_{L}^{4}-\frac{1}{2}M_{R}^{\prime }\right) ,  \label{newN1d4} \\
\begin{picture}(15,7)(-2,0) \put(0,1.5){\makebox(0,0)[c]{$d $}}
\put(11,1.5){\makebox(0,0)[c]{$\bar u$}} \put(0,0){\line(1,0){10}}
\put(2.5,0){\vector(1,0){0.1}} \put(7.5,0){\vector(1,0){0.1}}
\put(6,5){\makebox(0,0)[l]{$W^+_\mu$}} \multiput(5,0.25)(0,1){5}{\zvs}
\end{picture}\kern-0.5cm &=&-ie{\frac{1}{2\sqrt{2}s_{W}}}\gamma _{\mu
}\left( 2M_{L}^{1}+2M_{L}^{3}-2M_{R}^{1}-2M_{R}^{3}\right)  \notag \\
&+&ie{\frac{1}{2\sqrt{2}s_{W}}}\gamma _{\mu }\gamma _{5}\left(
2M_{L}^{1}+2M_{L}^{3}+2M_{R}^{1}+2M_{R}^{3}\right) .  \label{newN1d5}
\end{eqnarray}

The operators $\mathcal{L}_{L}^{4}$ and $\mathcal{L}_{R}^{\prime }$
contribute to two-point function. The relevant Feynman rules are %
\setlength{\unitlength}{15mm}
\begin{eqnarray}
\begin{picture}(3,1)(-0.4,0) \put(0,0.375){\makebox(0,0)[c]{$u$}}
\put(2,0.375){\makebox(0,0)[c]{$\bar u $}}
\put(1.0,0){\makebox(0,0)[c]{$\times$}} \put(0,0){\line(1,0){2}}
\put(0.5,0){\vector(1,0){0.02}} \put(1.5,0){\vector(1,0){0.02}} \end{picture}%
\kern-0.5cm &=&i(M_{L}^{4}+\frac{1}{2}M_{R}^{\prime })\not{p}+i(-M_{L}^{4}+%
\frac{1}{2}M_{R}^{\prime })\not{p}\gamma _{5},  \label{new-1} \\
\begin{picture}(3,1)(-0.4,0) \put(0,0.375){\makebox(0,0)[c]{$d$}}
\put(2,0.375){\makebox(0,0)[c]{$\bar d $}}
\put(1.0,0){\makebox(0,0)[c]{$\times$}} \put(0,0){\line(1,0){2}}
\put(0.5,0){\vector(1,0){0.02}} \put(1.5,0){\vector(1,0){0.02}} \end{picture}%
\kern-0.5cm &=&i(-M_{L}^{4}-\frac{1}{2}M_{R}^{\prime })\not{p}+i(M_{L}^{4}-%
\frac{1}{2}M_{R}^{\prime })\not{p}\gamma _{5}.  \label{new-2}
\end{eqnarray}

Rather than giving the actual Feynman rules in the unphysical basis, we
collect the various tensor structures that can result from the calculation
of the relevant diagrams in table~\ref{table-1}.
\begin{table}[!h]
\centering
\begin{equation*}
\begin{array}{l||c|c|c|c|c|c|}
Tensor~structure & M_{L}^{1} & M_{L}^{2} & M_{L}^{3} & M_{R}^{1} & M_{R}^{2}
& M_{R}^{3} \\ \hline
i\bar{q}_{L}\,g[\tau ^{1}\not{W}^{1}+\tau ^{2}\not{W}^{2}]q_{L} & 1 &  & 1 &
&  &  \\
i\bar{q}_{L}\,\tau ^{3}[g\not{W}^{3}-g^{\prime }\not{B}]q_{L} & 1 &  & -1 &
&  &  \\
i\bar{q}_{L}\,[g\not{W}^{3}-g^{\prime }\not{B}]q_{L} &  & -1 &  &  &  &  \\
i\bar{q}_{R}\,g[\tau ^{1}\not{W}^{1}+\tau ^{2}\not{W}^{2}]q_{R} &  &  &  & -1
&  & 1 \\
i\bar{q}_{R}\,\tau ^{3}[g\not{W}^{3}-g^{\prime }\not{B}]q_{R} &  &  &  & -1
&  & -1 \\
i\bar{q}_{R}\,[g\not{W}^{3}-g^{\prime }\not{B}]q_{R} &  &  &  &  & -1 &
\end{array}
\end{equation*}
\caption{Various structures appearing in the matching of the vertex and the
corresponding contributions to $\mathcal{L}_{L,R}^{1,2,3}$}
\label{table-1}
\end{table}
We include only those that can be matched to insertions of the operators $%
\mathcal{L}_{L,R}^{1,2,3}$ (the contributions to $\mathcal{L}_{L}^{4}$ and $%
\mathcal{L}_{R}^{\prime }$ can be determined from the matching of the
two-point functions). The corresponding contributions of these structures to
$M_{L,R}^{1},$ $M_{L,R}^{2}$ and $M_{L,R}^{3}$ are also given in table~\ref
{table-1}. Once $M_{L}^{4}$ has been replaced by its value, obtained in the
matching of the two-point functions, only the listed structures can show up
in the matching of the vertex, otherwise the $SU(2)\times U(1)$ symmetry
would not be preserved.

\section{Four-fermion operators}

\label{MsectorappC}The complete list of four-fermion operators relevant for
the discussion in section \ref{new-phys} of Chapter \ref{mattersectorchapter}
is in tables \ref{table-2} and \ref{table-3} of that section. It is also
explained in that section the convenience of fierzing the operators in the
last seven rows of table \ref{table-2} in order to write them in the form $%
\mathbf{J}\cdot \mathbf{j}$. Here we just give the list that comes out
naturally from our analysis, tables \ref{table-2} and \ref{table-3}, without
further physical interpretation. The list is given for fermions belonging to
the representation $\mathbf{3}$ of $SU(3)_{c}$ (techniquarks). By using
Fierz transformations one can easily find out relations among some of these
operators when the fermions are color singlet (technileptons), which is
telling us that some of these operators are not independent in this case. A
list of independent operators for technileptons is also given in section \ref
{new-phys} of Chapter \ref{mattersectorchapter}. In particular, the
independent chirality preserving operators for colorless fermions are in the
first six rows of table \ref{table-2} (those whose name we write in capital
letters) and, additionally, only the two operators $\left( \bar{Q}_{L}\gamma
_{\mu }q_{L}\right) \left( \bar{q}_{L}\gamma ^{\mu }Q_{L}\right) ,$ $\left(
\bar{Q}_{R}\gamma _{\mu }q_{R}\right) \left( \bar{q}_{R}\gamma ^{\mu
}Q_{R}\right) $ from the last seven rows.

Let us outline the procedure we have followed to obtain this basis in the
(more involved) case of colored fermions.

There are only two color singlet structures one can build out of four
fermions, namely ($\alpha ,$ $\beta ,$ ... are color indices)
\begin{eqnarray}
(\bar{\psi}\psi )(\bar{\psi}^{\prime }\psi ^{\prime }) &\equiv &\bar{\psi}%
_{\alpha }\psi _{\alpha }\;\bar{\psi}_{\beta }^{\prime }\psi _{\beta
}^{\prime }\;,  \label{26.5d} \\
(\bar{\psi}\vec{\lambda}\psi )\cdot (\bar{\psi}^{\prime }\vec{\lambda}\psi
^{\prime }) &\equiv &\bar{\psi}_{\alpha }(\vec{\lambda})_{\alpha \beta }\psi
_{\beta }\;\cdot \;\bar{\psi}_{\gamma }^{\prime }(\vec{\lambda})_{\gamma
\delta }\psi _{\delta }^{\prime }\;,  \label{26.5e}
\end{eqnarray}
where, $\psi $ stands for any field belonging to the representation $\mathbf{%
3}$ of $SU(3)_{c}$ ($\psi $ will be either $q$ or $Q$); $\alpha $, $\beta $%
,..., are color indices; and the primes ( $^{\prime }$) remind us that $\psi
$ and $\bar{\psi}$ carry the same additional indices (Dirac, $SU(2)$, ...).

Next we classify the Dirac structures. Since $\psi $ is either $\psi _{L}$
[it belongs to the representation $(\frac{1}{2},0)$ of the Lorentz group] or
$\psi _{R}$ [representation $(0,\frac{1}{2})$], we have five sets of fields
to analyze, namely
\begin{eqnarray}
&&\{\bar{\psi}_{L},\psi _{L},\bar{\psi}_{L}^{\prime },\psi _{L}^{\prime
}\},\quad \lbrack R\leftrightarrow L];\qquad \{\bar{\psi}_{L},\psi _{L},\bar{%
\psi}_{R},\psi _{R}\};  \label{26.5b} \\
&&\{\bar{\psi}_{L},\psi _{R},\bar{\psi}_{L}^{\prime },\psi _{R}^{\prime
}\},\qquad \lbrack R\leftrightarrow L].  \label{26.5c}
\end{eqnarray}
There is only an independent scalar we can build with each of the three sets
in~(\ref{26.5b}). Our choice is
\begin{eqnarray}
&&\bar{\psi}_{L}\gamma ^{\mu }\psi _{L}\;\bar{\psi}_{L}^{\prime }\gamma
_{\mu }\psi _{L}^{\prime },\qquad \lbrack R\leftrightarrow L];
\label{28.5-59} \\
&&\bar{\psi}_{L}\gamma ^{\mu }\psi _{L}\;\bar{\psi}_{R}\gamma _{\mu }\psi
_{R}.  \label{28.5-60}
\end{eqnarray}
where the prime is not necessary in the second equation because $R$ and $L$
suffice to remind us that the two $\psi $ and $\bar{\psi}$ may carry
different ($SU(2)$, technicolor, ...) indices. There appear to be four other
independent scalar operators: $\bar{\psi}_{L}\gamma ^{\mu }\psi _{L}^{\prime
}\;\bar{\psi}_{L}^{\prime }\gamma _{\mu }\psi _{L}$, $[R\leftrightarrow L]$;
$\allowbreak \bar{\psi}_{L}\psi _{R}\;\bar{\psi}_{R}\psi _{L}$; and $\bar{%
\psi}_{L}\sigma ^{\mu \nu }\psi _{R}\;\bar{\psi}_{R}\sigma _{\mu \nu }\psi
_{L}$. However, Fierz symmetry implies that the first three are not
independent, and the fourth one vanishes, as can be also seen using the
identity $2i\sigma ^{\mu \nu }\gamma ^{5}=\epsilon ^{\mu \nu \rho \lambda
}\sigma _{\rho \lambda }$. For each of the two operators in~(\ref{26.5c}),
two independent scalars can be constructed. Our choice is
\begin{eqnarray}
&&\bar{\psi}_{L}\psi _{R}\;\bar{\psi}_{L}^{\prime }\psi _{R}^{\prime
},\qquad \lbrack R\leftrightarrow L];  \label{28.5-68} \\
&&\bar{\psi}_{L}\psi _{R}^{\prime }\;\bar{\psi}_{L}^{\prime }\psi
_{R},\qquad \lbrack R\leftrightarrow L].  \label{28.5-69}
\end{eqnarray}
Again, there appear to be four other scalar operators:
$\bar{\psi}_{L}\sigma ^{\mu \nu }\psi _{R}\;\bar{\psi}_{L}^{\prime
}\sigma _{\mu \nu }\psi _{R}^{\prime }$, $[R\leftrightarrow
L]$;\break $\bar{\psi}_{L}\sigma ^{\mu \nu }\psi _{R}^{\prime }
\;\bar{\psi}_{L}^{\prime }\sigma _{\mu \nu }\psi _{R}$, $%
[R\leftrightarrow L]$; which, nevertheless, can be shown not to be
independent but related to (\ref{28.5-68}) and (\ref{28.5-69}) by Fierz
symmetry. To summarize, the independent scalar structures are (\ref{28.5-59}%
), (\ref{28.5-60}), (\ref{28.5-68}) and (\ref{28.5-69}).

Next, we combine the color and the Dirac structures. We do this for the
different cases (\ref{28.5-59}) to (\ref{28.5-69}) separately. For operators
of the form (\ref{28.5-59}), we have the two obvious possibilities
(Hereafter, color and Dirac indices will be implicit)
\begin{eqnarray}
&&(\bar{\psi}_{L}\gamma ^{\mu }\psi _{L})(\bar{\psi}_{L}^{\prime }\gamma
_{\mu }\psi _{L}^{\prime }),\qquad \lbrack R\leftrightarrow L];
\label{26.5f} \\
&&(\bar{\psi}_{L}\gamma ^{\mu }\psi _{L}^{\prime })(\bar{\psi}_{L}^{\prime
}\gamma _{\mu }\psi _{L}),\qquad \lbrack R\leftrightarrow L];  \label{26.5g}
\end{eqnarray}
where fields in parenthesis have their color indices contracted as in~(\ref
{26.5d}) and~(\ref{26.5e}). Note that the operator $(\bar{\psi}_{L}\gamma
^{\mu }\vec{\lambda}\psi _{L})\cdot (\bar{\psi}_{L}^{\prime }\gamma _{\mu }%
\vec{\lambda}\psi _{L}^{\prime })$, or its $R$ version, is not independent
(recall that $(\vec{\lambda})_{\alpha \beta }\cdot (\vec{\lambda})_{\gamma
\delta }=2\delta _{\alpha \delta }\delta _{\beta \gamma }-2/3\;\delta
_{\alpha \beta }\delta _{\gamma \delta }$). For operators of the form (\ref
{28.5-60}), we take
\begin{eqnarray}
&&(\bar{\psi}_{L}\gamma ^{\mu }\psi _{L})(\bar{\psi}_{R}\gamma _{\mu }\psi
_{R}),  \label{C1} \\
&&(\bar{\psi}_{L}\gamma ^{\mu }\vec{\lambda}\psi _{L})\cdot (\bar{\psi}%
_{R}\gamma _{\mu }\vec{\lambda}\psi _{R}),  \label{C2}
\end{eqnarray}
Finally, for operators of the form (\ref{28.5-68}) and~(\ref{28.5-69}), our
choice is
\begin{eqnarray}
&&(\bar{\psi}_{L}\psi _{R})(\bar{\psi}_{L}^{\prime }\psi _{R}^{\prime
}),\quad \lbrack R\leftrightarrow L];\quad (\bar{\psi}_{L}\vec{\lambda}\psi
_{R})\cdot (\bar{\psi}_{L}^{\prime }\vec{\lambda}\psi _{R}^{\prime }),\quad
\lbrack R\leftrightarrow L];  \label{26.5i} \\
&&(\bar{\psi}_{L}\psi _{R}^{\prime })(\bar{\psi}_{L}^{\prime }\psi
_{R}),\quad \lbrack R\leftrightarrow L];\quad (\bar{\psi}_{L}\vec{\lambda}%
\psi _{R}^{\prime })\cdot (\bar{\psi}_{L}^{\prime }\vec{\lambda}\psi
_{R}),\quad \lbrack R\leftrightarrow L].  \label{26.5k}
\end{eqnarray}
All them are independent unless further symmetries [e.g., $SU(2)_{L}\times
SU(2)_{R}$] are introduced.

To introduce the $SU(2)_{L}\times SU(2)_{R}$ symmetry one just assigns $%
SU(2) $ indices ($i$, $j$, $k$, ...) to each of the fields in~(\ref{26.5f}--%
\ref{26.5k}). We can drop the primes hereafter since there is no other
symmetry left but technicolor which for the present analysis is trivial
(recall that we are only interested in four fermion operators of the form $Q%
\bar{Q}q\bar{q}$, thus technicolor indices must necessarily be matched in
the obvious way: $Q^{A}\bar{Q}^{A}q\bar{q}$). For each of the operators in~(%
\ref{26.5f}) and~(\ref{26.5g}), there are two independent ways of
constructing $SU(2)_{L}\times SU(2)_{R}$ invariants. Only two of the four
resulting operators turn out to be independent (actually, the other two are
exactly equal to the first ones). The independent operators are chosen to be
\begin{eqnarray}
&&(\bar{\psi}_{L}^{i}\gamma ^{\mu }\psi _{L}^{i})(\bar{\psi}_{L}^{j}\gamma
_{\mu }\psi _{L}^{j})\equiv (\bar{\psi}_{L}\gamma ^{\mu }\psi _{L})(\bar{\psi%
}_{L}\gamma _{\mu }\psi _{L}),\quad \lbrack R\leftrightarrow L];  \label{D1}
\\
&&(\bar{\psi}_{L}^{i}\gamma ^{\mu }\psi _{L}^{j})(\bar{\psi}_{L}^{j}\gamma
_{\mu }\psi _{L}^{i}),\quad \lbrack R\leftrightarrow L];  \label{D2}
\end{eqnarray}
For each of the operators in~(\ref{C1}--\ref{26.5k}), the same
straightforward group analysis shows that there is only one way to construct
a $SU(2)_{L}\times SU(2)_{R}$ invariant. Discarding the redundant operators
and imposing hermiticity and $CP$ invariance one finally has, in addition to
the operators~(\ref{D1}) and~(\ref{D2}), those listed below (from now on, we
understand that fields in parenthesis have their Dirac, color and also
flavor indices contracted as in~(\ref{D1}))
\begin{eqnarray}
&&(\bar{\psi}_{L}\gamma ^{\mu }\psi _{L})(\bar{\psi}_{R}\gamma _{\mu }\psi
_{R}),  \label{1.6a} \\
&&(\bar{\psi}_{L}\gamma ^{\mu }\vec{\lambda}\psi _{L})\cdot (\bar{\psi}%
_{R}\gamma _{\mu }\vec{\lambda}\psi _{R}),  \label{1.6b} \\
&&(\bar{\psi}_{L}^{i}\psi _{R}^{j})(\bar{\psi}_{L}^{k}\psi _{R}^{l})\epsilon
_{ik}\epsilon _{jl}+(\bar{\psi}_{R}^{i}\psi _{L}^{j})(\bar{\psi}_{R}^{k}\psi
_{L}^{l})\epsilon _{ik}\epsilon _{jl},  \label{1.6c} \\
&&(\bar{\psi}_{L}^{i}\vec{\lambda}\psi _{R}^{j})\cdot (\bar{\psi}_{L}^{k}%
\vec{\lambda}\psi _{R}^{l})\epsilon _{ik}\epsilon _{jl}+(\bar{\psi}_{R}^{i}%
\vec{\lambda}\psi _{L}^{j})\cdot (\bar{\psi}_{R}^{k}\vec{\lambda}\psi
_{L}^{l})\epsilon _{ik}\epsilon _{jl}.  \label{1.6d}
\end{eqnarray}

We are now in a position to obtain very easily the custodially preserving
operators of tables~\ref{table-2} and~\ref{table-3} 
We simply replace $\psi$ by $q$ and $Q$ (a pair of each: a field and its
conjugate) in all possible independent ways.

To break the custodial symmetry we simply insert $\tau ^{3}$ matrices in the
$R$-sector of the custodially preserving operators we have just obtain (left
columns of tables~\ref{table-2} and~\ref{table-3}). However, not all the
operators obtained this way are independent since one can prove the
following relations
\begin{eqnarray}
(\bar{q}_{R}^{i}\gamma ^{\mu }Q_{R}^{j})(\bar{Q}_{R}^{j}\gamma _{\mu }[\tau
^{3}q_{R}]^{i}) &=&(\bar{q}_{R}\gamma ^{\mu }\tau ^{3}Q_{R})(\bar{Q}%
_{R}\gamma _{\mu }q_{R})+(\bar{q}_{R}\gamma ^{\mu }Q_{R})(\bar{Q}_{R}\gamma
_{\mu }\tau ^{3}q_{R})  \notag \\
&&\qquad -(\bar{q}_{R}^{i}\gamma ^{\mu }[\tau ^{3}Q_{R}]^{j})(\bar{Q}%
_{R}^{j}\gamma _{\mu }q_{R}^{i}), \\
(\bar{q}_{R}^{i}\gamma ^{\mu }[\tau ^{3}Q_{R}]^{j})(\bar{Q}_{R}^{j}\gamma
_{\mu }[\tau ^{3}q_{R}]^{i}) &=&(\bar{q}_{R}\gamma ^{\mu }Q_{R})(\bar{Q}%
_{R}\gamma _{\mu }q_{R})+(\bar{q}_{R}\gamma ^{\mu }\tau ^{3}Q_{R})(\bar{Q}%
_{R}\gamma _{\mu }\tau ^{3}q_{R})  \notag \\
&&\qquad -(\bar{q}_{R}^{i}\gamma ^{\mu }Q_{R}^{j})(\bar{Q}_{R}^{j}\gamma
_{\mu }q_{R}^{i}), \\
(\bar{q}_{R}^{i}\gamma ^{\mu }[\tau ^{3}q_{R}]^{j})(\bar{Q}_{R}^{j}\gamma
_{\mu }[\tau ^{3}Q_{R}]^{i}) &=&(\bar{q}_{R}\gamma ^{\mu }q_{R})(\bar{Q}%
_{R}\gamma _{\mu }Q_{R})+(\bar{q}_{R}\gamma ^{\mu }\tau ^{3}q_{R})(\bar{Q}%
_{R}\gamma _{\mu }\tau ^{3}Q_{R})  \notag \\
&&\qquad -(\bar{q}_{R}^{i}\gamma ^{\mu }q_{R}^{j})(\bar{Q}_{R}^{j}\gamma
_{\mu }Q_{R}^{i}), \\
&&(\bar{q}_{R}^{i}\gamma ^{\mu }[\tau ^{3}q_{R}]^{j})(\bar{Q}_{R}^{j}\gamma
_{\mu }Q_{R}^{i})\qquad \phantom{+}  \notag \\
+(\bar{q}_{R}^{i}\gamma ^{\mu }q_{R}^{j})(\bar{Q}_{R}^{j}\gamma _{\mu }[\tau
^{3}Q_{R}]^{i}) &=&(\bar{q}_{R}\gamma ^{\mu }q_{R})(\bar{Q}_{R}\gamma ^{\mu
}\tau ^{3}Q_{R})  \notag \\
&&\qquad +(\bar{q}_{R}\gamma ^{\mu }\tau ^{3}q_{R})(\bar{Q}_{R}\gamma ^{\mu
}Q_{R}).
\end{eqnarray}

Our final choice of custodially breaking operators is the one in the right
columns of tables~\ref{table-2} and~\ref{table-3}.

%

\section{Renormalization of the matter sector}

\label{MsectorappD}Although most of the material in this section is
standard, it is convenient to collect some of the important expressions, as
the renormalization of the fermion fields is somewhat involved, and also to
set up the notation. Let us introduce three wave-function renormalization
constants for the fermion fields
\begin{eqnarray}
\left(
\begin{array}{c}
u \\
d
\end{array}
\right) _{L} &\rightarrow &Z_{L}^{1/2}\left(
\begin{array}{c}
u \\
d
\end{array}
\right) _{L}, \\
u_{R} &\rightarrow &(Z_{R}^{u})^{1/2}u_{R}, \\
d_{R} &\rightarrow &(Z_{R}^{d})^{1/2}d_{R}.  \label{newglo4}
\end{eqnarray}
where $u$ ($d$) stands for the field of the up-type (down-type) fermion. We
write
\begin{equation}
Z_{i}=1+\delta Z_{i}
\end{equation}
We also renormalize the fermion masses according to
\begin{equation*}
m_{f}\rightarrow m_{f}+\delta m_{f},
\end{equation*}
where $f=u,\;d$. These substitutions generate the counterterms needed to
cancel the UV divergencies. The corresponding Feynman rules are
\begin{eqnarray}
\setlength{\unitlength}{3mm}\begin{picture}(15,4)(-2,0)
\put(0,1.5){\makebox(0,0)[c]{$q$}} \put(11,1.5){\makebox(0,0)[c]{$\bar q $}}
\put(5.12,0){\makebox(0,0)[c]{$\times$}} \put(0,0){\line(1,0){10}}
\put(2.5,0){\vector(1,0){0.1}} \put(7.5,0){\vector(1,0){0.1}} \end{picture} %
&=&i\delta Z_{V}^{f}\not{p}-i\delta Z_{A}^{f}\not{p}\gamma _{5}-i\left( {%
\frac{\delta m_{f}}{m_{f}}}+\delta Z_{V}^{f}\right) ,  \label{17.6a} \\
\setlength{\unitlength}{3mm}\begin{picture}(15,7)(-2,0)
\put(0,1.5){\makebox(0,0)[c]{$q $}} \put(11,1.5){\makebox(0,0)[c]{$\bar q$}}
\put(5.25,0){\makebox(0,0)[c]{$\times$}} \put(0,0){\line(1,0){10}}
\put(2.5,0){\vector(1,0){0.1}} \put(7.5,0){\vector(1,0){0.1}}
\put(6,5){\makebox(0,0)[l]{$Z_\mu$}} \multiput(4.8,0.25)(0,1){5}{\zvs}
\end{picture} &=&-ie\gamma _{\mu }(v_{f}-a_{f}\,\gamma _{5})(\delta
Z_{1}^{Z}-\delta Z_{2}^{Z})  \notag \\
&-&ie\gamma _{\mu }\,Q_{f}\,(\delta Z_{1}^{Z\gamma }-\delta Z_{2}^{Z\gamma })
\notag \\
&-&ie\gamma _{\mu }(v_{f}\,\delta Z_{V}^{f}+a_{f}\,\delta Z_{A}^{f})  \notag
\\
&+&ie\gamma _{\mu }\gamma _{5}(v_{f}\,\delta Z_{A}^{f}+a_{f}\,\delta
Z_{V}^{f})  \label{17.6h} \\
\setlength{\unitlength}{3mm}\begin{picture}(15,7)(-2,0)
\put(0,1.5){\makebox(0,0)[c]{$q $}} \put(11,1.5){\makebox(0,0)[c]{$\bar q$}}
\put(5.25,0){\makebox(0,0)[c]{$\times$}} \put(0,0){\line(1,0){10}}
\put(2.5,0){\vector(1,0){0.1}} \put(7.5,0){\vector(1,0){0.1}}
\put(6,5){\makebox(0,0)[l]{$A_\mu$}} \multiput(4.8,0.25)(0,1){5}{\zvs}
\end{picture} &=&-ie\gamma _{\mu }\,Q_{f}(\delta Z_{1}^{\gamma }-\delta
Z_{2}^{\gamma }+\delta Z_{V}^{f}-\delta Z_{A}^{f}\,\gamma _{5})  \notag \\
&-&ie\gamma _{\mu }(v_{f}-a_{f}\,\gamma _{5})(\delta Z_{1}^{Z\gamma }-\delta
Z_{2}^{Z\gamma }) \\
\setlength{\unitlength}{3mm}\begin{picture}(15,7)(-2,0)
\put(0,1.5){\makebox(0,0)[c]{$d $}} \put(11,1.5){\makebox(0,0)[c]{$\bar u$}}
\put(5.25,0){\makebox(0,0)[c]{$\times$}} \put(0,0){\line(1,0){10}}
\put(2.5,0){\vector(1,0){0.1}} \put(7.5,0){\vector(1,0){0.1}}
\put(6,5){\makebox(0,0)[l]{$W^+_\mu$}} \multiput(4.8,0.25)(0,1){5}{\zvs}
\end{picture} &=&-i\gamma _{\mu }(1-\gamma _{5})\,(\delta Z_{1}^{W}-\delta
Z_{2}^{W}+\delta Z_{L})  \label{17.6i}
\end{eqnarray}
Here we have introduced the notation
\begin{equation}
\delta Z_{L}=\delta Z_{V}^{u,d}+\delta Z_{A}^{u,d},\quad \delta
Z_{R}^{u,d}=\delta Z_{V}^{u,d}-\delta Z_{A}^{u,d},
\end{equation}
and
\begin{equation}
v_{f}={\frac{\tau ^{3}/2-2Q_{f}s_{W}^{2}}{2s_{W}c_{W}}},\qquad a_{f}={\frac{%
\tau ^{3}/2}{2s_{W}c_{W}}}.  \label{vfaf}
\end{equation}
Note that the Feynman rules for the vertices contain additional
renormalization constants which should be familiar from the oblique
corrections.

The fermion self-energies can be decomposed as
\begin{equation}
\Sigma ^{f}(p)=\not{p}\,\Sigma _{V}^{f}(p^{2})+\not{p}\gamma _{5}\,\Sigma
_{A}^{f}(p^{2})+m\,\Sigma _{S}^{f}(p^{2}).
\end{equation}

By adding the counterterms one obtains de renormalized self-energies, which
admit the same decomposition. One has
\begin{eqnarray}
\hat{\Sigma}_{V}^{f}(p^{2}) &=&\Sigma _{V}^{f}(p^{2})-\delta Z_{V}^{f},
\label{17.6d} \\
\hat{\Sigma}_{A}^{f}(p^{2}) &=&\Sigma _{A}^{f}(p^{2})+\delta Z_{A}^{f},
\label{17.6e} \\
\hat{\Sigma}_{S}^{f}(p^{2}) &=&\Sigma _{S}^{f}(p^{2})+{\frac{\delta m_{f}}{%
m_{f}}}+\delta Z_{V}^{f},  \label{17.6f}
\end{eqnarray}
where the hat denotes renormalized quantities. The on-shell renormalization
conditions amount to
\begin{eqnarray}
{\frac{\delta m_{u,d}}{m_{u,d}}} &=&-\Sigma _{V}^{u,d}(m_{u,d}^{2})-\Sigma
_{S}^{u,d}(m_{u,d}^{2}),  \label{9.6a} \\
\delta Z_{V}^{d} &=&\Sigma _{V}^{d}(m_{d}^{2})+2m_{d}^{2}[\Sigma
_{V}^{d}{}^{\prime }(m_{d}^{2})+\Sigma _{S}^{d}{}^{\prime }(m_{d}^{2})],
\label{17.6b} \\
\delta Z_{A}^{u,d} &=&-\Sigma _{A}^{u,d}(m_{u,d}^{2}),  \label{17.6c}
\end{eqnarray}
where $\Sigma ^{\prime }(m^{2})=[\partial \Sigma (p^{2})/\partial
p^{2}]_{p^{2}=m^{2}}$. Eq.~(\ref{9.6a}) guarantees that $m_{u}$, $m_{d}$ are
the physical fermion masses. The other two equations, come from requiring
that the residue of the down-type fermion be unity. One cannot
simultaneously impose this condition to both up- and down-type fermions.
Actually, one can easily work out the residue of the up-type fermions which
turns out to be $1+\delta _{res}$ with
\begin{equation}
\delta _{res}=\hat{\Sigma}_{V}^{u}(m_{u}^{2})+2m_{u}^{2}\left[ \hat{\Sigma}%
_{V}^{u}{}^{\prime }(m_{u}^{2})+\hat{\Sigma}_{S}^{u}{}^{\prime }(m_{u}^{2})%
\right] .  \label{17.6j}
\end{equation}

\section{Effective Lagrangian coefficients}

\label{MsectorappE}In this appendix we shall provide the general expressions
for the coefficients $a_{i}$ and $M_{L,R}^{i}$ in theories of the type we
have been considering in Chapter \ref{mattersectorchapter}. The results are
for the usual representations of $SU(2)\times SU(3)_{c}$. Extension to other
representations is possible using the prescriptions listed in section \ref
{heavyfermions} in Chapter \ref{mattersectorchapter}. The coefficients $%
a_{i} $ in theories with technifermion doublets with masses ($m_{1}$, $m_{2}$%
), are given by
\begin{eqnarray}
a_{0} &=&\frac{n_{TC}n_{D}}{64\pi ^{2}M_{Z}^{2}s_{W}^{2}}\left( \frac{%
m_{2}^{2}+m_{1}^{2}}{2}+\frac{m_{1}^{2}m_{2}^{2}\ln \frac{m_{1}^{2}}{%
m_{2}^{2}}}{m_{2}^{2}-m_{1}^{2}}\right) +\frac{1}{16\pi ^{2}}\frac{3}{8}(%
\frac{1}{\hat{\epsilon}}-\log \frac{\Lambda ^{2}}{\mu ^{2}}), \\
a_{1} &=&-\frac{n_{TC}n_{D}}{96\pi ^{2}}+\frac{n_{TC}\left(
n_{Q}-3n_{L}\right) }{3\times 96\pi ^{2}}\ln \frac{m_{1}^{2}}{m_{2}^{2}}+%
\frac{1}{16\pi ^{2}}\frac{1}{12}(\frac{1}{\hat{\epsilon}}-\log \frac{\Lambda
^{2}}{\mu ^{2}}), \\
a_{8} &=&-\frac{n_{TC}\left( n_{c}+1\right) }{96\pi ^{2}}\frac{1}{\left(
m_{2}^{2}-m_{1}^{2}\right) ^{2}}\left\{ \frac{5}{3}m_{1}^{4}-\frac{22}{3}%
m_{2}^{2}m_{1}^{2}+\frac{5}{3}m_{2}^{4}\right.  \notag \\
&&+\left. \left( m_{2}^{4}-4m_{2}^{2}m_{1}^{2}+m_{1}^{4}\right) \frac{%
m_{2}^{2}+m_{1}^{2}}{m_{2}^{2}-m_{1}^{2}}\ln \frac{m_{1}^{2}}{m_{2}^{2}}%
\right\} ,
\end{eqnarray}
where $n_{TC}$ the number of technicolors (taken equal to 2 in all numerical
discussions), $n_{D}$ is the number of technidoublets. It is interesting to
note that all effective Lagrangian coefficients (except for $a_{1}$) depend
on $n_{D}$ and are independent of the actual hypercharge (or charge)
assignment. $n_{Q}$ and $n_{L}$ are the actual number of techniquarks and
technileptons. In the one-generation model $n_{Q}=3$, $n_{L}=1$ and,
consequently, $n_{D}=4$. Furthermore in this model $a_{1}$ is mass
independent. For simplicity we have written $m_{1}$ for the dynamically
generated mass of the $u$-type technifermion and $m_{2}$ for the one of the $%
d$-type, and assumed that they are the same for all doublets. This is of
course quite questionable as a large splitting between the technielectron
and the technineutrino seems more likely and they should not necessarily
coincide with techniquark masses, but the appropriate expressions can be
easily inferred from the above formulae anyway. For the coefficients $%
M_{L,R}^{i}$ we have
\begin{eqnarray}
2M_{L}^{1} &=&{\frac{n_{D}n_{TC}G^{2}}{16\pi ^{2}M^{2}}}a_{\vec{L}%
^{2}}\left\{ {\frac{m_{1}^{2}+m_{2}^{2}}{2}}-m_{1}^{2}\left( 1+{\frac{%
m_{1}^{2}}{m_{1}^{2}-m_{2}^{2}}}\right) \log {\frac{m_{1}^{2}}{M^{2}}}\right.
\notag \\
&&-m_{2}^{2}\left. \left( 1+{\frac{m_{2}^{2}}{m_{2}^{2}-m_{1}^{2}}}\right)
\log {\frac{m_{2}^{2}}{M^{2}}}\right\} ,  \label{aa1} \\
M_{L}^{2} &=&{\frac{n_{D}n_{TC}G^{2}}{16\pi ^{2}M^{2}}}\left\{ \left(
a_{L^{2}}-a_{RL}\right) A_{-}+a_{R_{3}L}A_{+}\right\} ,  \label{aa3} \\
2M_{L}^{3} &=&{\frac{n_{D}n_{TC}G^{2}}{16\pi ^{2}M^{2}}}\;a_{\vec{L}%
^{2}}\left\{ {\frac{m_{1}^{2}+m_{2}^{2}}{2}}+m_{1}^{2}\left( 1-{\frac{%
m_{1}^{2}}{m_{1}^{2}-m_{2}^{2}}}\right) \log {\frac{m_{1}^{2}}{M^{2}}}\right.
\notag \\
&&+m_{2}^{2}\left. \left( 1-{\frac{m_{2}^{2}}{m_{2}^{2}-m_{1}^{2}}}\right)
\log {\frac{m_{2}^{2}}{M^{2}}}\right\} ,  \label{aa4} \\
M_{L}^{4} &=&0,  \label{aa7} \\
2M_{R}^{1} &=&{\frac{n_{D}n_{TC}G^{2}}{16\pi ^{2}M^{2}}}\left\{ \left(
a_{LR_{3}}-a_{RR_{3}}\right) A_{-}+a_{R_{3}^{2}}A_{+}+a_{\vec{R}%
^{2}}B_{+}\right\} ,  \label{aa2} \\
M_{R}^{2} &=&{\frac{n_{D}n_{TC}G^{2}}{16\pi ^{2}M^{2}}}\left\{ \left(
a_{LR}-a_{R^{2}}\right) A_{-}+a_{R_{3}R}A_{+}\right\} ,  \label{aa5} \\
2M_{R}^{3} &=&{\frac{n_{D}n_{TC}G^{2}}{16\pi ^{2}M^{2}}}\left\{ \left(
a_{LR_{3}}-a_{RR_{3}}\right) A_{-}+a_{R_{3}^{2}}A_{+}+a_{\vec{R}%
^{2}}B_{-}\right\} ,  \label{aa6}
\end{eqnarray}
where
\begin{eqnarray}
A_{\pm } &=&\mp m_{1}^{2}\log {\frac{m_{1}^{2}}{M^{2}}}-m_{2}^{2}\log {\frac{%
m_{2}^{2}}{M^{2}}}  \label{bb1} \\
B_{\pm } &=&\pm 2m_{1}m_{2}-m_{1}^{2}\left( 1\pm {\frac{2m_{1}m_{2}}{%
m_{1}^{2}-m_{2}^{2}}}\right) \log {\frac{m_{1}^{2}}{M^{2}}}  \notag \\
&-&m_{2}^{2}\left( 1\pm {\frac{2m_{2}m_{1}}{m_{2}^{2}-m_{1}^{2}}}\right)
\log {\frac{m_{2}^{2}}{M^{2}}}.  \label{bb2}
\end{eqnarray}
We have not bothered to write the chiral divergences counterterms in the
above expressions. They are identical to those of section \ref{heavyfermions}
in Chapter \ref{mattersectorchapter}. Although we have written the full
expressions obtained using chiral quark model methods, one should be well
aware of the approximations made in the text.

\chapter{Fermionic Self-Energy calculations in $R_{\protect\xi }$ gauges.}

\label{selfenergiesapp}In dimensional regularization we have
\begin{equation*}
\left[ dx^{\mathtt{d}}\bar{f}\not{\partial}f\right] =0=1-\mathtt{d}+2\left[ f%
\right] ,
\end{equation*}
and
\begin{equation*}
\left[ dx^{\mathtt{d}}\partial _{\mu }A_{\nu }\partial ^{\mu }A^{\nu }\right]
=0=2-\mathtt{d}+2\left[ A\right] ,
\end{equation*}
finally
\begin{equation*}
\left[ g\mu ^{\frac{\epsilon }{2}}dx^{\mathtt{d}}\bar{f}\not{A}f\right] =0=%
\frac{\epsilon }{2}-\mathtt{d}+2\left[ f\right] +\left[ A\right] ,
\end{equation*}
hence, for the following calculations we take
\begin{equation*}
\epsilon =4-\mathtt{d}.
\end{equation*}
We will use the naive prescription for $\gamma ^{5}$ in $\mathtt{d}$
dimensions, i.e. we will take it as anticommuting with $\gamma ^{\mu }$.
Since we do not need to calculate triangle diagrams, this easy-to-use
prescription is compatible with Ward identities \cite{Buras:1998ra}.

\section{Fermionic Self-Energies}

We want to calculate the 1-loop diagrams with Higgs and Goldstone bosons as
internal lines
\begin{equation}
\begin{array}{ccccccc}
-i\Sigma _{ij}^{u\phi } & \equiv & u_{j}\overset{\phi }{\widehat{\rightarrow
d\rightarrow }}u_{i}, & \qquad & -i\Sigma _{ij}^{d\phi } & \equiv & d_{j}%
\overset{\phi }{\widehat{\rightarrow u\rightarrow }}d_{i},
\end{array}
\label{goldstone}
\end{equation}
where $\phi $ can be the Higgs $\rho $ or the Goldstone bosons $\chi ^{i}$ ;
and the 1-loop diagrams with gauge bosons as internal lines
\begin{equation}
\begin{array}{ccccccc}
-i\Sigma _{ij}^{u\phi } & \equiv & u_{j}\overset{\phi }{\widehat{\rightarrow
d\rightarrow }}u_{i}, & \qquad & -i\Sigma _{ij}^{d\phi } & \equiv & d_{j}%
\overset{\phi }{\widehat{\rightarrow u\rightarrow }}d_{i},
\end{array}
\label{gauge}
\end{equation}
where $\phi $ can be $W^{\pm },$ $Z,$ a foton $A$ or a gluon $G$ according
to the notation of this appendix.

\section{Feynman rules}

\subsection{Vertices}

In the Standard Model we have the kinetic terms
\begin{eqnarray*}
\mathcal{L}_{R} &=&if^{\dagger }\gamma ^{0}\gamma ^{\mu }\left\{ \partial
_{\mu }+ig^{\prime }\left( \frac{\tau ^{3}}{2}+z\right) B_{\mu }+ig_{s}\frac{%
\mathbf{\lambda }}{2}\mathbf{\cdot G}_{\mu }\right\} Rf, \\
\mathcal{L}_{L} &=&if^{\dagger }\gamma ^{0}\gamma ^{\mu }\left\{ \partial
_{\mu }+ig^{\prime }zB_{\mu }+ig\frac{\tau ^{3}}{2}W_{\mu }^{3}\right. \\
&&\left. +ig\left( K_{-}\frac{\tau ^{-}}{2}W_{\mu }^{+}+K_{-}^{\dagger }%
\frac{\tau ^{+}}{2}W_{\mu }^{-}\right) +ig_{s}\frac{\mathbf{\lambda }}{2}%
\mathbf{\cdot G}_{\mu }\right\} Lf,
\end{eqnarray*}
where $L$ and $R$ are the left and right projectors and $z$ is a real
parameter which takes the value $\frac{1}{6}$ for quarks and $\frac{-1}{2}$
for leptons. In the non-linear representation we have for the mass term
\begin{eqnarray*}
\mathcal{L}_{m} &=&-f^{\dagger }\gamma ^{0}\left\{ \left( \tau
^{u}M+K^{\dagger }\tau ^{d}M\right) \tau ^{u}y^{u}\right. \\
&&\left. +\left( \tau ^{d}M+K\tau ^{u}M\right) \tau ^{d}y^{d}\right\}
Rf+h.c.,
\end{eqnarray*}
where $y^{u}$ and $y^{u}$ are the diagonal Yukawa matrices
\begin{eqnarray*}
y_{ij}^{u} &=&\delta _{ij}\frac{m_{i}^{u}}{v}, \\
y_{ij}^{d} &=&\delta _{ij}\frac{m_{i}^{d}}{v}, \\
i,j &=&1,2,3\quad \mathrm{(family\ indices)}
\end{eqnarray*}
$K$ is the CKM matrix and $M$ is given by
\begin{eqnarray*}
M &=&\left( v+\rho \right) U=\left( v+\rho \right) e^{i\tau ^{i}\chi
^{i}/v}=v+\rho +i\tau ^{i}\chi ^{i}+i\tau ^{i}\frac{\rho }{v}\chi
^{i}+O\left( \chi ^{2}\right) \\
&=&v+\rho +\left( 1+\frac{\rho }{v}\right) \left( i\tau ^{3}\chi ^{3}+i\tau
^{-}\chi ^{+}+i\tau ^{+}\chi ^{-}\right) +O\left( \chi ^{2}\right) ,
\end{eqnarray*}
where $\rho $ and the $\chi ^{i}$ are the non-linear Higgs and Goldstone
bosons fields respectively and
\begin{equation*}
\chi ^{\pm }\equiv \frac{\chi ^{1}\mp i\chi ^{2}}{\sqrt{2}}.
\end{equation*}
Then
\begin{eqnarray*}
\mathcal{L}_{m} &=&-f^{\dagger }\gamma ^{0}\left\{ \left( v+\rho +i\left( 1+%
\frac{\rho }{v}\right) \chi ^{3}\right) \tau ^{u}y^{u}+iK^{\dagger }\tau
^{+}\left( 1+\frac{\rho }{v}\right) \chi ^{-}y^{u}\right. \\
&&+\left. \left( v+\rho -i\left( 1+\frac{\rho }{v}\right) \chi ^{3}\right)
\tau ^{d}y^{d}+iK\tau ^{-}\left( 1+\frac{\rho }{v}\right) \chi
^{+}y^{d}\right\} Rf \\
&&+h.c.+O\left( \chi ^{2}\right) ,
\end{eqnarray*}
or
\begin{eqnarray*}
\mathcal{L}_{m} &=&-\bar{u}\left( v+\rho +i\left( R-L\right) \left( 1+\frac{%
\rho }{v}\right) \chi ^{3}\right) y^{u}u \\
&&-\bar{d}\left( v+\rho -i\left( R-L\right) \left( 1+\frac{\rho }{v}\right)
\chi ^{3}\right) y^{d}d \\
&&-i\sqrt{2}\bar{u}\left( 1+\frac{\rho }{v}\right) \chi ^{+}\left(
RKy^{d}-y^{u}KL\right) d \\
&&-i\sqrt{2}\bar{d}\left( 1+\frac{\rho }{v}\right) \chi ^{-}\left(
RK^{\dagger }y^{u}-y^{d}K^{\dagger }L\right) u+O\left( \chi ^{2}\right) ,
\end{eqnarray*}
from where we can read the vertices
\begin{eqnarray}
\bar{u}_{i}\rho u_{j} &=&-i\delta _{ij}\frac{m_{i}^{u}}{v},  \notag \\
\bar{d}_{i}\rho d_{j} &=&-i\delta _{ij}\frac{m_{i}^{d}}{v},  \notag \\
\bar{u}_{i}\chi ^{3}u_{j} &=&\delta _{ij}\frac{m_{i}^{u}}{v}\left(
R-L\right) ,  \notag \\
\bar{d}_{i}\chi ^{3}d_{j} &=&\delta _{ij}\frac{m_{i}^{d}}{v}\left(
L-R\right) ,  \notag \\
\bar{u}_{i}\chi ^{+}d_{j} &=&\frac{\sqrt{2}}{v}\left(
K_{ij}m_{j}^{d}R-m_{i}^{u}K_{ij}L\right) ,  \notag \\
\bar{d}_{i}\chi ^{-}u_{j} &=&\frac{\sqrt{2}}{v}\left( K_{ij}^{\dagger
}m_{j}^{u}R-m_{i}^{d}K_{ij}^{\dagger }L\right) ,  \label{e1}
\end{eqnarray}
And the four leg vertices including the Higgs are
\begin{eqnarray}
\bar{u}_{i}\rho \chi ^{3}u_{j} &=&\delta _{ij}\frac{m_{i}^{u}}{v^{2}}\left(
R-L\right) ,  \notag \\
\bar{d}_{i}\rho \chi ^{3}d_{j} &=&\delta _{ij}\frac{m_{i}^{d}}{v^{2}}\left(
L-R\right) ,  \notag \\
\bar{u}_{i}\rho \chi ^{+}d_{j} &=&\frac{\sqrt{2}}{v^{2}}\left(
K_{ij}m_{j}^{d}R-m_{i}^{u}K_{ij}L\right) ,  \notag \\
\bar{d}_{i}\rho \chi ^{-}u_{j} &=&\frac{\sqrt{2}}{v^{2}}\left(
m_{j}^{u}K_{ji}^{\ast }R-K_{ji}^{\ast }m_{i}^{d}L\right) .  \label{flv}
\end{eqnarray}
While from the kinetic terms we obtain
\begin{eqnarray*}
\mathcal{L}_{R} &=&i\bar{u}\gamma ^{\mu }\left\{ \partial _{\mu }+ig^{\prime
}\left( \frac{1}{2}+z\right) B_{\mu }+ig_{s}\frac{\mathbf{\lambda }}{2}%
\mathbf{\cdot G}_{\mu }\right\} Ru \\
&&+i\bar{d}\gamma ^{\mu }\left\{ \partial _{\mu }+ig^{\prime }\left( \frac{-1%
}{2}+z\right) B_{\mu }+ig_{s}\frac{\mathbf{\lambda }}{2}\mathbf{\cdot G}%
_{\mu }\right\} Rd,
\end{eqnarray*}
which, using
\begin{eqnarray*}
s_{W} &\equiv &\sin \theta _{W}\equiv \frac{g^{\prime }}{\sqrt{%
g^{2}+g^{\prime \,2}}}, \\
c_{W} &\equiv &\cos \theta _{W}\equiv \frac{g}{\sqrt{g^{2}+g^{\prime \,2}}},
\\
e &\equiv &gs_{W}=g^{\prime }c_{W}, \\
W_{\mu }^{3} &=&s_{W}A_{\mu }+c_{W}Z_{\mu }, \\
B_{\mu } &=&c_{W}A_{\mu }-s_{W}Z_{\mu },
\end{eqnarray*}
becomes
\begin{eqnarray}
\mathcal{L}_{R} &=&i\bar{u}\gamma ^{\mu }\left\{ \partial _{\mu }+ie\left(
\frac{1}{2}+z\right) \left( A_{\mu }-\frac{s_{W}}{c_{W}}Z_{\mu }\right)
+ig_{s}\frac{\mathbf{\lambda }}{2}\mathbf{\cdot G}_{\mu }\right\} Ru  \notag
\\
&&+i\bar{d}\gamma ^{\mu }\left\{ \partial _{\mu }+ie\left( \frac{-1}{2}%
+z\right) \left( A_{\mu }-\frac{s_{W}}{c_{W}}Z_{\mu }\right) +ig_{s}\frac{%
\mathbf{\lambda }}{2}\mathbf{\cdot G}_{\mu }\right\} Rd,  \label{kinRphys}
\end{eqnarray}
and
\begin{eqnarray*}
\mathcal{L}_{L} &=&i\bar{u}\gamma ^{\mu }\left\{ \partial _{\mu }+ig^{\prime
}zB_{\mu }+ig\frac{1}{2}W_{\mu }^{3}+ig_{s}\frac{\mathbf{\lambda }}{2}%
\mathbf{\cdot G}_{\mu }\right\} Lu \\
&&+i\bar{d}\gamma ^{\mu }\left\{ \partial _{\mu }+ig^{\prime }zB_{\mu }-ig%
\frac{1}{2}W_{\mu }^{3}+ig_{s}\frac{\mathbf{\lambda }}{2}\mathbf{\cdot G}%
_{\mu }\right\} Ld \\
&&-\frac{g}{\sqrt{2}}\left[ \bar{u}\gamma ^{\mu }KW_{\mu }^{+}Ld+\bar{d}%
\gamma ^{\mu }K^{\dagger }W_{\mu }^{-}Lu\right] ,
\end{eqnarray*}
and therefore
\begin{eqnarray}
\mathcal{L}_{kin} &=&\mathcal{L}_{L}+\mathcal{L}_{R}=i\bar{f}\gamma ^{\mu
}\left\{ \partial _{\mu }+ig_{s}\frac{\mathbf{\lambda }}{2}\mathbf{\cdot G}%
_{\mu }+ie\left( \frac{\tau ^{3}}{2}+z\right) A_{\mu }\right.  \notag \\
&&+\left. \frac{ie}{s_{W}c_{W}}\left[ \left( \frac{\tau ^{3}}{2}%
c_{W}^{2}-zs_{W}^{2}\right) L-\left( \frac{\tau ^{3}}{2}+z\right) s_{W}^{2}R%
\right] Z_{\mu }\right\} f  \notag \\
&&-\frac{e}{\sqrt{2}s_{W}}\left[ \bar{u}\gamma ^{\mu }KW_{\mu }^{+}Ld+\bar{d}%
\gamma ^{\mu }K^{\dagger }W_{\mu }^{-}Lu\right] ,  \label{kinphys}
\end{eqnarray}
So from Eq. (\ref{kinphys}) we can read the vertices
\begin{eqnarray}
\bar{u}_{i}G_{\mu }^{a}u_{j} &=&\bar{d}_{i}G_{\mu }^{a}d_{j}=-i\delta
_{ij}g_{s}\frac{\lambda }{2}^{a}\gamma _{\mu },  \notag \\
\bar{u}_{i}A_{\mu }u_{j} &=&-i\delta _{ij}e\left( z+\frac{1}{2}\right)
\gamma _{\mu },  \notag \\
\bar{d}_{i}A_{\mu }d_{j} &=&-i\delta _{ij}e\left( z-\frac{1}{2}\right)
\gamma _{\mu },  \notag \\
\bar{u}_{i}Z_{\mu }u_{j} &=&i\delta _{ij}\gamma _{\mu }\frac{e}{c_{W}s_{W}}%
\left[ s_{W}^{2}\left( z+\frac{1}{2}\right) R+\left( zs_{W}^{2}-\frac{1}{2}%
c_{W}^{2}\right) L\right] ,  \notag \\
\bar{d}_{i}Z_{\mu }d_{j} &=&i\delta _{ij}\gamma _{\mu }\frac{e}{c_{W}s_{W}}%
\left[ s_{W}^{2}\left( z-\frac{1}{2}\right) R+\left( zs_{W}^{2}+\frac{1}{2}%
c_{W}^{2}\right) L\right] ,  \notag \\
\bar{u}_{i}W_{\mu }^{+}d_{j} &=&-i\gamma _{\mu }\frac{e}{\sqrt{2}s_{W}}%
K_{ij}L,  \notag \\
\bar{d}_{i}W_{\mu }^{-}u_{j} &=&-i\gamma _{\mu }\frac{e}{\sqrt{2}s_{W}}%
K_{ij}^{\dagger }L,  \label{gaugevert}
\end{eqnarray}

\subsection{Propagators}

Defining
\begin{equation*}
\left\langle \varphi _{1}\varphi _{2}\right\rangle \equiv \int
d^{4}x\left\langle T\varphi _{1}\left( x\right) \varphi _{2}\left( y\right)
\right\rangle _{tree}e^{ik\left( x-y\right) },
\end{equation*}
then after gauge fixing, for the propagators we have the following Feynman
rules
\begin{eqnarray}
\left\langle \rho \rho \right\rangle &=&\frac{i}{k^{2}-M_{\rho
}^{2}+i\varepsilon },  \notag \\
\left\langle \chi ^{3}\chi ^{3}\right\rangle &=&\frac{i}{k^{2}-\xi
M_{Z}^{2}+i\varepsilon },  \notag \\
\left\langle \chi ^{+}\chi ^{-}\right\rangle &=&\frac{i}{k^{2}-\xi
M_{W}^{2}+i\varepsilon },  \notag \\
\left\langle W_{\mu }^{+}W_{\nu }^{-}\right\rangle &=&\frac{-i}{%
k^{2}-M_{W}^{2}+i\varepsilon }\left( g_{\mu \nu }+\left( \xi -1\right) \frac{%
k_{\mu }k_{\nu }}{k^{2}-\xi M_{W}^{2}}\right) ,  \notag \\
\left\langle Z_{\mu }Z_{\nu }\right\rangle &=&\frac{-i}{k^{2}-M_{Z}^{2}+i%
\varepsilon }\left( g_{\mu \nu }+\left( \xi -1\right) \frac{k_{\mu }k_{\nu }%
}{k^{2}-\xi M_{Z}^{2}}\right) ,  \notag \\
\left\langle A_{\mu }A_{\nu }\right\rangle &=&\frac{-i}{k^{2}+i\varepsilon }%
\left( g_{\mu \nu }+\left( \xi -1\right) \frac{k_{\mu }k_{\nu }}{k^{2}}%
\right) ,  \notag \\
\left\langle G_{\mu }^{a}G_{\nu }^{b}\right\rangle &=&\frac{-i\delta ^{ab}}{%
k^{2}+i\varepsilon }\left( g_{\mu \nu }+\left( \xi -1\right) \frac{k_{\mu
}k_{\nu }}{k^{2}}\right) ,  \notag \\
\left\langle f\bar{f}\right\rangle &=&\frac{i\left( \not{k}+m_{f}\right) }{%
k^{2}-m_{f}^{2}+i\varepsilon },  \label{e2}
\end{eqnarray}
where the same gauge fixing parameter $\xi $ has been taken for all gauge
bosons (one can easily take $\xi _{W}\neq \xi _{Z}\neq \xi _{A}$ if
necessary).

\section{Higgs and Goldstone bosons as internal lines}

\begin{itemize}
\item  {\huge $-i\Sigma _{ij}^{u\chi ^{+}}$}
\end{itemize}

Using Eqs. (\ref{e1}) and (\ref{e2}) we obtain
\begin{eqnarray}
-i\Sigma _{ij}^{u\chi ^{+}} &=&\sum_{h}\frac{2\mu ^{\epsilon }}{\left( 2\pi
\right) ^{\mathtt{d}}v^{2}}\int d^{\mathtt{d}}k\left(
m_{h}^{d}R-m_{i}^{u}L\right) K_{ih}\frac{i\left( \not{k}+m_{h}^{d}\right) }{%
k^{2}-m_{h}^{d2}+i\varepsilon }  \notag \\
&&\times \left( m_{j}^{u}R-m_{h}^{d}L\right) K_{hj}^{\dagger }\frac{i}{%
\left( p-k\right) ^{2}-\xi M_{W}^{2}+i\varepsilon }  \notag \\
&=&\sum_{h}\frac{2\mu ^{\epsilon }K_{ih}K_{hj}^{\dagger }}{\left( 2\pi
\right) ^{\mathtt{d}}v^{2}}  \notag \\
&&\times \int d^{\mathtt{d}}k\frac{\not{k}\left(
m_{i}^{u}m_{j}^{u}R+m_{h}^{d2}L\right) -m_{h}^{d2}\left(
m_{j}^{u}R+m_{i}^{u}L\right) }{\left( k^{2}-m_{h}^{d2}+i\varepsilon \right)
\left( \left( p-k\right) ^{2}-\xi M_{W}^{2}+i\varepsilon \right) },
\label{sux+}
\end{eqnarray}
Introducing a Feynman parameter we have
\begin{eqnarray*}
-i\Sigma _{ij}^{u\chi ^{+}} &=&\sum_{h}\left\{ \frac{2\mu ^{\epsilon
}K_{ih}K_{hj}^{\dagger }}{\left( 2\pi \right) ^{\mathtt{d}}v^{2}}%
\int_{0}^{1}dx\int d^{\mathtt{d}}k\right. \\
&&\times \left. \frac{\not{k}\left( m_{i}^{u}m_{j}^{u}R+m_{h}^{d2}L\right)
-m_{h}^{d2}\left( m_{j}^{u}R+m_{i}^{u}L\right) }{\left[ x\left(
k^{2}-m_{h}^{d2}+i\varepsilon \right) +\left( 1-x\right) \left( \left(
p-k\right) ^{2}-\xi M_{W}^{2}+i\varepsilon \right) \right] ^{2}}\right\}
\end{eqnarray*}
but
\begin{eqnarray*}
&&x\left( k^{2}-m_{h}^{d2}+i\varepsilon \right) +\left( 1-x\right) \left(
\left( p-k\right) ^{2}-\xi M_{W}^{2}+i\varepsilon \right) \\
&=&k^{2}-2kp\left( 1-x\right) +\left( p^{2}-\xi M_{W}^{2}\right) \left(
1-x\right) -m_{h}^{d2}x+i\varepsilon \\
&=&\left( k-p\left( 1-x\right) \right) ^{2}+p^{2}x\left( 1-x\right) -\xi
M_{W}^{2}\left( 1-x\right) -m_{h}^{d2}x+i\varepsilon
\end{eqnarray*}
so
\begin{eqnarray*}
-i\Sigma _{ij}^{u\chi ^{+}} &=&\sum_{h}\left\{ \frac{2\mu ^{\epsilon
}K_{ih}K_{hj}^{\dagger }}{\left( 2\pi \right) ^{\mathtt{d}}v^{2}}%
\int_{0}^{1}dx\int d^{\mathtt{d}}k\right. \\
&&\times \left. \frac{\left( \not{k}+\not{p}\left( 1-x\right) \right) \left(
m_{i}^{u}m_{j}^{u}R+m_{h}^{d2}L\right) -m_{h}^{d2}\left(
m_{j}^{u}R+m_{i}^{u}L\right) }{\left[ k^{2}+p^{2}x\left( 1-x\right) -\xi
M_{W}^{2}\left( 1-x\right) -m_{h}^{d2}x+i\varepsilon \right] ^{2}}\right\} \\
&=&\sum_{h}\left\{ \frac{2K_{ih}K_{hj}^{\dagger }}{v^{2}}%
\int_{0}^{1}dxA_{h}^{Wd}\left( \mu ,x,p^{2},\epsilon \right) \right. \\
&&\times \left. \left[ \not{p}\left( m_{i}^{u}m_{j}^{u}R+m_{h}^{d2}L\right)
\left( 1-x\right) -m_{h}^{d2}\left( m_{j}^{u}R+m_{i}^{u}L\right) \right]
\right\} ,
\end{eqnarray*}
where
\begin{equation*}
A_{h}^{Wd}\left( x,p^{2},\epsilon \right) \equiv \int \frac{d^{\mathtt{d}}k}{%
\left( 2\pi \right) ^{\mathtt{d}}}\frac{\mu ^{\epsilon }}{\left[
k^{2}-\Delta _{h}^{Wd}\right] ^{2}},
\end{equation*}
with
\begin{equation*}
\Delta _{h}^{Wd}\equiv p^{2}x\left( x-1\right) +\xi M_{W}^{2}\left(
1-x\right) +m_{h}^{d2}x-i\varepsilon ,
\end{equation*}
but
\begin{eqnarray}
A_{h}^{W}\left( x,p^{2},\epsilon \right) &=&\frac{i\mu ^{\epsilon }}{\left(
4\pi \right) ^{2-\frac{\epsilon }{2}}}\frac{\Gamma \left( \frac{\epsilon }{2}%
\right) }{\Gamma \left( 2\right) }\left( \frac{1}{\Delta _{h}^{Wd}}\right) ^{%
\frac{\epsilon }{2}}  \notag \\
&=&\frac{i}{\left( 4\pi \right) ^{2}}\left( \hat{\epsilon}^{-1}-\ln \left(
\frac{\Delta _{h}^{Wd}}{\mu ^{2}}\right) \right) +O\left( \epsilon \right) ,
\label{facil}
\end{eqnarray}
where
\begin{equation*}
\hat{\epsilon}^{-1}\equiv 2\epsilon ^{-1}+\ln \left( 4\pi \right) -\gamma
_{E},
\end{equation*}
so, finally
\begin{eqnarray}
-i\Sigma _{ij}^{u\chi ^{+}} &=&-i\check{\Sigma}_{ij}^{u\chi ^{+}}-\sum_{h}%
\frac{i2K_{ih}K_{hj}^{\dagger }}{\left( 4\pi \right) ^{2}v^{2}}%
\int_{0}^{1}dx\ln \left( \frac{\Delta _{h}^{Wd}}{\mu ^{2}}\right)  \notag \\
&&\times \left[ \not{p}\left( m_{i}^{u}m_{j}^{u}R+m_{h}^{d2}L\right) \left(
1-x\right) -m_{h}^{d2}\left( m_{j}^{u}R+m_{i}^{u}L\right) \right] ,
\label{e3}
\end{eqnarray}
where the divergent part is given by
\begin{equation*}
-i\check{\Sigma}_{ij}^{u\chi ^{+}}=\sum_{h}\frac{i2K_{ih}K_{hj}^{\dagger }}{%
\left( 4\pi \right) ^{2}v^{2}}\hat{\epsilon}^{-1}\left( \frac{1}{2}\left(
m_{i}^{u}m_{j}^{u}L+m_{h}^{d2}R\right) \not{p}-m_{h}^{d2}\left(
m_{j}^{u}R+m_{i}^{u}L\right) \right) .
\end{equation*}

\begin{itemize}
\item  {\huge $-i\Sigma _{ij}^{d\chi ^{-}}$}
\end{itemize}

Using Eqs. (\ref{e1}) and (\ref{e2}) we obtain
\begin{eqnarray*}
-i\Sigma _{ij}^{d\chi ^{-}} &=&\sum_{h}\frac{2\mu ^{\epsilon }}{\left( 2\pi
\right) ^{\mathtt{d}}v^{2}}\int d^{\mathtt{d}}k\left(
m_{h}^{u}R-m_{i}^{d}L\right) K_{ih}^{\dagger }\frac{i\left( \not{k}%
+m_{h}^{u}\right) }{k^{2}-m_{h}^{u2}+i\varepsilon } \\
&&\times \left( m_{j}^{d}R-m_{h}^{u}L\right) K_{hj}\frac{i}{\left(
p-k\right) ^{2}-\xi M_{W}^{2}+i\varepsilon },
\end{eqnarray*}
which is identical to the expression for $\Sigma _{ij}^{u\chi ^{+}}$ given
by Eq. (\ref{sux+}) performing the changes ($u\leftrightarrow d$ and $%
K\leftrightarrow K^{\dagger }$). So
\begin{equation*}
\Delta _{h}^{Wd}\equiv p^{2}x\left( x-1\right) +\xi M_{W}^{2}\left(
1-x\right) +m_{h}^{d2}x=0,
\end{equation*}
\begin{eqnarray}
-i\Sigma _{ij}^{d\chi ^{-}} &=&-i\check{\Sigma}_{ij}^{d\chi ^{-}}-\sum_{h}%
\frac{i2K_{ih}^{\dagger }K_{hj}}{\left( 4\pi \right) ^{2}v^{2}}%
\int_{0}^{1}dx\ln \left( \frac{\Delta _{h}^{Wu}}{\mu ^{2}}\right)  \notag \\
&&\times \left[ \not{p}\left( m_{i}^{d}m_{j}^{d}R+m_{h}^{u2}L\right) \left(
1-x\right) -m_{h}^{u2}\left( m_{j}^{d}R+m_{i}^{d}L\right) \right] ,
\label{e4}
\end{eqnarray}
where the divergent part is given by
\begin{equation*}
-i\check{\Sigma}_{ij}^{d\chi ^{-}}=\sum_{h}\frac{i2K_{ih}^{\dagger }K_{hj}}{%
\left( 4\pi \right) ^{2}v^{2}}\hat{\epsilon}^{-1}\left( \frac{1}{2}\not{p}%
\left( m_{i}^{d}m_{j}^{d}R+m_{h}^{u2}L\right) -m_{h}^{u2}\left(
m_{j}^{d}R+m_{i}^{d}L\right) \right) .
\end{equation*}

\begin{itemize}
\item  {\huge $-i\Sigma _{ij}^{u\chi ^{3}}$}
\end{itemize}

Using Eqs. (\ref{e1}) and (\ref{e2}) we obtain
\begin{eqnarray}
-i\Sigma _{ij}^{u\chi ^{3}} &=&\frac{m_{i}^{u2}\mu ^{\epsilon }\delta _{ij}}{%
\left( 2\pi \right) ^{\mathtt{d}}v^{2}}\int d^{\mathtt{d}}k\left( R-L\right)
\frac{i\left( \not{k}+m_{i}^{u}\right) }{k^{2}-m_{i}^{u2}+i\varepsilon }
\notag \\
&&\times \left( R-L\right) \frac{i}{\left( p-k\right) ^{2}-\xi
M_{Z}^{2}+i\varepsilon }  \notag \\
&=&\frac{m_{i}^{u2}\mu ^{\epsilon }\delta _{ij}}{\left( 2\pi \right) ^{%
\mathtt{d}}v^{2}}\int d^{\mathtt{d}}k\frac{\not{k}-m_{i}^{u}}{\left(
k^{2}-m_{i}^{u2}+i\varepsilon \right) \left( \left( p-k\right) ^{2}-\xi
M_{Z}^{2}+i\varepsilon \right) }.  \label{sux3}
\end{eqnarray}
Introducing a Feynman parameter we have
\begin{eqnarray*}
-i\Sigma _{ij}^{u\chi ^{3}} &=&\frac{m_{i}^{u2}\mu ^{\epsilon }\delta _{ij}}{%
\left( 2\pi \right) ^{\mathtt{d}}v^{2}}\int_{0}^{1}dx\int d^{\mathtt{d}}k \\
&&\times \frac{\not{k}-m_{i}^{u}}{\left[ x\left(
k^{2}-m_{i}^{u2}+i\varepsilon \right) +\left( 1-x\right) \left( \left(
p-k\right) ^{2}-\xi M_{Z}^{2}+i\varepsilon \right) \right] ^{2}},
\end{eqnarray*}
but
\begin{eqnarray*}
&&x\left( k^{2}-m_{i}^{u2}\right) +\left( 1-x\right) \left( \left(
p-k\right) ^{2}-\xi M_{Z}^{2}\right) \\
&=&k^{2}-2kp\left( 1-x\right) +\left( p^{2}-\xi M_{Z}^{2}\right) \left(
1-x\right) -m_{i}^{u2}x \\
&=&\left( k-p\left( 1-x\right) \right) ^{2}+p^{2}x\left( 1-x\right) -\xi
M_{Z}^{2}\left( 1-x\right) -m_{i}^{u2}x
\end{eqnarray*}
so
\begin{eqnarray*}
-i\Sigma _{ij}^{u\chi ^{3}} &=&\left\{ \frac{m_{i}^{u2}\mu ^{\epsilon
}\delta _{ij}}{\left( 2\pi \right) ^{\mathtt{d}}v^{2}}\int_{0}^{1}dx\int d^{%
\mathtt{d}}k\right. \\
&&\times \left. \frac{\not{k}+\not{p}\left( 1-x\right) -m_{i}^{u}}{\left[
k^{2}+p^{2}x\left( 1-x\right) -\xi M_{Z}^{2}\left( 1-x\right) -m_{i}^{u2}x%
\right] ^{2}}\right\} \\
&=&\frac{m_{i}^{u2}\delta _{ij}}{v^{2}}\int_{0}^{1}dxA_{i}^{Zu}\left( \mu
,x,p^{2},\epsilon \right) \left( \not{p}\left( 1-x\right) -m_{i}^{u}\right) ,
\end{eqnarray*}
where
\begin{equation*}
A_{i}^{Zu}\left( x,p^{2},\epsilon \right) \equiv \int \frac{d^{\mathtt{d}}k}{%
\left( 2\pi \right) ^{\mathtt{d}}}\frac{\mu ^{\epsilon }}{\left[
k^{2}-\Delta _{i}^{Zu}\right] ^{2}},
\end{equation*}
with
\begin{equation*}
\Delta _{i}^{Zu}\equiv p^{2}x\left( x-1\right) +\xi M_{Z}^{2}\left(
1-x\right) +m_{i}^{u2}x.
\end{equation*}
So finally we obtain
\begin{equation}
-i\Sigma _{ij}^{u\chi ^{3}}=-i\check{\Sigma}_{ij}^{u\chi ^{3}}-i\frac{%
m_{i}^{u2}\delta _{ij}}{\left( 4\pi \right) ^{2}v^{2}}\int_{0}^{1}dx\ln
\left( \frac{\Delta _{i}^{Zu}}{\mu ^{2}}\right) \left( \not{p}\left(
1-x\right) -m_{i}^{u}\right) ,  \label{e5}
\end{equation}
where
\begin{equation*}
-i\check{\Sigma}_{ij}^{u\chi ^{3}}=i\frac{m_{i}^{u2}\delta _{ij}}{\left(
4\pi \right) ^{2}v^{2}}\hat{\epsilon}^{-1}\left( \frac{1}{2}\not{p}%
-m_{i}^{u}\right) ,
\end{equation*}

\begin{itemize}
\item  {\huge $-i\Sigma _{ij}^{d\chi ^{3}}$}
\end{itemize}

Using Eqs. (\ref{e1}) and (\ref{e2}) we obtain
\begin{eqnarray*}
-i\Sigma _{ij}^{d\chi ^{3}} &=&\frac{m_{i}^{d2}\mu ^{\epsilon }\delta _{ij}}{%
\left( 2\pi \right) ^{\mathtt{d}}v^{2}}\int d^{\mathtt{d}}k\left( L-R\right)
\frac{i\left( \not{k}+m_{i}^{d}\right) }{k^{2}-m_{i}^{d2}+i\varepsilon } \\
&&\times \left( L-R\right) \frac{i}{\left( p-k\right) ^{2}-\xi
M_{Z}^{2}+i\varepsilon },
\end{eqnarray*}
which is identical to the expression for $\Sigma _{ij}^{u\chi ^{3}}$ given
by Eq. (\ref{sux3}) performing the change ($u\leftrightarrow d$). So
\begin{equation}
-i\Sigma _{ij}^{d\chi ^{3}}=-i\check{\Sigma}_{ij}^{d\chi ^{3}}-i\frac{%
m_{i}^{d2}\delta _{ij}}{\left( 4\pi \right) ^{2}v^{2}}\int_{0}^{1}dx\ln
\left( \frac{\Delta _{i}^{Zd}}{\mu ^{2}}\right) \left( \not{p}\left(
1-x\right) -m_{i}^{d}\right) ,  \label{e6}
\end{equation}
where
\begin{equation*}
-i\check{\Sigma}_{ij}^{d\chi ^{3}}=i\frac{m_{i}^{d2}\delta _{ij}}{\left(
4\pi \right) ^{2}v^{2}}\hat{\epsilon}^{-1}\left( \frac{1}{2}\not{p}%
-m_{i}^{d}\right) ,
\end{equation*}

\begin{itemize}
\item  {\huge $-i\Sigma _{ij}^{u\rho }$}
\end{itemize}

Using Eqs. (\ref{e1}) and (\ref{e2}) we obtain
\begin{eqnarray}
-i\Sigma _{ij}^{u\rho } &=&\frac{-m_{i}^{u2}\mu ^{\epsilon }\delta _{ij}}{%
\left( 2\pi \right) ^{\mathtt{d}}v^{2}}\int d^{\mathtt{d}}k\frac{i\left(
\not{k}+m_{i}^{u}\right) }{k^{2}-m_{i}^{u2}+i\varepsilon }\frac{i}{\left(
p-k\right) ^{2}-M_{\rho }^{2}+i\varepsilon }  \notag \\
&=&\frac{m_{i}^{u2}\mu ^{\epsilon }\delta _{ij}}{\left( 2\pi \right) ^{%
\mathtt{d}}v^{2}}\int d^{\mathtt{d}}k\frac{\not{k}+m_{i}^{u}}{\left(
k^{2}-m_{i}^{u2}+i\varepsilon \right) \left( \left( p-k\right) ^{2}-\xi
M_{Z}^{2}+i\varepsilon \right) },  \label{suro}
\end{eqnarray}
which is identical to the expression for $\Sigma _{ij}^{u\chi ^{3}}$ given
by Eq. (\ref{sux3}) performing the changes ($m_{i}^{u}\rightarrow -m_{i}^{u}$
and $\xi M_{Z}^{2}\rightarrow M_{\rho }^{2}$). So we have
\begin{equation}
-i\Sigma _{ij}^{u\rho }=-i\check{\Sigma}_{ij}^{u\rho }-i\frac{%
m_{i}^{u2}\delta _{ij}}{\left( 4\pi \right) ^{2}v^{2}}\int_{0}^{1}dx\ln
\left( \frac{\Delta _{i}^{\rho u}}{\mu ^{2}}\right) \left( \not{p}\left(
1-x\right) +m_{i}^{u}\right) ,  \label{e7}
\end{equation}
where
\begin{equation*}
-i\check{\Sigma}_{ij}^{u\chi ^{3}}=i\frac{m_{i}^{u2}\delta _{ij}}{\left(
4\pi \right) ^{2}v^{2}}\hat{\epsilon}^{-1}\left( \frac{1}{2}\not{p}%
+m_{i}^{u}\right) ,
\end{equation*}
and
\begin{equation*}
\Delta _{i}^{\rho u}\equiv p^{2}x\left( x-1\right) +M_{\rho }^{2}\left(
1-x\right) +m_{i}^{u2}x.
\end{equation*}

\begin{itemize}
\item  {\huge $-i\Sigma _{ij}^{d\rho }$}
\end{itemize}

Using Eqs. (\ref{e1}) and (\ref{e2}) we obtain
\begin{equation*}
-i\Sigma _{ij}^{d\rho }=\frac{-m_{i}^{d2}\mu ^{\epsilon }\delta _{ij}}{%
\left( 2\pi \right) ^{\mathtt{d}}v^{2}}\int d^{\mathtt{d}}k\frac{i\left(
\not{k}+m_{i}^{u}\right) }{k^{2}-m_{i}^{d2}+i\varepsilon }\frac{i}{\left(
p-k\right) ^{2}-M_{\rho }^{2}+i\varepsilon },.
\end{equation*}
which is identical to the expression for $\Sigma _{ij}^{u\rho }$ given by
Eq. (\ref{suro}) performing the change ($u\rightarrow d$). So we have
\begin{equation}
-i\Sigma _{ij}^{d\rho }=-i\check{\Sigma}_{ij}^{d\rho }-i\frac{%
m_{i}^{d2}\delta _{ij}}{\left( 4\pi \right) ^{2}v^{2}}\int_{0}^{1}dx\ln
\left( \frac{\Delta _{i}^{\rho d}}{\mu ^{2}}\right) \left( \not{p}\left(
1-x\right) +m_{i}^{d}\right) ,  \label{e8}
\end{equation}
where
\begin{equation*}
-i\check{\Sigma}_{ij}^{d\chi ^{3}}=i\frac{m_{i}^{d2}\delta _{ij}}{\left(
4\pi \right) ^{2}v^{2}}\hat{\epsilon}^{-1}\left( \frac{1}{2}\not{p}%
+m_{i}^{d}\right) .
\end{equation*}

\section{Gauge bosons as internal lines}

Here we will calculate the 1-loop fermion self energies given by Eq. (\ref
{gauge}). All the integrals that will appear are of the form
\begin{eqnarray}
-i\Sigma _{ij} &=&\sum_{h}\frac{\mu ^{\epsilon }}{\left( 2\pi \right) ^{%
\mathtt{d}}}\int d^{\mathtt{d}}kS_{ih}\gamma ^{\mu }\left(
a_{L}L+a_{R}R\right) \frac{i\left( \not{p}-\not{k}+m_{h}\right) }{\left(
p-k\right) ^{2}-m_{h}+i\varepsilon }S_{hj}^{\dagger }\gamma ^{\nu }\left(
a_{L}L+a_{R}R\right)  \notag \\
&&\times \frac{-i}{k^{2}-M^{2}+i\varepsilon }\left( g_{\mu \nu }+\left( \xi
-1\right) \frac{k_{\mu }k_{\nu }}{k^{2}-\xi M^{2}}\right) ,  \label{scheme}
\end{eqnarray}
so let us calculate it
\begin{eqnarray}
-i\Sigma _{ij} &=&\sum_{h}\frac{\mu ^{\epsilon }S_{ih}S_{hj}^{\dagger }}{%
\left( 2\pi \right) ^{\mathtt{d}}}\int d^{\mathtt{d}}k\frac{\gamma ^{\mu
}\left( \not{p}-\not{k}\right) \gamma ^{\nu }\left(
a_{L}^{2}L+a_{R}^{2}R\right) +\gamma ^{\mu }\gamma ^{\nu }m_{h}a_{L}a_{R}}{%
\left( p-k\right) ^{2}-m_{h}+i\varepsilon }  \notag \\
&&\times \frac{1}{k^{2}-M^{2}+i\varepsilon }\left( g_{\mu \nu }+\left( \xi
-1\right) \frac{k_{\mu }k_{\nu }}{k^{2}-\xi M^{2}}\right) ,  \notag
\end{eqnarray}
or
\begin{equation}
-i\Sigma _{ij}=\sum_{h}S_{ih}S_{hj}^{\dagger }\left[ \left(
A_{h}+B_{h}\right) \left( a_{L}^{2}L+a_{R}^{2}R\right)
+m_{h}a_{L}a_{R}\left( C_{h}+D_{h}\right) \right] ,  \label{fscheme}
\end{equation}
where
\begin{eqnarray*}
A_{h} &\equiv &\frac{\mu ^{\epsilon }}{\left( 2\pi \right) ^{\mathtt{d}}}%
\int d^{\mathtt{d}}k\frac{\gamma ^{\mu }\left( \not{p}-\not{k}\right) \gamma
^{\nu }g_{\mu \nu }}{\left[ \left( p-k\right) ^{2}-m_{h}^{2}\right] \left(
k^{2}-M^{2}\right) }, \\
B_{h} &\equiv &\frac{\mu ^{\epsilon }}{\left( 2\pi \right) ^{\mathtt{d}}}%
\int d^{\mathtt{d}}k\frac{\left( \xi -1\right) \not{k}\left( \not{p}-\not{k}%
\right) \not{k}}{\left[ \left( p-k\right) ^{2}-m_{h}^{2}\right] \left(
k^{2}-M^{2}\right) \left( k^{2}-\xi M^{2}\right) }, \\
C_{h} &\equiv &\frac{\mu ^{\epsilon }}{\left( 2\pi \right) ^{\mathtt{d}}}%
\int d^{\mathtt{d}}k\frac{\gamma ^{\mu }\gamma ^{\nu }g_{\mu \nu }}{\left[
\left( p-k\right) ^{2}-m_{h}^{2}\right] \left( k^{2}-M^{2}\right) }, \\
D_{h} &\equiv &\frac{\mu ^{\epsilon }}{\left( 2\pi \right) ^{\mathtt{d}}}%
\int d^{\mathtt{d}}k\frac{\left( \xi -1\right) k^{2}}{\left[ \left(
p-k\right) ^{2}-m_{h}^{2}\right] \left( k^{2}-M^{2}\right) \left( k^{2}-\xi
M^{2}\right) },
\end{eqnarray*}
Let us calculate $B_{h}$ and $D_{h}$ firstly. Introducing two Feynman
parameters we have
\begin{eqnarray*}
B_{h} &=&\frac{2\mu ^{\epsilon }\left( \xi -1\right) }{\left( 2\pi \right) ^{%
\mathtt{d}}}\int_{0}^{1}dx\int_{0}^{1-x}dy\int d^{\mathtt{d}}k\not{k}\left(
\not{p}-\not{k}\right) \not{k} \\
&&\times \left\{ x\left( \left( p-k\right) ^{2}-m_{h}^{2}\right) +\left(
1-x-y\right) \left( k^{2}-\xi M^{2}\right) +y\left( k^{2}-M^{2}\right)
\right\} ^{-3}
\end{eqnarray*}
but
\begin{eqnarray*}
&&x\left( \left( p-k\right) ^{2}-m_{h}^{2}\right) +\left( 1-x-y\right)
\left( k^{2}-\xi M^{2}\right) +y\left( k^{2}-M^{2}\right) \\
&=&\left( k-xp\right) ^{2}+x\left( 1-x\right) p^{2}-yM^{2}-\left(
1-x-y\right) \xi M^{2}-xm_{h}^{2},
\end{eqnarray*}
so defining
\begin{equation*}
\Omega _{h}\equiv xm_{h}^{2}+yM^{2}+\left( 1-x-y\right) \xi M^{2}-x\left(
1-x\right) p^{2},
\end{equation*}
we have
\begin{eqnarray*}
B_{h} &=&\frac{2\mu ^{\epsilon }\left( \xi -1\right) }{\left( 2\pi \right) ^{%
\mathtt{d}}}\int_{0}^{1}dx\int_{0}^{1-x}dy\int d^{\mathtt{d}}k\frac{\left(
\not{k}+x\not{p}\right) \left( \not{p}\left( 1-x\right) -\not{k}\right)
\left( \not{k}+x\not{p}\right) }{\left( k^{2}-\Omega _{h}\right) ^{3}} \\
&=&\frac{2\mu ^{\epsilon }\left( \xi -1\right) }{\left( 2\pi \right) ^{%
\mathtt{d}}}\int_{0}^{1}dx\int_{0}^{1-x}dy\int d^{\mathtt{d}}k \\
&&\times \frac{\not{k}\not{p}\not{k}\left( 1-x\right) -x\not{k}\not{k}\not{p}%
+x^{2}\not{p}\not{p}\not{p}\left( 1-x\right) -x\not{p}\not{k}\not{k}}{\left(
k^{2}-\Omega _{h}\right) ^{3}},
\end{eqnarray*}
but
\begin{eqnarray*}
\not{k}\not{k} &=&k^{2}, \\
\not{p}\not{p} &=&p^{2}, \\
\not{k}\not{p}\not{k} &=&2kp\not{k}-k^{2}\not{p},
\end{eqnarray*}
so
\begin{equation*}
B_{h}=\frac{2\mu ^{\epsilon }\left( \xi -1\right) }{\left( 2\pi \right) ^{%
\mathtt{d}}}\int_{0}^{1}dx\int_{0}^{1-x}dy\int d^{\mathtt{d}}k\frac{2\left(
1-x\right) pk\not{k}-\left( 1+x\right) \not{p}k^{2}+x^{2}\left( 1-x\right)
p^{2}\not{p}}{\left( k^{2}-\Omega _{h}\right) ^{3}},
\end{equation*}
but
\begin{eqnarray}
\int \frac{d^{\mathtt{d}}k}{\left( 2\pi \right) ^{\mathtt{d}}}\frac{\mu
^{\epsilon }}{\left( k^{2}-\Omega _{h}\right) ^{3}} &=&\frac{-i\mu
^{\epsilon }}{\left( 4\pi \right) ^{2-\frac{\epsilon }{2}}}\frac{\Gamma
\left( 1+\frac{\epsilon }{2}\right) }{\Gamma \left( 3\right) }\left( \frac{1%
}{\Omega _{h}}\right) ^{1+\frac{\epsilon }{2}}  \notag \\
&=&\frac{-i}{2\left( 4\pi \right) ^{2}\Omega _{h}}+O\left( \epsilon \right) ,
\label{int1}
\end{eqnarray}
and
\begin{eqnarray}
\int \frac{d^{\mathtt{d}}k}{\left( 2\pi \right) ^{\mathtt{d}}}\frac{\mu
^{\epsilon }k^{\mu }k^{\nu }}{\left( k^{2}-\Omega _{h}\right) ^{3}} &=&\frac{%
i\mu ^{\epsilon }g^{\mu \nu }}{2\left( 4\pi \right) ^{2-\frac{\epsilon }{2}}}%
\frac{\Gamma \left( \frac{\epsilon }{2}\right) }{\Gamma \left( 3\right) }%
\left( \frac{1}{\Omega _{h}}\right) ^{\frac{\epsilon }{2}}  \notag \\
&=&\frac{ig^{\mu \nu }}{4\left( 4\pi \right) ^{2}}\left( \hat{\epsilon}%
^{-1}-\ln \frac{\Omega _{h}}{\mu ^{2}}\right) +O\left( \epsilon \right) ,
\label{int2}
\end{eqnarray}
with
\begin{eqnarray}
\int \frac{d^{\mathtt{d}}k}{\left( 2\pi \right) ^{\mathtt{d}}}\frac{\mu
^{\epsilon }k^{2}}{\left( k^{2}-\Omega _{h}\right) ^{3}} &=&\frac{i\mu
^{\epsilon }\left( 4-\epsilon \right) }{2\left( 4\pi \right) ^{2-\frac{%
\epsilon }{2}}}\frac{\Gamma \left( \frac{\epsilon }{2}\right) }{\Gamma
\left( 3\right) }\left( \frac{1}{\Omega _{h}}\right) ^{\frac{\epsilon }{2}}
\notag \\
&=&\frac{i}{\left( 4\pi \right) ^{2}}\left( \hat{\epsilon}^{-1}-\frac{1}{2}%
-\ln \frac{\Omega _{h}}{\mu ^{2}}\right) +O\left( \epsilon \right) ,
\label{int3}
\end{eqnarray}
we obtain
\begin{eqnarray*}
B_{h} &=&2\left( \xi -1\right) \int_{0}^{1}dx\int_{0}^{1-x}dy\left\{ \frac{%
-ix^{2}\left( 1-x\right) }{2\left( 4\pi \right) ^{2}\Omega _{h}}p^{2}\not{p}%
\right. \\
&&-\left. \frac{i\left( 1+x\right) }{\left( 4\pi \right) ^{2}}\left( \hat{%
\epsilon}^{-1}-\frac{1}{2}-\ln \frac{\Omega _{h}}{\mu ^{2}}\right) \not{p}+2%
\frac{i\left( 1-x\right) }{4\left( 4\pi \right) ^{2}}\left( \hat{\epsilon}%
^{-1}-\ln \frac{\Omega _{h}}{\mu ^{2}}\right) \not{p}\right\} ,
\end{eqnarray*}
or
\begin{equation}
B_{h}=\frac{-i\left( \xi -1\right) \not{p}}{\left( 4\pi \right) ^{2}}%
\int_{0}^{1}\int_{0}^{1-x}\left\{ \left( 1+3x\right) \left( \hat{\epsilon}%
^{-1}-\ln \frac{\Omega _{h}}{\mu ^{2}}\right) +\frac{x^{2}\left( 1-x\right)
}{\Omega _{h}}p^{2}-\left( 1+x\right) \right\} dydx,  \label{bhm}
\end{equation}
with
\begin{equation}
\Omega _{h}\equiv xm_{h}^{2}+yM^{2}+\left( 1-x-y\right) \xi M^{2}-x\left(
1-x\right) p^{2},  \label{def1}
\end{equation}
defining
\begin{eqnarray}
\Delta _{h} &\equiv &xm_{h}^{2}+\left( 1-x\right) \xi M^{2}+x\left(
x-1\right) p^{2},  \notag \\
\eta _{h} &\equiv &xm_{h}^{2}+\left( 1-x\right) M^{2}+x\left( x-1\right)
p^{2},  \label{def2}
\end{eqnarray}
we obtain
\begin{eqnarray*}
&&\left( \xi -1\right) \int_{0}^{1-x}\frac{1}{xm_{h}^{2}+yM^{2}+\left(
1-x-y\right) \xi M^{2}-x\left( 1-x\right) p^{2}}dy \\
&=&\frac{1}{M^{2}}\ln \left( \frac{p^{2}x\left( x-1\right) +\xi M^{2}\left(
1-x\right) +m_{h}^{2}x}{xm_{h}^{2}+\left( 1-x\right) \left(
M^{2}-xp^{2}\right) }\right) \quad M\neq 0,
\end{eqnarray*}
or
\begin{equation}
\left( \xi -1\right) \int_{0}^{1-x}\frac{1}{\Omega _{h}}dy=\left\{
\begin{array}{cc}
\frac{1}{M^{2}}\ln \left( \frac{\Delta _{h}}{\eta _{h}}\right) & M\neq 0 \\
\left( \xi -1\right) \frac{1-x}{\eta _{h}} & M=0
\end{array}
\right. ,  \label{great1}
\end{equation}
we also have
\begin{eqnarray*}
&&\left( \xi -1\right) \int_{0}^{1-x}\ln \left( \frac{xm_{h}^{2}+yM^{2}+%
\left( 1-x-y\right) \xi M^{2}-x\left( 1-x\right) p^{2}}{\mu ^{2}}\right) dy
\\
&=&\left( \xi -1\right) \left( x-1\right) \left( 1-\ln \left( \frac{M^{2}}{%
\mu ^{2}}\right) \right) -\frac{xm_{h}^{2}+\left( 1-x\right) \left(
M^{2}-xp^{2}\right) }{M^{2}}\ln \left( \frac{xm_{h}^{2}+\left( 1-x\right)
\left( M^{2}-xp^{2}\right) }{M^{2}}\right) \\
&&+\frac{p^{2}x\left( x-1\right) +\xi M^{2}\left( 1-x\right) +m_{h}^{2}x}{%
M^{2}}\ln \left( \frac{p^{2}x\left( x-1\right) +\xi M^{2}\left( 1-x\right)
+m_{h}^{2}x}{M^{2}}\right) \quad M\neq 0,
\end{eqnarray*}
or
\begin{equation}
\left( \xi -1\right) \int_{0}^{1-x}\ln \left( \frac{\Omega _{h}}{\mu ^{2}}%
\right) dy=\left\{
\begin{array}{cc}
\left( \xi -1\right) \left( x-1\right) \left( 1-\ln \frac{M^{2}}{\mu ^{2}}%
\right) -\frac{\eta _{h}}{M^{2}}\ln \frac{\eta _{h}}{M^{2}}+\frac{\Delta _{h}%
}{M^{2}}\ln \frac{\Delta _{h}}{M^{2}} & M\neq 0 \\
\left( \xi -1\right) \left( 1-x\right) \ln \frac{\eta _{h}}{\mu ^{2}} & M=0
\end{array}
\right. ,  \label{great2}
\end{equation}
using Eqs. (\ref{def2}), (\ref{great1}) and (\ref{great2}) Eq. (\ref{bhm})
becomes
\begin{eqnarray}
B_{h} &=&\check{B}_{h}+\frac{i\not{p}}{\left( 4\pi \right) ^{2}}%
\int_{0}^{1}\left\{ \left( 1+3x\right) \left[ \left( \xi -1\right) \left(
x-1\right) \left( 1-\ln \frac{M^{2}}{\mu ^{2}}\right) -\frac{\eta _{h}}{M^{2}%
}\ln \frac{\eta _{h}}{M^{2}}+\frac{\Delta _{h}}{M^{2}}\ln \frac{\Delta _{h}}{%
M^{2}}\right] \right.  \notag \\
&&+\left. x^{2}\left( x-1\right) \frac{p^{2}}{M^{2}}\ln \frac{\Delta _{h}}{%
\eta _{h}}+\left( \xi -1\right) \left( 1-x^{2}\right) \right\} dx\quad M\neq
0,  \label{bhM}
\end{eqnarray}
or
\begin{equation}
B_{h}=\check{B}_{h}+\frac{i\not{p}\left( \xi -1\right) }{\left( 4\pi \right)
^{2}}\int_{0}^{1}\left( 1-x\right) \left( \left( 1+3x\right) \ln \frac{\eta
_{h}}{\mu ^{2}}+x^{2}p^{2}\frac{x-1}{\eta _{h}}+1+x\right) dx\quad M=0,
\label{bh0}
\end{equation}
with the divergent part given by
\begin{equation}
\check{B}_{h}=\frac{-i\left( \xi -1\right) \not{p}\hat{\epsilon}^{-1}}{%
\left( 4\pi \right) ^{2}}\int_{0}^{1}\int_{0}^{1-x}\left( 1+3x\right) dydx=%
\frac{-i\left( \xi -1\right) \not{p}}{\left( 4\pi \right) ^{2}}\hat{\epsilon}%
^{-1},  \label{bhdiv}
\end{equation}
Analogously we have
\begin{eqnarray*}
D_{h} &=&\frac{2\mu ^{\epsilon }\left( \xi -1\right) }{\left( 2\pi \right) ^{%
\mathtt{d}}}\int_{0}^{1}dx\int_{0}^{1-x}dy\int d^{\mathtt{d}}k\frac{\left(
k+xp\right) ^{2}}{\left( k^{2}-\Omega _{h}\right) ^{3}} \\
&=&\frac{2\mu ^{\epsilon }\left( \xi -1\right) }{\left( 2\pi \right) ^{%
\mathtt{d}}}\int_{0}^{1}dx\int_{0}^{1-x}dy\int d^{\mathtt{d}}k\frac{%
k^{2}+x^{2}p^{2}}{\left( k^{2}-\Omega _{h}\right) ^{3}},
\end{eqnarray*}
using Eqs. (\ref{int1}) and (\ref{int3}) we obtain
\begin{equation}
D_{h}=\frac{2i\left( \xi -1\right) }{\left( 4\pi \right) ^{2}}%
\int_{0}^{1}dx\int_{0}^{1-x}dy\left( \hat{\epsilon}^{-1}-\frac{1}{2}-\ln
\frac{\Omega _{h}}{\mu ^{2}}-\frac{x^{2}p^{2}}{2\Omega _{h}}\right) ,
\label{dhm}
\end{equation}
and using Eqs. (\ref{def2}), (\ref{great1}) and (\ref{great2}) we obtain
\begin{eqnarray}
D_{h} &=&\check{D}_{h}+\frac{2i}{\left( 4\pi \right) ^{2}}%
\int_{0}^{1}dx\left\{ \left( \xi -1\right) \frac{x-1}{2}-\frac{x^{2}p^{2}}{%
2M^{2}}\ln \left( \frac{\Delta _{h}}{\eta _{h}}\right) \right.  \notag \\
&&-\left. \left( \xi -1\right) \left( x-1\right) \left( 1-\ln \frac{M^{2}}{%
\mu ^{2}}\right) +\frac{\eta _{h}}{M^{2}}\ln \frac{\eta _{h}}{M^{2}}-\frac{%
\Delta _{h}}{M^{2}}\ln \frac{\Delta _{h}}{M^{2}}\right\} \quad M\neq 0,
\label{dhM}
\end{eqnarray}
and
\begin{equation}
D_{h}=\check{D}_{h}+\frac{i\left( \xi -1\right) }{\left( 4\pi \right) ^{2}}%
\int_{0}^{1}\left( x-1\right) \left( 1+2\ln \frac{\eta _{h}}{\mu ^{2}}+\frac{%
x^{2}p^{2}}{\eta _{h}}\right) dx\quad M=0,  \label{dh0}
\end{equation}
with the divergent part given by
\begin{equation}
\check{D}_{h}=\frac{i\left( \xi -1\right) \hat{\epsilon}^{-1}}{\left( 4\pi
\right) ^{2}},  \label{dhdiv}
\end{equation}
Let us now calculate $A_{h}$ and $C_{h}$. Introducing a Feynman parameter we
have
\begin{eqnarray*}
A_{h} &=&\frac{\mu ^{\epsilon }}{\left( 2\pi \right) ^{\mathtt{d}}}%
\int_{0}^{1}dx\int d^{\mathtt{d}}k\gamma ^{\mu }\left( \not{p}-\not{k}%
\right) \gamma ^{\nu }g_{\mu \nu } \\
&&\times \left\{ x\left[ \left( p-k\right) ^{2}-m_{h}^{2}\right] +\left(
1-x\right) \left( k^{2}-M^{2}\right) \right\} ^{-2},
\end{eqnarray*}
but
\begin{eqnarray*}
&&x\left[ \left( p-k\right) ^{2}-m_{h}^{2}\right] +\left( 1-x\right) \left(
k^{2}-M^{2}\right) \\
&=&k^{2}-2xpk+xp^{2}-xm_{h}^{2}-\left( 1-x\right) M^{2} \\
&=&\left( k-xp\right) ^{2}+xp^{2}\left( 1-x\right) -xm_{h}^{2}-\left(
1-x\right) M^{2},
\end{eqnarray*}
so using Eq.(\ref{def2}) we obtain
\begin{eqnarray*}
A_{h} &=&\frac{\mu ^{\epsilon }}{\left( 2\pi \right) ^{\mathtt{d}}}%
\int_{0}^{1}dx\int d^{\mathtt{d}}k\frac{\gamma ^{\mu }\left( \left(
1-x\right) \not{p}-\not{k}\right) \gamma ^{\nu }g_{\mu \nu }}{\left(
k^{2}-\eta _{h}\right) ^{2}} \\
&=&\frac{\mu ^{\epsilon }}{\left( 2\pi \right) ^{\mathtt{d}}}%
\int_{0}^{1}dx\int d^{\mathtt{d}}k\frac{\left( 1-x\right) \gamma ^{\mu }\not{%
p}\gamma ^{\nu }g_{\mu \nu }}{\left( k^{2}-\eta _{h}\right) ^{2}},
\end{eqnarray*}
hence using Eq. (\ref{facil}) we have
\begin{eqnarray*}
A_{h} &=&\frac{i}{\left( 4\pi \right) ^{2}}\int_{0}^{1}dx\left( 1-x\right)
\gamma ^{\mu }\not{p}\gamma ^{\nu }g_{\mu \nu }\left( \hat{\epsilon}%
^{-1}-\ln \frac{\eta _{h}}{\mu ^{2}}\right) \\
&=&\frac{i}{\left( 4\pi \right) ^{2}}\int_{0}^{1}dx\left( 1-x\right) \not{p}%
\left( \epsilon -2\right) \left( \hat{\epsilon}^{-1}-\ln \frac{\eta _{h}}{%
\mu ^{2}}\right) \\
&=&\frac{-2i\not{p}}{\left( 4\pi \right) ^{2}}\int_{0}^{1}dx\left(
1-x\right) \left( \hat{\epsilon}^{-1}-1-\ln \frac{\eta _{h}}{\mu ^{2}}%
\right) ,
\end{eqnarray*}
and finally
\begin{equation}
A_{h}=\frac{-i\not{p}}{\left( 4\pi \right) ^{2}}\hat{\epsilon}^{-1}+\frac{2i%
\not{p}}{\left( 4\pi \right) ^{2}}\int_{0}^{1}dx\left( 1-x\right) \left(
1+\ln \frac{\eta _{h}}{\mu ^{2}}\right) ,  \label{ah}
\end{equation}
analogously for $C_{h}$ we have
\begin{eqnarray*}
C_{h} &=&\frac{\mu ^{\epsilon }}{\left( 2\pi \right) ^{\mathtt{d}}}%
\int_{0}^{1}dx\int d^{\mathtt{d}}k\frac{\gamma ^{\mu }\gamma ^{\nu }g_{\mu
\nu }}{\left( k^{2}-\eta _{h}\right) ^{2}} \\
&=&\frac{\mu ^{\epsilon }}{\left( 2\pi \right) ^{\mathtt{d}}}%
\int_{0}^{1}dx\int d^{\mathtt{d}}k\frac{4-\epsilon }{\left( k^{2}-\eta
_{h}\right) ^{2}},
\end{eqnarray*}
hence using Eq. (\ref{facil}) we have
\begin{eqnarray}
C_{h} &=&\frac{i}{\left( 4\pi \right) ^{2}}\int_{0}^{1}dx\left( 4-\epsilon
\right) \left( \hat{\epsilon}^{-1}-\ln \frac{\eta _{h}}{\mu ^{2}}\right)
\notag \\
&=&\frac{4i}{\left( 4\pi \right) ^{2}}\int_{0}^{1}dx\left( \hat{\epsilon}%
^{-1}-\frac{1}{2}-\ln \frac{\eta _{h}}{\mu ^{2}}\right) ,  \label{ch}
\end{eqnarray}
Now, let us finally calculate $A_{h}+B_{h}$ and $C_{h}+D_{h}$. From Eqs. (%
\ref{bhM}), (\ref{bhdiv}) and (\ref{ah}) we obtain
\begin{eqnarray}
A_{h}+B_{h} &=&-\frac{i\xi \not{p}\hat{\epsilon}^{-1}}{\left( 4\pi \right)
^{2}}+\frac{i\not{p}}{\left( 4\pi \right) ^{2}}\int_{0}^{1}\left\{ 2\left(
1-x\right) \left( 1+\ln \frac{\eta _{h}}{\mu ^{2}}\right) \right.  \notag \\
&&+\left( 1+3x\right) \left[ \left( \xi -1\right) \left( x-1\right) \left(
1-\ln \frac{M^{2}}{\mu ^{2}}\right) -\frac{\eta _{h}}{M^{2}}\ln \frac{\eta
_{h}}{M^{2}}+\frac{\Delta _{h}}{M^{2}}\ln \frac{\Delta _{h}}{M^{2}}\right]
\notag \\
&&+\left. x^{2}\left( x-1\right) \frac{p^{2}}{M^{2}}\ln \frac{\Delta _{h}}{%
\eta _{h}}+\left( \xi -1\right) \left( 1-x^{2}\right) \right\} dx\quad M\neq
0,  \label{a+bM}
\end{eqnarray}
and from Eqs. (\ref{bh0}), (\ref{bhdiv}) and (\ref{ah}) we obtain
\begin{eqnarray}
A_{h}+B_{h} &=&-\frac{i\xi \not{p}\hat{\epsilon}^{-1}}{\left( 4\pi \right)
^{2}}+\frac{i\not{p}}{\left( 4\pi \right) ^{2}}\int_{0}^{1}\left( 1-x\right)
\left\{ 2+2\ln \frac{\eta _{h}}{\mu ^{2}}+\left( \xi -1\right) \right.
\notag \\
&&\times \left. \left( \left( 1+3x\right) \ln \frac{\eta _{h}}{\mu ^{2}}%
+x^{2}p^{2}\frac{x-1}{\eta _{h}}+1+x\right) \right\} dx\quad M=0.
\label{a+b0}
\end{eqnarray}
From Eqs. (\ref{dhM}), (\ref{dhdiv}) and (\ref{ch}) we obtain
\begin{eqnarray}
C_{h}+D_{h} &=&\frac{i\left( \xi +3\right) \hat{\epsilon}^{-1}}{\left( 4\pi
\right) ^{2}}-\frac{2i}{\left( 4\pi \right) ^{2}}\int_{0}^{1}dx\left\{
1+2\ln \frac{\eta _{h}}{\mu ^{2}}+\left( \xi -1\right) \right.  \notag \\
&&\times \left. \left( x-1\right) \left( \frac{1}{2}-\ln \frac{M^{2}}{\mu
^{2}}\right) +\frac{\Delta _{h}}{M^{2}}\ln \frac{\Delta _{h}}{M^{2}}-\frac{%
\eta _{h}}{M^{2}}\ln \frac{\eta _{h}}{M^{2}}\right\} ,  \label{c+dM}
\end{eqnarray}
and from Eqs. (\ref{dh0}), (\ref{dhdiv}) and (\ref{ah}) we obtain
\begin{eqnarray}
C_{h}+D_{h} &=&\frac{i\left( \xi +3\right) \hat{\epsilon}^{-1}}{\left( 4\pi
\right) ^{2}}-\frac{2i}{\left( 4\pi \right) ^{2}}\int_{0}^{1}dx\left\{
1+2\ln \frac{\eta _{h}}{\mu ^{2}}+\left( \xi -1\right) \right.  \notag \\
&&\times \left. \left( 1-x\right) \left( \frac{1}{2}+\ln \frac{\eta _{h}}{%
\mu ^{2}}+\frac{x^{2}p^{2}}{2\eta _{h}}\right) \right\} \quad M=0,
\label{c+d0}
\end{eqnarray}
With these results let us calculate the concrete bare self energies

\begin{itemize}
\item  {\huge $-i\Sigma _{ij}^{uW^{+}}$}
\end{itemize}

Using Eqs. (\ref{gaugevert}) and (\ref{e2}) we obtain
\begin{eqnarray}
-i\Sigma _{ij}^{uW^{+}} &=&\sum_{h}\frac{\mu ^{\epsilon }}{\left( 2\pi
\right) ^{\mathtt{d}}}\int d^{\mathtt{d}}k\gamma ^{\mu }\left( -i\frac{e}{%
\sqrt{2}s_{W}}K_{ih}L\right) \frac{i\left( \not{p}-\not{k}+m_{h}^{d}\right)
}{\left( p-k\right) ^{2}-m_{h}^{d2}+i\varepsilon }\gamma ^{\nu }\left( -i%
\frac{e}{\sqrt{2}s_{W}}K_{hj}^{\dagger }L\right)  \notag \\
&&\times \frac{-i}{k^{2}-M_{W}^{2}+i\varepsilon }\left( g_{\mu \nu }+\left(
\xi -1\right) \frac{k_{\mu }k_{\nu }}{k^{2}-\xi M_{W}^{2}}\right) ,
\label{suw+ij}
\end{eqnarray}
from Eq. (\ref{scheme}) we obtain
\begin{eqnarray}
S_{ih} &=&K_{ih},  \notag \\
a_{L} &=&-i\frac{e}{\sqrt{2}s_{W}},  \notag \\
a_{R} &=&0,  \notag \\
m_{h} &=&m_{h}^{d},  \notag \\
M &=&M_{W},  \label{asuw+}
\end{eqnarray}
hence replacing in Eq. (\ref{fscheme}) we obtain
\begin{equation*}
-i\Sigma _{ij}^{uW^{+}}=\sum_{h}\frac{-e^{2}K_{ih}K_{hj}^{\dagger }}{%
2s_{W}^{2}}\left( A_{h}+B_{h}\right) L,
\end{equation*}
so using Eq. (\ref{a+bM}) we obtain
\begin{eqnarray}
-i\Sigma _{ij}^{uW^{+}} &=&\frac{ie^{2}\delta _{ij}\xi \not{p}L}{%
2s_{W}^{2}\left( 4\pi \right) ^{2}}\hat{\epsilon}^{-1}-\sum_{h}\frac{%
ie^{2}K_{ih}K_{hj}^{\dagger }\not{p}L}{2s_{W}^{2}\left( 4\pi \right) ^{2}}%
\int_{0}^{1}dx\left\{ 2\left( 1-x\right) \left( 1+\ln \frac{\eta _{h}^{dW}}{%
\mu ^{2}}\right) \right.  \notag \\
&&+\left( 1+3x\right) \left[ \left( \xi -1\right) \left( x-1\right) \left(
1-\ln \frac{M_{W}^{2}}{\mu ^{2}}\right) -\frac{\eta _{h}^{dW}}{M_{W}^{2}}\ln
\frac{\eta _{h}^{dW}}{M_{W}^{2}}+\frac{\Delta _{h}^{dW}}{M_{W}^{2}}\ln \frac{%
\Delta _{h}^{dW}}{M_{W}^{2}}\right]  \notag \\
&&+\left. x^{2}\left( x-1\right) \frac{p^{2}}{M_{W}^{2}}\ln \frac{\Delta
_{h}^{dW}}{\eta _{h}^{dW}}+\left( \xi -1\right) \left( 1-x^{2}\right)
\right\} ,  \label{suw+}
\end{eqnarray}
where from Eqs. (\ref{def2}) and (\ref{asuw+}) we have
\begin{eqnarray*}
\Delta _{h}^{dW} &\equiv &xm_{h}^{d2}+\left( 1-x\right) \xi
M_{W}^{2}+x\left( x-1\right) p^{2}, \\
\eta _{h}^{dW} &\equiv &xm_{h}^{d2}+\left( 1-x\right) M_{W}^{2}+x\left(
x-1\right) p^{2},
\end{eqnarray*}

\begin{itemize}
\item  {\huge $-i\Sigma _{ij}^{dW^{-}}$}
\end{itemize}

Using Eqs. (\ref{gaugevert}) and (\ref{e2}) we obtain
\begin{eqnarray*}
-i\Sigma _{ij}^{dW^{-}} &=&\sum_{h}\frac{\mu ^{\epsilon }}{\left( 2\pi
\right) ^{\mathtt{d}}}\int d^{\mathtt{d}}k\gamma ^{\mu }\left( -i\frac{e}{%
\sqrt{2}s_{W}}K_{ih}^{\dagger }L\right) \frac{i\left( \not{p}-\not{k}%
+m_{h}^{u}\right) }{\left( p-k\right) ^{2}-m_{h}^{u2}+i\varepsilon } \\
&&\times \gamma ^{\nu }\left( -i\frac{e}{\sqrt{2}s_{W}}K_{hj}L\right) \frac{%
-i}{k^{2}-M_{W}^{2}+i\varepsilon }\left( g_{\mu \nu }+\left( \xi -1\right)
\frac{k_{\mu }k_{\nu }}{k^{2}-\xi M_{W}^{2}}\right) ,
\end{eqnarray*}
which is identical to the expression for $\Sigma _{ij}^{uW^{+}}$ given by
Eq. (\ref{suw+ij}) performing the changes ($u\leftrightarrow d$ and $%
K\leftrightarrow K^{\dagger }$) so from Eq. (\ref{suw+}) we obtain
\begin{eqnarray}
-i\Sigma _{ij}^{dW^{-}} &=&\frac{ie^{2}\delta _{ij}\xi \not{p}L}{%
2s_{W}^{2}\left( 4\pi \right) ^{2}}\hat{\epsilon}^{-1}-\sum_{h}\frac{%
ie^{2}K_{ih}^{\dagger }K_{hj}\not{p}L}{2s_{W}^{2}\left( 4\pi \right) ^{2}}%
\int_{0}^{1}dx\left\{ 2\left( 1-x\right) \left( 1+\ln \frac{\eta _{h}^{uW}}{%
\mu ^{2}}\right) \right.  \notag \\
&&+\left( 1+3x\right) \left[ \left( \xi -1\right) \left( x-1\right) \left(
1-\ln \frac{M_{W}^{2}}{\mu ^{2}}\right) -\frac{\eta _{h}^{uW}}{M_{W}^{2}}\ln
\frac{\eta _{h}^{uW}}{M_{W}^{2}}+\frac{\Delta _{h}^{uW}}{M_{W}^{2}}\ln \frac{%
\Delta _{h}^{uW}}{M_{W}^{2}}\right]  \notag \\
&&+\left. x^{2}\left( x-1\right) \frac{p^{2}}{M_{W}^{2}}\ln \frac{\Delta
_{h}^{uW}}{\eta _{h}^{uW}}+\left( \xi -1\right) \left( 1-x^{2}\right)
\right\} ,  \label{sdw-}
\end{eqnarray}
where from Eq. (\ref{def2}) we have
\begin{eqnarray*}
\Delta _{h}^{uW} &\equiv &xm_{h}^{u2}+\left( 1-x\right) \xi
M_{W}^{2}+x\left( x-1\right) p^{2}, \\
\eta _{h}^{uW} &\equiv &xm_{h}^{u2}+\left( 1-x\right) M_{W}^{2}+x\left(
x-1\right) p^{2},
\end{eqnarray*}

\begin{itemize}
\item  {\huge $-i\Sigma _{ij}^{uZ}$}
\end{itemize}

Using Eqs. (\ref{gaugevert}) and (\ref{e2}) we obtain
\begin{eqnarray}
S_{ih} &=&\delta _{ih},  \notag \\
a_{L} &=&i\frac{e}{c_{W}s_{W}}\left( zs_{W}^{2}-\frac{1}{2}c_{W}^{2}\right) ,
\notag \\
a_{R} &=&i\frac{e}{c_{W}s_{W}}s_{W}^{2}\left( z+\frac{1}{2}\right) ,  \notag
\\
m_{h} &=&m_{h}^{u},  \notag \\
M &=&M_{Z},  \label{asuz}
\end{eqnarray}
where $z$ is a real parameter which takes the value $\frac{1}{6}$ for quarks
and $\frac{-1}{2}$ for leptons wherefrom the hypercharge $Y$ is obtained by
\begin{equation*}
Y=\left\{
\begin{tabular}{ll}
$z$ & $\mathrm{for\ lefts}$ \\
$\frac{\tau ^{3}}{2}+z$ & $\mathrm{for\ rights}$%
\end{tabular}
\right.
\end{equation*}
and the electric charge by
\begin{equation*}
Q=\frac{\tau ^{3}}{2}+z,
\end{equation*}
From Eq. (\ref{fscheme}) we obtain
\begin{equation*}
-i\Sigma _{ij}^{uZ}=\delta _{ij}\left[ \left( A_{i}+B_{i}\right) \left(
a_{L}^{2}L+a_{R}^{2}R\right) +m_{i}^{u}a_{L}a_{R}\left( C_{i}+D_{i}\right)
\right] ,
\end{equation*}
hence from Eqs. (\ref{a+bM}) and (\ref{c+dM}) we obtain
\begin{eqnarray}
-i\Sigma _{ij}^{uZ} &=&-\frac{\delta _{ij}i\xi \not{p}}{\left( 4\pi \right)
^{2}}\left( a_{L}^{2}L+a_{R}^{2}R\right) \hat{\epsilon}^{-1}+\frac{\delta
_{ij}i\not{p}}{\left( 4\pi \right) ^{2}}\int_{0}^{1}dx\left\{ 2\left(
1-x\right) \left( 1+\ln \frac{\eta _{i}^{uZ}}{\mu ^{2}}\right) \right.
\notag \\
&&+\left( 1+3x\right) \left[ \left( \xi -1\right) \left( x-1\right) \left(
1-\ln \frac{M_{Z}^{2}}{\mu ^{2}}\right) -\frac{\eta _{i}^{uZ}}{M_{Z}^{2}}\ln
\frac{\eta _{i}^{uZ}}{M_{Z}^{2}}+\frac{\Delta _{i}^{uZ}}{M_{Z}^{2}}\ln \frac{%
\Delta _{i}^{uZ}}{M_{Z}^{2}}\right]  \notag \\
&&+\left. x^{2}\left( x-1\right) \frac{p^{2}}{M_{Z}^{2}}\ln \frac{\Delta
_{i}^{uZ}}{\eta _{i}^{uZ}}+\left( \xi -1\right) \left( 1-x^{2}\right)
\right\} \left( a_{L}^{2}L+a_{R}^{2}R\right)  \notag \\
&&+\frac{i\left( \xi +3\right) \delta _{ij}m_{i}^{u}a_{L}a_{R}}{\left( 4\pi
\right) ^{2}}\hat{\epsilon}^{-1}-\frac{2i\delta _{ij}m_{i}^{u}a_{L}a_{R}}{%
\left( 4\pi \right) ^{2}}\int_{0}^{1}dx\left\{ 1+2\ln \frac{\eta _{i}^{uZ}}{%
\mu ^{2}}\right.  \notag \\
&&+\left. \left( \xi -1\right) \left( x-1\right) \left( \frac{1}{2}-\ln
\frac{M_{Z}^{2}}{\mu ^{2}}\right) +\frac{\Delta _{i}^{uZ}}{M_{Z}^{2}}\ln
\frac{\Delta _{i}^{uZ}}{M_{Z}^{2}}-\frac{\eta _{i}^{uZ}}{M_{Z}^{2}}\ln \frac{%
\eta _{i}^{uZ}}{M_{Z}^{2}}\right\} ,  \label{suz}
\end{eqnarray}
where from Eq. (\ref{def2}) we have
\begin{eqnarray*}
\Delta _{i}^{uZ} &\equiv &xm_{i}^{u2}+\left( 1-x\right) \xi
M_{Z}^{2}+x\left( x-1\right) p^{2}, \\
\eta _{i}^{uZ} &\equiv &xm_{i}^{u2}+\left( 1-x\right) M_{Z}^{2}+x\left(
x-1\right) p,
\end{eqnarray*}

\begin{itemize}
\item  {\huge $-i\Sigma _{ij}^{dZ}$}
\end{itemize}

Using Eqs. (\ref{gaugevert}) and (\ref{e2}) we obtain
\begin{eqnarray}
S_{ih} &=&\delta _{ih},  \notag \\
a_{L} &=&i\frac{e}{c_{W}s_{W}}\left( zs_{W}^{2}+\frac{1}{2}c_{W}^{2}\right) ,
\notag \\
a_{R} &=&i\frac{e}{c_{W}s_{W}}s_{W}^{2}\left( z-\frac{1}{2}\right) ,  \notag
\\
m_{h} &=&m_{h}^{d},  \notag \\
M &=&M_{Z},  \label{asdz}
\end{eqnarray}
From Eq. (\ref{fscheme}) we obtain
\begin{equation*}
-i\Sigma _{ij}^{dZ}=\delta _{ij}\left[ \left( A_{i}+B_{i}\right) \left(
a_{L}^{2}L+a_{R}^{2}R\right) +m_{i}^{d}a_{L}a_{R}\left( C_{i}+D_{i}\right) %
\right] ,
\end{equation*}
hence from Eqs. (\ref{a+bM}) and (\ref{c+dM}) we obtain
\begin{eqnarray}
-i\Sigma _{ij}^{dZ} &=&-\frac{\delta _{ij}i\xi \not{p}}{\left( 4\pi \right)
^{2}}\left( a_{L}^{2}L+a_{R}^{2}R\right) \hat{\epsilon}^{-1}+\frac{\delta
_{ij}i\not{p}}{\left( 4\pi \right) ^{2}}\int_{0}^{1}dx\left\{ 2\left(
1-x\right) \left( 1+\ln \frac{\eta _{i}^{dZ}}{\mu ^{2}}\right) \right.
\notag \\
&&+\left( 1+3x\right) \left[ \left( \xi -1\right) \left( x-1\right) \left(
1-\ln \frac{M_{Z}^{2}}{\mu ^{2}}\right) -\frac{\eta _{i}^{dZ}}{M_{Z}^{2}}\ln
\frac{\eta _{i}^{dZ}}{M_{Z}^{2}}+\frac{\Delta _{i}^{dZ}}{M_{Z}^{2}}\ln \frac{%
\Delta _{i}^{dZ}}{M_{Z}^{2}}\right]  \notag \\
&&+\left. x^{2}\left( x-1\right) \frac{p^{2}}{M_{Z}^{2}}\ln \frac{\Delta
_{i}^{dZ}}{\eta _{i}^{dZ}}+\left( \xi -1\right) \left( 1-x^{2}\right)
\right\} \left( a_{L}^{2}L+a_{R}^{2}R\right)  \notag \\
&&+\frac{i\left( \xi +3\right) \delta _{ij}m_{i}^{d}a_{L}a_{R}}{\left( 4\pi
\right) ^{2}}\hat{\epsilon}^{-1}-\frac{2i\delta _{ij}m_{i}^{d}a_{L}a_{R}}{%
\left( 4\pi \right) ^{2}}\int_{0}^{1}dx\left\{ 1+2\ln \frac{\eta _{i}^{dZ}}{%
\mu ^{2}}\right.  \notag \\
&&+\left. \left( \xi -1\right) \left( x-1\right) \left( \frac{1}{2}-\ln
\frac{M_{Z}^{2}}{\mu ^{2}}\right) +\frac{\Delta _{i}^{dZ}}{M_{Z}^{2}}\ln
\frac{\Delta _{i}^{dZ}}{M_{Z}^{2}}-\frac{\eta _{i}^{dZ}}{M_{Z}^{2}}\ln \frac{%
\eta _{i}^{dZ}}{M_{Z}^{2}}\right\} ,  \label{sdz}
\end{eqnarray}
where from Eq. (\ref{def2}) we have
\begin{eqnarray*}
\Delta _{i}^{dZ} &\equiv &xm_{i}^{d2}+\left( 1-x\right) \xi
M_{Z}^{2}+x\left( x-1\right) p^{2}, \\
\eta _{i}^{dZ} &\equiv &xm_{i}^{d2}+\left( 1-x\right) M_{Z}^{2}+x\left(
x-1\right) p,
\end{eqnarray*}

\begin{itemize}
\item  {\huge $-i\Sigma _{ij}^{uA}$}
\end{itemize}

Using Eqs. (\ref{gaugevert}) and (\ref{e2}) we obtain
\begin{eqnarray}
S_{ih} &=&\delta _{ih},  \notag \\
a_{L} &=&-ie\left( z+\frac{1}{2}\right) ,  \notag \\
a_{R} &=&-ie\left( z+\frac{1}{2}\right) ,  \notag \\
m_{h} &=&m_{h}^{u},  \notag \\
M &=&0,  \label{asua}
\end{eqnarray}
From Eq. (\ref{fscheme}) we obtain
\begin{equation*}
-i\Sigma _{ij}^{uA}=\delta _{ij}\left[ \left( A_{i}+B_{i}\right) \left(
a_{L}^{2}L+a_{R}^{2}R\right) +m_{i}^{u}a_{L}a_{R}\left( C_{i}+D_{i}\right) %
\right] ,
\end{equation*}
hence from Eqs. (\ref{a+b0}) and (\ref{c+d0}) we obtain
\begin{eqnarray}
-i\Sigma _{ij}^{uA} &=&-\frac{\delta _{ij}i\xi \not{p}}{\left( 4\pi \right)
^{2}}a_{L}^{2}\hat{\epsilon}^{-1}+\frac{\delta _{ij}i\not{p}}{\left( 4\pi
\right) ^{2}}a_{L}^{2}\int_{0}^{1}dx\left\{ 2\left( 1-x\right) \left( 1+\ln
\frac{\eta _{i}^{u}}{\mu ^{2}}\right) +\left( \xi -1\right) \right.  \notag
\\
&&\times \left. \left( 1-x\right) \left[ \left( 1+x\right) +\left(
1+3x\right) \ln \frac{\eta _{i}^{u}}{\mu ^{2}}-x^{2}\left( 1-x\right) \frac{%
p^{2}}{\eta _{i}^{u}}\right] \right\}  \notag \\
&&+\frac{i\left( \xi +3\right) \delta _{ij}m_{i}^{u}a_{L}^{2}}{\left( 4\pi
\right) ^{2}}\hat{\epsilon}^{-1}-\frac{2i\delta _{ij}m_{i}^{u}a_{L}^{2}}{%
\left( 4\pi \right) ^{2}}\int_{0}^{1}dx\left\{ 1+2\ln \frac{\eta _{i}^{u}}{%
\mu ^{2}}\right.  \notag \\
&&+\left. \left( \xi -1\right) \left( 1-x\right) \left( \frac{1}{2}+\ln
\frac{\eta _{i}^{u}}{\mu ^{2}}+\frac{x^{2}p^{2}}{2\eta _{i}^{u}}\right)
\right\} ,  \label{sua}
\end{eqnarray}
where from Eq. (\ref{def2}) we have
\begin{equation*}
\eta _{i}^{u}\equiv xm_{i}^{u2}-x\left( 1-x\right) p^{2},
\end{equation*}

\begin{itemize}
\item  {\huge $-i\Sigma _{ij}^{dA}$}
\end{itemize}

Using Eqs. (\ref{gaugevert}) and (\ref{e2}) we obtain
\begin{eqnarray}
S_{ih} &=&\delta _{ih},  \notag \\
a_{L} &=&-ie\left( z-\frac{1}{2}\right) ,  \notag \\
a_{R} &=&-ie\left( z-\frac{1}{2}\right) ,  \notag \\
m_{h} &=&m_{h}^{d},  \notag \\
M &=&0,  \label{asda}
\end{eqnarray}
From Eq. (\ref{fscheme}) we obtain
\begin{equation*}
-i\Sigma _{ij}^{dA}=\delta _{ij}\left[ \left( A_{i}+B_{i}\right) \left(
a_{L}^{2}L+a_{R}^{2}R\right) +m_{i}^{d}a_{L}a_{R}\left( C_{i}+D_{i}\right) %
\right] ,
\end{equation*}
hence from Eqs. (\ref{a+b0}) and (\ref{c+d0}) we obtain
\begin{eqnarray}
-i\Sigma _{ij}^{dA} &=&-\frac{\delta _{ij}i\xi \not{p}}{\left( 4\pi \right)
^{2}}a_{L}^{2}\hat{\epsilon}^{-1}+\frac{\delta _{ij}i\not{p}}{\left( 4\pi
\right) ^{2}}a_{L}^{2}\int_{0}^{1}dx\left\{ 2\left( 1-x\right) \left( 1+\ln
\frac{\eta _{i}^{d}}{\mu ^{2}}\right) +\left( \xi -1\right) \right.  \notag
\\
&&\times \left. \left( 1-x\right) \left[ \left( 1+x\right) +\left(
1+3x\right) \ln \frac{\eta _{i}^{d}}{\mu ^{2}}-x^{2}\left( 1-x\right) \frac{%
p^{2}}{\eta _{i}^{d}}\right] \right\}  \notag \\
&&+\frac{i\left( \xi +3\right) \delta _{ij}m_{i}^{d}a_{L}^{2}}{\left( 4\pi
\right) ^{2}}\hat{\epsilon}^{-1}-\frac{2i\delta _{ij}m_{i}^{d}a_{L}^{2}}{%
\left( 4\pi \right) ^{2}}\int_{0}^{1}dx\left\{ \left( 1+2\ln \frac{\eta
_{i}^{d}}{\mu ^{2}}\right) \right.  \notag \\
&&+\left. \left( \xi -1\right) \left( 1-x\right) \left( \frac{1}{2}+\ln
\frac{\eta _{i}^{d}}{\mu ^{2}}+\frac{x^{2}p^{2}}{2\eta _{i}^{d}}\right)
\right\} ,  \label{sda}
\end{eqnarray}
where from Eq. (\ref{def2}) we have
\begin{eqnarray*}
\Omega _{h}^{d} &\equiv &xm_{h}^{d2}-x\left( 1-x\right) p^{2}, \\
\eta _{h}^{d} &\equiv &xm_{h}^{d2}-x\left( 1-x\right) p^{2},
\end{eqnarray*}

\begin{itemize}
\item  {\huge $-i\Sigma _{ij}^{uG}$}
\end{itemize}

Using Eqs. (\ref{gaugevert}) and (\ref{e2}) we obtain
\begin{eqnarray}
S_{ih} &=&\delta _{ih},  \notag \\
a_{L} &=&-ig_{s}\frac{\lambda }{2}^{a},  \notag \\
a_{R} &=&-ig_{s}\frac{\lambda }{2}^{a},  \notag \\
m_{h} &=&m_{h}^{u},  \notag \\
M &=&0,  \label{asug}
\end{eqnarray}
From Eq. (\ref{fscheme}) we obtain
\begin{equation*}
-i\Sigma _{ij}^{uG}=\delta _{ij}\left[ \left( A_{i}+B_{i}\right) \left(
a_{L}^{2}L+a_{R}^{2}R\right) +m_{i}^{u}a_{L}a_{R}\left( C_{i}+D_{i}\right) %
\right] ,
\end{equation*}
hence from Eqs. (\ref{a+b0}) and (\ref{c+d0}) we obtain
\begin{eqnarray}
-i\Sigma _{ij}^{uG} &=&-\frac{\delta _{ij}i\xi \not{p}}{\left( 4\pi \right)
^{2}}a_{L}^{2}\hat{\epsilon}^{-1}+\frac{\delta _{ij}i\not{p}a_{L}^{2}}{%
\left( 4\pi \right) ^{2}}\int_{0}^{1}dx\left\{ 2\left( 1-x\right) \left(
1+\ln \frac{\eta _{i}^{u}}{\mu ^{2}}\right) +\left( \xi -1\right) \right.
\notag \\
&&\times \left. \left( 1-x\right) \left[ \left( 1+x\right) +\left(
1+3x\right) \ln \frac{\eta _{i}^{u}}{\mu ^{2}}-x^{2}\left( 1-x\right) \frac{%
p^{2}}{\eta _{i}^{u}}\right] \right\}  \notag \\
&&+\frac{i\left( \xi +3\right) \delta _{ij}m_{i}^{u}a_{L}^{2}}{\left( 4\pi
\right) ^{2}}\hat{\epsilon}^{-1}-\frac{2i\delta _{ij}m_{i}^{u}a_{L}^{2}}{%
\left( 4\pi \right) ^{2}}\int_{0}^{1}dx\left\{ \left( 1+2\ln \frac{\eta
_{i}^{u}}{\mu ^{2}}\right) \right.  \notag \\
&&+\left. \left( \xi -1\right) \left( 1-x\right) \left( \frac{1}{2}+\ln
\frac{\eta _{i}^{u}}{\mu ^{2}}+\frac{x^{2}p^{2}}{2\eta _{i}^{u}}\right)
\right\} ,  \label{sug}
\end{eqnarray}
where from Eq. (\ref{def2}) we have
\begin{equation*}
\eta _{i}^{u}\equiv xm_{i}^{u2}-x\left( 1-x\right) p^{2},
\end{equation*}
and we remember that
\begin{equation*}
\lambda ^{a}\lambda ^{a}=\frac{16}{3}I,
\end{equation*}
where in this case $I$ is the $3\times 3$ identity (color space).

\begin{itemize}
\item  {\huge $-i\Sigma _{ij}^{dG}$}
\end{itemize}

Using Eqs. (\ref{gaugevert}) and (\ref{e2}) we obtain
\begin{eqnarray}
S_{ih} &=&\delta _{ih},  \notag \\
a_{L} &=&-ig_{s}\frac{\lambda }{2}^{a},  \notag \\
a_{R} &=&-ig_{s}\frac{\lambda }{2}^{a},  \notag \\
m_{h} &=&m_{h}^{d},  \notag \\
M &=&0,  \label{asdg}
\end{eqnarray}
From Eq. (\ref{fscheme}) we obtain
\begin{equation*}
-i\Sigma _{ij}^{dG}=\delta _{ij}\left[ \left( A_{i}+B_{i}\right) \left(
a_{L}^{2}L+a_{R}^{2}R\right) +m_{i}^{d}a_{L}a_{R}\left( C_{i}+D_{i}\right) %
\right] ,
\end{equation*}
hence from Eqs. (\ref{a+b0}) and (\ref{c+d0}) we obtain
\begin{eqnarray}
-i\Sigma _{ij}^{dG} &=&-\frac{\delta _{ij}i\xi \not{p}}{\left( 4\pi \right)
^{2}}a_{L}^{2}\hat{\epsilon}^{-1}+\frac{\delta _{ij}i\not{p}a_{L}^{2}}{%
\left( 4\pi \right) ^{2}}\int_{0}^{1}dx\left\{ 2\left( 1-x\right) \left(
1+\ln \frac{\eta _{i}^{d}}{\mu ^{2}}\right) +\left( \xi -1\right) \right.
\notag \\
&&\times \left. \left( 1-x\right) \left[ \left( 1+x\right) +\left(
1+3x\right) \ln \frac{\eta _{i}^{d}}{\mu ^{2}}-x^{2}\left( 1-x\right) \frac{%
p^{2}}{\eta _{i}^{d}}\right] \right\}  \notag \\
&&+\frac{i\left( \xi +3\right) \delta _{ij}m_{i}^{d}a_{L}^{2}}{\left( 4\pi
\right) ^{2}}\hat{\epsilon}^{-1}-\frac{2i\delta _{ij}m_{i}^{d}a_{L}^{2}}{%
\left( 4\pi \right) ^{2}}\int_{0}^{1}dx\left\{ \left( 1+2\ln \frac{\eta
_{i}^{d}}{\mu ^{2}}\right) \right.  \notag \\
&&+\left. \left( \xi -1\right) \left( 1-x\right) \left( \frac{1}{2}+\ln
\frac{\eta _{i}^{d}}{\mu ^{2}}+\frac{x^{2}p^{2}}{2\eta _{i}^{d}}\right)
\right\} ,  \label{sdg}
\end{eqnarray}
where from Eq. (\ref{def2}) we have
\begin{equation*}
\eta _{i}^{d}\equiv xm_{i}^{d2}-x\left( 1-x\right) p^{2},
\end{equation*}
and we remember that
\begin{equation*}
\lambda ^{a}\lambda ^{a}=\frac{16}{3}I,
\end{equation*}
where in this case $I$ is the $3\times 3$ identity (color space).

\section{Self energy divergent parts}

From Eq. (\ref{fscheme}) we have that the general form for the 1-loop
fermion 1PI diagrams containing a gauge propagator is

\begin{equation*}
-i\Sigma _{ij}=\sum_{h}S_{ih}S_{hj}^{\dagger }\left[ \left(
A_{h}+B_{h}\right) \left( a_{L}^{2}L+a_{R}^{2}R\right)
+m_{h}a_{L}a_{R}\left( C_{h}+D_{h}\right) \right] ,
\end{equation*}
If we sum up the 1-loop contributions to the self energies we obtain
\begin{eqnarray*}
\Sigma _{ij}^{u} &=&\Sigma _{ij}^{ugold}+\Sigma _{ij}^{ugaug}, \\
\Sigma _{ij}^{d} &=&\Sigma _{ij}^{dgold}+\Sigma _{ij}^{dgaug},
\end{eqnarray*}
where
\begin{eqnarray*}
\Sigma _{ij}^{ugold} &\equiv &\Sigma _{ij}^{u\chi ^{+}}+\Sigma _{ij}^{u\chi
^{3}}+\Sigma _{ij}^{u\rho }, \\
\Sigma _{ij}^{ugaug} &\equiv &\Sigma _{ij}^{uW^{+}}+\Sigma _{ij}^{uZ}+\Sigma
_{ij}^{uA}+\Sigma _{ij}^{uG}, \\
\Sigma _{ij}^{dgold} &\equiv &\Sigma _{ij}^{d\chi ^{-}}+\Sigma _{ij}^{d\chi
^{3}}+\Sigma _{ij}^{d\rho }, \\
\Sigma _{ij}^{dgaug} &\equiv &\Sigma _{ij}^{dW^{-}}+\Sigma _{ij}^{dZ}+\Sigma
_{ij}^{dA}+\Sigma _{ij}^{dG},
\end{eqnarray*}
If we want only the divergent part of the above expressions ($-i\check{\Sigma%
}_{ij}$) we can perform a series of simplifications. First of all, from Eqs.
(\ref{a+bM}-\ref{a+b0}) and (\ref{c+dM}-\ref{c+d0}) we have that the
divergencies appearing in $\left( A_{h}+B_{h}\right) $ and $\left(
C_{h}+D_{h}\right) $ are
\begin{eqnarray*}
\left( A_{h}+B_{h}\right) ^{div} &=&-\frac{i\xi \not{p}\hat{\epsilon}^{-1}}{%
\left( 4\pi \right) ^{2}}, \\
\left( C_{h}+D_{h}\right) ^{div} &=&\frac{i\left( \xi +3\right) \hat{\epsilon%
}^{-1}}{\left( 4\pi \right) ^{2}},
\end{eqnarray*}
hence
\begin{equation*}
-i\check{\Sigma}_{ij}^{gauge}=\frac{i\hat{\epsilon}^{-1}}{\left( 4\pi
\right) ^{2}}\sum_{h}S_{ih}S_{hj}^{\dagger }\left[ -\xi \not{p}\left(
a_{L}^{2}L+a_{R}^{2}R\right) +m_{h}a_{L}a_{R}\left( 3+\xi \right) \right] ,
\end{equation*}
Another fact is that $a_{R}=0$ for the $W$ boson and $S_{ih}=K_{ih}$ in this
case and $S_{ih}=\delta _{ih}$ in the others, hence we have
\begin{equation*}
-i\check{\Sigma}_{ij}^{gauge}=\frac{i\hat{\epsilon}^{-1}\delta _{ij}}{\left(
4\pi \right) ^{2}}\left[ -\xi \not{p}\left( a_{L}^{2}L+a_{R}^{2}R\right)
+m_{i}a_{L}a_{R}\left( 3+\xi \right) \right] ,
\end{equation*}
so the divergent part of the 1-loop fermion self energies is
\begin{eqnarray}
-i\check{\Sigma}_{ij}^{gauge} &=&\frac{i\hat{\epsilon}^{-1}\delta _{ij}}{%
\left( 4\pi \right) ^{2}}\left[ -\xi \not{p}\left(
L\sum_{WZAG}a_{L}^{2}+R\sum_{WZAG}a_{R}^{2}\right) \right.  \notag \\
&&+\left. m_{i}\left( 3+\xi \right) \sum_{WZAG}a_{L}a_{R}\right] ,
\label{sgdiv}
\end{eqnarray}
where $\sum_{WZAG}$ means the sum over the values corresponding to the
different gauge bosons. But from Eq. (\ref{gaugevert}) we have
\begin{eqnarray}
\sum_{WZAG}a_{Lu}^{2} &=&-e^{2}\left[ \frac{1}{2s_{W}^{2}}+\frac{\left(
zs_{W}^{2}-\frac{1}{2}c_{W}^{2}\right) ^{2}}{c_{W}^{2}s_{W}^{2}}+\left( z+%
\frac{1}{2}\right) ^{2}\right] -\frac{4}{3}g_{s}^{2}=-e^{2}\frac{%
3c_{W}^{2}+4z^{2}s_{W}^{2}}{4c_{W}^{2}s_{W}^{2}}-\frac{4}{3}g_{s}^{2},
\notag \\
\sum_{WZAG}a_{Ld}^{2} &=&-e^{2}\left[ \frac{1}{2s_{W}^{2}}+\frac{\left(
zs_{W}^{2}+\frac{1}{2}c_{W}^{2}\right) ^{2}}{c_{W}^{2}s_{W}^{2}}+\left( z-%
\frac{1}{2}\right) ^{2}\right] -\frac{4}{3}g_{s}^{2}=-e^{2}\frac{%
3c_{W}^{2}+4z^{2}s_{W}^{2}}{4c_{W}^{2}s_{W}^{2}}-\frac{4}{3}g_{s}^{2},
\notag \\
\sum_{WZAG}a_{Ru}^{2} &=&-e^{2}\left[ \frac{s_{W}^{2}\left( z+\frac{1}{2}%
\right) ^{2}}{c_{W}^{2}}+\left( z+\frac{1}{2}\right) ^{2}\right] -\frac{4}{3}%
g_{s}^{2}=-e^{2}\frac{\left( 2z+1\right) ^{2}}{4c_{W}^{2}}-\frac{4}{3}%
g_{s}^{2},  \notag \\
\sum_{WZAG}a_{Rd}^{2} &=&-e^{2}\left[ \frac{s_{W}^{2}\left( z-\frac{1}{2}%
\right) ^{2}}{c_{W}^{2}}+\left( z-\frac{1}{2}\right) ^{2}\right] -\frac{4}{3}%
g_{s}^{2}=-e^{2}\frac{\left( 2z-1\right) ^{2}}{4c_{W}^{2}}-\frac{4}{3}%
g_{s}^{2},  \notag \\
\sum_{WZAG}a_{Lu}a_{Ru} &=&-e^{2}\left[ \frac{\left( zs_{W}^{2}-\frac{1}{2}%
c_{W}^{2}\right) \left( z+\frac{1}{2}\right) }{c_{W}^{2}}+\left( z+\frac{1}{2%
}\right) ^{2}\right] -\frac{4}{3}g_{s}^{2}=-e^{2}\frac{\left( 2z+1\right) z}{%
2c_{W}^{2}}-\frac{4}{3}g_{s}^{2},  \notag \\
\sum_{WZAG}a_{Ld}a_{Rd} &=&-e^{2}\left[ \frac{\left( zs_{W}^{2}+\frac{1}{2}%
c_{W}^{2}\right) \left( z-\frac{1}{2}\right) }{c_{W}^{2}}+\left( z-\frac{1}{2%
}\right) ^{2}\right] -\frac{4}{3}g_{s}^{2}=-e^{2}\frac{\left( 2z-1\right) z}{%
2c_{W}^{2}}-\frac{4}{3}g_{s}^{2},  \notag \\
&&  \label{gaugediv}
\end{eqnarray}
that is from Eqs. (\ref{sgdiv}) and (\ref{gaugediv}) we obtain
\begin{eqnarray*}
-i\check{\Sigma}_{ij}^{ugaug} &=&\frac{i\delta _{ij}2\epsilon ^{-1}}{\left(
4\pi \right) ^{2}}\left[ \xi \not{p}\left( e^{2}\frac{%
3c_{W}^{2}+4z^{2}s_{W}^{2}}{4c_{W}^{2}s_{W}^{2}}L+e^{2}\frac{\left(
2z+1\right) ^{2}}{4c_{W}^{2}}R+\frac{4}{3}g_{s}^{2}\right) \right. \\
&&-\left. m_{i}^{u}\left( 3+\xi \right) \left( e^{2}\frac{\left( 2z+1\right)
z}{2c_{W}^{2}}+\frac{4}{3}g_{s}^{2}\right) \right] ,
\end{eqnarray*}
and the same interchanging $u\leftrightarrow d$ and $z\leftrightarrow -z$.
Regarding the Higgs and goldstone bosons contribution from Eqs. (\ref{e3}) (%
\ref{e5}) (\ref{e7}) we obtain
\begin{eqnarray*}
-i\check{\Sigma}_{ij}^{ugold} &=&\sum_{h}\frac{i4K_{ih}K_{hj}^{\dagger }}{%
\left( 4\pi \right) ^{2}v^{2}}\epsilon ^{-1}\left( \frac{1}{2}\left(
m_{i}^{u}m_{j}^{u}L+m_{h}^{d2}R\right) \not{p}-m_{h}^{d2}\left(
m_{j}^{u}R+m_{i}^{u}L\right) \right) \\
&&+\frac{i2m_{i}^{u2}\delta _{ij}}{\left( 4\pi \right) ^{2}v^{2}}\epsilon
^{-1}\not{p},
\end{eqnarray*}
or
\begin{eqnarray*}
-i\check{\Sigma}_{ij}^{ugold} &=&\sum_{h}\frac{i4K_{ih}m_{h}^{d2}K_{hj}^{%
\dagger }}{\left( 4\pi \right) ^{2}v^{2}}\epsilon ^{-1}\left[ \left( \frac{1%
}{2}\not{p}-m_{i}^{u}\right) L-m_{j}^{u}R\right] \\
&&+\frac{i2\delta _{ij}m_{i}^{u2}\epsilon ^{-1}\not{p}}{\left( 4\pi \right)
^{2}v^{2}}\left( L+2R\right) ,
\end{eqnarray*}
and from Eqs. (\ref{e4}) (\ref{e6}) and (\ref{e8}) we have the same
interchanging $u\leftrightarrow d$ and $K\leftrightarrow K^{\dagger }$.

\chapter{\textit{t}-channel subprocess cross sections}

\label{TchannelappA}In this appendix we present the analytical results
obtained for the matrix elements $M_{+}^{d}$ and $M_{+}^{\bar{u}}$
corresponding to the processes of Figs. \ref{u+gt+b-d+tot} and \ref
{d-gt+b-u-tot} respectively and the ones corresponding to anti-top
production $M_{-}^{u}$ and $M_{-}^{\bar{d}}$. Defining
\begin{eqnarray*}
g_{+} &=&g_{R}, \\
g_{-} &=&g_{L},
\end{eqnarray*}
we have the square modulus
\begin{equation}
\left| M_{-}^{u}\right| ^{2}=g_{s}^{2}\left(
O_{11}A_{11}+O_{22}A_{22}+O_{c}\left( A_{p}^{\left( +\right) }+A_{p}^{\left(
-\right) }+A_{m_{t}}^{\left( +\right) }+A_{m_{t}}^{\left( -\right)
}+A_{m_{b}}^{\left( +\right) }+A_{m_{b}}^{\left( -\right) }\right) \right) ,
\label{casifinal}
\end{equation}
with
\begin{eqnarray}
O_{11} &=&\frac{1}{4\left( k_{1}\cdot p_{1}\right) ^{2}},  \notag \\
O_{22} &=&\frac{1}{4\left( k_{1}\cdot p_{2}\right) ^{2}},  \notag \\
O_{c} &=&\frac{1}{4\left( k_{1}\cdot p_{1}\right) \left( k_{1}\cdot
p_{2}\right) },  \label{oij}
\end{eqnarray}
and
\begin{eqnarray*}
A_{11} &=&\frac{\left| g\right| ^{4}\left| K_{ud}\right| ^{2}}{\left(
k_{2}^{2}-M_{W}^{2}\right) ^{2}}\left\{ im_{t}^{2}m_{b}\frac{g_{L}^{\ast
}g_{R}-g_{R}^{\ast }g_{L}}{2}\varepsilon ^{\mu \nu \alpha \beta }\left(
k_{1}-p_{1}\right) _{\mu }n_{\nu }q_{2\alpha }q_{1\beta }\right. \\
&&+m_{t}m_{b}\frac{g_{R}^{\ast }g_{L}+g_{L}^{\ast }g_{R}}{2}\left[
m_{t}\left( q_{2}\cdot \left( k_{1}-p_{1}\right) \right) \left( q_{1}\cdot
n\right) \right. \\
&&-\left. m_{t}\left( q_{1}\cdot \left( k_{1}-p_{1}\right) \right) \left(
q_{2}\cdot n\right) -\left( q_{1}\cdot q_{2}\right) \left( m_{t}^{2}-\left(
k_{1}\cdot p_{1}\right) \right) \right] \\
&&+2\left| g_{L}\right| ^{2}\left( q_{2}\cdot p_{2}\right) \left[ \left(
m_{t}^{2}+\frac{p_{1}+m_{t}n}{2}\cdot \left( k_{1}-p_{1}\right) \right)
\left( q_{1}\cdot \left( k_{1}-p_{1}\right) \right) \right. \\
&&-\left. \frac{1}{2}m_{t}^{3}\left( n\cdot q_{1}\right) +\left( \frac{%
p_{1}+m_{t}n}{2}\cdot q_{1}\right) \left( k_{1}\cdot p_{1}\right) \right] \\
&&+2\left| g_{R}\right| ^{2}\left( q_{1}\cdot p_{2}\right) \left[ \left(
m_{t}^{2}+\frac{p_{1}-m_{t}n}{2}\cdot \left( k_{1}-p_{1}\right) \right)
\left( q_{2}\cdot \left( k_{1}-p_{1}\right) \right) \right. \\
&&+\left. \left. \frac{1}{2}m_{t}^{3}\left( n\cdot q_{2}\right) +\left(
\frac{p_{1}-m_{t}n}{2}\cdot q_{2}\right) \left( k_{1}\cdot p_{1}\right) %
\right] \right\} ,
\end{eqnarray*}
and
\begin{eqnarray*}
A_{22} &=&\frac{\left| g\right| ^{4}\left| K_{ud}\right| ^{2}}{\left(
k_{2}^{2}-M_{W}^{2}\right) ^{2}}\left\{ \left( k_{1}\cdot p_{2}\right) \left[
2\left| g_{R}\right| ^{2}\left( q_{1}\cdot k_{1}\right) \left( q_{2}\cdot
\frac{p_{1}-m_{t}n}{2}\right) \right. \right. \\
&&+\left. 2\left| g_{L}\right| ^{2}\left( q_{2}\cdot k_{1}\right) \left(
q_{1}\cdot \frac{p_{1}+m_{t}n}{2}\right) \right] \\
&&+m_{b}^{2}\left[ 2\left| g_{R}\right| ^{2}\left( q_{1}\cdot \left(
k_{1}-p_{2}\right) \right) \left( q_{2}\cdot \frac{p_{1}-m_{t}n}{2}\right)
\right. \\
&&+\left. 2\left| g_{L}\right| ^{2}\left( q_{2}\cdot \left(
k_{1}-p_{2}\right) \right) \left( q_{1}\cdot \frac{p_{1}+m_{t}n}{2}\right) %
\right] \\
&&+m_{b}\frac{g_{L}^{\ast }g_{R}+g_{R}^{\ast }g_{L}}{2}\left(
m_{b}^{2}-\left( k_{1}\cdot p_{2}\right) \right) \left[ -m_{t}\left(
q_{1}\cdot q_{2}\right) \right. \\
&&-\left. \left( q_{1}\cdot n\right) \left( q_{2}\cdot p_{1}\right) +\left(
q_{2}\cdot n\right) \left( q_{1}\cdot p_{1}\right) \right] \\
&&-\left. im_{b}\frac{g_{L}^{\ast }g_{R}-g_{R}^{\ast }g_{L}}{2}\left(
m_{b}^{2}-\left( k_{1}\cdot p_{2}\right) \right) \varepsilon ^{\mu \nu
\alpha \beta }n_{\mu }p_{1\nu }q_{2\alpha }q_{1\beta }\right\} ,
\end{eqnarray*}
and
\begin{eqnarray*}
A_{p}^{\left( \pm \right) } &=&-\frac{\left| g\right| ^{4}\left|
K_{ud}\right| ^{2}}{\left( k_{2}^{2}-M_{W}^{2}\right) ^{2}}\left| g_{\pm
}\right| ^{2}\left\{ \left( q_{1}\cdot q_{2}\right) \left[ \left( \left(
k_{1}-p_{1}\right) \cdot \left( k_{2}-p_{1}\right) \right) \left( \frac{%
p_{1}\mp m_{t}n}{2}\cdot p_{2}\right) \right. \right. \\
&&+\left. \left( \left( k_{1}-p_{1}\right) \cdot \frac{p_{1}\mp m_{t}n}{2}%
\right) \left( \left( k_{2}-p_{1}\right) \cdot p_{2}\right) -\left( \left(
k_{1}-p_{1}\right) \cdot p_{2}\right) \left( \frac{p_{1}\mp m_{t}n}{2}\cdot
\left( k_{2}-p_{1}\right) \right) \right] \\
&&+\left( \left( k_{2}-p_{1}\right) \cdot q_{2}\right) \left[ \left(
p_{2}\cdot \left( k_{1}-p_{1}\right) \right) \left( q_{1}\cdot \frac{%
p_{1}\mp m_{t}n}{2}\right) -\left( q_{1}\cdot p_{2}\right) \left( \left(
k_{1}-p_{1}\right) \cdot \frac{p_{1}\mp m_{t}n}{2}\right) \right] \\
&&-\left( \left( k_{1}-p_{1}\right) \cdot q_{2}\right) \left[ \left(
p_{2}\cdot \left( k_{2}-p_{1}\right) \right) \left( q_{1}\cdot \frac{%
p_{1}\mp m_{t}n}{2}\right) -\left( q_{1}\cdot p_{2}\right) \left( \left(
k_{2}-p_{1}\right) \cdot \frac{p_{1}\mp m_{t}n}{2}\right) \right] \\
&&+\left( \left( k_{2}-p_{1}\right) \cdot q_{1}\right) \left[ \left(
p_{2}\cdot \left( k_{1}-p_{1}\right) \right) \left( q_{2}\cdot \frac{%
p_{1}\mp m_{t}n}{2}\right) -\left( q_{2}\cdot p_{2}\right) \left( \left(
k_{1}-p_{1}\right) \cdot \frac{p_{1}\mp m_{t}n}{2}\right) \right] \\
&&-\left( \left( k_{1}-p_{1}\right) \cdot q_{1}\right) \left[ \left(
p_{2}\cdot \left( k_{2}-p_{1}\right) \right) \left( q_{2}\cdot \frac{%
p_{1}\mp m_{t}n}{2}\right) -\left( q_{2}\cdot p_{2}\right) \left( \left(
k_{2}-p_{1}\right) \cdot \frac{p_{1}\mp m_{t}n}{2}\right) \right] \\
&&\pm \left( \left( k_{1}-p_{1}\right) \cdot \left( k_{2}-p_{1}\right)
\right) \left[ \left( \left( \frac{p_{1}\mp m_{t}n}{2}\right) \cdot
q_{2}\right) \left( p_{2}\cdot q_{1}\right) -\left( \left( \frac{p_{1}\mp
m_{t}n}{2}\right) \cdot q_{1}\right) \left( p_{2}\cdot q_{2}\right) \right]
\\
&&\pm \left. \left( \frac{p_{1}\mp m_{t}n}{2}\cdot p_{2}\right) \left[
\left( \left( k_{1}-p_{1}\right) \cdot q_{2}\right) \left( \left(
k_{2}-p_{1}\right) \cdot q_{1}\right) -\left( \left( k_{1}-p_{1}\right)
\cdot q_{1}\right) \left( \left( k_{2}-p_{1}\right) \cdot q_{2}\right) %
\right] \right\} ,
\end{eqnarray*}
and
\begin{eqnarray*}
A_{m_{t}}^{\left( \pm \right) } &=&\frac{\left| g\right| ^{4}\left|
K_{ud}\right| ^{2}}{\left( k_{2}^{2}-M_{W}^{2}\right) ^{2}}\frac{\left|
g_{\pm }\right| ^{2}}{2}\left\{ \left( m_{t}n\cdot p_{2}\right) \left[
\left( p_{1}\cdot q_{2}\right) \left( \left( k_{2}-p_{1}\right) \cdot
q_{1}\right) -\left( \left( k_{2}-p_{1}\right) \cdot q_{2}\right) \left(
p_{1}\cdot q_{1}\right) \right] \right. \\
&&-\left( m_{t}n\cdot q_{2}\right) \left[ \left( p_{1}\cdot p_{2}\right)
\left( \left( k_{2}-p_{1}\right) \cdot q_{1}\right) -\left( \left(
k_{2}-p_{1}\right) \cdot p_{2}\right) \left( p_{1}\cdot q_{1}\right) \right]
\\
&&+\left( m_{t}n\cdot q_{1}\right) \left[ \left( p_{1}\cdot p_{2}\right)
\left( \left( k_{2}-p_{1}\right) \cdot q_{2}\right) -\left( \left(
k_{2}-p_{1}\right) \cdot p_{2}\right) \left( p_{1}\cdot q_{2}\right) \right]
\\
&&+m_{t}^{2}\left[ \left( q_{2}\cdot p_{2}\right) \left( q_{1}\cdot \left(
k_{2}-p_{1}\right) \right) +\left( q_{1}\cdot p_{2}\right) \left( q_{2}\cdot
\left( k_{2}-p_{1}\right) \right) -\left( q_{1}\cdot q_{2}\right) \left(
p_{2}\cdot \left( k_{2}-p_{1}\right) \right) \right] \\
&&\pm m_{t}\left( n\cdot \left( k_{2}-p_{1}\right) \right) \left[ \left(
q_{2}\cdot p_{2}\right) \left( q_{1}\cdot p_{1}\right) +\left( q_{1}\cdot
p_{2}\right) \left( q_{2}\cdot p_{1}\right) -\left( q_{1}\cdot q_{2}\right)
\left( p_{2}\cdot p_{1}\right) \right] \\
&&\mp \left. m_{t}\left( p_{1}\cdot \left( k_{2}-p_{1}\right) \right) \left[
\left( q_{2}\cdot p_{2}\right) \left( q_{1}\cdot n\right) +\left( q_{1}\cdot
p_{2}\right) \left( q_{2}\cdot n\right) -\left( q_{1}\cdot q_{2}\right)
\left( p_{2}\cdot n\right) \right] \right\} ,
\end{eqnarray*}
and
\begin{eqnarray*}
A_{m_{b}}^{\left( \pm \right) } &=&\frac{m_{b}\left| g\right| ^{4}\left|
K_{ud}\right| ^{2}}{\left( k_{2}^{2}-M_{W}^{2}\right) ^{2}}\frac{g_{\pm
}^{\ast }g_{\mp }}{2}\left\{ 2\left( p_{1}\cdot p_{2}\right) \left[ \left(
n\cdot q_{2}\right) \left( \left( k_{1}-p_{1}\right) \cdot q_{1}\right)
-\left( n\cdot q_{1}\right) \left( \left( k_{1}-p_{1}\right) \cdot
q_{2}\right) \right] \right. \\
&&-2\left( n\cdot p_{2}\right) \left[ \left( p_{1}\cdot q_{2}\right) \left(
\left( k_{1}-p_{1}\right) \cdot q_{1}\right) -\left( p_{1}\cdot q_{1}\right)
\left( \left( k_{1}-p_{1}\right) \cdot q_{2}\right) \right] \\
&&\pm i\varepsilon ^{\mu \nu \alpha \beta }q_{2\alpha }q_{1\beta }\left(
n_{\mu }p_{1\nu }\left( k_{1}-p_{1}\right) \cdot p_{2}+p_{2\mu }n_{\nu
}\left( k_{1}-p_{1}\right) \cdot p_{1}+p_{1\mu }p_{2\nu }\left(
k_{1}-p_{1}\right) \cdot n\right) \\
&&\mp i\varepsilon ^{\mu \nu \alpha \beta }q_{2\alpha }q_{1\beta }\left(
k_{1}-p_{1}\right) _{\mu }\left[ n_{\nu }\left( p_{1}\cdot \left(
k_{2}-p_{1}\right) \right) +\left( k_{2}-p_{1}\right) _{\nu }\left(
p_{1}\cdot n\right) \right] \\
&&+\left( n\cdot \left( k_{1}-p_{1}\right) \right) \left[ \left( p_{1}\cdot
q_{2}\right) \left( \left( k_{2}-p_{1}\right) \cdot q_{1}\right) -\left(
\left( k_{2}-p_{1}\right) \cdot q_{2}\right) \left( p_{1}\cdot q_{1}\right) %
\right] \\
&&-\left( n\cdot q_{2}\right) \left[ \left( p_{1}\cdot \left(
k_{1}-p_{1}\right) \right) \left( \left( k_{2}-p_{1}\right) \cdot
q_{1}\right) -\left( \left( k_{2}-p_{1}\right) \cdot \left(
k_{1}-p_{1}\right) \right) \left( p_{1}\cdot q_{1}\right) \right] \\
&&+\left( n\cdot q_{1}\right) \left[ \left( p_{1}\cdot \left(
k_{1}-p_{1}\right) \right) \left( \left( k_{2}-p_{1}\right) \cdot
q_{2}\right) -\left( \left( k_{2}-p_{1}\right) \cdot \left(
k_{1}-p_{1}\right) \right) \left( p_{1}\cdot q_{2}\right) \right] \\
&&+2m_{t}\left[ \left( q_{2}\cdot \left( k_{1}-p_{1}\right) \right) \left(
q_{1}\cdot p_{2}\right) +\left( q_{1}\cdot \left( k_{1}-p_{1}\right) \right)
\left( q_{2}\cdot p_{2}\right) -\left( q_{1}\cdot k_{1}\right) \left(
q_{2}\cdot k_{1}\right) \right] \\
&&+\left. m_{t}\left( q_{1}\cdot q_{2}\right) \left[ \left( p_{2}\cdot
p_{1}\right) +\left( \left( k_{1}-p_{1}\right) \cdot \left(
k_{1}-p_{2}\right) \right) \right] \right\} \\
&&+m_{b}^{2}\frac{\left| g_{\pm }\right| ^{2}}{2}\frac{\left| g\right|
^{4}\left| K_{ud}\right| ^{2}\left| K_{tb}\right| ^{2}}{\left(
k_{2}^{2}-M_{W}^{2}\right) ^{2}}\left\{ -m_{t}\left[ \left( n\cdot
q_{2}\right) \left( p_{1}\cdot q_{1}\right) -\left( n\cdot q_{1}\right)
\left( p_{1}\cdot q_{2}\right) \right] \right. \\
&&+m_{t}^{2}\left( q_{1}\cdot q_{2}\right) -2\left[ \left( q_{2}\cdot \left(
k_{1}-p_{1}\right) \right) \left( q_{1}\cdot \frac{p_{1}\mp m_{t}n}{2}%
\right) \right. \\
&&+\left. \left. \left( q_{1}\cdot \left( k_{1}-p_{1}\right) \right) \left(
q_{2}\cdot \frac{p_{1}\mp m_{t}n}{2}\right) -\left( q_{1}\cdot q_{2}\right)
\left( \left( k_{1}-p_{1}\right) \cdot \frac{p_{1}\mp m_{t}n}{2}\right) %
\right] \right\} ,
\end{eqnarray*}
Finally, it can be shown that we can obtain the other matrix elements from
the above expressions performing the following changes
\begin{equation}
\begin{array}{lcr}
\left| M_{-}^{u}\right| ^{2}\longleftrightarrow \left| M_{+}^{\bar{u}%
}\right| ^{2} & \quad \Leftrightarrow \quad & n\longleftrightarrow -n, \\
\left| M_{-}^{u}\right| ^{2}\longleftrightarrow \left| M_{+}^{d}\right| ^{2}
& \quad \Leftrightarrow \quad & g_{L}\leftrightarrow g_{R}^{\ast }, \\
\left| M_{-}^{u}\right| ^{2}\longleftrightarrow \left| M_{-}^{\bar{d}%
}\right| ^{2} & \quad \Leftrightarrow \quad & q_{1}\leftrightarrow q_{2},
\end{array}
\label{change}
\end{equation}
it is useful to note also that all matrix elements are symmetric under the
change
\begin{equation}
\left( n,g_{L},q_{1}\right) \leftrightarrow \left( -n,g_{R}^{\ast
},q_{2}\right) ,  \label{sym}
\end{equation}

\end{document}